\documentclass[onecolumn]{IEEEtran}
\usepackage{amsmath,amsthm, amssymb, epsfig, verbatim, color, amsfonts} % spconf
\usepackage{graphicx}
\usepackage{ifpdf}
\title{Communication over Individual Channels -- a general framework}
\author{Yuval Lomnitz, Meir Feder \\
Tel Aviv University, Dept. of EE-Systems \\
Email: \{yuvall,meir\}@eng.tau.ac.il}

\ifx\theorem\undefined
\theoremstyle{plain}
\newtheorem{theorem}{Theorem}
\newtheorem{lemma}{Lemma}%[theorem] % enable the remark in order to number the lemmas as subnumbering of theorems
\newtheorem{corollary_in_theorem}{Corollary}[theorem]
\theoremstyle{definition}
\newtheorem{definition}{Definition}%[section]
\newtheorem{example}{Example}%[section]
\fi

\def\vr{\mathbf}
\def\mt{\mathbf}

\def\cov{\text{cov}}
\def\tr{\text{tr}}
\def\Pr{\mathrm{Pr}}
\def\E{\mathbb{E}}
\def\Ber{\mathrm{Ber}}
\def\Normal{\mathcal{N}}
\def\const{\mathrm{const}}

\def\Remp{{R_{\mathrm{emp}}}}
\newcommand\Rempname[1]{{R_{\mathrm{emp}}^{#1}}}
\def\Ract{{R_{\mathrm{act}}}}
\def\Ind{\mathrm{Ind}} % another option: uspackage{bbm}, \mathbbm{1}
\def\Ber{\mathrm{Ber}}

\def\half{\tfrac{1}{2}}

\def\msg{\mathrm{\mathbf{m}}} % message in a communication system
\def\msg{\mathbf{m}} % message in a communication system

\def\endofproof{\hspace{\stretch{1}}$\Box$}
\def\defeq{\triangleq} % \equiv

\def\etal{\textit{et al}}

\def\unif{\mathbb{U}}
\def\verdu{Verd\'u~}

\newcommand{\tsubs}[1]{{\scriptscriptstyle \mathrm{#1}}}

\def\ntoinfty{\arrowexpl{n \to \infty}}

\newcommand{\argmax}[1] {\underset{#1}{\textstyle \mathrm{argmax}}\hspace{0.5ex}}

\newcommand{\arrowexpl}[1] {\underset{#1}{\textstyle \longrightarrow}}
\newcommand{\placeunder}[2] {\underset{#2}{#1}}

\ifx\note\undefined
\newenvironment{note}{\comment}{\endcomment} % uncomment this line in order to remove all "notes" paragraphs
\fi

\newcommand{\datemark}[1]{}
\newcommand{\todo}[1]{}

\newcommand{\selector}[2]{\onlypaper{#1}\onlyphd{#2}} % e.g. \selector{paper}{work}. These macros are defined in the header of each article.

\newenvironment{inputpath}[1]
{ \let\origtexinput\input \renewcommand{\input}[1]{\origtexinput{#1/##1}}
\let\origtexincludegraphics\includegraphics \renewcommand{\includegraphics}[2][]{\origtexincludegraphics[##1]{#1/##2}} }
{ \let\input\origtexinput \let\includegraphics\origtexincludegraphics}

\allowdisplaybreaks[1]

\newcommand{\unfinished}[1]{}
\newcommand{\onlyphd}[1]{}
\newcommand{\onlypaper}[1]{#1}
\newcommand{\excluded}[1]{}

\def\ML{\scriptscriptstyle \mathrm{ML}}
\def\NML{\scriptscriptstyle \mathrm{NML}}
\def\dfb{d_{\scriptscriptstyle \mathrm{FB}}}

\def\dalf{\frac{d}{2}}

\begin{document}
\maketitle

\begin{abstract}
We consider the problem of communicating over a channel for which no mathematical model is specified, and the achievable rates are determined as a function of the channel input and output sequences known a-posteriori, without assuming any a-priori relation between them. In a previous paper we have shown that the empirical mutual information between the input and output sequences is achievable without specifying the channel model, by using feedback and common randomness, and a similar result for real-valued input and output alphabets. In this paper, we present a unifying framework which includes the two previous results as particular cases. We characterize the region of rate functions which are achievable, and show that asymptotically the rate function is equivalent to a conditional distribution of the channel input given the output. We present a scheme that achieves these rates with asymptotically vanishing overheads.
\end{abstract}

\todo{For the paper version, consider (for shortening)to split into two papers: one will discuss the asymptotical limits and rate adaptation (i.e. $P/Q$ and attainability of $P/Q$ in the case of causality, and the other paper will discuss the constructions, and will include the chapter on rate adaptivity of the ML construction). Also, remove the $\hat H$ notation and only mention it.}

\section{Introduction}\label{chap:intro}

\unfinished{[The introduction is currently only for the paper, not for thesis]}

\begin{figure}[h]
\setlength{\unitlength}{0.8mm}
\center
\begin{picture}(140, 30)
\put(22,16){Encoder}\put(20,10){\line(1,0){20}}\put(40,10){\line(0,1){15}}\put(40,25){\line(-1,0){20}}\put(20,25){\line(0,-1){15}}
\put(70,20){?}
\put(102,16){Decoder}\put(100,10){\line(1,0){20}}\put(120,10){\line(0,1){15}}\put(120,25){\line(-1,0){20}}\put(100,25){\line(0,-1){15}}
\put(40,21.5){\vector(1,0){20}}\put(45,22.5){$x_i$}
\put(80,21.5){\vector(1,0){20}}\put(85,22.5){$y_i$}
\put(100,15){\vector(-1,0){60}}
\end{picture}
\vspace{1ex}
\caption{The individual channel communication setup}\label{fig:individual_channel_model}
\end{figure}
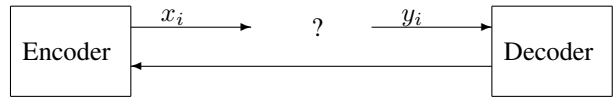

This paper revisits the ``individual channel'' communication model \cite{YL_individual_full}, which provides an alternative framework for communication over unknown channels. The communication setup is illustrated in Figure~\ref{fig:individual_channel_model}. An encoder sends an input sequence $\vr x \in \mathcal{X}^n$ into the channel. The output of the channel $\vr y \in \mathcal{Y}^n$ is determined in a completely arbitrary way which is unknown to the encoder and the decoder. However, there is a perfect feedback link from the decoder to the encoder, and we also assume the existence of common randomness. Under these assumptions we would like to characterize a communication rate for the channel. Clearly, since nothing is guaranteed with respect to the output, one cannot guarantee any positive communication rate a-priori, and achieve a vanishing error probability. Therefore, instead, we define a rate as a function of the specific input and output sequences ($\Remp(\vr x, \vr y)$, termed a \textit{rate function}).

The motivations for this communication model are elaborated upon in our initial paper \cite{YL_individual_full}, and will be briefly explained here through an example. We consider the example of the binary channel $y_i=x_i \oplus e_i$, where $e_i$ is an arbitrary sequence. The traditional way to deal with this channel would be by using the arbitrarily varying channels (AVC) framework \cite{Lapidoth_AVC}. In this framework feedback is not considered, and the AVC capacity is the maximum reliable communication rate that can be attained irrespective of the choice, or distribution, of the state sequence (in this case $e_i$). However, in order to obtain a positive capacity, it is necessary to place a constraint on $e_i$. Suppose that we limit the maximum rate of errors to $\frac{1}{n} \sum e_i \defeq \hat \epsilon \leq \epsilon_0 \leq \half$, then by applying common randomness the AVC capacity becomes $1 - h_b(\epsilon_0)$. This result requires placing an a-priori constraint. Furthermore, because of the worst-case nature of the AVC capacity, the communication rate will not improve if $\epsilon < \epsilon_0$, i.e. the channel is actually better than we have assumed. Shayevitz and Feder \cite{Ofer_ModuloAdditive} proposed to deal with this issue by using feedback, and have presented a scheme that without assuming any prior constraint on $\hat \epsilon$, achieves the rate $1 - h_b(\hat \epsilon)$.

This result, and its extensions \cite{Eswaran} allows us to replace a-priori constraints by the empirical distribution of the noise (or state) sequence that actually occurred, thus alleviating the worst-case assumptions. The result is that the rate is defined by the sequence (i.e. the channel). Still, we need to assume a channel model relating the input and the output. Since channel models are in many cases a coarse abstraction of reality, and in some cases may be completely unknown, the next step is to ask: can we do without the model, by, so to speak ``extracting'' this model from the empirical data? In doing so, we define the empirical rate function by using both the input and the output. This is a fundamental change with respect to the previous models, since the input is determined by the scheme itself.

In the \selector{previous paper}{initial work} \cite{YL_individual_full} we have shown that it is possible to attain the empirical mutual information $\Remp(\vr x, \vr y) = \hat I(\vr x; \vr y)$, as well as the function $\Remp(\vr x, \vr y) = \half \log \frac{1}{1- \hat\rho^2(\vr x, \vr y)}$, where $\hat\rho$ is the empirical correlation factor. The later function is suitable for channels with real-valued inputs and outputs. These rate functions are appealing since they are direct counterparts of statistical information measures. For the case of a discrete memoryless channel, the empirical mutual information over the sequences tends in probability to the statistical mutual information over the input and output random variables. The second function tends to the mutual information between two Gaussian random variables with the same correlation factor, and thus is optimal for Gaussian  channels. These results generalize achievability results for compound channels and AVCs, and enable to easily re-derive the previously mentioned results \cite{Ofer_ModuloAdditive,Eswaran}, and even extend them \cite[Section VII.B]{YL_individual_full}. However many questions are left open. For example, how can these functions be modified to include memory or take into account MIMO channels, and what is the set of achievable rate functions? Is there a general way to extend the concept of ``empirical mutual information''? In addition, in the \selector{previous paper}{initial work} we have separated the discussion on the discrete and the continuous cases, from technical reasons, and the natural question that raises to mind is whether the two results can be put into a unified theory.

The main objective of this paper is to define such a unifying theory, by first characterizing the set of achievable rate functions, presenting general communication schemes for achieving these rates with, and without feedback (where only in the first case, the communication rate is adaptive), and presenting a tighter analysis of the overheads related to universally achieving these rate functions. The new techniques used in this paper enable us to derive various rate functions and analyze the overhead (or rate loss) required for attaining them in a finite block length. We present refined proof techniques that lead to tighter bounds and re-derive, and improve over the previous results \cite{YL_individual_full,YL_MIMO_ITW2010,YL_ModuloAdditiveEilat}. However note that the different proof techniques used in the previous work \cite{YL_individual_full} are interesting on their own, and sometimes more intuitive. We will highlight the connections between the results in the sequel.

\section{Overview\onlyphd{~of Part~I}}\label{chap:overview}
Following is a high level overview of the ideas and results presented in this \selector{paper}{part}. As mentioned above we would like to refrain from stating the channel model. We define the rate of a system using a ``rate function'' $\Remp(\vr x, \vr y)$ of the input sequence $\vr x$ and output sequence $\vr y$. We would like to find systems which guarantee attaining certain rate functions.

The first step is to define what ``attaining'' a rate function means. We refer to two kinds of systems: fixed rate systems without feedback, and adaptive-rate systems using feedback. The adaptive rate systems guarantee that the transmitted rate would be at least $\Remp(\vr x, \vr y)$ while keeping a small probability of error, for any input and output sequence. I.e. this guarantee holds irrespective of the channel model. In the fixed rate case, since we cannot guarantee any positive rate a-priori (the Shannon capacity of the channel in Figure~\ref{fig:individual_channel_model} is $0$), the system only guarantees reliable communication when $\Remp \geq R$ (the event $\Remp < R$ can be considered as ``outage''). Therefore the adaptive case is of more interest from a practical perspective. We allow unlimited common randomness between the encoder and the decoder, and in order to avoid circular definitions, we constrain the input distribution to a given prior $Q$. These definitions are stated formally and discussed in \selector{Section}{Chapter}~\ref{chap:definitions}.

In classical communication and information theory, one only considers the average error probability over the channel law and requires a certain static rate of communication, whereas here we require that the rate function would be specified per input-output pair $\vr x, \vr y$, and that a certain error probability would be achieved. This may be seen as an over-requirement, however note that every system has, in effect, a rate function: one can always look at all the cases where the input was a specific $\vr x$ and the output was a specific $\vr y$ and ask what was the actual rate of error free bits that was received in this case. Thus, we can consider the ``rate function'' as way for characterizing communication systems which is ``channel independent''. On the other hand, as we will see in Section~\ref{sec:good_put_bound}, with a small overhead, the rate function of any system can be attained with a \emph{fixed} error probability.

\onlyphd{
For example, it can be shown \cite{YL_individual_full} that the empirical mutual information $\Remp(\vr x, \vr y) = \hat I(\vr x; \vr y)$ is attainable (adaptively and non adaptively) under these definitions. This result was shown at the beginning of the current research and supplies motivation for this model. The results of \cite{YL_individual_full} will be reproduced in this report in a general way, yielding tighter bounds. The proofs used in \cite{YL_individual_full} are not repeated here, but the techniques are sometimes more intuitive, and are briefly reviewed in Chapter~\ref{chap:phd_related_results}. The empirical mutual information can be thought of as the prototype of rate functions: it is appealing since it is a direct counterpart of the statistical mutual information, and for the case of a discrete memoryless channel, the empirical mutual information over the sequences tends in probability to the mutual information over the input and output random variables. The achievability of $ \hat I(\vr x; \vr y)$ generalizes achievability results for compound channels and AVCs, and enable to easily re-derive the previously mentioned results \cite{Ofer_ModuloAdditive,Eswaran}, and even extend them \cite[Section VII.B]{YL_individual_full}. In this part more general rate functions are developed.
}

The first question we ask is -- which rate functions are achievable (\selector{Section}{Chapter}~\ref{chap:fundamental_limits_on_remp})? Theorem~\ref{theorem:remp_ublb} gives a necessary and a sufficient condition for the achievability of a rate function (in the non-adaptive case), which are tight in the sense of the achieved rate for large block size $n \to \infty$. In an analogy to universal source coding, this theorem is equivalent to the Kraft inequality, stating which source encoders are feasible (in terms of the set of word lengths).
Based on this result, we can characterize the ``intrinsic redundancy'', which is a property of any rate function, determining the redundancy that would be needed to achieve it (Theorem~\ref{theorem:remp_achievability_upto}).
\onlyphd{We also analyze the reason for the gap between the necessary and sufficient condition which relates to the codebook distribution.}
Then, considering more general systems, it is shown that the good-put associated with a specific choice of $\vr x, \vr y$ in any system, is in-fact an achievable rate function, and therefore can be achieved with an error probability as low as desired, per sequence, up to a small overhead in rate.

The characterization of Theorem~\ref{theorem:remp_ublb} is based on the CDF of the rate function with respect to the input distribution $Q$, which is inconvenient to handle. In \selector{Section}{Chapter}~\ref{chap:asymptotic_limit_remp} we deal with the asymptotic behavior of rate functions, and show that asymptotically, the achievability of rate functions can be determined based on a simpler condition similar to the Chernoff bound (Theorem~\ref{theorem:chernoff_tightness}). The main result of this \selector{section}{chapter} is Theorem~\ref{theorem:remp_conditional_form_asymptotic} which shows that the maximum rate functions are asymptotically of the form $\Remp = \frac{1}{n} \log \left( \frac{P(\vr x | \vr y)}{Q(\vr x)} \right)$ for some conditional probability  $P(\vr x | \vr y)$. Thus, selecting rate functions is asymptotically equivalent to selecting conditional distributions $P(\vr x | \vr y)$. Returning again to the analogy to source coding, this claim is similar to the claim that, due to Kraft inequality, every source encoder is defined by a probability distribution on the set of possible messages \cite{Barron_MDL}.

The set of achievable rate functions is rather arbitrary (like the set of possible encoders, in the analogy). In \selector{Section}{Chapter}~\ref{chap:useful_constructions} we discuss the problem of selecting the rate function, using several possible constructions. Each construction has a certain justification and results in a certain form. The first construction that we term ``maximum likelihood construction'' (Section~\ref{sec:ML_rate_functions}) is based on taking the maximum of the form $\frac{1}{n} \log \left( \frac{P_{\theta}(\vr x | \vr y)}{Q(\vr x)} \right)$ over a class of models $\theta$. Achieving this rate function guarantees matching (or surpassing) the rate of any system operating over any of the channels in the model class. Another way to remove the arbitrariness (Section~\ref{sec:rate_func_by_parameters}) is to limit the scope to rate functions defined based on a predefined set of parameters (for example the empirical second order moments, or zero order joint statistics). When the parameters can take only a sub-exponential number of values, the input and output sequences can be grouped into ``types'' of sequences having the same values of the empirical parameters. Theorem~\ref{theorem:optimal_type_based} determines the optimal rate function that can be obtained in this case. We particularize the result to the memoryless case, and present the best rate function that can be defined by zero order statistics (Lemma~\ref{lemma:maximum_iid_type_based_Remp}). This rate function can be also stated in terms of the ``maximum likelihood'' construction, and on the other hand is close to the empirical mutual information, which means that the empirical mutual information is essentially optimal (in terms of using the zero order statistics). A third way to define a rate function (Section~\ref{sec:rate_of_given_metric}) is by taking another system as a reference and asking what is the maximum rate that can be achieved with a given decoding metric and a given prior, when the number of messages is allowed to vary -- i.e. conditioned on a certain pair of input and output, how many messages can one send while still maintaining a small probability of error? In the rest of the \selector{paper}{work} we focus mainly on the ``maximum likelihood' construction.

The main strength of the ``individual channel'' approach is when the rate function can be obtained adaptively, without outage. \selector{Section}{Chapter}~\ref{chap:rate_adaptivity} focuses on rate adaptivity. In Section~\ref{sec:rate_adaptive_scheme} we present a communication scheme that attains an adaptive rate using multiple iterations of rateless coding. Theorem~\ref{theorem:framework} and its corollaries characterize the performance of the proposed rate adaptive scheme. The scheme is based on a decoding metric that must satisfy some conditions and needs to be specified later, and the rate function is given as function of this metric. In what follows we substitute various metrics to obtain various rate functions. In Section~\ref{sec:Remp_adaptive_conditional} we show that under a ``causality'' condition, the rate function $\Remp = \frac{1}{n} \log \left( \frac{P(\vr x | \vr y)}{Q(\vr x)} \right)$ (which is the asymptotical bound for all rate functions) can be \emph{adaptively} achieved (Theorem~\ref{theorem:adaptive_causal_distribution}).

Next we focus on ``maximum likelihood'' rate functions (Section~\ref{sec:rate_adaptivity_RempML}). In Theorem~\ref{theorem:RempML_adaptive_redundancy_by_weighting} we show the achievability of such rate functions when the ``maximum likelihood'' probability $\max_{\theta} P_{\theta}(\vr x | \vr y)$ can be given as a weighted sum of $P_{\theta}(\vr x | \vr y)$ (which always holds when the number of $\theta$-s is subexponential in $n$). We particularize this result for rate functions based on empirical probabilities (Theorem~\ref{theorem:RempML_adaptive_redundancy_empirical}) and present bounds on the redundancy for the adaptive and non-adaptive case. In the more general case where $\theta$ belongs to an infinite class, we do not have a general result on adaptivity, however we show that some properties required for the application of Theorem~\ref{theorem:framework} hold in general for the ``maximum likelihood'' construction (Lemma~\ref{lemma:summability_continuous}).

The rate adaptive scheme presented in Section~\ref{sec:rate_adaptive_scheme} is finite horizon, i.e. it requires prior knowledge of the block length $n$. In Section~\ref{sec:rate_adaptive_inf_horizon} we present an infinite horizon extension of the scheme, based on a simple ``doubling trick''. The modified scheme attains the results of Theorem~\ref{theorem:framework} under some assumptions. Unfortunately the results regarding rate adaptivity in \selector{Section}{Chapter}~\ref{chap:rate_adaptivity} are not as tight and elegant as the results in the non-adaptive case -- this manifests itself in the relatively high redundancy of the scheme (which generally behaves like $O \left( \sqrt{\frac{\log n}{n}} \right)$ in the block length), as well as its complexity, and the fact we do not have a tight lower bound (necessary condition) on the redundancy.

In \selector{Section}{Chapter}~\ref{chap:examples} we present examples for rate functions, which include as particular cases the \selector{previous results \cite{YL_individual_full}}{results of the early work \cite{YL_individual_full}}. The rate functions include the empirical mutual information (Section~\ref{sec:examples_eMI}), an extension that uses memory in the channel (which is optimal for stationary ergodic channels, Section~\ref{sec:examples_markov}), a discussion on extensions that include time variation (Section~\ref{sec:examples_time_variation}), the modulo-additive rate function presented by Shayevitz and Feder \cite{Ofer_ModuloAdditive} (Section~\ref{sec:examples_modadditive}), rate functions based on compression (Section~\ref{sec:examples_compression}), and a second-order rate function for the MIMO channel (Section~\ref{sec:GaussianMIMO}, Theorem~\ref{theorem:GaussianMIMO} and Lemma~\ref{lemma:GaussianMIMO_opt}). \onlyphd{We also present results showing the strength of the individual channel model in achieving capacity for probabilistic and semi-probabilistic channels (Section~\ref{sec:examples_probabilistic})}.
%These rate functions are summarized in Table~\ref{tbl:rate_functions} (refer to Section~\ref{sec:empirical_distributions_and_information_measures} for notation).

\selector{Section}{Chapter}~\ref{chap:comments_research} is devoted to comments and further research. In Section~\ref{sec:comparison_with_prev_paper} we compare with the results of the \selector{previous paper}{initial work} \cite{YL_individual_full}.
\todo{continue further research chapter}
\onlyphd{At the end of the appendix we give a table of the Nomenclature (Table~\ref{tbl:nomenclature}) and a list of theorems and lemmas (Table~\ref{tbl:theorems_and_lemmas}).}

Before beginning the formal parts, several comments are due on the general approach taken in this \selector{paper}{work}. First, this work is theoretical in nature. No effort is made to improve the decoder complexity, or reduce the amount of common randomness required. The reason behind this is that we are mainly interested in examining this communication concept. If we see the concept is fruitful, the next step should be trying to make it the implementation practical. Also, while we do not attempt to be practical regarding the implementation, the requirements from the system do need to be related to practical targets. The second comment is that in this work we focus on transmission rate rather than on error exponents. The theoretical reason is that the discussion around error exponents is based on the fact the error probability with a fixed rate and a known, stationary ergodic channel, decreases exponentially. Here, the rate is not fixed, and the channel is not specified, so this does not necessarily hold true. The second reason is practical -- from a practical perspective of requirements, there is no reason to require the system's error probability to decrease exponentially fast (if at all, the block error rate should be allowed to increase with $n$). Rather, it makes sense to require a small, but fixed, error probability.

\selector{
\section{Definitions}}{\section{Individual channels -- definitions}}\label{chap:definitions}
\onlypaper{The definitions in this section almost identical to the ones stated in the previous paper \cite{YL_individual_full}, and are repeated here for completeness. The main difference is the absence of the set $J$.}
We define the channel, adaptive and non-adaptive systems and achievability in the adaptive and non-adaptive sense. If the motivation for these definitions is not immediately clear, the asymptotically achievable rate functions $\hat I(\vr x; \vr y)$ and $\half \log \frac{1}{{1 - \hat\rho^2}}$ can be regarded as motivating examples.

\onlypaper{
\subsection{Notation}
Uppercase letters denote random variables, and respective lowercase letters denote their sample values. Boldface letters are used to denote vectors, which are by default of length $n$. Superscript and subscript indices are applied to vectors to define subsequences in the standard way, i.e. $\vr x_i^j \defeq (x_i, x_{i+1}, ... , x_j)$, $\vr x^i \defeq \vr x_1^i$. The indices $i,j$ are allowed to exceed the range of indices where $\vr x$ is defined (for example be negative), in which case only the indices in the definition range will be considered (e.g. $\vr x_{-1}^{n+2} = \vr x_1^n$, $\vr x^{-1} = \emptyset$).
 The indicator function $\Ind(E)$ where $E$ is a set or a probabilistic event is defined as $1$ over the set (or when the event occurs) and $0$ otherwise. $P \circ Q$ denotes the product of conditional probability functions e.g. $(P \circ Q)(x,y) = P(x) \cdot Q(y|x)$. $\unif(A)$ denotes a uniform distribution over the set $A$.

$\mathbb{R}$ denotes the set of real numbers, and $\Normal(\mu,\sigma^2)$ denotes a Gaussian distribution with mean $\mu$ and variance $\sigma^2$. $\lVert \vr x \rVert \defeq \sqrt{\vr x^T \vr x}$ denotes $L_2$ norm. $\Ber(p)$ denotes the Bernoulli distribution, and $h_b(p) \defeq H(\Ber(p)) = -p \log p - (1-p) \log (1-p) $ denotes the binary entropy function.

A hat ($\hat{\square}$) denotes an estimated value. The empirical mutual information of two vectors $\hat I(\vr x; \vr y)$ is the mutual information between two random variables $X,Y$ whose joint distribution equals the empirical distribution of $\vr x, \vr y$ \cite[Section II]{MethodOfTypes}. An exact definition of empirical mutual information and other empirical information measures is delayed to sections \ref{sec:def_emp_entropy} and \ref{sec:def_eMI}. We denote $I(P,W)$ the mutual information $I(X;Y)$ when $(X,Y) \sim P(x)\cdot W(y|x)$.

The functions $\log(\cdot)$ and $\exp(\cdot)$ as well as information theoretic quantities $H(\cdot), I(\cdot;\cdot), D(\cdot || \cdot)$ are in base 2 (bits) (and can be interpreted as other information units by changing the base of the $\log$). We use $\ln(\cdot)$ to denote the natural logarithm.

Bachmann \& Landau notations are used for orders of magnitude. Specifically, $f_n = \Theta(g_n)$, means $\exists n_0,\alpha,\beta > 0: \forall n > n_0: \alpha g_n \leq f_n \leq \beta g_n$, $f_n \in o(g_n)$ or $f_n = o(g_n)$ means $\frac{f_n}{g_n} \ntoinfty 0$ and $f_n \in \omega(g_n)$ means $\frac{f_n}{g_n} \ntoinfty \infty$.

Most of the results apply both to the case where the input is discrete, and characterized by a probability mass function, and to the case it is continuous and characterized by density function. When denoting $p(\vr x)$ as the probability of $\vr x$ without specifying whether $\vr x$ is continuous or discrete, it means that $p(\vr x)$ may be substituted by either the probability mass function or a density function, as applicable.

Note that proofs are given sometimes after the Theorem/Lemma is stated, and sometimes before it, as seems easier to read. In the later case the Theorem/Lemma summarizes a conclusion from a discussion.
} % onlypaper

\subsection{Individual channels and rate functions}
\begin{definition}[Channel]\label{def:individual_channel}
A channel is defined by a pair of input and output alphabets $\mathcal{X,Y}$, and is denoted $\mathcal{X} \to \mathcal{Y}$
\end{definition}

\begin{definition}[Rate function]\label{def:rate_func}
A rate function $\Remp : \mathcal{X}^n \times \mathcal{Y}^n \to \mathbb{R}$ for the channel $\mathcal{X} \to \mathcal{Y}$ may be any real valued function of $\vr x \in \mathcal{X}^n, \vr y \in \mathcal{Y}^n$.
\end{definition}
Note that we do not preclude negative values, for reasons of notational convenience. Also, we have defined the set of possible outputs as $n$ length vectors $\mathcal{Y}^n$ mainly for the sake of concreteness; many of the results in the paper do not assume anything about the structure of $\vr y$, and thus in general, the output does not have to be a vector of the same length of the input.

\subsection{Fixed rate communication without feedback}
\begin{definition}[Fixed rate encoder, decoder, error probability]\label{def:sys_nonadaptive}
A randomized block encoder and decoder pair for the channel $\mathcal{X \to Y}$ with block length $n$ and rate $R$ without feedback is defined by a random variable $S$ distributed over the set $\mathcal{S}$, a mapping $\vr X
 : \{1,2,\ldots\exp(nR)\} \times \mathcal{S} \to \mathcal{X}^n$ and a mapping $\hat\msg : \mathcal{Y}^n \times \mathcal{S} \to \{1,2,
\ldots\exp(nR)\}$. The error probability for message $\msg \in \{1,2,\ldots\exp(nR)\}$ is defined as
\begin{equation}\label{eq:A228}
P_e^{(\msg)}(\vr x, \vr y) = \Pr \left(\hat\msg(\vr y, S) \neq \msg \big\vert \vr X(\msg,S) = \vr x \right)
\end{equation}
where for $\vr x$ such that the conditioning in \eqref{eq:A228} cannot hold, we define $P_e^{(\msg)}(\vr x, \vr y)=0$.
\end{definition}

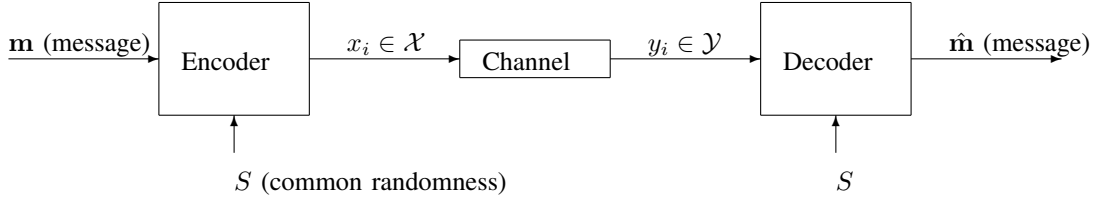
\begin{figure*}[t]
\setlength{\unitlength}{1mm}
\hspace{\stretch{1}}
\begin{picture}(140,30)
\put(23,16){Encoder}\put(20,10){\line(1,0){20}}\put(40,10){\line(0,1){15}}\put(40,25){\line(-1,0){20}}\put(20,25){\line(0,-1){15}}
\put(63,16){Channel}\put(60,15){\line(1,0){20}}\put(80,15){\line(0,1){5}}\put(80,20){\line(-1,0){20}}\put(60,20){\line(0,-1){5}}
\put(103,16){Decoder}\put(100,10){\line(1,0){20}}\put(120,10){\line(0,1){15}}\put(120,25){\line(-1,0){20}}\put(100,25){\line(0,-1){15}}
\put(0,17.5){\vector(1,0){20}}\put(0,18.5){$\msg$ (message)}
\put(40,17.5){\vector(1,0){20}}\put(45,18.5){$x_i \in \mathcal{X}$}
\put(80,17.5){\vector(1,0){20}}\put(85,18.5){$y_i \in \mathcal{Y}$}
\put(120,17.5){\vector(1,0){20}}\put(125,18.5){$\hat{\msg}$ (message)}
\put(30,5){\vector(0,1){5}}\put(30,0){$S$ (common randomness)}
\put(110,5){\vector(0,1){5}}\put(110,0){$S$}
\end{picture}
\hspace{\stretch{1}}
\caption{Non rate adaptive encoder-decoder pair without feedback}\label{fig:system_non_adaptive}
\end{figure*}

This system is illustrated in Figure~\ref{fig:system_non_adaptive}. We treat $\vr x$ as a random variable and $\vr y$ as a deterministic sequence. This does not preclude applying the results to a channel whose output $\vr y$ is a random variable and depends on $\vr x$, since all results are conditioned on both $\vr x$ and $\vr y$. Note that the encoder rate must pertain to a discrete number of messages $\exp(nR) \in \mathbb{Z}_+$, but the empirical rates we refer to in the sequel may be any positive real numbers. In the sequel, $\msg$ is treated sometimes as a series of bits and sometimes as an index of the message.

\begin{definition}[Achievability]\label{def:achievability_nonadaptive}
A rate function $\Remp : \mathcal{X}^n \times \mathcal{Y}^n \to \mathbb{R}$ is \emph{achievable} with a prior $Q(\vr x)$ defined over $\mathcal{X}^n$ and error probability $\epsilon$ if for any $R > 0$, there exist a pair of randomized encoder and decoder, with a rate of at least $R$ such that for any message $\msg$: $\vr X \sim Q$ and for any $\vr x, \vr y$ where $\Remp(\vr x, \vr y) \geq R$, $P_e^{(\msg)}(\vr x, \vr y) \leq \epsilon$.
\end{definition}

We sometimes term this kind of achievability ``non-adaptive achievability'' to separate it from the adaptive achievability defined below. The usage of the notation $\Remp$ does not immediately imply the rate function is achievable (or adaptively, or asymptotically achievable, by the definitions below). We sometimes place an superscript asterisk $\Remp^*$ to specify that the given function is indeed achievable. Note that the definition requires that the conditions hold for all $R > 0$, however this is done mainly for convenience, and if we are interested in the achievability of $\Remp$ at a specific $R$ we can always define a new rate function $\Remp' = \begin{cases} R & \Remp \geq R \\ 0 & \Remp < R \end{cases}$ whose achievability indicates that the achievability conditions are met for $\Remp$ for the specific $R$.

\subsection{Adaptive rate communication with feedback}
\begin{figure*}[t]
\setlength{\unitlength}{1mm}
\hspace{\stretch{1}}
\begin{picture}(140, 30)
\put(23,16){Encoder}\put(20,10){\line(1,0){20}}\put(40,10){\line(0,1){15}}\put(40,25){\line(-1,0){20}}\put(20,25){\line(0,-1){15}}
\put(63,20){Channel}\put(60,19){\line(1,0){20}}\put(80,19){\line(0,1){5}}\put(80,24){\line(-1,0){20}}\put(60,24){\line(0,-1){5}}
\put(103,16){Decoder}\put(100,10){\line(1,0){20}}\put(120,10){\line(0,1){15}}\put(120,25){\line(-1,0){20}}\put(100,25){\line(0,-1){15}}
\put(0,17.5){\vector(1,0){20}}\put(6,18.5){$\msg$}\put(0,14){(message)}
\put(40,21.5){\vector(1,0){20}}\put(42,22.5){$x_i \in \mathcal{X}$}
\put(80,21.5){\vector(1,0){20}}\put(82,22.5){$y_i \in \mathcal{Y}$}
\put(100,15){\vector(-1,0){60}}\put(55,10){$f_i \in \mathcal{F}$ (feedback)}
\put(120,21.5){\vector(1,0){20}}\put(125,22.5){$R$ (rate)}
\put(120,15){\vector(1,0){20}}\put(125,16){$\hat{\msg}$ (message)}
\put(30,5){\vector(0,1){5}}\put(30,0){$S$ (common randomness)}
\put(110,5){\vector(0,1){5}}\put(110,0){$S$}
\end{picture}
\hspace{\stretch{1}}
\caption{Rate adaptive encoder-decoder pair with feedback}\label{fig:system_adaptive}
\end{figure*}
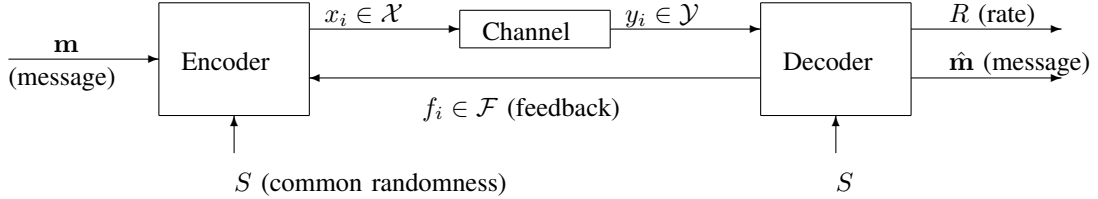

\begin{definition}[Adaptive rate encoder, decoder, error probability]\label{def:sys_adaptive}
A randomized block encoder and decoder pair for the channel $\mathcal{X \to Y}$ with block length $n$, adaptive rate and feedback is defined as follows:
\begin{itemize}
\item The message $\msg$ is expressed by the infinite sequence ${\msg}_1^{\infty} \in \{0,1\}^{\infty}$
\item The common randomness is defined as a random variable $S$ distributed over the set $\mathcal{S}$
\item The feedback alphabet is denoted $\mathcal{F}$
\item The encoder is defined by a series of mappings $X_k = X_k(\msg, S, \vr f^{k-1})$
\item The decoder is defined by the feedback function $f_k = \varphi_k(\vr y^k, S)$, the decoding function $\hat\msg (\vr y, S)$ and the rate function $R(\vr y, S)$.
\end{itemize}
The random variables $\vr X$, $\hat\msg$ and $R$ denote the outcomes of the respective functions. The error probability for message $\msg$ is defined as
\begin{equation}\label{eq:A227}
P_e^{(\msg)}(\vr x, \vr y) = \Pr \left({\hat {\msg}}_1^{\lceil nR \rceil} \neq  {{\msg}}_1^{\lceil nR \rceil} \big\vert \vr X = \vr x, \vr y \right)
\end{equation}

In other words, a recovery of the first $\lceil nR \rceil$ bits by the decoder is considered a successful reception. For $\vr x$ such that the conditioning in \eqref{eq:A227} cannot hold, we define $P_e^{(\msg)}(\vr x, \vr y)=0$.  The conditioning on $\vr y$ is mainly for clarification, since it is treated as a fixed vector. This system is illustrated in Figure~\ref{fig:system_adaptive}.

\end{definition}
In all cases discussed in this paper the feedback is binary $\mathcal{F} = \{0,1\}$. Furthermore we sometime consider reducing the feedback rate below $1$ bit/use. In this case some of the feedback values $f_k$ will be fixed to $0$, and the feedback rate is the ratio of unconstrained feedback bits.

\begin{definition}[Adaptive achievability]\label{def:achievability_adaptive}
A rate function $\Remp : \mathcal{X}^n \times \mathcal{Y}^n \to \mathbb{R}$ is \emph{adaptively achievable} with a prior $Q(\vr x)$ defined over $\mathcal{X}^n$ and error probability $\epsilon$, if there exist randomized encoder and decoder with feedback, such that $\vr x \sim Q$ and for all $\vr x \in \mathcal{X}^n, \vr y \in \mathcal{Y}^n$:
\begin{equation}\label{eq:A253}
\Pr \left\{ \left({\hat {\msg}}_1^{\lceil nR \rceil} \neq  {{\msg}}_1^{\lceil nR \rceil} \right) \cup (R < \Remp(\vr x, \vr y)) \Big| \vr X = \vr x, \vr y \right\} \leq \epsilon
\end{equation}
In other words, with probability at least $1-\epsilon$, a message with a rate of at least $\Remp$ is decoded correctly.
\end{definition}

The model in which the decoder determines the transmission rate is lenient in the sense that it gives the flexibility to exchange rate for error probability: the decoder may estimate the error probability and decrease it by reducing the decoding rate.

\subsection{Approximate achievability}
\begin{definition}[Achievability up to a gap]\label{def:achievability_upto}
We say that $\Remp(\vr x, \vr y)$ is achievable (adaptively/non adaptively)  \textit{up to $\mu$} (or with a gap of $\mu$) with a certain $Q$ and $\epsilon$, if $\Remp(\vr x, \vr y)' = \Remp(\vr x, \vr y) - \mu$ is achievable (adaptively/non adaptively, resp.)
\end{definition}
Note that $\mu$ can be translated to a loss in rate. This is clear in the adaptive case where the rate is a function of $\Remp$. In the non adaptive case the definition above means there is a system that transmits at rate $R-\mu$ and achieves error probability of less than $\epsilon$ whenever $\Remp \geq R$ (which is equivalent to $\Remp - \mu \geq R - \mu$).

\begin{definition}[Asymptotic achievability]\label{def:asymptotic_achievability}
A sequence of rate functions defined for $n=1,2,\ldots$ is \emph{asymptotically achievable} (adaptively / non adaptively) with a prior $Q(\vr x)$ defined for vectors $\vr \in \mathcal{X}^n$ of increasing size, if for all $\epsilon > 0$ there exists a sequence of functions $F_n(t)$, $n=1,2,\ldots$ with $F_n(t) \ntoinfty t$, such that $F_n(\Remp(\vr x, \vr y))$ is achievable (adaptively / non adaptively, resp.) with the given $\epsilon$ and $Q(\vr x)$.
\end{definition}

Note that relating the rate function to the achievable function through $F_n(t) \ntoinfty t$ is in general weaker than requiring that their ratio would tend to 1, since $F_n(t) \ntoinfty t$ does not necessarily uniformly converge. As an example, consider $F_n(t) = \min(t, n)$, and the two (equal) functions $f_n = g_n = 2n$ then although $\frac{f_n}{g_n} \ntoinfty 1$, $\frac{F_n(f_n)}{g_n} \ntoinfty \half$. The reason to use this definition is that indeed in many cases of interest, the convergence of the rate function is non uniform. However the results are useful since $t$ has a meaning of rate, and the slow convergence occurs only at high rates.

\subsection{Discussion}
Note that achievability is defined with respect to a fixed prior $Q(\vr x)$. Although the rate function depends on specific sequences, for actual \emph{communication} to happen it is necessary to select input sequences, and $Q(\vr x)$ defines the main property of this selection needed for our purpose, i.e. the input distribution.

The reason for fixing $Q$ is that the achievable rates are a function of the channel input, which is determined by the scheme itself. This is an opening for possible falsity -- the encoder may choose sequences for which the rate is attained more easily. For example, by setting $\vr x = 0$ one can attain $\Remp = \hat I(\vr x; \vr y)$ in a void way, since the rate function will always be $0$. We circumvent this difficulty by constraining an input distribution, and by using common randomness, requiring that the encoder emits input symbols that are random and distributed according to the defined prior. This breaks the circular dependence that might have been created, by specifying the input behavior together with the rate function.

In a high level view we can say that the individual channel framework does not contain any tools to modify the input behavior -- since nothing is assumed on the effect of a change in the input, and therefore the input prior is constrained. From this reason, in the current framework we only gain rate adaptivity from feedback, and but do not improve the communication rate. In channels with memory, it is possible to improve the channel capacity using feedback, but this improvement is due to modification of the input distribution (conditioned on the output). This gain cannot be obtained in the current framework due to the constraint on the input distribution.

Note that these results hold under the theoretical assumption that one may have access to a random variable of any desired distribution, which is in some cases un-feasible to generate in an exact manner -- see further discussion in \selector{our previous paper \cite{YL_individual_full}}{Section~***}.

\section{Fundamental limitations on rate functions}\label{chap:fundamental_limits_on_remp}
The selection of rate functions is rather arbitrary. This could be seen by the following example: suppose $\Remp(\vr x, \vr y)$ is achievable, and let $\pi : \mathcal{Y}^n \to \mathcal{Y}^n $ be a permutation of the output values, then clearly $\Remp(\vr x, \pi(\vr y))$ is also achievable, by
placing the permutation $\pi$ before the decoder (so that the effective channel output seen by the system is $\vr y' = \pi(\vr y)$). In general none of the rate functions generated by various values of $\pi$ is uniformly better than the others. In the sequel we will discuss possible reasonable ways to choose rate functions, that may eliminate some of these choices. However we start with the more basic question: what is the set of achievable rate functions?

In this section and the following ones we focus only on the non-adaptive case, and characterize the set of achievable rate functions. The role of this bound is similar to the role of Kraft's inequality in source encoding -- it does not indicate a \emph{preference} to specific encoders, but merely states which encoding lengths are \emph{possible} (can be implemented by uniquely decodable encoders) and which are not. The rate function $\Remp$ takes the role of encoding lengths in Kraft's inequality.

\subsection{A characterization of the set of achievable rate functions}\label{sec:achievability_conditions}
The following theorem presents a necessary and a sufficient conditions for a rate function $\Remp$ to be achievable, in the fixed sense.
\begin{theorem}\label{theorem:remp_ublb}
Consider communication over block length $n$, with a prior $Q$ and error probability $\epsilon$.
If $\Remp(\vr x, \vr y)$ is achievable, or adaptively achievable, then:
\begin{equation}\label{eq:A434}
\forall y \in \mathcal{Y}^n, R \in \mathbb{R} : \qquad
Q \left\{ \Remp(\vr X, \vr y) \geq R \right\}  \leq \frac{1}{1 - \epsilon} \exp(-nR)
\end{equation}
Conversely, if
\begin{equation}\label{eq:A435}
\forall y \in \mathcal{Y}^n, R \geq 0 : \qquad
Q \left\{ \Remp(\vr X, \vr y) \geq R \right\}  \leq \epsilon \exp(-nR)
\end{equation}
then $\Remp(\vr x, \vr y)$ is achievable.
\end{theorem}

\begin{figure}
\centering
\ifpdf
  \setlength{\unitlength}{1bp}%
  \begin{picture}(190.02, 186.01)(0,0)
  \put(0,0){\includegraphics{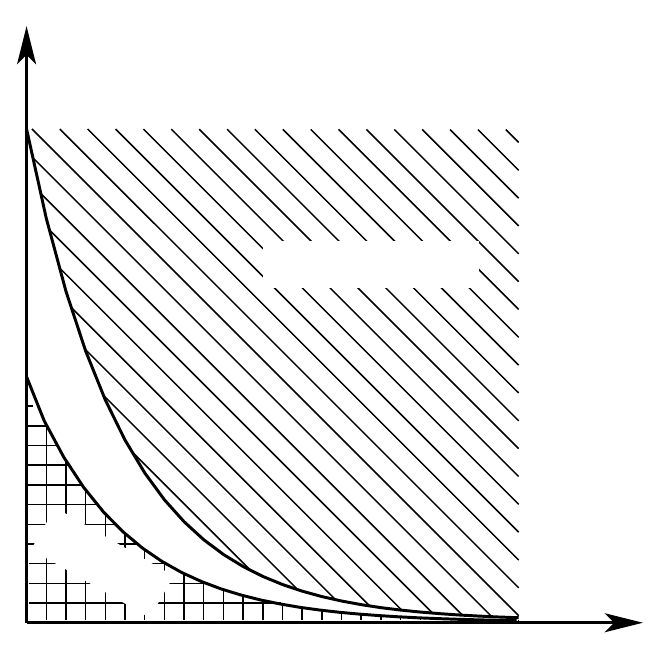}}
  \put(174.90,16.16){\fontsize{14.23}{17.07}\selectfont $R$}
  \put(10.49,169.23){\fontsize{14.23}{17.07}\selectfont $Q \{ \Remp \geq R \}$}
  \put(11.28,32.65){\rotatebox{330.00}{\fontsize{8.54}{10.24}\selectfont \smash{\makebox[0pt][l]{Achievable}}}}
  \put(13.48,78.41){\rotatebox{315.00}{\fontsize{8.54}{10.24}\selectfont \smash{\makebox[0pt][l]{Gap}}}}
  \put(78.52,109.70){\rotatebox{0.00}{\fontsize{8.54}{10.24}\selectfont \smash{\makebox[0pt][l]{Non Achievable}}}}
  \end{picture}%
\else
  \setlength{\unitlength}{1bp}%
  \begin{picture}(190.02, 186.01)(0,0)
  \put(0,0){\includegraphics{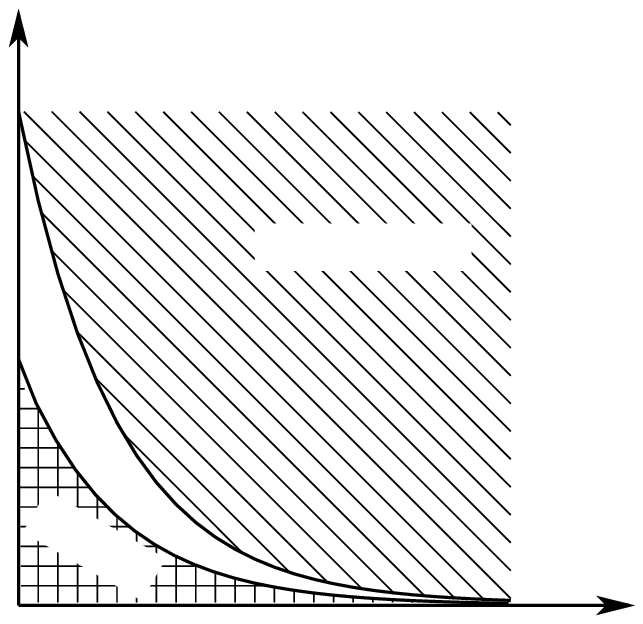}}
  \put(174.90,16.16){\fontsize{14.23}{17.07}\selectfont $R$}
  \put(10.49,169.23){\fontsize{14.23}{17.07}\selectfont $Q \{ \Remp \geq R \}$}
  \put(11.28,32.65){\rotatebox{330.00}{\fontsize{8.54}{10.24}\selectfont \smash{\makebox[0pt][l]{Achievable}}}}
  \put(13.48,78.41){\rotatebox{315.00}{\fontsize{8.54}{10.24}\selectfont \smash{\makebox[0pt][l]{Gap}}}}
  \put(78.52,109.70){\rotatebox{0.00}{\fontsize{8.54}{10.24}\selectfont \smash{\makebox[0pt][l]{Non Achievable}}}}
  \end{picture}%
\fi
\caption{\label{fig:Q_Remp_Exponential_Decay}%
 Achievable and unachivable regions in Theorem \ref{theorem:remp_ublb}}
\end{figure}

Where $Q \left\{ \Remp(\vr X, \vr y) \geq R \right\}$ means the probability with respect to $\vr X$ distributed $Q$ of the event $\Remp(\vr X, \vr y) \geq R$. The necessary condition refers to both achievable and adaptively achievable rate functions, whereas the sufficient condition only refers to achievable rate function (adaptive achievability is discussed in \selector{Section}{Chapter}~\ref{chap:rate_adaptivity}). Note that the necessary condition holds trivially for $R \leq 0$ (the definition is extended to negative $R$-s for matters of convenience, which will become clear later on). These conditions are depicted graphically in Figure~\ref{fig:Q_Remp_Exponential_Decay} where the horizontal axis is the rate and the vertical axis is the probability $Q \left\{ \Remp \geq R \right\}$.

Both bounds characterize the achievability of $\Remp$ based on the probability of $\Remp$ to exceed a threshold for a fixed value of $\vr y$ (its CCDF). The rationale behind this characterization is as follows. Consider the system of Definition~\ref{def:sys_nonadaptive}, and fix the output $\vr y$. Clearly, no information can be transmitted in this case. At each block, there is a codebook of input sequences $\vr X_i$, $i=1,2,\ldots,\exp(nR)$ that would be transmitted if the input message is $\msg=i$. The decoder does not know which of these words was chosen but only knows the codebook. However, it guarantees that in high probability it will decode the correct word, if this word has $\Remp(\vr X_i, \vr y) \geq R$. This is possible only if in most codebooks, only one word satisfies the condition. This leads to the bound on the probability of $\Remp(\vr X_i, \vr y) \geq R$.

Note that if a rate function satisfies the sufficient condition with strict inequality (for all or some $\vr y$-s and $R$-s), then it can be modified to a larger function meeting the condition with equality, by using the inverse transform theorem, i.e. by passing the random variable $\Remp(\vr X, \vr y)$ through its CDF to obtain a uniform random variable and then through the desired CDF satisfying \eqref{eq:A435} with equality.
A remarkable property in the necessary and sufficient conditions is that, since they are given per value of $\vr y$, there is no tradeoff between different $\vr y$ (i.e. one can decide on a rate function separately for each $\vr y$). Indeed, these are only bounds, and in an accurate characterization of the domain of achievable rate functions there is a tradeoff between different $\vr y$-s. But later on we shall see that this property, of separation between $\vr y$-s holds also in the asymptotical form of the bound (Theorem~\ref{theorem:remp_conditional_form_asymptotic}).

Following Theorem~\ref{theorem:remp_ublb}, it is convenient to make the following definition:  define the \textit{intrinsic redundancy} of a rate function $\Remp$ with respect to a prior $Q$ as:
\begin{equation}\label{eq:Adef_intrinsic_redundancy}
\mu_Q(\Remp) \defeq
\sup_{\vr y,R \in \mathbb{R}} \left\{ \frac{1}{n} \log Q \{\Remp(\vr X, \vr y) \geq R\}  + R \right\}
\end{equation}
This definition simply extracts the normalized coefficient before the $\exp(-nR)$ in Theorem~\ref{theorem:remp_ublb}, i.e. it is the minimum value $\mu_Q$ such that:
\begin{equation}\label{eq:A519}
\forall \vr y,R: \qquad Q \{\Remp(\vr X, \vr y ) \geq R\} \leq \exp(n \cdot \mu_Q) \cdot \exp(-n R)
\end{equation}

Theorem~\ref{theorem:remp_ublb} can now be stated as follows:
\begin{enumerate}
\item A rate function $\Remp$ is achievable if $\mu_Q(\Remp) \leq \frac{1}{n} \log \epsilon$
\item A rate function $\Remp$ is achievable only if $\mu_Q(\Remp) \leq \frac{1}{n} \log \frac{1}{1-\epsilon}$
\end{enumerate}
It is easy to see that the inequalities above together with the definition of $\mu_Q$ directly imply the inequalities in Theorem~\ref{theorem:remp_ublb}. Note that the two bounds on $\mu_Q(\Remp)$ converge to $0$ for fixed $\epsilon$ as $n \to \infty$.

Intuitively the intrinsic redundancy characterizes an overhead that exists in $\Remp$ and will be expressed in a loss when trying to achieve this rate function. The more ``ambitious'' the rate function, the larger the redundancy. We note the following two properties of $\mu_Q$:
\label{sec:irp524}
\begin{enumerate}
\item When an offset $\delta \in \mathbb{R}$ is added to (or subtracted from) the rate function:
\begin{equation}\label{eq:A516}
\mu_Q(\Remp + \delta) = \mu_Q(\Remp) + \delta
\end{equation}
\item When taking the maximum of several rate functions $\Remp(\vr x, \vr y) = \max_{k \in \{1,\ldots,K\}} \Remp_k(\vr x, \vr y)$, we have:
\begin{equation}\label{eq:A521}
\mu_Q \left(\max_{k \in \{1,\ldots,K\}} \Remp_k \right) \leq \max_{k \in \{1,\ldots,K\}} \mu_Q(\Remp_k)  + \frac{\log(K)}{n}
\end{equation}
$\frac{\log(K)}{n}$ can be regarded as the price payed for ``universality'', in the sense of exceeding several rate functions.
\end{enumerate}
The proof of these properties is straightforward and is deferred to Section~\ref{sec:int_redundancy_properties}.

Suppose that a rate function $\Remp$ has a given intrinsic redundancy $\mu_Q(\Remp)$, we may reduce it by an offset $\delta$ to make this rate function achievable. Denote $\Remp^* = \Remp - \delta$, then $\Remp^*$ will be achievable if $\mu_Q(\Remp^*) = \mu_Q(\Remp) - \delta \leq \frac{1}{n} \log \epsilon$, i.e. if $\delta \geq \mu_Q(\Remp) + \frac{1}{n} \log \frac{1}{\epsilon}$. Conversely, it will not be achievable if $\mu_Q(\Remp^*) = \mu_Q(\Remp) - \delta > \frac{1}{n} \log \frac{1}{1-\epsilon}$, i.e. if $\delta < \mu_Q(\Remp) - \frac{1}{n}  \log \frac{1}{1-\epsilon}$. Using this argument, we can characterize the achievability of rate functions by specifying what value of $\delta$ (overhead) turns them into achievable. This is formalized in the following theorem:
\begin{theorem}\label{theorem:remp_achievability_upto}
For a rate function $\Remp$ to be achievable up to $\delta$, with prior $Q$ and error probability $\epsilon$, it is necessary that $\delta \geq \mu_Q(\Remp) - \frac{1}{n} \log \frac{1}{1-\epsilon}$ and sufficient that $\delta \geq \mu_Q(\Remp) + \frac{1}{n} \log \frac{1}{\epsilon}$.
\end{theorem}
This theorem gives a meaning to the term ``intrinsic redundancy'' and we can see how it affects the actual redundancy. The actual redundancy is comprised of a term depending on the intrinsic redundancy and a term depending on the desired error probability. The proof is given by the discussion above. Using this theorem we can see more clearly the rate penalty for decreasing the error probability. Supposing that we know a rate function $\Remp$ is achievable with an error probability $\epsilon_1$, then we may use the theorem to bound the redundancy required to achieve it with an error probability $\epsilon_2$. Furthermore, \eqref{eq:A521} implies that competing against $K$ competitors who attain the rate functions $\Remp_i$, incurs a small asymptotical price.

Up to the gap between the necessary and sufficient conditions in Theorems~\ref{theorem:remp_ublb},\ref{theorem:remp_achievability_upto}, these conditions are the equivalent of Kraft inequality for rate functions. If a rate function meets them, it is tight in the sense that it cannot be improved uniformly. In some sense however they are weaker than Kraft inequality, since the later applies to each uniquely decodable fixed to variable code, while our conditions apply only to communication systems which attain the error probability individually for each $\vr x, \vr y$ \onlyphd{(see also the comments in Section~\ref{sec:minimal_reqs_for_sufficient})}. In general, when comparing to information theoretic results pertaining to probabilistic channel settings, because the requirements we make are stricter (we require a rate and error probability guarantee per $\vr x, \vr y$ rather than on average), our achievability results are stronger, while our necessary conditions (converse) are weaker, since they hold for a restricted class of systems.

Theorems~\ref{theorem:remp_ublb}-\ref{theorem:remp_achievability_upto} also bring another observation: any rate function which is achievable (by any system), is also achievable using random coding (the system achieving the sufficient condition), up to a small overhead.

The gap between the upper and lower bounds of Theorem~\ref{theorem:remp_ublb},\ref{theorem:remp_achievability_upto} is equivalent to an overhead of $\log \frac{1-\epsilon}{\epsilon}$ bits over the entire transmission. This overhead is $20$ bits for $\epsilon = 10^{-6}$, so in the scope of working with a fixed but small $\epsilon$ (rather than $\epsilon \ntoinfty 0$), the difference between the bounds is small. An analysis for the reasons of this gap can be found in \selector{\cite{YL_PhdThesis}}{\ref{sec:gap_in_ublb}}. It is shown that the necessary condition can be reduced by almost one bit at the price of complicating the decoder and the proof, and cannot be further reduced in the current form of the bound. It appears by that analysis that for most rate functions, the required redundancy is close to the one required by the sufficient condition.

\subsection{Proof of Theorem~\ref{theorem:remp_ublb}}\label{sec:proof_remp_ublb}
\subsubsection{Necessary condition (converse)}\label{sec:proof_remp_ublb_NC}
In this section we prove the first part of Theorem~\ref{theorem:remp_ublb}. We need to show that the condition \eqref{eq:A434} holds for achievable, and adaptively achievable rate functions. We begin with the case of achievable rate functions (non adaptively).

Suppose $\Remp$ is achievable with $Q$, $\epsilon$. Consider and encoder and a decoder designed for rate $R$ over block size $n$ and satisfying Definition~\ref{def:achievability_nonadaptive}. There are $M \geq \exp(nR)$ input messages. Each input message $\msg=i \in \{1,\ldots,M\}$ is translated by the encoder into the random sequence $\vr X_i$, which is a random variable distributed in $\mathcal{X}^n$ (implemented by the common randomness $S$), and is known to the decoder.

According to the requirements of Definition~\ref{def:achievability_nonadaptive}, the distribution of $\vr X_i$ should be $Q(\vr x)$, since the definition requires the input distribution to be $Q(\vr x)$ for any input message. However for the converse we assume a milder condition: we only assume that the scheme achieves $Q$ on average, i.e. that the input distribution is $Q$ when $i$ is chosen uniformly over $\{1,\ldots,M\}$, in other words:
\begin{equation}\label{eq:A476}
\forall \vr x: \frac{1}{M} \sum_{i=1}^M \Pr(\vr X_i = \vr x) = Q(\vr x)
\end{equation}
Note that the codewords may be statistically dependent.

Denoting by $\hat \msg$ the decoded message, then according to Definition~\ref{def:achievability_nonadaptive}, we have:
\begin{equation}\label{eq:A483}
\forall \vr y, i \in \{1,\ldots,M\}: \Pr \left\{ \hat \msg \neq i \Big| \Remp(\vr X_i, \vr y) \geq R \right\} \leq \epsilon
\end{equation}
Note that the definition implies that \eqref{eq:A483} holds with respect to the transmitted message. However, since $\hat \msg$ is a function of $\vr y$ and $S$, for a fixed $\vr y$ it does not depend on the transmitted message, and therefore, by considering that any of the possible messages may be input to the encoder, and using Definition~\ref{def:achievability_nonadaptive} with respect to this message, we have that \eqref{eq:A483} holds for any $i$. Therefore the following holds for any $\vr y$ (where probabilities are over the randomness in the codebook):
\begin{equation}\label{eq:Aconverse_derivation}
\begin{split}
1
&=
\sum_{i=1}^M \Pr \left\{ \hat \msg = i \right\}
\geq
\sum_i \Pr \left\{ (\hat \msg = i) \cap (\Remp(\vr X_i, \vr y) \geq R) \right\}
\\&=
\sum_i \Pr \left\{ \hat \msg = i \Big| \Remp(\vr X_i, \vr y) \geq R \right\} \Pr \left\{ \Remp(\vr X_i, \vr y) \geq R \right\}
\\& \stackrel{\eqref{eq:A483}}{\geq}
\sum_i (1 - \epsilon) \Pr \left\{ \Remp(\vr X_i, \vr y) \geq R \right\}
=
(1 - \epsilon) \sum_i \sum_{\vr x: \Remp(\vr x, \vr y) \geq R} \Pr(\vr X_i = \vr x)
\\&=
(1 - \epsilon) \sum_{\vr x: \Remp(\vr x, \vr y) \geq R}  \sum_i \Pr(\vr X_i = \vr x)
\stackrel{\eqref{eq:A476}}{=}
(1 - \epsilon) \sum_{\vr x: \Remp(\vr x, \vr y) \geq R}  M Q(\vr x)
\\&=
(1 - \epsilon) M Q \left\{ \Remp(\vr X, \vr y) \geq R \right\}
\end{split}
\end{equation}
Therefore
\begin{equation}\label{eq:A518}
Q \left\{ \Remp(\vr X, \vr y) \geq R \right\} \leq \frac{1}{(1 - \epsilon) M} \leq \frac{1}{1 - \epsilon} \exp(-nR)
\end{equation}
This holds for any $\vr y$. In addition Definition~\ref{def:achievability_nonadaptive} requires that such a system will exist for any $R$, therefore \eqref{eq:A518} holds for any $R$ as well. This proves the claim for the case of achievable $\Remp$.

The case of adaptively achievable $\Remp$ follows from the same argument. First, one may convert the adaptive rate system with feedback into a non-adaptive rate system with feedback: fix a rate $R$ and let the decoder output only $nR$ bits, and an error if the rate is $\Remp < R$. Therefore whenever $\Remp > R$ in probability $1-\epsilon$ the message will be decoded correctly. Now, note that \eqref{eq:Aconverse_derivation} refers to any fixed value of $\vr y$. Therefore \eqref{eq:Aconverse_derivation} holds even if the encoder knows the value of $\vr y$, and particularly it holds also in the presence of feedback (partial and sequential knowledge of $\vr y$). Hence the results holds also for $\Remp$ which is adaptively achievable.

\subsubsection{Sufficient condition (direct)}\label{sec:proof_remp_ublb_SC}
The direct side is shown by generating the $M = \lceil \exp(nR) \rceil$ codewords $\vr X_i$ i.i.d. with distribution $Q$. Thus, the condition on the input distribution is met. The decoder, after observing $\vr y$, chooses $\hat \msg$ to be the index of the word with the maximum value of $\Remp(\vr X_i, \vr y)$ (breaking ties arbitrarily), i.e.
\begin{equation}\label{eq:A531}
\hat \msg = \argmax{i} \left[ \Remp(\vr X_i, \vr y) \right]
\end{equation}
We assume a given message $\msg$, and a given $\vr X_\msg = \vr x$. Since the codewords are independent, conditioning on $\vr x$ does not change the distribution of the other codewords. By the union bound, the probability of error is bounded by:
\begin{equation}\begin{split}\label{eq:A9090p}
P_e^{(\msg)}(\vr x, \vr y)
& \leq
\Pr \left\{ \bigcup_{i \neq \msg} \left( \Remp (\vr X_i , \vr y)  \geq  \Remp (\vr X_\msg , \vr y) \right) \Bigg| \vr X_\msg = \vr x \right\}
\\ & \leq
(M-1) \cdot Q \left\{ \Remp (\vr X , \vr y)  \geq  \Remp (\vr x , \vr y) \right\}
\\& \stackrel{\eqref{eq:A435}}{\leq}
(M-1) \cdot \epsilon \cdot \exp(- n \Remp (\vr x , \vr y))
\\& \leq
\epsilon \cdot \exp[n (R-\Remp (\vr x , \vr y))]
\end{split}\end{equation}
where in the last inequality we substituted $M \leq \exp(nR) + 1$.
Therefore if $\Remp (\vr x , \vr y) \geq R$, we will have $P_e^{(\msg)}(\vr x, \vr y) \leq \epsilon$ as required. \endofproof

\subsection{Comments on the proof of Theorem~\ref{theorem:remp_ublb}}
\begin{itemize}
\item To understand the proof of the necessary condition, it is useful to think that the channel output $\vr y$ is set to a constant. Thus, the decoder is isolated from the encoder, and is required to decide on the message $\hat \msg$ based solely on its knowledge of the codebook.

\item The proof of the theorem teaches something about the way rate functions are achieved: conditioning on $\vr x$ and $\vr y$, the different codebooks generated all include $\vr x$, and in addition other codeword. If $\Remp(\vr x, \vr y)$ is large, then in most codebooks, the other codewords will have a smaller value of $\Remp$, due to the constraint on its distribution. Therefore, by choosing the word with the maximum $\Remp$, the decoder would usually be correct. The necessary condition means that this is actually required to happen in order for $\Remp$ to be achievable: as the decoder is ``isolated'' from the encoder, and still committed to \eqref{eq:A483}. If there are several words with $\Remp(\vr X_i, \vr y) \geq R$ the decoder will need to toss a coin and split the distribution in some way between them, with a large probability to be in error.
\onlyphd{Section~\ref{sec:gap_in_ublb} sheds more light on this topic. }
\onlypaper{The analysis of the gap between the necessary and sufficient condition in \cite{YL_PhdThesis} sheds more light on this topic.}

\item By the current definitions, it is assumed that the input distribution $Q$ does not depend on $\vr y$. However note that since the proofs of the necessary and sufficient conditions both consider a fixed value of $\vr y$, the results hold, under a suitable formulation, also for the case where the input distribution depends on $\vr y$.

\item We can adopt two point of views when considering systems satisfying Theorem~\ref{theorem:remp_ublb} (the achievability of rate functions): one is as communication systems trying to convey messages over an unknown channel; another is a cynical perspective in which we do not assume the input and output are related (and thus it is impossible to convey information), but we are only trying to design systems that satisfy the promises of the theorems, and the question is viewed as a game between the encoder and decoder, and the environment choosing $\vr y$ and the message. The first point of view gives us the motivation and application of the theorems; the second is more suitable for the design and analysis. This is similar to the case of prediction and learning with expert advice \cite{Nicolo}\cite{Vovk97} -- when designing these learning algorithms the assumption is that the information supplied by the experts is completely arbitrary, and therefore the target is not to ``learn'' but just to compete; but the application of the results is for learning (where we assume there is some information at least in some of the experts advice). \todo{wording}
\end{itemize}

\subsection{Examples}
\begin{example}[A wire]
Consider the binary input -- binary output channel $\mathcal{X}=\mathcal{Y}=\{0,1\}$ with the rate function $\Remp = \Ind(\vr x = \vr y)$, i.e. $\Remp=1$ iff the output is identical to the input. This function is easily achievable, with $Q(\vr x) = \unif(\mathcal{X}^n)$. To attain this rate function without error $\epsilon=0$, one simply transmits the message un-coded, at a rate $R=1$. If the channel output happened to equal the input, the communication had succeeded. If it happened to be different, $\Remp=0<R$ and thus no guarantee was made. $Q(\vr x)$ needs to be uniform in order to achieve rate 1. For this rate function and any $R \leq 1$, the condition $\Remp \geq R$ is satisfied by one sequence, and therefore $Q \{ \Remp \geq R \} = \frac{1}{2^n}$. This satisfies the necessary condition in Theorem~\ref{theorem:remp_ublb} with equality for $\epsilon=0$, and thus the sufficient condition is not tight here.

Note that the codebook that achieves this rate function is not a random i.i.d. codebook - the codewords are fixed, or, in order to achieve the input distribution condition, should be generated by randomly permuting the $2^n$ possible sequences. Therefore the codewords are correlated, which is necessary in order to obtain the necessary condition. Furthermore, the regions of $\vr x \in \mathcal{X}^n$ for which $\Remp(\vr x, \vr y) > 0$ obtained for different $\vr y$-s are disjoint, in which case, as we have noted, the necessary condition could be tight. If we had insisted on generating the codewords independently, then this rate function could not be achieved without some loss, due to the probability of two codewords being equal, therefore in that case the maximum rate would be closer to rate determined by the sufficient condition.
\end{example}

\begin{example}[A fixed codebook]
Similarly, consider transmission using a fixed codebook of $M = \exp(nR_0)$ codewords, and an arbitrary fixed decoder. We may randomly permute the messages in order to guarantee a fixed input distribution for any message. In this case $Q(\vr x) = \frac{1}{M}$ when $\vr x$ is in the codebook and $0$ otherwise. Define the rate function $\Remp(\vr x, \vr y)$ as $R_0$ if $\vr y$ is decoded by the decoder to the message represented by $\vr x$, and $0$ otherwise. Then for $R \leq R_0$, $Q \{ \Remp \geq R \} = \frac{1}{M} = \exp(-n R_0) \leq \exp(-nR)$, and as before the necessary condition is satisfied with equality for $\epsilon = 0$.
\end{example}

\begin{example}[The empirical mutual information]
Lemma~1 in our \selector{previous paper}{initial work} \cite{YL_individual_full} states that for any i.i.d. prior $Q$,  $Q \left( \hat{I}(\vr x; \vr y) \geq R \right) \leq \exp \left( -n \left(R - \delta_n \right) \right)$ with $\delta_n = |\mathcal{X}||\mathcal{Y}|\frac{\log(n+1)}{n} \ntoinfty 0$. Therefore $\mu_Q(\hat I) \leq \delta_n$, and the conclusion from Theorem~\ref{theorem:remp_achievability_upto} is that this function is achievable up to $\delta_n + \frac{1}{n} \log \frac{1}{\epsilon}$. Note that the actual intrinsic redundancy is about half of this bound (see Section~\ref{sec:examples_eMI}).
\end{example}

\begin{example}[A second order rate function]\label{example:ir_of_siso}
The rate function $\Remp = \half \log \frac{1}{1 - \hat\rho^2}$ presented in the \selector{previous paper}{initial work} \cite{YL_individual_full} has an intrinsic redundancy $\mu_Q(\Remp) = \infty$. This results from the factor $n-1$ instead of $n$ in Lemma~4 there, which causes the fact $-\frac{1}{n} \log \Pr(\Remp \geq R)$ grows slower than $R$ for large values of $R$. The implication is that this rate function cannot be attained with a fixed loss, but the loss must grow with $R$. So for example one cannot attain $\Remp-\delta$, but one can attain $\gamma \cdot \Remp$ (with $\gamma \ntoinfty 1$). The proof is technical and is deferred to Appendix~\ref{sec:ir_of_siso_proof}.
\end{example}

\subsection{General systems and Good-put functions}\label{sec:good_put_bound}
The requirement to attain a fixed error probability for every $\vr x, \vr y$ releases the characterization of the communication system from dependence on the channel. On the other hand, it may seem as an over-requirement, since from application perspective requiring low \emph{average} error probability may be sufficient. In this section it is shown that this over-requirement is not as strong as may seem: any communication system may be converted to a system guaranteeing a small error probability, with a small price in the rate.\footnote{Practically, the later system may be more complex to implement.} This result holds in full generality only for the non-adaptive case, however considering the sub-set of adaptively achievable rate functions presented in \selector{Section}{Chapter}~\ref{chap:rate_adaptivity}, it makes sense to believe that for many systems of interest, this will hold also adaptively. Thus, the concept of attainable rate functions is not as esoteric as it would initially seem.

Let us consider a system delivering a rate $R_\tsubs{sys}$ with an error probability $\epsilon_\tsubs{sys}$. This system may be quite general. To fix thoughts, it may be useful to consider the two examples of a practical (Turbo/LDPC) encoder and a decoder, perhaps combined within a more complex system involving channel estimation, feedback, scrambling, etc, and on the other hand, a theoretical random coding system. Each system generates a certain input distribution $Q(\vr x) = Q_\tsubs{sys}(\vr x)$, which is assumed to be independent of the channel output.

In order to characterize the system with a single number, consider the rate of error-free bits delivered by the system, sometimes referred to as ``good-put'' (in contrast to throughput):
\begin{equation}\label{eq:A1040}
R_\tsubs{good} = (1 - \epsilon_\tsubs{sys}) R_\tsubs{sys}
.
\end{equation}
This value is a little optimistic, because it ignores the need to detect the errors. As an example, delivering one bit per second with error probability half is not the equivalent of half a bit per second. This additional gap is related to the factor $h_b(\epsilon_\tsubs{sys})$ in Fano's inequality\onlyphd{~(see Lemma~\ref{lemma:fano_rateadaptive})}, and is asymptotically negligible. Now, assuming that $\epsilon_\tsubs{sys}$ and $R_\tsubs{sys}$ are not fixed but may change (depending, e.g. on the channel, on common randomness), the good-put is the average of the above, i.e.
\begin{equation}\label{eq:A1040b}
R_\tsubs{good} = \E \left[ (1 - \epsilon_\tsubs{sys}) R_\tsubs{sys} \right]
.
\end{equation}
To obtain a characterization of a system, which is independent of the channel, the above may be conditioned on the channel input and output $\vr x, \vr y$. Define
\begin{equation}\label{eq:A1040c}
R_\tsubs{good}(\vr x, \vr y) \defeq \E \left[ (1 - \epsilon_\tsubs{sys}) R_\tsubs{sys} \Big| \vr x, \vr y \right]
.
\end{equation}
In other words, $R_\tsubs{good}(\vr x, \vr y)$ is the average good-put obtained with the system when the input and output happened to be $\vr x, \vr y$. For a deterministic block encoder/decoder, the conditional error probability is either $0$ or $1$, and the good-put is, respectively, either $R_\tsubs{sys}$ or $0$. The function $R_\tsubs{good}(\vr x, \vr y)$ is only a function of the system and not of the channel, and when a specific probabilistic channel is known, the average good put may be computed as $R_\tsubs{good} = \E \left[ R_\tsubs{good}(\vr X, \vr Y) \right]$.

Next, let us show that for any system, $R_\tsubs{good}(\vr x, \vr y)$ is an asymptotically achievable rate function (with the prior $Q(\vr x) = Q_\tsubs{sys}(\vr x)$). Initially, it is assumed that $R_\tsubs{sys}$ is a constant, i.e. the system delivers a constant rate, with a varying error probability $\epsilon_\tsubs{sys}(\vr x, \vr y)$. Assume the message $\msg$ is a uniform random variable $\unif\{1,\ldots,M \}$, $M=\exp(n R_\tsubs{sys})$. The system is defined by common randomness $S$ (possibly), a transmission function $\vr X(S,\msg)$ and a decoding function $\hat\msg(S,\vr y)$ (see Definition~\ref{def:sys_nonadaptive}). Now, consider the system's operation when $\vr y$ set as a constant. Any feedback the system might have, can be ignored, as it conveys constant information. In this case, $\hat\msg(S,\vr y)$ and $\msg$ are independent, and:
\begin{equation}\label{eq:A731}
\Pr \{ \hat\msg(S,\vr y) = \msg \} = \frac{1}{M} = \exp(-n R_\tsubs{sys})
.
\end{equation}
The error probability is
\begin{equation}\label{eq:A735}
\epsilon_\tsubs{sys}(\vr x, \vr y)
=
\Pr \left\{ \hat\msg(S,\vr y) \neq \msg \Big| \vr X(S,\msg) = \vr x \right\}
.
\end{equation}
Now,
\begin{equation}\begin{split}\label{eq:A743}
\exp(-n R_\tsubs{sys})
& =
\Pr \left\{ \hat\msg(S,\vr y) = \msg \right\}
\\&=
\sum_{\vr x} \Pr \left\{ \hat\msg(S,\vr y) = \msg \cap \vr X(S,\msg) = \vr x \right\}
\\&=
\sum_{\vr x} \Pr \left\{ \hat\msg(S,\vr y) = \msg | \vr X(S,\msg) = \vr x \right\} \cdot \Pr \left\{ \vr X(S,\msg) = \vr x \right\}
\\&=
\sum_{\vr x} (1 - \epsilon_\tsubs{sys}(\vr x, \vr y)) Q(\vr x)
\\&=
\sum_{\vr x} \frac{R_\tsubs{good}(\vr x, \vr y)}{R_\tsubs{sys}} Q(\vr x)
.
\end{split}\end{equation}
For any $R \leq R_\tsubs{sys}$, the sum above is bounded by :
\begin{equation}\begin{split}\label{eq:A758}
&
\sum_{\vr x} \frac{R_\tsubs{good}(\vr x, \vr y)}{R_\tsubs{sys}} Q(\vr x)
\\& \geq
\sum_{\vr x: R_\tsubs{good}(\vr x, \vr y) \geq R} \frac{R_\tsubs{good}(\vr x, \vr y)}{R_\tsubs{sys}} Q(\vr x)
\\& \geq
\frac{R}{R_\tsubs{sys}} \sum_{\vr x: R_\tsubs{good}(\vr x, \vr y) \geq R} Q(\vr x)
\\& =
\frac{R}{R_\tsubs{sys}} \Pr \left\{ R_\tsubs{good}(\vr X, \vr y) \geq R\right\}
\end{split}\end{equation}
Combining \eqref{eq:A743} and \eqref{eq:A758}, yields:
\begin{equation}\label{eq:A768}
\Pr \left\{ R_\tsubs{good}(\vr X, \vr y) \geq R \right\} \leq \frac{R_\tsubs{sys}}{R} \exp(-n R_\tsubs{sys})
.
\end{equation}
For $x \geq 1$ the function $x e^{-x}$ is decreasing. Substituting $\log(e) x = n R$, yields that $R \exp(-nR)$ is decreasing with $R$ for $R \geq \frac{\log(e)}{n}$, and therefore
$\frac{R_\tsubs{sys}}{R} \exp(-n R_\tsubs{sys}) \leq \exp(-n R)$. For $R < \frac{\log(e)}{n}$ (where $\exp(-nR) > e^{-1}$), the probability above \eqref{eq:A768} can be simply upper bounded by $1$. This yields the following simple bound:
\begin{equation}\label{eq:A775}
\Pr \left\{ R_\tsubs{good}(\vr X, \vr y) \geq R \right\} \leq e \cdot \exp(-n R)
.
\end{equation}
For the case of $R \geq R_\tsubs{sys}$, the above holds trivially. The bound above corresponds to the sufficient condition of Theorem~\ref{theorem:remp_ublb}, with an intrinsic redundancy of $\mu_Q(R_\tsubs{good}) \leq \frac{\log(e)}{n}$, and is therefore it is asymptotically achievable (Theorem~\ref{theorem:remp_achievability_upto}). Notice that the system achieving this rate (Section~\ref{sec:proof_remp_ublb_SC}) is potentially very different than the original system. Furthermore, the bound leading from \eqref{eq:A768} to \eqref{eq:A775} is very coarse, which implies the good-put is a very pessimistic bound on the rate that can be achieved. This is because the error probability can be exponentially improved with a decrease in the rate, while in the good-put function, there is only a linear decrease (e.g. the error probability when attaining $R_\tsubs{good}=\half R_\tsubs{sys}$ is $\half$ with the original system, whereas it could have been significantly better). The extension to rate adaptive systems appears in Appendix~\ref{sec:good_put_bound_rateadaptive} \todo{Currently unfinished}. This is summarized by the following Lemma:
\begin{theorem}\label{theorem:goodput_converse}
The good-put function \eqref{eq:A1040c} of any fixed-rate or adaptive rate system (Definitions~\ref{def:sys_nonadaptive},\ref{def:sys_adaptive}), possibly including common randomness and feedback, is an asymptotically achievable rate function, with the prior generated by the system's codebook distribution, and has an intrinsic redundancy of $\mu_Q(R_\tsubs{good}) \leq \frac{\log(e)}{n}$.
\end{theorem}
An interesting and insightful resulting of the combination of Theorem~\ref{theorem:goodput_converse} and Theorem~\ref{theorem:remp_conditional_form_asymptotic} which is proven in \selector{Section}{Chapter}~\ref{chap:asymptotic_limit_remp}, is that the rate of any system can be characterized by two probability functions $P(\vr x | \vr y)$ and $Q(\vr x)$ (where the second is the input distribution).

\todo{this raises the question: what is $P$ of a given system? is it $P(\vr x | \vr y) = \Pr(\hat{\vr x} = \vr x | \vr X = \vr x, \vr y)$ ?}

If, furthermore, this achievable rate function satisfies the structure defined in \selector{Section}{Chapter}~\ref{chap:rate_adaptivity}, then it is also asymptotically adaptively achievable. I.e. there exists a system attaining the same rates, but with an error probability as small as desired, per any pair of sequences.

\section{An asymptotical characterization of achievable rate functions}\label{chap:asymptotic_limit_remp}
In Theorem~\ref{theorem:remp_ublb} we have shown that achievable rate function have a CCDF upper bounded by a decaying exponential function. Therefore it stands to reason that the Chernoff bound for the probability $Q(\Remp(\vr X, \vr y) \geq R)$ may be rather tight. From this observation we derive asymptotical necessary and sufficient conditions which are easier to calculate. The main result of this section is that asymptotically achievable rate functions are bounded by the form $\frac{1}{n} \log \frac{f(\vr x | \vr y)}{Q(\vr x)}$ for some conditional probability assignment $f(\vr x | \vr y)$. As a result this form can be used as a prototype for rate functions.

\subsection{The Chernoff and Markov inequalities}\label{sec:chernoff_markov_exposition}
The Chernoff and Markov inequalities are useful tools in the following analysis. The Markov inequality simply states that for any non-negative random variable $A$,
\begin{equation}\label{eq:Amarkov_bound}
\Pr \{ A \geq t \} \leq \frac{\E[A]}{t}
\end{equation}
The proof is simple, by applying the expected value operator to both sides of $\Ind( A \geq t ) \leq \frac{A}{t}$. From this simple bound, many useful bounds can be derived, for example the Chebyshev inequality is obtained by substituting $A=(X - \E[X])^2$. The Chernoff upper bound for $\Pr(X \geq \tau)$ is obtained by substituting $A = \exp(\beta X), t = \exp(\beta \tau)$ for some constant $\beta>0$, and then optimizing over $\beta$. The main strength of Chernoff bound results from the fact that when $X$ is a sum of independent random variables $X = \sum_i X_i$, then $\E[A] = \E[\exp(\beta X)] = \E[\prod_i \exp(\beta X_i)] = \prod_i \E[\exp(\beta X_i)]$ breaks into a product of terms associated with each individual element, which is in most cases simpler to calculate. Since information theoretic values are associated with log-probabilities, the Markov and Chernoff bounds are virtually the same in our context (the Chernoff bound when applied to the log-probabilities is equivalent to the Markov inequality applied to the probabilities). \onlyphd{In *** we present a simple proof of the achievability of Shannon capacity, using the Markov inequality as a main tool}

\subsection{Application of the Chernoff bound}\label{sec:chernoff_application}
Consider a sequence of rate functions $\Remp(\vr x^n, \vr y^n)$ for $n=1,2,\ldots$. We would like to find out whether $\Remp$ is asymptotically attainable. Although $\Remp$ may be asymptotically attainable, the intrinsic redundancy associated with it may not tend to zero. In other words, it may be possible to attain $F_n(\Remp(\vr x^n, \vr y^n))$ (with $F_n(t) \ntoinfty t$), but $F_n$ is not necessarily of the form $F_n(t) = t - \delta_n$ with $\delta_n \ntoinfty 0$. Therefore it is useful to consider more general functions $F_n(t)$. As an example for such a case see the rate function for the continuous MIMO channel presented in Section~\ref{sec:GaussianMIMO}, which is achieved up to $F_n(t) = \gamma_n t - \delta_n$.

We consider the rate function $F_n( \Remp(\vr x, \vr y) )$. Using the Chernoff/Markov inequality to bound the probabilities in Theorem~\ref{theorem:remp_ublb}, we have:
\begin{equation}\begin{split}\label{eq:A726}
Q \{ F_n( \Remp(\vr X, \vr y) ) \geq R \}
& =
Q \{ \exp(n F_n(\Remp(\vr X, \vr y))) \geq \exp(n R) \}
\\&\stackrel{\eqref{eq:Amarkov_bound}}{\leq}
\underset{Q}{\E} \left[ \exp(n F_n[\Remp(\vr X, \vr y)]) \right] \exp(- n R)
=
L_{F,n} \cdot \exp(- n R)
\end{split}\end{equation}
where
\begin{equation}\label{eq:Adef_L_Fn}
L_{F,n} \defeq \underset{Q}{\E} \left[ \exp(n F_n[\Remp(\vr X, \vr y)]) \right]
\end{equation}
In many cases, for a suitable choice of $F$, such as $F_n = \gamma t$, calculating $L_{F,n}$ is simpler than calculating the probability $Q \{ \Remp(\vr X, \vr y) \geq R \}$.
From this bound we have that the intrinsic redundancy \eqref{eq:Adef_intrinsic_redundancy} of $F_n[\Remp]$ satisfies
\begin{equation}\label{eq:A756}
\mu_Q(F_n[\Remp]) \stackrel{\eqref{eq:A726}\eqref{eq:Adef_intrinsic_redundancy}}{\leq} \frac{1}{n} \log L_{F,n}
\end{equation}
by Theorem~\ref{theorem:remp_achievability_upto}, this implies that $F_n[\Remp]$ is achievable up to $\delta_n = \frac{1}{n} \log L_{F,n} + \frac{1}{n} \log \frac{1}{\epsilon}$. If for any sequence $F_n(t) \ntoinfty t$, we have $\frac{1}{n} \log L_{F,n} \ntoinfty 0$ (in other words, $L_{F,n}$ increases subexponentially with $n$), this implies that $F_n[\Remp] - \delta_n$ is achievable where $\delta_n \ntoinfty 0$ and therefore $\Remp$ is asymptotically achievable. On the other hand, as we show below, this condition is also necessary. This manifests the claim that the use of the Chernoff bound is asymptotically tight.

\subsection{Asymptotic tightness of the Chernoff bound}\label{sec:chernoff_tightness}
\begin{theorem}\label{theorem:chernoff_tightness}
A sequence of rate functions $\Remp(\vr x^n, \vr y^n)$ is asymptotically achievable with a sequence of priors $Q(\vr x^n)$, iff there exists a sequence of functions $F_n(t) \ntoinfty t$, such that for all $\vr y$:
\begin{equation}\label{eq:A784}
\limsup_{n \to \infty} \frac{1}{n} \log L_{F,n} \leq 0
\end{equation}
where $L_{F,n}$ is defined in \eqref{eq:Adef_L_Fn}.
\end{theorem}
Note that comparing with the conditions of Theorem~\ref{theorem:remp_ublb}, which are conditions on the CCDF of $\Remp(\vr X, \vr y)$ and must be satisfied per $R$, the condition above is a simpler condition on an expected value, which doesn't explicitly refer to the rate $R$.

Let us begin with the following lemma which is the heart of the reverse part.
\begin{lemma}\label{lemma:chernoff_tightness_lemma}
Any achievable rate function $\Remp$ (with $\epsilon, Q$) satisfies for $\gamma < 1$:
\begin{equation}\label{eq:A957}
\forall \vr y: \underset{\vr X \sim Q}{\E} \left[ \exp(n \gamma \Remp(\vr X, \vr y)) \right] \leq \frac{1}{(1-\epsilon)(1-\gamma)}
\end{equation}
\end{lemma}

\textit{Proof:} Suppose that $\Remp$ achievable, by Theorem~\ref{theorem:remp_ublb} this implies
\begin{equation}\label{eq:A796}
\forall y \in \mathcal{Y}^n, R \in \mathbb{R} : Q \left\{ \Remp(\vr X, \vr y) \geq R \right\}  \leq \frac{1}{1 - \epsilon} \exp(-nR)
\end{equation}

Intuitively it is clear that this constraint on the CCDF of $\Remp$ implies the exponential factor in \eqref{eq:A957} is canceled out by the exponential decay of the distribution. For a fixed $\vr y$, define the random variable $V \defeq \exp(-n \Remp(\vr X, \vr y))$ and substitute $r \defeq \exp(-nR)$. Then the above can be written as a condition on the CDF of $V$, $F_V(r)$:
\begin{equation}\begin{split}\label{eq:A796b}
\forall r > 0 :
F_V(r)
& \defeq
\Pr(V \leq r)
=
Q \left\{ \exp(-n \Remp(\vr X, \vr y)) \leq \exp(-n R) \right\}
\\&=
Q \left\{ \Remp(\vr X, \vr y) \geq R \right\}
\leq
\frac{1}{1 - \epsilon} \exp(-nR)
\\& =
\frac{r}{1 - \epsilon}
\end{split}\end{equation}
Next, this condition on the CDF is translated to a conclusion on the expected value. Since by definition $F_V(r) \in [0,1]$ we can write the bound as $F_V(r) \leq F_U(r) \defeq \min \left( \frac{r}{1 - \epsilon}, 1 \right)$, i.e. $F_V(r)$ is bounded by the CDF of a uniform random variable $U \sim \unif[0,1-\epsilon]$. This implies that we can bound $V \geq U$, as formulated in the following Lemma:
\begin{lemma}[CDF inequality]\label{lemma:cdf_inequality}
Let $V$ be a random variable and let the probability function of $V$ be bounded by $F_V(x) \leq F_U(x)$, where $F_U(x)$ is a probability function and is monotonically increasing for all $x$ such that $0 < F_U(x) < 1$, then there exists a random variable $U \sim F_U$ such that $V \geq U$.
\end{lemma}

\textit{Proof:} Since $F_U(x)$ is monotonically increasing it is invertible for values in the region $(0,1)$. Let $U = F_U^{-1} ( F_V( V ) )$. Then by the well known inverse transform theorem \unfinished{[? Missing Ref]} $F_V( V )$ is uniform $\unif[0,1]$ and therefore by applying $F_U^{-1}$ we obtain that $U$ is distributed according to $F_U$. Since $F_U$ is monotonically increasing, so is its inverse. Thus by applying $F_U^{-1}$ to both sides of the inequality $F_V(V) \leq F_U(V)$ we obtain $U \leq V$. \endofproof

Returning to the proof of Lemma~\ref{lemma:chernoff_tightness_lemma}, let $U \sim \unif[0,1-\epsilon]$ be a random variable that satisfies $U \leq V$, then
\begin{equation}\begin{split}\label{eq:A837}
\underset{Q}{\E} \left[ \exp(n \gamma \Remp(\vr X, \vr y)) \right]
\\&=
\E \left[ \frac{1}{V^{\gamma}} \right]
\leq
\E \left[ \frac{1}{U^{\gamma}} \right]
\\&=
\int_{0}^{1-\epsilon} \frac{1}{u^{\gamma}} \frac{1}{1-\epsilon} d u
=
\frac{(1-\epsilon)^{1-\gamma}}{(1-\epsilon)(1-\gamma)}
\\&\leq
\frac{1}{(1-\epsilon)(1-\gamma)}
\end{split}\end{equation}
The condition $\gamma < 1$ is required for the integral to exist.
\endofproof
Notice that it is possible to prove the result by using integration in parts, however the current proof technique avoids any continuity/integrability assumptions.

\textit{Proof of Theorem~\ref{theorem:chernoff_tightness}:}\\

\textit{Direct part:} if \eqref{eq:A784} holds for some sequence $F_n(t)$, then there exists an upper bounding sequence $\overline \delta_n \ntoinfty 0$ such that $\frac{1}{n} \log L_{F_n} \leq \overline \delta_n$, therefore by Theorem~\ref{theorem:remp_achievability_upto} and \eqref{eq:A756}, we have that $F_n[\Remp]$ is achievable up to
\begin{equation}\label{eq:A795}
\mu_Q(F_n(\Remp)) + \frac{1}{n} \log \frac{1}{\epsilon}
=
\frac{1}{n} \log L_{F,n} + \frac{1}{n} \log \frac{1}{\epsilon}
\leq
\overline \delta_n + \frac{1}{n} \log \frac{1}{\epsilon}
\end{equation}
Therefore defining $G_n(t) = F_n(t) - \left( \overline \delta_n + \frac{1}{n} \log \frac{1}{\epsilon} \right)$, we have that $G_n(t) \ntoinfty t$, and $G_n(\Remp) = F_n(\Remp) - \left( \overline \delta_n + \frac{1}{n} \log \frac{1}{\epsilon} \right) $ is achievable, and therefore by definition $\Remp$ is asymptotically achievable.

\textit{Reverse part:}
Suppose that $\Remp$ is asymptotically achievable. Then by definition for any $\epsilon$, there exists a sequence of functions $F_n(t) \ntoinfty t$ such that $F_n[\Remp]$ is achievable. By Lemma~\ref{lemma:chernoff_tightness_lemma} this implies (for $\gamma_n < 1$):
\begin{equation}\label{eq:A947b}
\underset{Q}{\E} \left[ \exp(n \gamma_n F_n[\Remp(\vr X, \vr y)]) \right] \leq \frac{1}{(1-\epsilon)(1-\gamma)}
.
\end{equation}
Defining $G_n(t) \defeq \gamma_n \cdot F_n(t)$, then by definition \eqref{eq:Adef_L_Fn} the LHS equals $L_{G,n}$. Choosing $\gamma_n = 1 - \frac{1}{n}$ we have that $G_n(t) \ntoinfty t$, while
\begin{equation}\label{eq:A837}
L_{G,n}
=
\underset{Q}{\E} \left[ \exp(n G_n[\Remp(\vr X, \vr y)]) \right]
\stackrel{\eqref{eq:A947b}}{\leq}
\frac{n}{(1-\epsilon)}
,
\end{equation}
and therefore
\begin{equation}\begin{split}\label{eq:A856}
\limsup_{n \to \infty} \frac{1}{n} \log L_{G,n}
& \leq
\limsup_{n \to \infty} \frac{1}{n} \log \frac{n}{1-\epsilon}
\\& =
\lim_{n \to \infty} \frac{1}{n} \log \frac{n}{1-\epsilon}
=
0
\end{split}\end{equation}
which satisfies \eqref{eq:A784}.
\endofproof

\subsection{Conditional probabilities and rate functions}\label{sec:chernoff_cond_probability}
We now apply Theorem~\ref{theorem:chernoff_tightness} to obtain a more intuitive form for the asymptotical rate functions. We assume that the conditions of Theorem~\ref{theorem:chernoff_tightness} hold. For the sake of discussion, let us for the moment replace the limits with equalities, i.e. assume that $\frac{1}{n} \log L_{F,n} = 0$ (i.e. $L_{F,n} = 1$) and $F_n(t) = t$. Then by definition \eqref{eq:Adef_L_Fn} we have:
\begin{equation}\label{eq:A797}
L_{F,n}
=
\underset{Q}{\E} \left[ \exp(n \Remp(\vr X, \vr y)) \right]
=
\sum_{\vr x \in \mathcal{X}^n} Q(\vr x) \exp(n \Remp(\vr x, \vr y))
=
1
\end{equation}
Denote the summand:
\begin{equation}\label{eq:A809}
f(\vr x | \vr y) = Q(\vr x) \exp(n \Remp(\vr x, \vr y))
\end{equation}
then \eqref{eq:A797} implies $\sum_{\vr x} f(\vr x | \vr y) = 1$ for every $\vr y$. Therefore $f(\vr x | \vr y)$ is a legitimate conditional distribution on $\vr x$. By inverting the relation \eqref{eq:A809}, $\Remp$ is written as:
\begin{equation}\label{eq:Aremp_conditional_form}
\Remp(\vr x, \vr y) = \frac{1}{n} \log \frac{f(\vr x | \vr y)}{Q(\vr x)}
\end{equation}
The considerations above remain the same for continuous input, by replacing the sum with an integral. Note that this rate function is not defined for $\vr x$ with $Q(\vr x)=0$, however by the definitions of achievability, the values of $\Remp$ for such $\vr x$ have no consequence, and therefore we may leave them ``undefined''.  This form \eqref{eq:Aremp_conditional_form} provides a general way to obtain rate functions which are achievable up to a small factor. Specifically, since rate functions of the form \eqref{eq:Aremp_conditional_form} have by definition $L_{F=t,n} = 1$, they have $\mu_Q(\Remp) \leq 0$ \eqref{eq:A756}, and are therefore, achievable up to $\delta_n = \frac{1}{n} \log \frac{1}{\epsilon}$ (Theorem~\ref{theorem:remp_achievability_upto}). This observation is formalized below.
\begin{lemma}\label{lemma:remp_conditional_form}
For any conditional distribution $f(\vr x | \vr y)$, the rate function defined in \eqref{eq:Aremp_conditional_form} has $\mu_Q(\Remp) \leq 0$ and is achievable (with a prior $Q$ and error probability $\epsilon$) up to $\delta_n = \frac{1}{n} \log \frac{1}{\epsilon}$.
\end{lemma}
On the other hand, it is also possible to give a lower bound on the redundancy of this rate function (the reverse of Lemma~\ref{lemma:remp_conditional_form}) by using the proof technique from Theorem~\ref{theorem:chernoff_tightness}. The following Lemma is proven in Appendix~\ref{sec:proof_remp_conditional_form_reverse}:
\begin{lemma}\label{lemma:remp_conditional_form_reverse}
If the rate function defined in \eqref{eq:Aremp_conditional_form} satisfies $\Remp \leq R_{\max} \in \mathbb{R}^+$, then this function is achievable (with a prior $Q$ and error probability $\epsilon$) up to $\delta$, only if $\delta \geq - \frac{\log (n) + \log \frac{e \cdot R_{\max}}{1-\epsilon}}{n - R_{\max}^{-1}}$
\end{lemma}
The fact the bound is negative is not surprising, since this rate function has a non-positive intrinsic redundancy. Using both Lemmas we can bound the redundancy $\delta$ up to an order of $O(\frac{\log n}{n})$.
\todo{Consider inserting these lemmas into the previous theorem, using the expectation notation, together with the intermediate result on the expected value: i.e. a result on expected value given achievability, a result on redundancy given expected value}.

The main result of this section states that all rate functions are asymptotically bounded by the form of \eqref{eq:Aremp_conditional_form} (for some $f$). I.e. this is a general way to construct all asymptotically achievable rate functions.

\begin{theorem}\label{theorem:remp_conditional_form_asymptotic}
A sequence of rate functions $\Remp(\vr x^n, \vr y^n)$ is asymptotically achievable (with a sequence of priors $Q(\vr x^n)$), iff there exist a sequence of functions $F_n(t) \ntoinfty t$ and a sequence of conditional distributions $f(\vr x^n | \vr y^n)$ such that
\begin{equation}\label{eq:A950}
F_n[\Remp(\vr x^n, \vr y^n)] \leq \frac{1}{n} \log \left( \frac{f(\vr x^n | \vr y^n)}{Q(\vr x^n)} \right)
\end{equation}
\end{theorem}

\textit{Proof:}
Direct part: if \eqref{eq:A950} holds, then $F_n[\Remp(\vr x^n, \vr y^n)]$ is upper bounded by the rate function \eqref{eq:Aremp_conditional_form}, which is asymptotically achievable by Lemma~\ref{lemma:remp_conditional_form}, and therefore by definition $\Remp(\vr x^n, \vr y^n)$ is asymptotically achievable.

Reverse part: suppose $\Remp$ is asymptotically achievable, then by Theorem~\ref{theorem:chernoff_tightness},
for some $F_n$ and a bounding sequence $\delta_n$:
\begin{equation}\label{eq:A951}
\frac{1}{n} \log L_{F,n} \leq \delta_n \ntoinfty 0
\end{equation}
Define
\begin{equation}\label{eq:A830}
f(\vr x^n | \vr y^n) = \frac{Q(\vr x) \cdot \exp(n F_n[\Remp(\vr x^n, \vr y^n)])}{L_{F,n}}
\end{equation}
by definition of $L_{F,n}$ \eqref{eq:Adef_L_Fn}, the denominator is the sum over $\vr x$ of the numerator therefore $f(\vr x^n | \vr y^n)$ is a conditional distribution. Extracting $L_{F,n}$ from \eqref{eq:A830} and substituting in \eqref{eq:A951} we have:
\begin{equation}\begin{split}\label{eq:A950b}
\frac{1}{n} \log L_{F,n}
&=
\frac{1}{n} \log \left( \frac{Q(\vr x) \cdot \exp(n F_n[\Remp(\vr x^n, \vr y^n)])}{f(\vr x^n | \vr y^n)} \right)
\\&=
F_n[\Remp(\vr x^n, \vr y^n)] - \frac{1}{n} \log \left( \frac{f(\vr x^n | \vr y^n)}{Q(\vr x)} \right)
\\&\leq
\delta_n
\end{split}\end{equation}

Defining $G_n(t) = F_n(t) - \delta_n$ we have that
\begin{equation}\label{eq:A981}
G_n[\Remp(\vr x^n, \vr y^n)] \leq \frac{1}{n} \log \left( \frac{f(\vr x^n | \vr y^n)}{Q(\vr x)} \right)
\end{equation}
Therefore $G_n$ satisfies the conditions of the theorem.
\endofproof

\subsection{Manipulating rate functions}
Following the results of this and the previous \selector{section}{chapter} we can consider various manipulations of rate functions.
\begin{itemize}
\item In Section~\ref{sec:irp524} we have seen that when taking the maximum over $K$ rate functions, the increase in the intrinsic redundancy is at most $\frac{\log K}{n}$.
\item Theorem~\ref{theorem:remp_ublb} states the achievability conditions separately per $\vr y$. Therefore if we have two rate functions that satisfy the sufficient condition, and we mix them by arbitrarily choosing for each $\vr y$ one of the rate functions, the resulting rate function is achievable.
\item Suppose that we have $K$ sequences of rate functions of the form
\begin{equation}\label{eq:A1524}
\Remp^{(k)}(\vr x, \vr y) = \frac{1}{n} \log \frac{P_k(\vr x | \vr y)}{Q(\vr x)}
\qquad
k=1,\ldots,K
\end{equation}
By definition this rate function has a non-positive intrinsic redundancy. Then the following rate function:
\begin{equation}\label{eq:A1524b}
\Remp(\vr x^n, \vr y^n) = \frac{1}{n} \log \frac{\sum_k P_k(\vr x | \vr y)}{Q(\vr x)}
= \frac{1}{n} \log \frac{\frac{1}{K} \sum_k P_k(\vr x | \vr y)}{Q(\vr x)} + \frac{\log K}{n}
\end{equation}
satisfies $\Remp \geq \Remp^{(k)}$ (as visible from the first expression in \eqref{eq:A1524b}), and has intrinsic redundancy at most $\frac{\log K}{n}$ (as visible from the second expression in \eqref{eq:A1524b}).
\end{itemize}
These results have analogs in universal source coding. In source coding, given $K$ encoders with encoding lengths $l_k(\vr x) = -\log(p_k(\vr x))$ (for the source sequence $\vr x$), by defining the universal distribution $p(\vr x) = \frac{1}{K} \sum p_k(\vr x)$, one obtains the encoding lengths $l(\vr x) = -\log(p(\vr x))$, which satisfy $l(\vr x) \leq l_k(\vr x) + \log(K)$, i.e. there is a regret of at most $\log(K)$ compared to the $K$ encoders. This fact, that stems from the logarithmic relation between probabilities and encoding lengths is the basis for universal encoding (since the normalized penalty $\frac{\log(K)}{n}$ vanishes as $n \to \infty$). Similarly in our case, the logarithmic loss in the number of competitors will be the basis for universally competing with multiple models.

\todo{Here - or somewhere else - note the relation to NML and Rissanen's result.}
\todo{also in some place i think we can connect the result of having small regret without adaptation and larger regret with adaptation to the case of probability assignment with known family versus experts. In our case the fact $y$ is not known means the family is not known}

\subsection{Discussion}
\textit{The definition of asymptotical achievability}: As we have noted, the definition of asymptotical achievability is rather loose, by allowing any $F_n(t) \ntoinfty t$ that translates the rate function to a strictly achievable one. This is done mainly for the sake of the adaptive case, in which, as we shall see, $F_n$ takes various forms, usually non linear. However for the non adaptive case, the definition could have been narrowed by considering only $F_n(t)$ of the linear form $F_n(t) = \gamma_n \cdot t - \delta_n$ with $\gamma_n \ntoinfty 1$, $\delta_n \ntoinfty 0$. All results in this section would be true also under this restricted form of $F_n(t)$.

\section{Constructions for rate functions}\label{chap:useful_constructions}
In the last two sections we have defined the conditions for achievability of rate functions, but haven't dealt with the selection of the rate function out of all achievable functions. In this section, we deal with the problem of selecting the rate function. We define constructions for rate functions which have meaningful structure. This is similar to choosing, from all encoders which comply with Kraft inequality, those that compete well with all encoders based on a family of models. We propose two main constructions:
\begin{enumerate}
\item ML construction: Rate functions that guarantee achieving the mutual information rate over a family of potential channel distributions.
\item Rate functions that are defined via a certain parameterization or classification of sequences.
\end{enumerate}

These constructions supply reasoning for choosing a specific rate function, give a uniform way to construct several rate functions that seem to be of interest, and will allow us later to prove general claims referring to the construction (rather than specific to a certain rate function).

\subsection{Empirical distributions and information measures}\label{sec:empirical_distributions_and_information_measures}
We begin with some definitions that will be useful in the sequel. The definitions below are applicable to probability distributions or probability density functions, unless stated otherwise.

\subsubsection{Empirical distribution}
Given sequences (or equivalently vectors or ordered tuples) $\vr a = (a_i)_{i=1}^n$, $\vr b = (b_i)_{i=1}^n$ where $a_i \in A, b_i \in B$ and $A,B$ are discrete alphabet sets, we define the empirical distribution:
\begin{equation}\label{eq:A1562}
\hat P_{\vr a}(a) = \hat P_{(a_i)_{i=1}^n}(a) = \frac{\sum_{i=1}^n \Ind(a_i = a)}{n}
\qquad
a \in A
\end{equation}
and the conditional empirical distribution
\begin{equation}\label{eq:A1562b}
\hat P_{(a_i | b_i)_{i=1}^n}(a|b) = \frac{\hat P_{(a_i, b_i)_{i=1}^n}(a,b)}{\hat P_{(b_i)_{i=1}^n}(b)}
\qquad
a \in A, b \in B
\end{equation}
For example $\hat P_{(x_i | x_{i-1},x_{i-2})_{i=2}^{10}}(\tilde{x}_0 | \tilde{x}_{-1}, \tilde{x}_{-2})$ yields the empirical distribution of each value in the sequence $\vr x_2^{10}$ given the two previous values. The empirical distribution of a sequence $\vr x$ denoted $\hat P_{\vr x}(x)$ is just the zero order empirical distribution.

\subsubsection{Empirical probability}
Given a probability law $Q(\vr x)$, the probability of the sequence $\vr x$ is $Q(\vr x)$. The empirical \emph{probability} of the discrete sequence $\vr x$, is the probability of the sequence under the i.i.d. empirical distribution of itself, and denoted $\hat p(\vr x)$. I.e.:
\begin{equation}\begin{split}\label{eq:A1585}
\hat p(\vr x)
&=
(\hat P_{\vr x})^n (\vr x)
=
\prod_{i=1}^n \hat P_{\vr x}(x_i)
\\&=
\prod_{\tilde{x} \in \mathcal{X}} \prod_{i: x_i = \tilde{x}} \hat P_{\vr x}(\tilde{x})
=
\prod_{\tilde{x} \in \mathcal{X}} \hat P_{\vr x}(\tilde{x})^{n \hat P_{\vr x}(\tilde{x})}
\end{split}\end{equation}
Note that the empirical probability is, in general, not a legitimate probability distribution (but a super-distribution, i.e. it has $\sum_{\vr x} \hat p(\vr x) \geq 1$), as we shall see below.

Similarly, we define the conditional empirical probability, as the probability of the sequence under the conditional empirical distribution of itself (induced by another sequence). To keep the definitions general we denote the conditioning sequence by $\vr z \in \mathcal{Z}^n$ (here and in the sequel). This conditioning sequence may include $\vr y$ or possibly delayed or modified versions of $\vr x$ and $\vr y$. For the purpose of this section it does not matter whether $\vr z$ is derived from $\vr x$ since all sequences are fixed. The conditional empirical probability means that for each set of symbols in $\vr x$ for which a certain symbol in $\vr z$ appears, i.e. $z_i = \tilde z$, we separately measure the empirical probability.

\begin{equation}\begin{split}\label{eq:A1585b}
\hat p(\vr x | \vr z)
&=
\prod_{i=1}^n \hat P_{\vr x|\vr z}(x_i|z_i)
\\&=
\prod_{\tilde{x} \in \mathcal{X}, \tilde{z} \in \mathcal{Z}} \prod_{i: x_i = \tilde{x}, z_i = \tilde z} \hat P_{\vr x|\vr z}(x_i|z_i)
=
\prod_{\tilde{x} \in \mathcal{X}, \tilde{z} \in \mathcal{Z}} \hat P_{\vr x|\vr z}(\tilde{x}|\tilde z)^{n \hat P_{\vr x \vr z}(\tilde{x}, \tilde{z})}
\end{split}\end{equation}

\subsubsection{Maximum likelihood probability}\label{sec:ML_probability}
In structuring universal schemes, we many times base a universal model on a wide class of probabilistic models \cite{Barron_MDL} (attempting to beat each model in the class). The definition of maximum likelihood probability generalizes the definition of empirical probability above, and provides a useful tool for constructing rate functions.

Denote by $p_{\theta}(\vr x)$ a class of distributions over the sequence $\vr x$, with the index $\theta \in \Theta$ (the class $\Theta$ not necessarily finite or countable). The maximum likelihood estimate of $\theta$ from $\vr x$ is
\begin{equation}\label{eq:A1601}
\hat \theta_{\ML}(\vr x) \defeq \argmax{\theta} p_{\theta}(\vr x)
\end{equation}
The maximum likelihood distribution defined by $\vr x$ is the distribution defined by the parameter $\theta = \hat \theta_{\ML}(\vr x)$. The maximum likelihood \emph{probability} of the sequence $\vr x$, is the maximum probability given to $\vr x$ by any member in $p_{\theta}(\vr x)$, or can be alternatively written as the probability of $\vr x$ under the maximum likelihood distribution:
\begin{equation}\label{eq:A1610}
\hat p_{\ML}(\vr x) \defeq \max_{\theta} p_{\theta}(\vr x) = p_{\hat \theta_{\ML}(\vr x)}(\vr x)
\end{equation}
By definition $\hat p_{\ML}(\vr x)$ satisfies $\hat p_{\ML}(\vr x) \geq p_{\theta}(\vr x)$. Except in degenerate cases, $\hat p_{\ML}(\vr x)$ is not a probability distribution, but a (strict) super-probability. Specifically, if we have two different distributions $p_1(\vr x), p_2(\vr x)$, then at least at one point $p_1(\vr x) > p_2(\vr x)$ (or equivalently $p_2 > p_1$) therefore the sum $\sum_{\vr x \in \mathcal{X}^n} \hat p_{\ML}(\vr x) = \sum_{\vr x \in \mathcal{X}^n} \max (p_1(\vr x), p_2(\vr x)) > \sum_{\vr x \in \mathcal{X}^n} p_1(\vr x) = 1$, since the summand is at least $p_1$ and larger than $p_1$ at at least one point.

The definition extends trivially to the conditional case. Using a class of conditional distributions $p_{\theta}(\vr x | \vr z)$ with respect to the generic sequence $\vr z \in \mathcal{Z}^n$, every fixed value of $\vr z$ induces a set of probabilities on $\vr x$. We define
\begin{equation}\label{eq:A1610c}
\hat p_{\ML}(\vr x | \vr z) \defeq \max_{\theta} p_{\theta}(\vr x | \vr z)
\end{equation}
Note that the class of conditional distributions $p_{\theta}(\vr x | \vr z)$ may be derived from a class of joint distributions $p_{\theta}(\vr x, \vr z)$, but this is not necessary.

For discrete sequences, taking $\Theta$ to be the class of i.i.d. distributions (defined by the probability $\theta(x), x \in \mathcal{X}$ for each value of $x$)
\begin{equation}\label{eq:A1625}
p_{\theta}(\vr x) = \prod_{i=1}^n \theta(x_i) \defeq \theta^n(\vr x)
\end{equation}
we have that the maximum likelihood distribution is the empirical distribution of $\vr x$, i.e.
\begin{equation}\label{eq:A1298}
\hat p_{\ML}(\vr x) = \max_{\theta} \theta^n(\vr x) = \hat p(\vr x)
\end{equation}
This is shown below:
\begin{equation}\begin{split}\label{eq:A1625b}
\log p_{\theta}(\vr x)
&=
\sum_{i=1}^n \log \theta(x_i)
=
\sum_{\tilde{x} \in \mathcal{X}} n \hat P_{\vr x}(\tilde{x}) \log \theta(\tilde{x})
\\& =
n \sum_{\tilde{x} \in \mathcal{X}} \hat P_{\vr x}(\tilde{x}) \log \hat P_{\vr x}(\tilde{x})
- n \sum_{\tilde{x} \in \mathcal{X}} \hat P_{\vr x}(\tilde{x}) \log \frac{\hat P_{\vr x}(\tilde{x})}{\theta(\tilde{x})}
\\&=
\log p_{\theta = \hat P_{\vr x}}(\vr x) - n D(\hat P_{\vr x} \| \theta)
\leq
\log p_{\theta = \hat P_{\vr x}}(\vr x)
\end{split}\end{equation}
Therefore $\hat \theta_{\ML}(\vr x) = \argmax{\theta} p_{\theta}(\vr x) = \hat P_{\vr x}$. As a result, the empirical probability of $\vr x$, $\hat p(\vr x)$, equals the maximum likelihood probability of $\vr x$ under the i.i.d. model class. Therefore the maximum likelihood probability is a generalization of empirical probability, which is not limited to discrete sequences, and can be applied to continuous sequences, and include time structure.

Another consequence of the fact that $\hat p_{\ML}(\vr x) = \hat p(\vr x)$ for the class of memoryless models is that for any i.i.d. distribution $Q^n(\vr x)$, and every sequence: $\hat p(\vr x) = \max_{\theta} p_{\theta}(\vr x) \geq Q^n(\vr x)$ (since $Q \in \Theta$).

The same result holds for the conditional case, i.e. defining the class $\Theta$ as the class of conditionally memoryless models $p_{\theta}(\vr x | \vr z) = \prod_{i=1}^n \theta(x_i | z_i)$, we have that $\hat p_{\ML}(\vr x | \vr z) = \hat p(\vr x | \vr z)$. To see that, note that the distribution $p_{\theta}(\vr x | \vr z)$ can be written as a product of the distribution of sub-vectors of $\vr x$ which have constant $z_i$ (i.e. all indices for which $z_i = \tilde z$). Each of these sub-vectors has an independent set of parameters $\theta(\cdot | \tilde z)$, and maximizing the probability over $\theta$ implies maximizing the probability of each sub-vector separately. As we have seen above, this maximization yields the empirical probability of $\vr x$ over the sub-vector. Therefore the maximum is obtained for $\theta(\tilde x | \tilde z) = \hat P_{\vr x | \vr z}(\tilde x | \tilde z)$.

\subsubsection{Maximum likelihood, empirical and quazi-empirical entropies}\label{sec:def_emp_entropy}
\onlypaper{\todo{Suggestion: not include anything about empirical entropies in the paper, just in the PhD. In the places I used this for the examples, replace them with $\log p$-s}}

Given a probability distribution $p(x)$, the self information of the element $x$ is defined as
\begin{equation}\label{eq:A1654}
\log \frac{1}{p(x)}
\end{equation}
and the entropy is the expected value of the self information:
\begin{equation}\label{eq:A1658}
H(X) = \E \left[ \log \frac{1}{p(X)} \right] = - \sum_{x} p(x) \log p(x)
\end{equation}

We define the \textit{quazi-empirical} entropy of a sequence $\vr x$ with respect to a model $p(x)$ as above expression, where the expected value is replaced by the empirical expectation:
\begin{equation}\begin{split}\label{eq:A1667}
\hat H_p(\vr x)
& \defeq
\hat \E \left[ \log \frac{1}{p(x_i)} \right]
=
- \sum_{\tilde{x}} \hat P_{\vr x}(\tilde{x}) \log p(\tilde{x})
=
- \frac{1}{n} \sum_{i=1}^n \log p(x_i)
\\ & =
- \frac{1}{n} \log \prod_{i=1}^n p(x_i)
=
- \frac{1}{n} \log p^n(\vr x)
\end{split}\end{equation}
The last expression implies that the quazi-empirical entropy is the normalized self information of the sequence $\vr x$, with the i.i.d. probability $p$.

For discrete sequences, the empirical entropy of a sequence $\vr x, \vr y$ is defined as the entropy of the random variable with the distribution $X \sim \hat P_{\vr x}(x)$
\cite[Section II]{MethodOfTypes}. The empirical entropy of a sequence $\vr x$ is obtained from \eqref{eq:A1658} by replacing the distribution $p(x)$ with the empirical distribution $\hat P_{\vr x}(x)$:
\begin{equation}\label{eq:A1663}
\hat H(\vr x)  = - \sum_{\tilde{x}} \hat P_{\vr x}(\tilde{x}) \log \hat P_{\vr x}(\tilde{x})
\end{equation}
Equivalently using \eqref{eq:A1585} we may relate $\hat H(\vr x)$ to the empirical probability:
\begin{equation}\label{eq:A1363}
\hat H(\vr x) = - \frac{1}{n} \log \hat p(\vr x)
\end{equation}
This supplies an intuitively appealing way to understand $\hat H$ as the normalized self information of the sequence, under its estimated i.i.d. probability $\hat P_{\vr x}$. Equivalently we may write the empirical entropy as the quazi-empirical entropy using the empirical distribution $\hat H(\vr x) = H_{\hat P_x}(\vr x)$. From the relation between the empirical probability and the maximum likelihood probability $\hat p(\vr x) = \max_p p^n(\vr x)$, we have that
\begin{equation}\label{eq:A1369b}
\hat H(\vr x) = - \frac{1}{n} \log \hat p(\vr x) = - \max_p \frac{1}{n} \log p^n(\vr x) = \min_p \hat H_p(\vr x)
\end{equation}
I.e. in extracting the i.i.d. model extracted from $\vr x$ (rather than using an arbitrary $p$) we minimize its quazi-empirical entropy.

As an extension, given a class of models $P_{\theta}(\vr x), \theta \in \Theta$, we may define the maximum likelihood entropy of a sequence as the normalized self information of the sequence under the maximum-likelihood distribution.
\begin{equation}\label{eq:A1707}
\hat H_{\ML}(\vr x) = - \frac{1}{n} \log \hat p_{\ML}(\vr x)
\end{equation}

As before, all relations extend trivially to the conditional case (conditioned on the generic sequence $\vr z$), by simply considering each sub-vector of $\vr x$ related to a specific value in $\vr z$. I.e.
\begin{equation}\label{eq:A1667c}
\hat H_p(\vr x | \vr z)
=
- \frac{1}{n} \sum_{i=1}^n \log p(x_i | z_i)
=
- \frac{1}{n} \log p^n(\vr x | \vr z)
\end{equation}
\begin{equation}\label{eq:A1389}
\hat H(\vr x | \vr z)
=
- \sum_{\tilde{x},\tilde{z}} \hat P_{\vr x \vr z}(\tilde{x},\tilde{z}) \log \hat P_{\vr x|\vr z}(\tilde{x}|\tilde{z})
=
- \frac{1}{n} \log \hat p(\vr x | \vr z)
=
\min_p \hat H_p(\vr x | \vr z)
\end{equation}

While the standard chain rule holds for empirical entropies (being entropies of dummy random variables), it does not, in general, hold for entropies defined by maximum likelihood probabilities. Since, in general, we have:
\begin{equation}\begin{split}\label{eq:A1430}
\hat p_{\ML}(\vr x, \vr z)
&=
\max_{\theta \in \Theta} P_{\theta} (\vr x, \vr z)
=
\max_{\theta \in \Theta} \left[ P_{\theta} (\vr z) P_{\theta} (\vr x | \vr z) \right]
\\&\leq
\max_{\theta \in \Theta} P_{\theta} (\vr z) \cdot \max_{\theta \in \Theta} P_{\theta} (\vr x | \vr z)
=
\hat p_{\ML}(\vr z) \cdot \hat p_{\ML}(\vr x | \vr z)
\end{split}\end{equation}
Then
\begin{equation}\label{eq:A1442}
\hat H_{\ML}(\vr x, \vr z)
=
- \frac{1}{n} \log \hat p_{\ML}(\vr x, \vr z)
\geq
- \frac{1}{n} \hat p_{\ML}(\vr z) - \frac{1}{n} \hat p_{\ML}(\vr x | \vr z)
=
\hat H_{\ML}(\vr z) + \hat H_{\ML}(\vr x | \vr z)
\end{equation}
However, equality holds in \eqref{eq:A1430}, \eqref{eq:A1442} when the parameters $\theta$ can be separated into a set of parameters $\theta_z$ controlling $P_{\theta} (\vr z)$ and a set $\theta_{x|z}$ controlling $P_{\theta} (\vr x | \vr z)$. This occurs for example in the discrete memoryless case (where $\hat H_{\ML}$ is the empirical entropy), since the single letter distribution $\theta(x,z)$ can be separated into $\theta(z)$ and $\theta(x|z)$, and therefore we have equality in this case.

\subsubsection{Empirical mutual information}\label{sec:def_eMI}
Similarly to the empirical entropy, the empirical mutual information of two vectors $\hat I(\vr x; \vr y)$ is defined as the mutual information between two random variables $X,Y$ with the joint distribution $(X,Y) \sim \hat P_{\vr x, \vr y}(x,y)$, i.e. whose joint distribution equals the empirical distribution of $\vr x, \vr y$ \cite[Section II]{MethodOfTypes}. This way of defining the empirical mutual information and empirical entropy as mutual information/entropy of alternative random variables, can be extended to conditional forms. In general, all expressions such as $\hat H(\vr x)$, $\hat H(\vr x | \vr y)$, $\hat I (\vr x; \vr y)$, $\hat I (\vr x; \vr y | \vr z)$, $\hat I (\vr x; \vr y | \vr z = z_0)$ are interpreted as their respective probabilistic counterparts $H(X)$, $H(X|Y)$, $I (X;Y)$, $I (X;Y|Z)$, $I (X;Y|Z=z_0)$ where $(X,Y,Z)$ are random variables distributed according to the empirical distribution of the vectors $\hat{P}_{(\vr x, \vr y, \vr z)}$. Equivalently  $(X,Y,Z)$ can be defined as a random selection of an element of the vectors i.e. $(X,Y,Z)=(x_i, y_i, z_i), i \sim \unif \{1,\ldots,n\}$. It is clear from this equivalence that known properties of these values, such as relations between mutual information and entropy, non-negativity, chain rules, etc, are directly translated to relations on their empirical counterparts.

In particular, we can write the empirical mutual information as:
\begin{equation}\label{eq:A1724}
\hat I(\vr x; \vr y) = \hat H(\vr x) - \hat H(\vr x | \vr y) = \hat H(\vr x)  +  \hat H(\vr y) - \hat H(\vr x , \vr y)
\end{equation}
Writing the entropies as the self information under the empirical distribution we have:
\begin{equation}\begin{split}\label{eq:A1729}
\hat I(\vr x; \vr y)
&=
\hat H(\vr x)  +  \hat H(\vr y) - \hat H(\vr x , \vr y)
\\&=
- \frac{1}{n} \log \hat p(\vr x)  - \frac{1}{n} \log \hat p(\vr y) + \frac{1}{n} \log \hat p(\vr x, \vr y)
\\&=
\frac{1}{n} \log \frac{\hat p(\vr x, \vr y)}{\hat p(\vr x) \hat p(\vr y)}
=
\frac{1}{n} \log \frac{\hat p(\vr x | \vr y)}{\hat p(\vr x)}
\end{split}\end{equation}
Note the similarity to the form \eqref{eq:Aremp_conditional_form}.

\subsection{Maximum likelihood based rate functions}\label{sec:ML_rate_functions}
\subsubsection{Rationale}
In Section~\ref{sec:chernoff_cond_probability} we observed that attainable rate functions are asymptotically limited by the form
\begin{equation}\label{eq:A1616}
\Remp(\vr x, \vr y) = \frac{1}{n} \log \frac{P(\vr x | \vr y)}{Q(\vr x)}
\end{equation}
Let us assume that there is a probabilistic model relating $\vr y$ to $\vr x$, and $P(\vr x | \vr y)$ is the true conditional probability resulting from this model. In this case the value $i(\vr x, \vr y) = \log \frac{P(\vr x | \vr y)}{Q(\vr x)}$ is termed the \textit{information spectrum} or \textit{information density} \cite[(1.5)]{HanVerdu}, and we have that the mutual information between the input and output vectors is
\begin{equation}\label{eq:A1622}
I(\vr X; \vr Y) = \underset{\vr X, \vr Y}{\E} i(\vr X, \vr Y)
\end{equation}
As noted by Han and \verdu \cite{HanVerdu}, for general models (not necessarily i.i.d. or ergodic), the mutual information $I(\vr X; \vr Y)$ is not necessarily an achievable rate, and their characterization of channel capacity in this case relies on the ``$\liminf$ in probability'' of $\frac{1}{n} \cdot i(\vr X, \vr Y)$, which means the maximum value $\alpha$ such that the probability that $\frac{1}{n} \cdot  i(\vr X, \vr Y) \leq \alpha$ tends to $0$ as $n \to \infty$. In other words, achieving a rate $R$ requires that in high probability $i(\vr X, \vr Y) \geq n R$.

Setting the rate function as the normalized information density of a specific probabilistic model, i.e. $\Remp(\vr x, \vr y) = \frac{1}{n} i(\vr x, \vr y)$, is advantageous, especially when this rate function is attained adaptively, since this means that on average, the communication rate would be $\E \Remp = \frac{1}{n} \E i(\vr X, \vr Y) = \frac{1}{n}  I(\vr X, \vr Y)$. For general models, and with the suitable prior $Q(\vr x)$, this value may be is larger than the Han-\verdu capacity (which a lower bound in probability of $i$ rather than its mean). This occurs due to the use of feedback for rate adaptation. As an example, suppose a non-ergodic binary channel may be in one of two states, which are determined by a single random drawing with equal probabilities -- either the output equals the input for $j=1,\ldots,n$, or it is independent of the input. Clearly, no positive rate can be guaranteed on this channel, but if one allows the rate to vary, we may achieve a rate of $1$ [bit/use], $\half$ the time, and thus a rate of $\half$ [bit/use] on average. \onlyphd{To complete the example, the information density and the Han-Verd\'u capacity  of this channel are analyzed in Appendix~\ref{sec:binary_onoff_example}}

If we attain the normalized information density $\Remp(\vr x, \vr y) = \frac{1}{n} \cdot i(\vr x, \vr y)$ adaptively, then not only we attain the mutual information on average, but we also attain a rate of at least the liminf in probability of $i(\vr x, \vr y)$ with high probability (the later value becomes the channel capacity if the input distribution $Q(\vr x)$ is optimized). Another rationale for choosing $\frac{1}{n} \log \frac{P(\vr x | \vr y)}{Q(\vr x)}$ as the rate function, is that we know from Theorem~\ref{theorem:remp_conditional_form_asymptotic} that asymptotically the rate function is bounded by $\Rempname{(f)}=\frac{1}{n} \log \frac{f(\vr x | \vr y)}{Q(\vr x)}$ for some conditional distribution $f(\vr x | \vr y)$. If one assumes that the channel model truly induces the conditional probability $P(\vr x | \vr y)$, then the average rate would be $\E \Rempname{(f)} = \frac{1}{n} \sum_{\vr x, \vr y} P(\vr x | \vr y) P(\vr y) \log \frac{f(\vr x | \vr y)}{Q(\vr x)}$ which is maximized when $f(\vr x | \vr y) = P(\vr x | \vr y)$. I.e. when the channel induces $P$, any choice other than $P$ in the numerator will degrade the achieved rate, while choosing $P$ attains the mutual information. So far, we have justified why it makes sense to choose the rate function $\frac{1}{n} \log \frac{P(\vr x | \vr y)}{Q(\vr x)}$ if the channel is assumed to be known.

However, the main motivation for the individual channel framework is to avoid the probabilistic model. One possible approach is to guarantee a rate close to the information density, for a class of models. Let $P_{\theta}(\vr x, \vr y)$ $\theta \in \Theta$ be a class of models for joint probability of the vectors $\vr x, \vr y$. We denote by $P_{\theta}(\vr x)$, $P_{\theta}(\vr x | \vr y)$ the marginal and the conditional distribution resulting from $P_{\theta}(\vr x, \vr y)$. Then a possible rate function is the maximum normalized information density over all models in the family.
\begin{equation}\label{eq:A1674}
\Rempname{\ML} = \max_{\theta \in \Theta} \frac{1}{n} \log \frac{P_{\theta}(\vr x | \vr y)}{Q(\vr x)}
=
\frac{1}{n} \log \frac{\max_{\theta \in \Theta} P_{\theta}(\vr x | \vr y)}{Q(\vr x)}
=
\frac{1}{n} \log \frac{\hat p_{\ML}(\vr x | \vr y)}{Q(\vr x)}
\end{equation}
Clearly, attaining this rate function guarantees attaining the above properties (the mutual information rate on average and the liminf in high probability) for all channels in the family. The family of distributions may be constrained to have $\forall \theta: P_{\theta}(\vr x) = Q(\vr x)$ but this is not necessary, and it is sometimes more convenient to avoid this constraint. However we assume that there exist $\theta$ such that $P_{\theta}(\vr x) = Q(\vr x)$ and therefore \eqref{eq:A1674} includes maximization over information densities (and possibly other values which are not legitimate information densities, but are still achievable rate functions). In this case the $\theta$ achieving the maximization in the numerator would not necessary yield the ``correct'' marginal $P_{\theta}(\vr x) = Q(\vr x)$.
\todo{change to $\sup$ instead of $\max$ in all ML-s}

To summarize, we have seen that attaining the ML-based rate function \eqref{eq:A1674} is advantageous. In the sequel we analyze the intrinsic redundancy associated with this rate function, and show how it can be achieved adaptively in many cases of interest. However we must note that there is a gap between the justification for this rate function, and what attaining it actually yields. In justifying this rate function we have analyzed the behavior in the case that the relation between $\vr x$ and $\vr y$ is governed by a probability law from a given class, however the system attaining $\Remp$ of \eqref{eq:A1674} will not only guarantee this behavior but guarantees a certain rate and error probability for each pair of sequences (which is more than required to obtain the target of achieving the mutual information rate for all channels in the class, using feedback). Therefore we should not treat this system as the best system attaining the mutual information rate, but rather as a system attaining the $\Remp$ of \eqref{eq:A1674} per each pair of sequences, where this $\Remp$ on one hand guarantees a certain behavior when $\vr x, \vr y$ are governed by a probability law from the class, but also guarantees some computable rate when a different probability law is applied. This may be compared against a system which attempts to learn $\theta$ by measuring the channel, and may also attain the mutual information rate, but does not give any guarantee on what occurs when another probability law is applied.

\subsubsection{Intrinsic redundancy}\label{sec:redundancy_of_Remp_ML}
For finite classes, it is easy to bound the intrinsic redundancy of \eqref{eq:A1674}. Since the intrinsic redundancy of $\frac{1}{n} \log \frac{P_{\theta}(\vr x | \vr y)}{Q(\vr x)}$ is non-positive (see Section~\ref{sec:chernoff_cond_probability}), according to Property~2 of the intrinsic redundancy (Section~\ref{sec:irp524}), the intrinsic redundancy of $\Rempname{\ML}$ is at most $\frac{\log |\Theta|}{n}$. Therefore we may allow the size of the class to increase with $n$, and as long as this increase is sub-exponential, the intrinsic redundancy $\mu_Q(\Rempname{\ML})$ would tend to $0$ with $n$, and therefore $\Rempname{\ML}$ of \eqref{eq:A1674} would be asymptotically achievable. However, as we shall see, \eqref{eq:A1674} may be asymptotically achievable even for infinite parametric classes as long as suitable smoothness conditions hold.

\todo{for Phd, consider taking this part to the background and just quickly referring to it. Or maybe the inverse: write the background in easy language and refer here for further details}
The size of the model class yields a coarse estimate for the intrinsic redundancy of \eqref{eq:A1674}. A finer analysis is by relating the intrinsic redundancy to the regret of a universal distribution representing the model class $\{P_{\theta}(\vr x | \vr y)\}$. In universal source coding of a family of sources with distributions $P_{\theta}(\vr x)$, one seeks a single distribution $P(\vr x)$, which approximates all distributions in the class, up to a certain loss $\mathcal{R}(\theta, \vr x, P) = \log \frac{P_{\theta}(\vr x)}{P(\vr x)}$, termed the ``regret'', which represents the difference in encoding lengths when $P$ is used, compared to when $P_{\theta}(\vr x)$ is used \cite{Barron_MDL}. The minimax regret $\mathcal{R}_{\mathrm{minimax}} \defeq \min_{P} \max_{\theta, \vr x} \mathcal{R}(\theta, \vr x, P)$ is the minimum value of the worst case regret over all models $\theta$ and sequences $\vr x$.

It is easy to show \cite{Barron_MDL} that the distribution $P$ which achieves the minimax regret is
\begin{equation}\label{eq:A1717}
P_{\NML}(\vr x) =
\frac{\max_{\theta} P_{\theta}(\vr x)}{\sum_{\vr {\tilde x}} \max_{\theta} P_{\theta}(\vr {\tilde x})}
=
\frac{\hat p_{\ML}(\vr x)}{\sum_{\vr {\tilde x}} \hat p_{\ML}(\vr {\tilde x})}
\end{equation}
This distribution is simply a normalization of the super-probability $\hat p_{\ML}(\vr x)$ (which we would like to approximate by a probability), and is termed ``Normalized Maximum Likelihood'' (NML). The regret is determined by the size of the normalization factor
\begin{equation}\label{eq:A1759}
\log \frac{\hat p_{\ML}(\vr x)}{P_{\NML}(\vr x)} = \log c_{\NML}
\end{equation}
where
\begin{equation}\label{eq:A1759b}
c_{\NML} = \sum_{\vr {\tilde x}} \hat p_{\ML}(\vr {\tilde x})
\end{equation}

The fact $P_{\NML}$ is minimax optimal is evident by observing, that $P_{\NML}$ is required to be the closet probability that approximates the superprobability $\hat p_{\ML}$ (in a logarithmic minimax regret sense), and a normalization by a constant factor, which yields a constant regret is best, since decreasing the factor at any point would necessarily require increasing it at other points, thus increasing the maximum regret. The resulting regret was analyzed by Barron, Rissanen, Yu and others and is known up to negligible terms in many cases of interest. For continuous parametric families, where $\theta$ is a vector of size $k$ it was shown by Rissanen \cite[Theorem~1]{Rissanen_FisherInfo} that under certain conditions, there exists $\tilde P$ having the following regret, determined up to a vanishing factor:
\begin{equation}\label{eq:A1763}
\forall \vr x, \theta: \mathcal{R}(\theta, \vr x, \tilde P) = \log \frac{P_{\theta}(\vr x)}{\tilde P(\vr x)} = \frac{k}{2} \log \frac{n}{2 \pi} + \log \int_{\Theta} \sqrt{|I(\theta)|} d \theta + o_n(1)
\end{equation}
where $I(\theta) = \lim_{n \to \infty} \frac{1}{n} \E \left[ \frac{\partial^2}{\partial \theta^2} \ln P_{\theta}(\vr x) \right]$ is the limit of the normalized Fisher information matrix. Since this value does not grow with $n$ the main factor in the regret is $\frac{k}{2} \log n$, which is the penalty associated with the ``richness'' of the class. Rissanen's conditions are sometimes limiting. As an example, they do not hold for the class of memoryless sources where $\theta$ is the vector of letter probabilities, at the boundary of $\Theta$, i.e. when one of the element of $\theta$ is $0$ or $1$, since the Fisher information is infinite at these points. One solution is to apply the result only to the interior of $\Theta$ and account for the boundaries separately. However specifically for the class of memoryless sources, there are explicit expressions for the regret, with the same behavior as determined by \eqref{eq:A1763}. See Section~\ref{sec:RempML_adaptive_redundancy_empirical} in the following for a more detailed discussion of the memoryless and conditional cases. A conclusion from \eqref{eq:A1763} is that the minimax redundancy of the NML, which is optimal, satisfies
\begin{equation}\label{eq:A1781}
\mathcal{R}(\theta, \vr x, P_{\NML}) = \log c_{\NML} \leq \frac{k}{2} \log \frac{n}{2 \pi} + \log \int_{\Theta} \sqrt{|I(\theta)|} d \theta + o_n(1)
\end{equation}

Returning to our problem we begin with a general analysis of the intrinsic redundancy of $\Rempname{\ML}$ assuming that the conditions for \eqref{eq:A1763} hold. For each $\vr y$ separately, we form a distribution $P^*(\vr x | \vr y)$ on $\vr x$ which has a bounded regret with respect to the maximum likelihood probability $\hat p_{\ML}(\vr x | \vr y)$ (one option is the NML). By \eqref{eq:A1763} we have that
\begin{equation}\label{eq:A1763b}
\forall \vr x, \vr y: \log \frac{\sup_{\theta} P_{\theta}(\vr x | \vr y)}{P^*(\vr x | \vr y)}
\leq
\frac{k}{2} \log \frac{n}{2 \pi} + \log \int_{\Theta} \sqrt{|I_{\vr y}(\theta)|} d \theta + o_n(1)
=
\frac{k}{2} \log n + O_n(1)
\end{equation}
where here the asymptotical Fisher information matrix $I$ may, in general depend on $\vr y$. Now writing
\begin{equation}\label{eq:A1788}
\Rempname{\ML} =
\frac{1}{n} \log \frac{\hat p_{\ML}(\vr x | \vr y)}{Q(\vr x)}
=
\frac{1}{n} \log \frac{P^*(\vr x | \vr y)}{Q(\vr x)} + \frac{1}{n} \log \frac{\hat p_{\ML}(\vr x | \vr y)}{P^*(\vr x | \vr y)}
\leq
\frac{1}{n} \log \frac{P^*(\vr x | \vr y)}{Q(\vr x)} + \frac{k}{2} \cdot \frac{\log n}{n}  + O_n(1/n)
\end{equation}
Since $P^*$ is a probability distribution, the first term has a non-positive intrinsic redundancy (Lemma~\ref{lemma:remp_conditional_form}), and therefore by the additivity of intrinsic redundancy, $\Rempname{\ML}$ has intrinsic redundancy of $\mu_Q(\Rempname{\ML}) \leq  \frac{k}{2} \cdot \frac{\log n}{n}  + O_n(1/n)$.

Note that although the intrinsic redundancy obtained here has a similar form to the minimax regret in universal source coding, the number of parameters $k$ will be in most cases larger due to the conditioning on $\vr y$. As an example, to model all i.i.d. sources over alphabet $\mathcal{X}$ one needs $|\mathcal{X}|-1$ parameters to define the letter distribution ($|\mathcal{X}|$ letter distributions, and a constraint on the sum). To model all memoryless distributions $P(\vr x | \vr y) = \prod_{i=1}^n p(x_i | y_i)$, one needs $|\mathcal{X}|-1$ parameters for each value of $y_i$ therefore $k = (|\mathcal{X}|-1) \cdot |\mathcal{Y}|$ parameters.

\subsubsection{Universality over a set of probabilistic non-ergodic channels}
\todo{Universality over a set of probabilistic non-ergodic channels $P_i(\vr y|\vr x)$ for given $Q(\vr x)$: by Han-Verdu we know that if someone operates with rate $R_i$ over the channel $P_i$ then $\frac{1}{n} \log P_1(\vr x | \vr y)/Q(\vr x) > R$ with high probability, then if we achieve $\frac{1}{n} \log \sum_i P_i(\vr x | \vr y)/Q(\vr x)$ we exceed $R_i$. Another consequence is the optimality of the GLRT in terms of rate, for general channels. This consequence is interesting, since it is known that GLRT is not optimal in terms of error exponent, and most results on GLRT pertain to channels with specific structures}

\todo{From the other direction: it is very easy to prove Feinstein's Lemma \cite[Theorem 1]{HanVerdu} based on the tools here. We know that $\Remp = \frac{1}{n} i(\vr x; \vr y)$ has zero intrinsic redundancy and therefore is achievable with error probability $\epsilon_0$ up to $\delta = \frac{1}{n} \log \epsilon_0^{-1}$. If we operate this communiction system over the probabilistic channel, then we have two sources of error: either $\Remp - \delta < R$, or in case $\Remp - \delta \geq R$, the error probability $\epsilon_0$. Thus our overall error probability satisfies: $\epsilon \leq \Pr(\Remp - \delta \leq R) + \epsilon_0 = \Pr \left\{ \frac{1}{n} i(\vr X; \vr Y) \leq R + \delta \right\} + \exp(-n \delta)$, which is exactly the Lemma}.

\subsection{Variations on the maximum likelihood construction}
\subsubsection{The doubly maximum likelihood construction}\label{sec:MLML_construction}
In the maximum-likelihood construction proposed above \eqref{eq:A1674} the rate function depends on the prior $Q$. It is sometimes convenient to avoid the specific dependence on $Q$ by replacing $Q(\vr x)$ it with the maximum-likelihood probability $\hat p_{\ML}(\vr x)$ of the sequence $\vr x$.

Since we assumed there exists $\theta$ such that $P_{\theta}(\vr x) = Q(\vr x)$, we have $\hat p_{\ML}(\vr x) = \max_{\theta} P_{\theta}(\vr x) \geq Q(\vr x)$, therefore we have:
\begin{equation}\label{eq:A1696}
\Rempname{\ML*} = \frac{1}{n} \log \frac{\hat p_{\ML}(\vr x | \vr y)}{\hat p_{\ML}(\vr x)} \leq \frac{1}{n} \log \frac{\hat p_{\ML}(\vr x | \vr y)}{Q(\vr x)} = \Rempname{\ML}
\end{equation}

Therefore if $\Rempname{\ML}$ is achievable (in any of the senses), $\Rempname{\ML*}$ is achievable as well. $\Rempname{\ML*}$ is sometimes more convenient to use since it does not include the prior $Q$ in an explicit form, and may be suitable for a large class of priors. The empirical mutual information as well as other rate functions presented in \cite{YL_individual_full, YL_MIMO_ITW2010} are of this form. In the examples in \selector{Section}{Chapter}~\ref{chap:examples} we usually use the form $\Rempname{\ML}$ for analysis and present the two forms $\Rempname{\ML}, \Rempname{\ML*}$ for each case. It can be observed from Table~\unfinished{[? internal ref]} that $\Rempname{\ML*}$ has a more intuitively appealing form. The rate functions of this form are inherently sub-optimal, since they are in general uniformly inferior with respect to the respective $\Rempname{\ML}$, but this sub-optimality is insignificant since it is expressed only when the maximum likelihood probability significantly differs from the actual one. In most cases, if $\vr x$ is a typical sequence, then the maximum likelihood estimate will be close to the true value, and the empirical probability will be close to the true one, and therefore the difference is insignificant for typical sequences. As we argue in Section~\ref{sec:eMI_optimality}, the main interest should be on the values of the rate function for typical $\vr x$, therefore in many cases the difference between $\Rempname{\ML}$ and $\Rempname{\ML*}$ is immaterial.

\subsubsection{The use of universal distributions}
As we have seen, the maximum likelihood probability, after being normalized, yields the NML probability measure which is close to any distribution in the family. In general, one may define other such ``universal distributions'' based on similar or different criteria, and define the rate function as:
\begin{equation}\label{eq:A1626}
\Remp = \frac{1}{n} \log \frac{P_u(\vr x | \vr y)}{Q(\vr x)}
\end{equation}
where $P_u$ is universal conditional probability. Similarly as done in the previous section, $Q(\vr x)$ may be replaced by a ``close'' universal distribution $P_u(\vr x)$, however since in this case we do not have the inequality $P_u(\vr x) \geq Q(\vr x)$, a bound on $\frac{P_u}{Q}$ may be required to show the modified rate function is achievable.

\subsection{Entropy based notation for maximum likelihood rate functions}\label{sec:Remp_ML_as_entropy_diff}
It is intuitively appealing to write $\Rempname{\ML}$ and $\Rempname{\ML*}$ as a difference of entropies. Using the definitions from Section~\ref{sec:empirical_distributions_and_information_measures}:
\begin{eqnarray*}
\hat H_{\ML}(\vr x | \vr y) &=& -\frac{1}{n} \log \hat p_{\ML}(\vr x | \vr y) \\
\hat H_{\ML}(\vr x) &=& -\frac{1}{n} \log \hat p_{\ML}(\vr x) \\
\hat H_{Q}(\vr x) &=& -\frac{1}{n} \log Q(\vr x) \\
\end{eqnarray*}
we have in analogy to $I(X;Y) = H(X) - H(X|Y)$:
\begin{equation}\label{eq:A1666}
\Rempname{\ML} \stackrel{\eqref{eq:A1674}}{=} \hat H_{Q}(\vr x) - \hat H_{\ML}(\vr x | \vr y)
\end{equation}
\begin{equation}\label{eq:A1670}
\Rempname{\ML*} \stackrel{\eqref{eq:A1696}}{=} \hat H_{\ML}(\vr x) - \hat H_{\ML}(\vr x | \vr y)
\end{equation}
In many cases the empirical entropies above have an intuitive interpretation as a measure for the complexity of the vectors.
When the equality in \eqref{eq:A1430} holds, we can write the rate function $\Rempname{\ML*}$ in a symmetric form:
\begin{equation}\label{eq:A1675}
\Rempname{\ML*} = \frac{1}{n} \log \frac{\hat p_{\ML}(\vr x, \vr y)}{\hat p_{\ML}(\vr x) \cdot \hat p_{\ML}(\vr y)} = \hat H_{\ML}(\vr x) + \hat H_{\ML}(\vr y) - \hat H_{\ML}(\vr x, \vr y)
\end{equation}

\subsection{Rate functions defined by given empirical parameters}\label{sec:rate_func_by_parameters}
Now we examine a different construction for rate functions, relying on a parametric representation of the input and output sequences. For example, in \cite[Lemma 3]{YL_individual_full} we have justified the rate function $\half \log \frac{1}{1 - \hat \rho^2}$ for the continuous real-valued channel, as the best rate function defined by second order statistics, in a compound channel setting. More generally, suppose that we decide on a certain empirical parametrization of the sequences $\vr x, \vr y$ (e.g. zero order empirical statistics, empirical second order moments, etc), can we find the ``best'' rate function that can be defined using this parametrization?

Let $\hat \theta(\vr x, \vr y) \in \Theta$ be an predefined estimator of a parameter vector $\theta \in \Theta$, and let $Q$ be a predetermined prior. We limit our scope to rate functions defined as:
\begin{equation}\label{eq:A1582}
\Remp = R(\hat \theta(\vr x, \vr y))
\end{equation}
where $R(\theta)$ is a function of our choice. Given $\hat \theta, Q$ we would like to find the maximum $R(\theta)$ for which $\Remp$ would be achievable.

\subsubsection{Optimal rate functions over types}\label{sec:1588}
An alternative formulation of the problem is to say that the set of sequences $(\vr x, \vr y) \in \mathcal{X}^n \times \mathcal{Y}^n$ is separated into disjoint sets, termed \textit{types}, $\mathcal{T}_{xy} \subset (\mathcal{X}^n, \mathcal{Y}^n)$, and the rate function is required to be a function of the type $\Remp = R(\mathcal{T}_{xy})$. This formulation is equivalent to the former, since we may define the types as the sets of sequences that yield the same value of the parameter, i.e. $\mathcal{T}_{xy}(\theta) \defeq \{ (\vr x,\vr y): \hat \theta(\vr x, \vr y) = \theta \}$. However, we now further constrain ourselves to the case where the number of types is finite (equivalently, the set of possible parameter values is finite). This assumption is more suitable to the discrete case, since when the sequences $\vr x, \vr y$ are discrete, the number of possible parameter values, for a certain block length $n$ is finite.

As an example, suppose that the parametrization is by the zero order empirical statistics. In this case $\hat \theta$ is a vector comprised of the $|\mathcal{X}| \cdot |\mathcal{Y}|$ elements of the empirical probability $\hat P_{\vr x, \vr y}(\tilde x, \tilde y)$. Since each element of the empirical probability is in the set $\left\{ \frac{i}{n} \right\}_{i=0}^n$, there are at most $N_T \leq (n+1)^{|\mathcal{X}| \cdot |\mathcal{Y}|}$ values. Alternatively, the types defined by the sets of sequences with the same value of the parameter, i.e. the empirical distribution, are in this case the regular types defined by Csisz\'ar \cite{MethodOfTypes}\cite[Chapter 11]{CoverThomas_InfoTheoryBook}, and the number of types is bounded by $N_T$ above. The concept of types was generalized in various ways \cite[Sec. VII]{MethodOfTypes}\cite{Seroussi_UnivType}. However currently we do not assume anything about the structure of the type classes, and they can be arbitrary sets of pairs $(\vr x, \vr y)$. Our only assumption is that the number of type classes is finite and upper bounded by a given value, denoted $N_T$.

We begin with an upper bound on the rate function. We denote by $\mathcal{T}_{xy}(\vr x, \vr y)$ the type class associated with a specific pair of sequences. Consider a specific type $\mathcal{T}_{xy}^0$. If $(\vr x, \vr y) \in \mathcal{T}_{xy}^0$ (i.e. $\mathcal{T}_{xy}(\vr x, \vr y) = \mathcal{T}_{xy}^0$), then $\Remp(\vr x, \vr y) \defeq R(\mathcal{T}_{xy}(\vr x, \vr y)) = R(\mathcal{T}_{xy}^0)$, therefore for a specific $\vr y$,
\begin{equation}\label{eq:A1605}
\underset{Q}{\Pr} \left\{ \Remp(\vr X, \vr y) \geq R(\mathcal{T}_{xy}^0) \right\}
\geq
\underset{Q}{\Pr} \left\{ (\vr X, \vr y) \in \mathcal{T}_{xy}^0 \right\}
=
Q \left\{ \vr x: (\vr x, \vr y) \in \mathcal{T}_{xy}^0 \right\}
=
Q \left\{ \mathcal{T}_{x|y}^0(\vr y) \right\}
\end{equation}
where we have defined the conditional type $\mathcal{T}_{x|y}^0(\vr y) \defeq \{ \vr x: (\vr x, \vr y) \in \mathcal{T}_{xy}^0 \}$ (in an analogy to the regular definition of conditional types \cite[Lemma II.3]{MethodOfTypes}). On the other hand, by Theorem~\ref{theorem:remp_ublb}, if $\Remp$ is achievable then
\begin{equation}\label{eq:A1617}
\underset{Q}{\Pr} \left\{ \Remp(\vr X, \vr y) \geq R(\mathcal{T}_{xy}^0) \right\} \leq (1-\epsilon)^{-1} \exp(-n R(\mathcal{T}_{xy}^0))
\end{equation}
Combining the two inequalities \eqref{eq:A1605},\eqref{eq:A1617} we have that
\begin{equation}\label{eq:A1621}
\forall \vr y: Q \left\{ \mathcal{T}_{x|y}^0(\vr y) \right\} \leq (1-\epsilon)^{-1} \exp(-n R(\mathcal{T}_{xy}^0))
\end{equation}
i.e.
\begin{equation}\label{eq:A1624}
R(\mathcal{T}_{xy}^0) \leq -\frac{1}{n} \sup_{\vr y} \log Q \left\{ \mathcal{T}_{x|y}^0(\vr y) \right\} + \frac{1}{n} \log \frac{1}{1-\epsilon}
\end{equation}
For large $n$, the second term in the RHS of \eqref{eq:A1624} tends to $0$ and the first term is therefore the dominant one. Note that for vectors $\vr y$ that do not appear in $\mathcal{T}_{xy}^0$, $\mathcal{T}_{x|y}^0(\vr y)$ is an empty set, and therefore these do not affect the supremum and can be removed.

We now show that the first term in the RHS of \eqref{eq:A1624} indeed leads to an achievable rate function if the number of types is not too large. Let the rate function be defined as:
\begin{equation}\label{eq:A1624a}
R(\mathcal{T}_{xy}) = -\frac{1}{n} \sup_{\vr y} \log Q \left\{ \mathcal{T}_{x|y}(\vr y) \right\} - \delta
\end{equation}

From \eqref{eq:A1624a} we have for any $\vr y$:
\begin{equation}\label{eq:A1624b}
Q \left\{ \mathcal{T}_{x|y}(\vr y) \right\} \leq \exp[-n (R(\mathcal{T}_{xy}) + \delta)]
\end{equation}

For any $\vr y$ and $R \in \mathbb{R}$:
\begin{equation}\begin{split}\label{eq:A1636}
\underset{Q}{\Pr} \left\{ \Remp(\vr X, \vr y) \geq R \right\}
&=
\underset{Q}{\Pr} \left\{ R(\mathcal{T}_{xy}(\vr X, \vr y)) \geq R \right\}
\\&=
\sum_{\mathcal{T}_{xy}^0} \underset{Q}{\Pr} \left\{ \left( R(\mathcal{T}_{xy}(\vr X, \vr y)) \geq R  \right) \cap \left(\mathcal{T}_{xy}(\vr X, \vr y) = \mathcal{T}_{xy}^0 \right) \right\}
\\&=
\sum_{\mathcal{T}_{xy}^0: R(\mathcal{T}_{xy}^0) \geq R} \underset{Q}{\Pr} \left\{ (\vr X, \vr y) \in \mathcal{T}_{xy}^0 \right\}
=
\sum_{\mathcal{T}_{xy}^0: R(\mathcal{T}_{xy}^0) \geq R} Q \left\{ \mathcal{T}_{x|y}^0(\vr y)  \right\}
\\& \stackrel{\eqref{eq:A1624b}}{\leq}
\sum_{\mathcal{T}_{xy}^0: R(\mathcal{T}_{xy}^0) \geq R} \exp[-n (R(\mathcal{T}_{xy}) + \delta)]
\leq
\sum_{\mathcal{T}_{xy}^0: R(\mathcal{T}_{xy}^0) \geq R} \exp[-n (R + \delta)]
\\ & \leq
N_T \cdot \exp(-n \delta) \cdot \exp(-nR)
\end{split}
\end{equation}
Therefore in order to satisfy the sufficient condition of Theorem~\ref{theorem:remp_ublb}, it is sufficient to require $N_T \cdot \exp(-n \delta) \leq \epsilon$, i.e.  $\delta = \frac{1}{n} \log \frac{N_T}{\epsilon}$.

We summarize these results in the following theorem.
\begin{theorem}\label{theorem:optimal_type_based}
Let $\mathbb{T}_{\mathcal{XY}}$ denote a set of no more than $|\mathbb{T}_{\mathcal{XY}}| \leq N_T$ disjoint sets (types) covering the set of sequences $\mathcal{X}^n \times \mathcal{Y}^n$. For two sequences $(\vr x, \vr y)$, let $\mathcal{T}_{xy} \in \mathbb{T}_{\mathcal{XY}}$ denote the type containing these sequences, let $\mathcal{T}_{x|y}(\vr {\tilde y}) \defeq \{ \vr {\tilde x}: (\vr {\tilde x}, \vr {\tilde y}) \in \mathcal{T}_{xy} \}$ denote the respective conditional type. For a prior $Q$ on $\mathcal{X}^n$ define the following rate function:
\begin{equation}\label{eq:A1669}
\Remp(\vr x, \vr y) = -\frac{1}{n} \sup_{\vr {\tilde y}} \log Q \left\{ \mathcal{T}_{x|y}(\vr {\tilde y}) \right\}
\end{equation}
Then for a prior $Q$ and an error probability $\epsilon$:
\begin{enumerate}
\item Any achievable rate function which can be written as a function of the joint type $\mathcal{T}_{xy}$ of the sequences $\vr x, \vr y$ (i.e. the set $\mathcal{T}_{xy}$ such that $(\vr x, \vr y) \in \mathcal{T}_{xy}$), can exceed $\Remp$ by more than $\frac{1}{n} \log \frac{1}{1-\epsilon}$
\item $\Remp$ is achievable up to $\delta = \frac{1}{n} \log \frac{N_T}{\epsilon}$
\item Furthermore, if the number of types increases subexponentially with $n$, i.e. $\frac{1}{n} \log N_T \ntoinfty 0$, then $\Remp$ is asymptotically achievable.
\end{enumerate}
\end{theorem}
\textit{Proof:} the proof is given by the derivation above (the first claim in \eqref{eq:A1624} and the second in \eqref{eq:A1636}). The last claim results trivially from the second, since under the assumption, $\delta \ntoinfty 0$. \endofproof.

\begin{figure}
\centering
\ifpdf
  \setlength{\unitlength}{1bp}%
  \begin{picture}(384.42, 258.08)(0,0)
  \put(0,0){\includegraphics{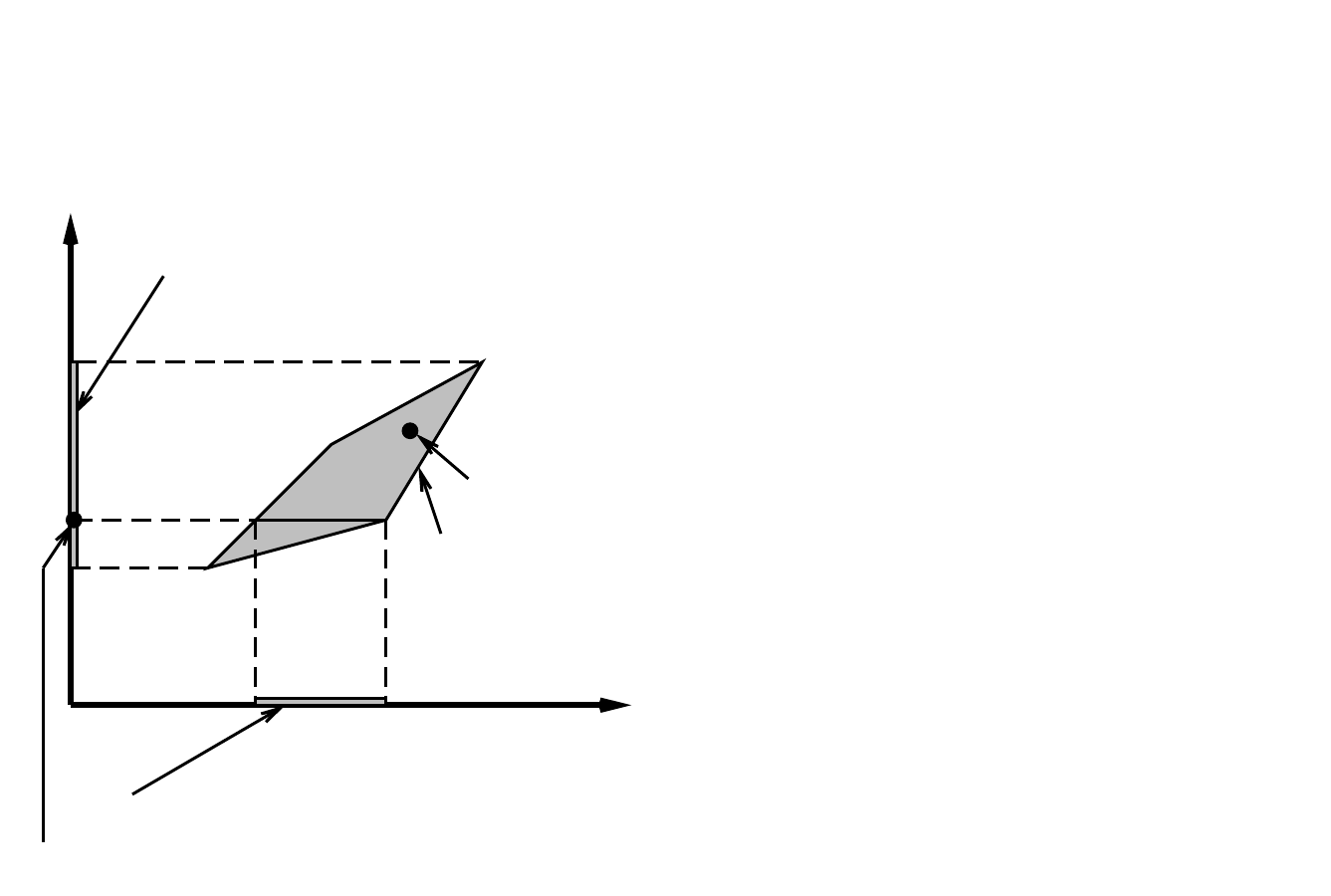}}
  \put(115.66,45.08){\fontsize{9.96}{11.95}\selectfont Input sequences $\mathcal{X}^n$}
  \put(13.45,109.28){\rotatebox{90.00}{\fontsize{9.96}{11.95}\selectfont \smash{\makebox[0pt][l]{Output sequences $\mathcal{Y}^n$}}}}
  \put(135.50,112.55){\fontsize{7.97}{9.56}\selectfont (a) The  specific pair $(\vr x, \vr y)$}
  \put(123.60,92.71){\fontsize{7.97}{9.56}\selectfont (b) The type class of $(\vr x, \vr y)$}
  \put(8.51,7.38){\fontsize{7.97}{9.56}\selectfont (e) The specific $\vr {\tilde y}$ yielding the largest probability conditional type $\mathcal{T}_{x|y}(\vr {\tilde y})$}
  \put(32.32,21.27){\fontsize{7.97}{9.56}\selectfont (d) The largest probability conditional type $\mathcal{T}_{x|y}(\vr {\tilde y})$}
  \put(48.20,178.03){\fontsize{7.97}{9.56}\selectfont (c) The set of all $\vr {\tilde y}$ in the class (projection)}
  \end{picture}%
\else
  \setlength{\unitlength}{1bp}%
  \begin{picture}(384.42, 258.08)(0,0)
  \put(0,0){\includegraphics{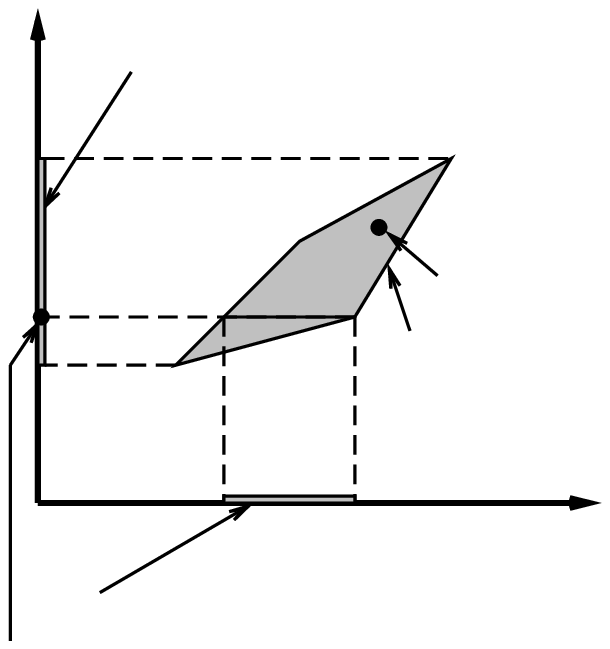}}
  \put(115.66,45.08){\fontsize{9.96}{11.95}\selectfont Input sequences $\mathcal{X}^n$}
  \put(13.45,109.28){\rotatebox{90.00}{\fontsize{9.96}{11.95}\selectfont \smash{\makebox[0pt][l]{Output sequences $\mathcal{Y}^n$}}}}
  \put(135.50,112.55){\fontsize{7.97}{9.56}\selectfont (a) The  specific pair $(\vr x, \vr y)$}
  \put(123.60,92.71){\fontsize{7.97}{9.56}\selectfont (b) The type class of $(\vr x, \vr y)$}
  \put(8.51,7.38){\fontsize{7.97}{9.56}\selectfont (e) The specific $\vr {\tilde y}$ yielding the largest probability conditional type $\mathcal{T}_{x|y}(\vr {\tilde y})$}
  \put(32.32,21.27){\fontsize{7.97}{9.56}\selectfont (d) The largest probability conditional type $\mathcal{T}_{x|y}(\vr {\tilde y})$}
  \put(48.20,178.03){\fontsize{7.97}{9.56}\selectfont (c) The set of all $\vr {\tilde y}$ in the class (projection)}
  \end{picture}%
\fi
\caption{\label{fig:Remp_type_based_illustration}%
 An illustration of the calculation of type-based rate function by Theorem~\ref{theorem:optimal_type_based}}
\end{figure}

The calculation of the $\Remp$ proposed above \eqref{eq:A1669} is illustrated in Figure~\ref{fig:Remp_type_based_illustration}. The axes lines denote the set of all sequences $\vr x \in \mathcal{X}^n$ (horizontal) and $\vr y \in \mathcal{Y}^n$ (vertical). For the specific pair $(\vr x, \vr y)$ for which the rate function is computed (a), the polygon (b) depicts the type class this pair belongs to (any arbitrary sub-group of pairs). All $\vr {\tilde y}$ in the type class (c) are scanned, to find the one (e) yielding the maximum-probability conditional type (d), illustrated by the maximum horizontal width in the figure.

The main gap between the upper bound and the lower bound of Theorem~\ref{theorem:optimal_type_based} is due to $N_T$ - the number of types. This gap is essentially unavoidable (when types are considered as general sets), since it is possible to construct a rate function that will nearly meet the necessary condition for one type (up to the gaps resulting from Theorem~\ref{theorem:remp_ublb}), by placing all the probability on that type (i.e. having $\Remp=0$ for all other types). In this case the bound of \eqref{eq:A1605} becomes tight for this type.

\subsubsection{On the optimality of the empirical mutual information}\label{sec:eMI_optimality}
We now particularize the result of Theorem~\ref{theorem:optimal_type_based} for the memoryless model, in which $\hat \theta$ is the joint empirical distribution, and the types $\mathcal{T}_{xy}$ are the standard types \cite{MethodOfTypes}. We use the coarse upper bound $N_T = (n+1)^{|\mathcal{X}| \cdot |\mathcal{Y}|}$ \cite[Theorem 11.1.1]{CoverThomas_InfoTheoryBook} (see also Section~\ref{sec:1588}). We assume that $Q(\vr x)$ is also memoryless, i.e. $Q(\vr x) = \prod_{i=1}^n Q(x_i)$.

Consider two sequences $(\vr x, \vr y)$ having an empirical distribution $\hat P_{\vr x, \vr y}$ and belonging to the type class $\mathcal{T}_{xy}$. For notational purposes, we denote by $\tilde{X},  \tilde{Y}$ dummy random variables, distributed according to $\hat P_{\vr x, \vr y}(\tilde x, \tilde y)$.

The size of the conditional type is $|\mathcal{T}_{x|y}(\vr y)| = c_n \exp(n H(\tilde{X} | \tilde{Y}))$ for any $\vr y \in \mathcal{T}_{y}$, where $c_n$ is a subexponential factor $\frac{\log c_n}{n} \ntoinfty 0$ \cite[Lemma II.3]{MethodOfTypes} (for other $\vr y$-s it is zero). All sequences in the conditional type are of the same type $\mathcal{T}_x$, and therefore have the same probability under $Q$, which is easily shown to equal $Q^n(\vr x) = \exp[-n (H(\tilde X) + D(\hat P_{\vr x} \| Q))]$ \cite[(II.1)]{MethodOfTypes}. Therefore we have for all $\vr {\tilde y} \in \mathcal{T}_y$:
\begin{equation}\begin{split}\label{eq:A1703}
Q \left\{ \mathcal{T}_{x|y}(\vr {\tilde y}) \right\}
&=
|\mathcal{T}_{x|y}(\vr {\tilde y})| \cdot Q^n(\vr x)
\\&=
c_n \exp(n H(\tilde{X} | \tilde{Y})) \exp[-n (H(\tilde X) + D(\hat P_{\vr x} \| Q))]
\\&=
c_n \exp[-n (H(\tilde X) - H(\tilde{X} | \tilde{Y}) + D(\hat P_{\vr x} \| Q))]
\\&=
c_n \exp[-n (I(\tilde X; \tilde{Y}) + D(\hat P_{\vr x} \| Q))]
\\&=
c_n \exp[-n (\hat I(\vr x; \vr y) + D(\hat P_{\vr x} \| Q))]
\end{split}\end{equation}

Hence, the rate function defined by Theorem~\ref{theorem:optimal_type_based} in our case is:
\begin{equation}\label{eq:A1719}
\Remp^{(\ref{theorem:optimal_type_based})}(\vr x, \vr y) = -\frac{1}{n} \sup_{\vr {\tilde y}} \log Q \left\{ \mathcal{T}_{x|y}(\vr {\tilde y}) \right\} = \hat I(\vr x; \vr y) + D(\hat P_{\vr x} \| Q) - \frac{\log c_n}{n}
\end{equation}
where $\frac{\log c_n}{n}$ is asymptotically vanishing. According to Theorem~\ref{theorem:optimal_type_based} this is the optimal rate function defined using types, up to asymptotically vanishing factors. Since $\frac{\log c_n}{n}$ is also asymptotically vanishing, the conclusion is:

\begin{lemma}\label{lemma:maximum_iid_type_based_Remp}
The following rate function:
\begin{equation}\label{eq:A1725}
\Remp(\vr x, \vr y)  = \hat I(\vr x; \vr y) + D(\hat P_{\vr x} \| Q)
\end{equation}
is the maximum rate function defined by zero-order statistics (equivalently, joint types) which is asymptotically achievable.
\end{lemma}
Note that we have used the term ``maximum rate function asymptotically achievable'' in a somewhat loose way. What it actually means is that $\Remp(\vr x, \vr y)$ can only be improved by asymptotically vanishing factors.

This result shows that formally, perhaps contrary to intuition, the empirical mutual information is \emph{not} the asymptotically optimal rate function defined by zero order statistics: the above rate function is uniformly better.

However considering this from another perspective, we may argue that this difference is immaterial. The rate function above \eqref{eq:A1725} significantly exceeds the empirical mutual information, due to its second term, only when $\vr x$ is non typical, i.e. when the empirical distribution of $\vr x$ significantly differs from the prior $Q$. Since $\vr x$ is fully controlled by the encoder and has a known probability distribution (as opposed to $\vr y$), increasing the rate for non-typical $\vr x$ does not give any actual gain, since we know in advance these events are rare \cite[Theorem III.3]{MethodOfTypes}, irrespective of the channel behavior. In other words, rate functions should be compared mainly based on their values for the typical set of $\vr x$ sequences. Considering this perspective, we may interpret the result above as essentially proving the optimality of the empirical mutual information, as it aligns with the above rate function for the typical $\vr x$. For any rate function asymptotically improving over $\hat I(\vr x; \vr y)$ (and still bounded by \eqref{eq:A1725}), the improvement may happen only for non-typical (and thus, low probability) $\vr x$. Furthermore, it is impossible to have a non-vanishing gain over $\hat I(\vr x, \vr y)$ for all sequences, since this would imply improving over \eqref{eq:A1725} for sequences with $\hat P_{\vr x} = Q$. Therefore we may conclude that the empirical mutual information is ``effectively'' optimal.

The fact that, strictly speaking, the empirical mutual information is not optimal is not surprising if one recalls \eqref{eq:A1729} that $\hat I(\vr x; \vr y) = \frac{1}{n} \log \frac{\hat p(\vr x | \vr y)}{\hat p(\vr x)}$, and therefore it is of the suboptimal form $\Rempname{\ML*}$ defined in Section~\ref{sec:MLML_construction}. Indeed, replacing $\hat p(\vr x)$ by $Q(\vr x)$ we obtain a rate function of the maximum likelihood form \eqref{eq:A1674} which equals the asymptotically optimal function presented above \eqref{eq:A1725}:
\begin{equation}\begin{split}\label{eq:A1739}
\frac{1}{n} \log \frac{\hat p(\vr x | \vr y)}{Q(\vr x)}
&=
\frac{1}{n} \log \frac{\hat p(\vr x | \vr y)}{\hat p(\vr x)} + \frac{1}{n} \log \frac{\hat p(\vr x)}{Q(\vr x)}
=
\hat I(\vr x; \vr y) + \frac{1}{n} \log \prod_{i=1}^n \frac{\hat P_{\vr x} (x_i)}{Q(x_i)}
\\&=
\hat I(\vr x; \vr y) + \frac{1}{n} \sum_{\tilde x \in \mathcal{X}} n \hat P_{\vr x} (\tilde x) \log \frac{\hat P_{\vr x} (\tilde x)}{Q(\tilde x)}
=
\hat I(\vr x; \vr y) + D(\hat P_{\vr x} \| Q)
\end{split}\end{equation}
This observation strengthens the motivation for the maximum likelihood construction \eqref{eq:A1674}, as we have now seen that in addition to the properties mentioned in Section~\ref{sec:ML_rate_functions} this construction yields an asymptotically optimal rate function in the memoryless case.

A way to understand the reason that $\frac{1}{n} \log \frac{\hat p(\vr x | \vr y)}{Q(\vr x)}$ is optimal is as follows: since we are looking for an asymptotically achievable form we consider only rate functions of the form $\Remp = \frac{1}{n} \log \frac{P(\vr x | \vr y)}{Q(\vr x)}$ (the asymptotically limiting form, by Theorem~\ref{theorem:remp_conditional_form_asymptotic}). Further constraining the rate function to be a function of the empirical statistics brings us to consider only memoryless $P$ and $Q$ (note that this is not a necessary condition !), i.e. we have
\begin{equation}\label{eq:A1974}
\Remp = \frac{1}{n} \sum_{i=1}^n \log \frac{P(x_i | y_i)}{Q(x_i)} = \sum_{x,y} \hat P_{\vr x | \vr y}(x|y) \hat P_{\vr y}(y) \log \frac{P(x | y)}{Q(x)}.
\end{equation}
This leaves us with the problem of choosing $P$. Since for every specific sequences $\vr x, \vr y$, $\sum_{x} \hat P_{\vr x | \vr y}(x|y) \log P(x | y) \leq \sum_{x} \hat P_{\vr x | \vr y}(x|y) \log \hat P_{\vr x | \vr y}(x|y)$, $\Remp$ is upper bounded by $\sum_{x,y} \hat P_{\vr x | \vr y}(x|y) \hat P_{\vr y}(y) \log \frac{P_{\vr x | \vr y}(x|y)}{Q(x)} = \frac{1}{n} \log \frac{\hat p(\vr x | \vr y)}{Q(\vr x)}$, and on the other hand as we have seen this rate function is achievable with an asymptotically vanishing redundancy.

\todo{Add the following observation: if one competes with the class of random encoders and decoders with i.i.d. codes, and with a metric based on type, then as we will show, the good-put of the reference for a given $\vr x, \vr y$ (i.e. conditioned), is an achievable rate function. On the other hand, it is only a function of the joint type, and therefore is upper bounded by the optimal type based rate function \eqref{eq:A1725}. Therefore if we fix $Q$, the system attaining this rate is competitively optimal in the sense of attaining a better rate than any of the reference systems}.

\todo{It is possible to make the same claims when, instead of i.i.d., $Q(\vr x)$ is a uniform selection from a type (constant composition) ?}

 \unfinished{
\subsubsection{Rate functions defined by empirical parameters}
discrete case - same as types
note that the factor $N_T$ is inherent since we could place all the region $\Remp \geq R$ on one type. For the continuous case, we could place it on a single value of $\hat \theta$.

continuous case - we need to make continuity assumption on $R(\theta)$.
One option is to use previous derivation but explicitly stating the offset due to continuity - i.e. we assume that if $\| \Delta \theta \| \leq \delta(\theta)$ then  $R(\theta + \Delta \theta) \geq R(\theta) - \epsilon(\theta)$. This will give more explicit tradeoff

Another approach: treat $\hat \theta$ as Gaussian vector. Look at the probability that $\hat \theta \in \{\theta: R_{\theta} \geq R \}$, by assuming some ``width'' of the region, and looking at closest point. This may relate us to the Fisher information matrix !? (naa. it is low probability area, hence off limits for central limit theorem..)

possible approach: look at the distribution of $\hat \theta$ when $\vr y$ is chosen. It seems that if we look at equi-probability stripes they should have equal rate. If a large portion of a certain stripe is in a certain rate, then it may be in the same rate just as well. On the other hand, we can put a limit to the rate in each stripe (which has a volume).
}

\subsection{The rate of a given decoding metric}\label{sec:rate_of_given_metric}
\todo{1. general on goodput given x,y, 2. specific for i.i.d. code, 3. tightness of union bound, 4. equivalence of metric and remp, 5. specific ML case - set of ML decoders, 6. constrain to memoryless metrics, 7. non i.i.d. codes}

\todo{Can we say that in general the good put of a system is an achievable rate function up to some overhead? i.e. if a system attains a high error probability for $\vr x, \vr y$ can we correct it? As example, if it transmits at rate $R$ and has error probability $\half$ for certain $\vr x, \vr y$, then it attains rate $\half R$. Can we fix it to attain this rate with a small error probability? for example, if it is indecisive between two messages, we can remove some codewords from the codebook (effectively, add more redundancy). But it seems that non-adaptively we cannot do it - we can only attain the rate function at the point $R$ (the rate of the orignial system). Adaptively -- it produces a different system, so we can't use operational arguments.}

\todo{Can we define a competitive model? Our competitor knows the sequences, but has a limitation that affects the range of possible rate function (can we put it as a limitation on $P(x|y)$ in the architype form?), so he chooses the best $\Remp$ than he can. Also, we need to be able to say that this is not only the best $\Remp$ but also the best good put. Then, can we attain this rate universally?}

As already mentioned, every single user communication system can be characterized by a rate function -- one could always ``freeze'' the channel and observe how the system performed (in terms of rate and error probability) over all instances in which a specific $\vr x$ was the input and a specific $\vr y$ was the output. Having characterized the system in this way, we may now consider how it operates over any channel of interest. In the particular case of random encoders and metric based decoders, explicit expressions for a rate function from a given metric and an input distribution can be given. Then, these expressions can be used in order to compete against a class of systems, defined by different decoding metrics.

We now consider the specific case of a random i.i.d. code and metric based decoder. This class of systems was selected since it allows a relatively simple analysis. On the other hand, this class of systems is able to attain the information theoretic bounds with respect to rate and error exponent (where the later is known to be tight over part of its domain \cite{MethodOfTypes}). This class was used as a comparison class by Ziv \cite{ZivUniversal}, and the current derivation is inspired by the analysis performed there.

The code is a random i.i.d. selection of $M = \exp(nR)$ codewords from a predetermined distribution $Q(\vr x)$. The decoder uses a decoding metric $u(\vr x, \vr y)$, and after seeing $\vr y$, chooses the word with the highest value of $u(\vr x, \vr y)$. Note that the system attaining the sufficient condition of Theorem~\ref{theorem:remp_ublb} also belongs to this class. With this metric and input distribution, under different channel assumptions and error probability requirements, one can obtain various feasible rates, i.e. the maximum rate in which the system can operate with the required error probability under the channel model. Now, we would like to avoid specifying the channel model and error probability, and say something about the rate possible with this metric for given input and output.

Given that the word $\vr x$ was transmitted and $\vr y$ was received, a decoding error would happen if any of the other words has a metric value higher than the metric of the transmitted word. The probability of any word having a metric value exceeding that of the transmitted word is
\begin{equation}\label{eq:A2132}
p(\vr x, \vr y) \defeq \underset{\vr {\tilde X} \sim Q}{\Pr} \left\{ u(\vr {\tilde X}, \vr y) > u(\vr x, \vr y) \right\}
\end{equation}
The probability of any of the $M-1$ competing words exceeding the correct word is $1 - (1 -p(\vr x, \vr y))^{M-1}$ and since this is a sufficient condition for an error we have that the conditional error probability is:
\begin{equation}\label{eq:A2136}
P_{e|xy}(\vr x, \vr y) \geq 1 - (1 -p(\vr x, \vr y))^{M-1}
\end{equation}
Note that this bound is tight up to the question of how ties are broken: it is an inequality only since we do not know if errors occur when $u(\vr {\tilde x}, \vr y) = u(\vr x, \vr y)$. If we had defined $p(\vr x, \vr y)$ by the event $u(\vr X, \vr y) \geq u(\vr x, \vr y)$ then we would have an inequality in the other direction. In the following, we sometimes omit the arguments $\vr x, \vr y$ and use $p, P_{e|xy}$ instead of $p(\vr x, \vr y), P_{e|xy}(\vr x, \vr y)$ (etc).

We may now ask the following question: given a specific pair $\vr x, \vr y$, how many codewords could one allow, while reaching a small probability of error? Using \eqref{eq:A2136} and requiring $P_{e|xy} \leq \epsilon$ we have:
\begin{equation}\label{eq:A2145}
1 - (1 -p)^{M-1} \leq \epsilon
\end{equation}
\begin{equation}\label{eq:A2148}
M \leq \frac{\log(1 - \epsilon)}{\log (1 -p)} + 1
\end{equation}
Note that both $\log(1 - \epsilon)$ and $\log (1 -p)$ are negative. Assuming $\epsilon \leq \half$, $-\log(1 - \epsilon) \leq \log(2) = 1$, and in order for $M$ to be large $M>>1$, $-\log (1 -p)$ is required to be small ($<<0$) and therefore $p$ needs to be close to $0$. Therefore we may approximate $-\log (1 -p) \approx p$. If one also assumes $\epsilon \approx 0$, the bound above can be written as $\approx \frac{\epsilon}{p} + 1$. Interestingly, if we had used the union bound to calculate the error probability, we would need to require $p \cdot (M-1) \leq \epsilon$, which would also mean $M = \frac{\epsilon}{p} + 1$. Here we can see in a simple way why for the purpose of determining the rate when the error probability is small and fixed, the union bound is tight.

Given $p,\epsilon$, it is possible to define the rate function $\frac{1}{n} \log M$ where $M$ satisfies \eqref{eq:A2148} with equality, i.e.
\begin{equation}\label{eq:A1854}
\Remp = \frac{1}{n} \log \left( \frac{\log(1 - \epsilon)}{\log (1 -p(\vr x, \vr y))} + 1 \right) \approx \frac{1}{n} \log \left( \frac{\epsilon}{p(\vr x, \vr y)}  \right)
.
\end{equation}

This yields a way of converting decoding metrics to rate functions. It is interesting to observe that $p(\vr X,\vr y)$ is uniformly $\unif[0,1]$ distributed. This is because since for each $\vr y$ it equals the inverse CDF $1-F_U(u)$ of the random variable $U$, defined as $U=u(\vr X, \vr y)$ (where $\vr X \sim Q$). Hence $F_U(U)$ is uniform $\unif[0,1]$. Also, per $\vr y$, $p(\vr x, \vr y)$ is decreasing in $u(\vr x, \vr y)$, and therefore decoding with the metric $\frac{1}{p(\vr x, \vr y)}$ is equivalent to decoding with $u(\vr x, \vr y)$. Similarly $\Remp$ can be used as a metric. It is interesting to note in this respect that if one wants to supercede the performance of $K$ systems with metrics $u_k(\vr x, \vr y), k=1,\ldots,K$, by taking the maximum over their respective $\Remp$ \eqref{eq:A1854}, the resulting metric is equivalently the minimum over $p_k(\vr x, \vr y)$. This yields the ``Merged decoder'' of Feder-Lapidoth \cite{FederLapidoth_UnivDecod98} ($p$ reflects the order over $\vr X$ defined there), and it is easy to see by the union bound (still, conditioning on $\vr x, \vr y$) that indeed the error probability is at most the sum of individual error probabilities. $p$ can be considered a canonization of $u$: while $u$ is of general form, $1/p$ is an equivalent metric, constrained to a specific distribution.

While \eqref{eq:A2148} gives us a relation between the probability $p(\vr x, \vr y)$ and the rate $R(\vr x, \vr y) = \frac{1}{n} \log M$, it requires a specification of $\epsilon$, the error probability. Most communication systems are not designed to yield a guaranteed same error probability per each $\vr x, \vr y$, but rather on average (actually, it is impossible to have fixed $M$ and obtain a given error probability uniformly). Therefore instead of considering the rate with a given error probability, we mix the two together and consider the ``goodput'', i.e. the average number of error-free bits per channel use, which is defined as $(1-P_e) \cdot R$ (see Section~\ref{sec:good_put_bound}). Given $\vr x, \vr y$ the question is, what is the maximum good-put that can be achieved.

Define
\begin{equation}\label{eq:A2166}
R_\tsubs{good}^*(\vr x, \vr y) = \sup_{M = 2,3,\ldots} (1-P_{e|xy}(\vr x, \vr y)) \cdot R
\end{equation}
where $R = \frac{1}{n} \log M$. I.e. for given $Q$ and $u(\vr x, \vr y)$, $R_\tsubs{good}(\vr x, \vr y)$ is the maximum error-free rate conditioned on $\vr x, \vr y$ which can be obtained with any number of codewords, considering the tradeoff between rate and error probability. Although $R_\tsubs{good}(\vr x, \vr y)$ is a function of $\vr x, \vr y$, the goodput of any fixed rate system (where $M$ is a constant) conditioned on seeing a specific pair $\vr x, \vr y$ is also bounded by  $R_\tsubs{good}(\vr x, \vr y)$ (due to the supremum with respect to $M$ above). Note that the original system possibly had a certain fixed rate, which we now ignore, since we look at all possible systems using the same metric.

We now compute an upper bound on $R_\tsubs{good}(\vr x, \vr y)$ by using the bound of \eqref{eq:A2136} and by relaxing the maximization over $M$ to $[2,\infty)$ (not necessarily integer).
\begin{equation}\label{eq:A2177}
R_\tsubs{good}^*(\vr x, \vr y) \stackrel{\eqref{eq:A2136}}{\leq} \sup_{M \in [2,\infty)} (1 -p)^{M-1} \cdot \frac{1}{n} \log M
.
\end{equation}
Writing $M$  as $M = \frac{\alpha}{-\ln (1-p)} + 1$, then $\ln \left[ (1-p)^{M-1} \right] = (M-1) \ln (1-p) = -\alpha$, and therefore
\begin{equation}\label{eq:A1890}
(1-p)^{M-1} \log M = e^{-\alpha} \left[ \log \alpha + \log \frac{1}{-\ln (1-p)} + \log (\frac{M}{M-1}) \right] \leq  \log \frac{1}{-\ln (1-p)} +  e^{-\alpha} \log \alpha + \log(2)
.
\end{equation}
It is easy to bound $f(\alpha) = e^{-\alpha} \ln \alpha \leq e^{-2}$, by writing  $f(\alpha) \leq e^{-\alpha} (\alpha -1 )$, and showing that the maximum of this bound is obtained for $\alpha = 2$. Defining $c = \log(2) + e^{-2}$, \eqref{eq:A1890} yields:
\begin{equation}\label{eq:A1896}
(1 -p)^{M-1} \cdot \frac{1}{n} \log M \leq \frac{1}{n} \log \left( \frac{1}{-\ln (1-p)} \right) +  \frac{c}{n}
,
\end{equation}
and therefore
\begin{equation}\label{eq:A1896b}
R_\tsubs{good}^*(\vr x, \vr y) \leq \frac{1}{n} \log \left( \frac{1}{-\ln (1-p)} \right) +  \frac{c}{n}
,
\end{equation}
hence, also the good-put function is bounded asymptotically like $\frac{1}{n} \log \frac{1}{p}$, and not far from $\Remp$. This is not surprising, since for any $\epsilon$, when trying to exceed the rate given by $\Remp$, the error probability quickly increases to close to $1$ and the rate drops. Therefore $\Remp$ cannot be exceeded significantly, even when the error probability constraint is removed. This implies that $R_\tsubs{good}^*(\vr x, \vr y)$ is asymptotically achievable as a rate function, which corresponds to what was shown in Section~\ref{sec:good_put_bound} (there it is shown for general systems, not necessarily with i.i.d. random-coding, but without the maximization on $M$).

The discussion above shows how to construct achievable rate functions from decoding metrics. Furthermore, if one has a set of reference decoders, with possible decoding metrics, by choosing the maximum resulting rate function, it is possible to guarantee a better rate than all the reference decoders. In the non-adaptive case, the meaning of guaranteeing a better rate is that if the universal system operates with rate $R$, then for any $\vr x, \vr y$ for which any of the reference systems yields a rate (or good-put) larger than $R$, the universal system will succeed, with high probability, to decode.

It is interesting in this respect to consider, for a specific i.i.d. input distribution $Q$, the family of decoders using memoryless additive metrics, i.e. $u(\vr x, \vr y) = \sum_{i=1}^n u(x_i,y_i)$. Clearly, the rate \eqref{eq:A1854} or good put \eqref{eq:A2166} functions attainable by this family of decoders is independent of the order of the letters in $(x_i,y_i)$, i.e. they depend only on the empirical distribution of $(\vr x, \vr y)$, and are therefore asymptotically limited by the rate function \eqref{eq:A1725} $\hat I(\vr x; \vr y) + D(\hat P_{\vr x} \| Q)$ defined in Lemma~\ref{lemma:maximum_iid_type_based_Remp}. Hence, if one considers the maximum rate over all decoders in this family, this rate would still be lower than the rate function of Lemma~\ref{lemma:maximum_iid_type_based_Remp}. On the other hand, as shall be seen, the rate function of Lemma~\ref{lemma:maximum_iid_type_based_Remp} is asymptotically adaptively achievable. As a result, there exists an adaptive rate system, that for any $\vr x, \vr y$ attains a rate at least as large as the rate that could be attained by with the best memoryless decoding metric. This universality would also hold true when $\vr y$ is determined by a probabilistic channel and the rate is taken on average over all pairs.

\section{Rate adaptivity}\label{chap:rate_adaptivity}
In Section~\ref{sec:chernoff_cond_probability} we have shown that, asymptotically, all attainable rate functions are limited by the form
\begin{equation}\label{eq:A1616b}
\Remp(\vr x, \vr y) = \frac{1}{n} \log \frac{P(\vr x | \vr y)}{Q(\vr x)}
\end{equation}
In this section we will present a rate-adaptive scheme that attains this rate function adaptively for many conditional distributions $P(\vr x | \vr y)$, but not for all. Generally, the requirement is that $P(\vr x | \vr y)$ could be computed sequentially, while $\vr y$ is gradually revealed to the decoder. Unlike in the non-adaptive case, we do not have an asymptotical characterization of all achievable rate functions. Furthermore, we do not have tight bounds on the redundancy required to achieve these rate functions. However, many rate functions of interest can be posed in a sequential form, and therefore, as we shall see, there are many examples of interesting rate functions which are adaptively achievable.

Before presenting the scheme, we would like to begin with a more fundamental question: why is feedback needed to yield rate adaptivity?

\subsection{A rate adaptive scheme}\label{sec:rate_adaptive_scheme}
The scheme proposed in order to achieve rate adaptivity (see definitions \ref{def:sys_adaptive},\ref{def:achievability_adaptive}) is based on an iterative application of rateless coding, and is similar in concept to the one used in \selector{the previous paper}{our initial work} \cite{YL_individual_full}. The idea of iterative rateless coding was first proposed by Eswaran~\etal~\cite{Eswaran}. We fix a number $K$ of bits per block. At each block, the encoder transmits symbols from the codeword selected based on the message bits. The decoder examines the channel output, and decides when it has ``enough information'' to decode, according to a termination condition. When this condition is satisfied, the decoder sends an indication through the feedback link, and a new block begins. In the new block, additional $K$ bits from the message string will be sent. The process ends at time $n$, and the last block is possibly not decoded. Thus, the rate varies by changing the number of blocks transmitted. Roughly speaking, as the rate function increases, the blocks become shorter, and the number of blocks increases.

We assume the feedback is completely reliable, but may have a limited rate and a delay. In order to model the effect of limiting the feedback rate, we define that a feedback of one bit is possible only once per $\dfb \geq 1$ symbols and has a delay of $\dfb$ symbols, i.e. the decoder may send a feedback bit only on symbol $i \cdot \dfb + 1$ ($i=1,2,\ldots$), and this bit will be seen by the encoder $\dfb$ symbols later at time $(i+1) \cdot \dfb + 1$.

Let $Q(\vr x)$ denote the input prior. Suppose a block ended at symbol $j$, then the codebook of $\exp(K)$ codewords for the new block starting after this symbol is generated by random i.i.d. selection of each codeword, according to the distribution $Q(\vr x_{j+1}^n | \vr x^j) = \frac{Q(\vr x^n)}{Q(\vr x^j)}$, where $\vr x^j$ are the symbols that had already been transmitted. This guarantees that irrespective of the message, the input distribution remains $Q$. The randomization is carried out by using the common randomness. Under the assumption that there are no decoding errors, the decoder knows $\vr x^j$ and using the this codebook is known at both sides of the link. If there are decoding errors, there may be unexpected behavior at the decoder side, however the input distribution is maintained $Q(\vr x)$ as required. For simplicity, we always treat the codewords as vectors of length $n$, where all the prefixes of the codewords will be fixed and equal $\vr x^j$. The codeword that encodes message $\msg$ ($\msg = 1,2,\ldots,\exp(K)$) is denoted $\vr x^{(\msg)}$. At each block $\msg$ is formed from new $K$ bits out of the input message sequence, and the encoder sends the symbols of $\vr x^{(\msg)}$ matching the time index, one by one.

The decoding is carried out by using a decoding metric $\psi(\vr x^k, \vr y^k, j)$ and a decoding threshold $\psi^*_{j,k}$, which are defined for all $0 \leq j < k \leq n$. $\psi(\vr x^k, \vr y^k, j)$ is interpreted as the decoding metric at time $k$ where the last block ended at time $j$. To prove the properties of this scheme that are given in Theorem~\ref{theorem:framework}, some assumptions on $\psi(\vr x^k, \vr y^k, j)$ are required. Potentially, these assumptions are satisfied only when $k-j$ is large enough $k-j > b_0$, and in this case $\psi^*$ will be defined as infinity for the first $b_0$ symbols in each block.

The decoder decides to decode the current block at time $k$ if
\begin{enumerate}
\item $k-1$ divides by $\dfb$ (i.e. there is a chance to send a feedback bit)
\item There exists a codeword $\msg \in \{1,2,\ldots,\exp(K)\}$ such that
\begin{equation}\label{eq:A1912}
\psi((\vr x^{(\msg)})_1^k, \vr y^k, j) \geq \psi^*_{j,k}
\end{equation}
Note that these $(\vr x^{(\msg)})_1^k$ include a common history of length $j$ and an unknown part of length $k-j$.
\end{enumerate}
If the decoder decided to terminate at symbol $k$, then the encoder will start a new block at symbol $k+\dfb$. Thus for the new block we will have $j' = k+\dfb - 1$ (the last symbol of the previous block). New blocks always start on symbols $i \cdot \dfb + 1$ (the first, at symbol $1$).

\begin{figure}
\centering
\ifpdf
  \setlength{\unitlength}{1bp}%
  \begin{picture}(369.64, 258.52)(0,0)
  \put(0,0){\includegraphics{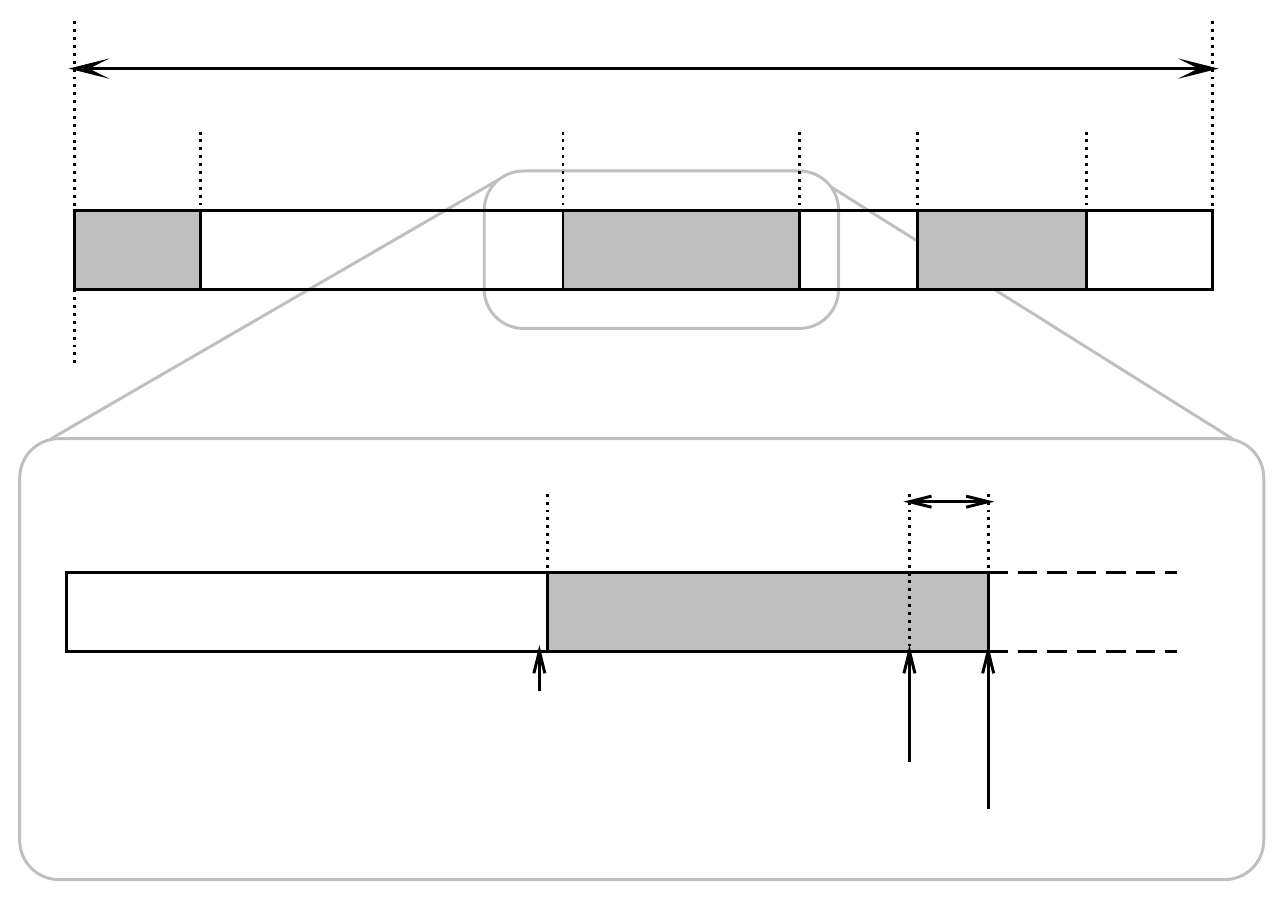}}
  \put(96.38,213.17){\fontsize{9.10}{10.93}\selectfont Block 2}
  \put(26.08,213.17){\fontsize{9.10}{10.93}\selectfont Block 1}
  \put(179.15,213.17){\fontsize{9.10}{10.93}\selectfont Block 3}
  \put(232.44,213.17){\fontsize{9.10}{10.93}\selectfont Block 4}
  \put(278.93,213.17){\fontsize{9.10}{10.93}\selectfont Block 5}
  \put(316.35,213.17){\fontsize{9.10}{10.93}\selectfont Block 6}
  \put(173.48,241.51){\fontsize{9.10}{10.93}\selectfont $n$}
  \put(174.61,108.85){\fontsize{9.10}{10.93}\selectfont Block 3}
  \put(293.67,106.58){\fontsize{9.10}{10.93}\selectfont Block 4}
  \put(74.83,108.85){\fontsize{9.10}{10.93}\selectfont Previous blocks}
  \put(119.06,46.49){\fontsize{9.10}{10.93}\selectfont $j$ (for block 3)}
  \put(60.09,80.50){\fontsize{9.10}{10.93}\selectfont $\vr x^j$ (fixed)}
  \put(173.48,80.50){\fontsize{9.10}{10.93}\selectfont $\vr x_{j+1}^n \sim Q(\cdot | \vr x^j)$}
  \put(264.19,119.06){\fontsize{9.10}{10.93}\selectfont $\dfb$}
  \put(173.48,32.88){\fontsize{9.10}{10.93}\selectfont Decoder decides to decode}
  \put(164.41,17.01){\fontsize{9.10}{10.93}\selectfont Encoder receives indication and starts a new block}
  \end{picture}%
\else
  \setlength{\unitlength}{1bp}%
  \begin{picture}(369.64, 258.52)(0,0)
  \put(0,0){\includegraphics{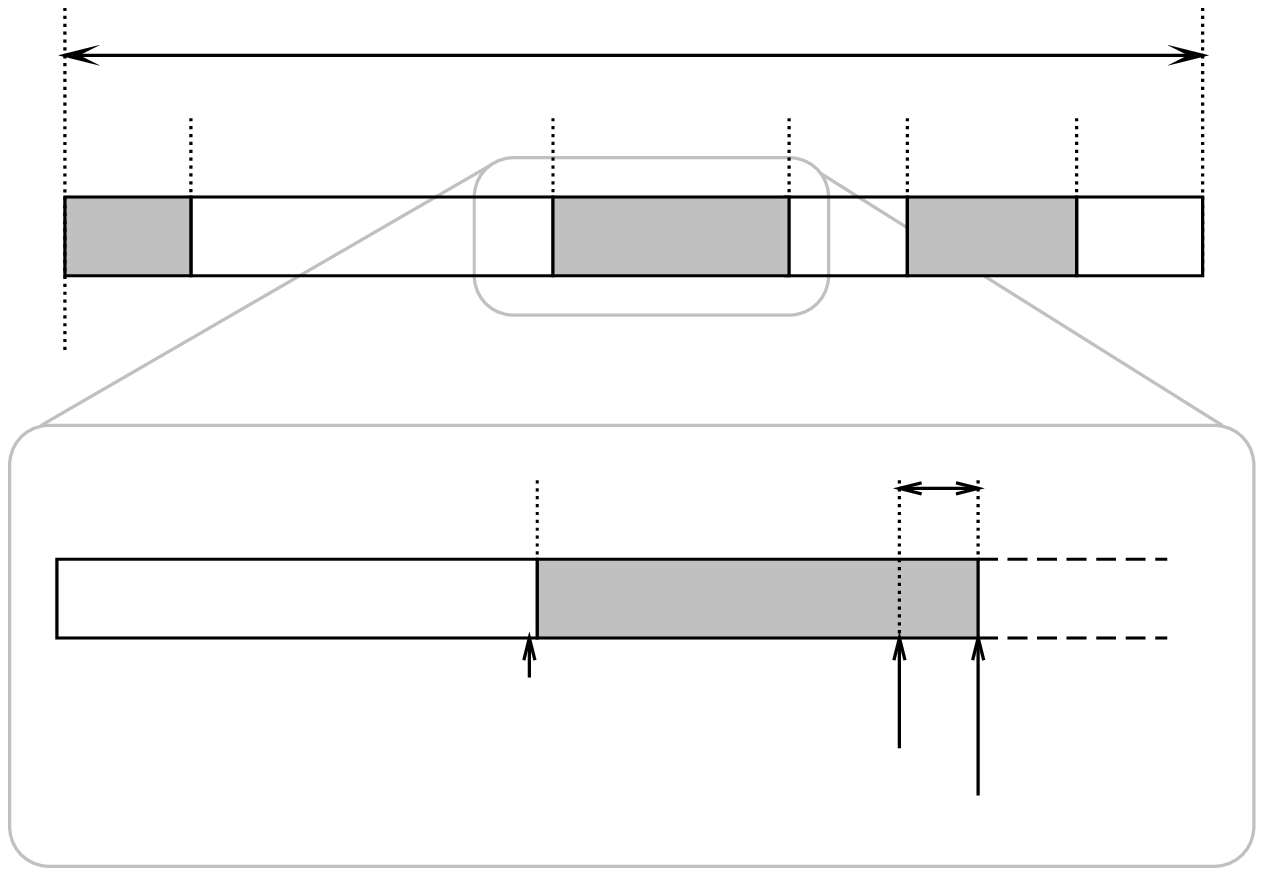}}
  \put(96.38,213.17){\fontsize{9.10}{10.93}\selectfont Block 2}
  \put(26.08,213.17){\fontsize{9.10}{10.93}\selectfont Block 1}
  \put(179.15,213.17){\fontsize{9.10}{10.93}\selectfont Block 3}
  \put(232.44,213.17){\fontsize{9.10}{10.93}\selectfont Block 4}
  \put(278.93,213.17){\fontsize{9.10}{10.93}\selectfont Block 5}
  \put(316.35,213.17){\fontsize{9.10}{10.93}\selectfont Block 6}
  \put(173.48,241.51){\fontsize{9.10}{10.93}\selectfont $n$}
  \put(174.61,108.85){\fontsize{9.10}{10.93}\selectfont Block 3}
  \put(293.67,106.58){\fontsize{9.10}{10.93}\selectfont Block 4}
  \put(74.83,108.85){\fontsize{9.10}{10.93}\selectfont Previous blocks}
  \put(119.06,46.49){\fontsize{9.10}{10.93}\selectfont $j$ (for block 3)}
  \put(60.09,80.50){\fontsize{9.10}{10.93}\selectfont $\vr x^j$ (fixed)}
  \put(173.48,80.50){\fontsize{9.10}{10.93}\selectfont $\vr x_{j+1}^n \sim Q(\cdot | \vr x^j)$}
  \put(264.19,119.06){\fontsize{9.10}{10.93}\selectfont $\dfb$}
  \put(173.48,32.88){\fontsize{9.10}{10.93}\selectfont Decoder decides to decode}
  \put(164.41,17.01){\fontsize{9.10}{10.93}\selectfont Encoder receives indication and starts a new block}
  \end{picture}%
\fi
\caption{\label{fig:Iterated_rateless_blocks_illustration_1}%
 An illustration of the rate adaptive scheme}
\end{figure}
The scheme is defined with respect to the parameters $K, \psi, \psi^*, b_0, \dfb$ and is performance will be a function of these factors. The scheme is illustrated in Figure~\ref{fig:Iterated_rateless_blocks_illustration_1}, where in the top of the figure, the division of the $n$ channel uses into blocks is depicted. The blocks have an arbitrary length. At the bottom of the figure, the process of decoding the third block is detailed, showing the transmitted word, and the feedback delay at the end of the block.

\subsection{The performance of the rate adaptive scheme}\label{sec:rate_adaptive_performance}
The following theorem formalizes a claim on the performance of the scheme presented above, under some assumptions on the parameters. The theorem gives the achieved rate as a function of the decoding metric, and shows that asymptotically $\Remp \approx \frac{1}{n} \log \psi(\vr x, \vr y, 0)$ is achievable. This relation, as well as the conditions we define on $\psi$, may appear a little cryptic at this point. This is mainly since, in order to keep the generality of the theorem, which will make it useful in several cases later on, we avoid specifying $\psi$. To better understand the theorem, it is useful at this point to think of the following substitution of $\psi$: $\psi(\vr x^k, \vr y^k, j) = \frac{P(\vr x_{j+1}^k | \vr y^k, \vr x^j)}{Q(\vr x_{j+1}^k | \vr x^j)}$ for some conditional probability law $P$. In this case, it is easy to see that the rate function defined above aligns with the generic conditional form of rate functions \eqref{eq:Aremp_conditional_form}, and the other conditions on $\psi$ will make sense.

\begin{theorem}\label{theorem:framework}
For the channel $\mathcal{X} \to \mathcal{Y}$, a given block length $n$, prior $Q(\vr x)$ and error probability $\epsilon$, and with respect to scheme of Section~\ref{sec:rate_adaptive_scheme} operating with $K$ bits/block, a decoding metric $\psi$, decoding thresholds $\psi^*$ and feedback delay $\dfb$, which satisfy the following conditions:
\begin{enumerate}
\item \textbf{CCDF condition}: The following bound holds for all $k-j > b_0 \geq 0$ (for $b_0 \in \mathbb{Z}^+$) and all $\vr y^k$:
\begin{equation}\label{eq:A1933}
\placeunder{\Pr}{Q} \left\{ \psi(\vr X^k, \vr y^k, j) \geq t | \vr x^j\right\} \leq \frac{L_{k-j}}{t}
\end{equation}
For some sequence $L_i \geq 0$. Alternatively, the following sufficient condition (due to Markov inequality) can be met:
\begin{equation}\label{eq:A1933b}
\placeunder{\E}{Q} \left[ \psi(\vr X^k, \vr y^k, j)  | \vr x^j \right] \leq L_{k-j}
\end{equation}
\item \textbf{Approximate summability (convexity)}: Let $\{j_b, k_b\}_{b=1}^B$ be a set of $B$ pairs of increasing indices indicating segments in time $j_1 < k_1 \leq j_2 < k_2, \ldots, j_b < k_b \leq j_{b+1}, \ldots, j_B < k_B \leq n$, where $(j_b,k_b)$ refers to symbols $j_b+1, \ldots, k_b$. Define $\psi_{0}^n \defeq \psi(\vr x, \vr y, 0)$ and $\psi_b \defeq \psi(\vr x^{k_b}, \vr y^{k_b}, j_b)$. I.e. one is the metric measured on the entire transmission, and the other is the metric for a specific segment. Let $m_0$ denote the number of symbols that are not included in any segment $m_0 \defeq n - \sum_{b=1}^B (k_b - j_b)$. Then there exists a function $f_0^{(n)}:\mathbb{R} \to \mathbb{R}^+$ such that the following is satisfied:
\begin{equation}\label{eq:A1945}
\log \psi_0^n - \sum_{b=1}^B \log \psi_b \leq f_0^{(n)}(\psi_0^n) \cdot m_0
\end{equation}
I.e. the difference between the $\log$-metric on the entire transmission and on the segments can be bounded as a function of the number of symbols not participating in the sum.
\item Technical assumptions:
\begin{itemize}
\item $L_i$ is non-decreasing in $i$ (for $i=1,2,\ldots$)
\item \unfinished{I think this should be removed: $d \cdot \epsilon \leq \exp(K)$}
\end{itemize}
\end{enumerate}
Define
\begin{equation}\label{eq:A1958}
\Remp = \frac{1}{n} \log \psi(\vr x^n, \vr y^n, 0) \defeq \frac{1}{n} \log \psi_0^n
\end{equation}
and
\begin{equation}\label{eq:A1962}
F_n(t) =  \left(1 + \frac{c_n  + b_1 \cdot f_0^{(n)}(\exp(n t))}{K} \right)^{-1} \cdot t - \frac{K}{n}
\end{equation}
with $c_n = \log \frac{n \cdot L_n}{\dfb \epsilon}$ and $b_1 = b_0 + 2\dfb - 1$. Then $F_n(\Remp)$ is adaptively achievable by the scheme of Section~\ref{sec:rate_adaptive_scheme}, using the threshold
\begin{equation}\label{eq:A1966}
\psi^*_{j,k} = \frac{n \cdot L_{k-j} \cdot \exp(K)}{d \epsilon}
\end{equation}
\end{theorem}
\todo{reword the theorem (it shouldn't be one-sentence page)!}

\begin{corollary_in_theorem}\label{corollary:framework_asymptote}
If $\frac{1}{n} \log L_n \ntoinfty 0$ and for some sequence $\delta_n \in [0,1], \delta_n \to 0$, $\forall t: \frac{f_0^{(n)}(\exp(n t))}{n \delta_n} \ntoinfty 0$ (this holds trivially if $f_0^{(n)}$ is upper bounded by a constant), then $\Remp$ is asymptotically adaptively achievable.
\end{corollary_in_theorem}

\begin{corollary_in_theorem}\label{corollary:framework_redundancy}
If the rate function $\Remp$ defined in \eqref{eq:A1958} is bounded $\Remp \leq R_{\max}$, then it is achievable up to $\delta_n = 3 \sqrt{\frac{R_{\max} \cdot (c_n  + b_1 \cdot f_0^{(n)*})}{n}}$, where $\displaystyle f_0^{(n)*} \defeq \max_{t \leq R_{\max}} f_0^{(n)}(\exp(n t))$. In other words, in this case we can bound the additive loss and have $F_n(t)$ of the form $t - \delta_n$. Furthermore, for small $\epsilon$ and large $n$, if $f_0^{(n)*}$ is upper bounded for all $n$, this rate function is achievable up to $\approx 2 \sqrt{\frac{\log \frac{n}{\epsilon}}{n}}$.
\end{corollary_in_theorem}

\begin{corollary_in_theorem}\label{corollary:framework_nonadaptive}
Under the conditions of Theorem~\ref{theorem:framework}, the above rate function \eqref{eq:A1958} is also non-adaptively achievable, with an intrinsic redundancy of $\mu_Q(\Remp) \leq \frac{1}{n} \log L_{n}$
\end{corollary_in_theorem}

Note that the Theorem refers to decoding metrics satisfying specific conditions. In some cases one can modify a given decoding metric by adding constants that will enable satisfying these conditions (see for example Section~\ref{sec:examples_compression_attanability}). A note regarding Corollary~\ref{corollary:framework_redundancy}: note that in the non-adaptive case the redundancy was of the order of $\Theta \left(\frac{1}{n} \right)$, whereas here it is larger by more than a square root $\Theta \left( \sqrt{\frac{\log n}{n}} \right)$. This relatively large redundancy is due to the fact we have divided the transmission into blocks and there are approximately $\Theta(\sqrt{n})$ blocks.

\subsection{An intuitive explanation}
\todo{Add a simpler explanation to the case of conditional probabilities.}

\subsection{Proof of Theorem~\ref{theorem:framework}}
For brevity we denote $d \defeq \dfb$. We begin by determining the decoding thresholds that allow us to bound the error probability by $\epsilon$. We require that for any symbol in which a decision is made $i \cdot d, 1 \leq i \leq n/d$, the probability of deciding in favor of a different codeword than the one that is transmitted is at most $\frac{d \epsilon}{n}$, conditioned on the input sequence, and on the assumption there were no errors up to this point. Since there are no more than $n/d$ such events, then by the union bound this would guarantee that the probability of any of these events, conditioned on the input sequence, is at most $\epsilon$.\footnote{the elements in the union are the following event: an error in the first decision, an error in the second decision given that the first is correct, etc. The union of these events is the event of any error occurring} When any of these events happens, there is an error, and we do not give any guarantee on the decoding rate. When none of these events happen, the message is perfectly decoded, and we will be able to give a deterministic lower bound on the rate. The probabilities are conditioned on the input sequence since Definition~\ref{def:achievability_adaptive} requires an error probability guarantee for any input and output sequence (the output sequence is treated as a deterministic sequence).

We consider a decoding at time $k$ where the previous block ended at time $j$. The true codeword is denoted $\msg$ and the channel input is therefore $\vr X^k = \left(\vr X^{(\msg)} \right)_1^k$. The alternative codeword is denoted $\tilde{\msg}$. By our definition, the two codewords are equal up to time $j$ (common history) and independent from time $j+1$ on. Therefore in terms of the probability of the decoding metric to exceed the threshold for codeword $\tilde{\msg}$, knowing the channel input $\vr X$ is equivalent to knowing the first $j$ elements of $\vr X^{(\tilde \msg)}$. In other words, given $\vr X$, $\vr X^{(\tilde \msg)}$ equals $\vr X^j$ to up time $j$ and is distributed $Q(\cdot| \vr X_j)$ from that time on. By our assumption that there are no decoding errors so far, the codebook used by the decoder is correct.

If $k - j < b_0$ then there is no guarantee on the distribution of $\psi$, and therefore we set $\psi^* = \infty$, i.e. do not decode regardless of the channel output. Assuming $k - j \geq b_0$, the probability of any codeword to exceed the threshold is:
\begin{equation}\label{eq:A2001}
\Pr \left\{ \psi((\vr X^{(\tilde \msg)})_1^k, \vr y^k, j) \geq \psi^*_{j,k} | \vr X \right\} = \underset{Q}{\Pr} \left\{ \psi(\vr X_1^k, \vr y^k, j) \geq \psi^*_{j,k} | \vr X^j \right\} \stackrel{\eqref{eq:A1933}}{\leq} \frac{L_{k-j}}{\psi^*_{j,k}}
\end{equation}
Since there are $\exp(K)-1$ competing codewords, using the union bound, the probability that any codeword will exceed the threshold is upper bounded by
\begin{equation}\label{eq:A2006}
\Pr \left\{ \exists \tilde \msg: \psi((\vr X^{(\tilde \msg)})_1^k, \vr y^k, j) \geq \psi^*_{j,k} | \vr X \right\} \leq \exp(K) \frac{L_{k-j}}{\psi^*_{j,k}} \stackrel{\text{Req.}}{\leq} \frac{d \epsilon}{n}
\end{equation}
Setting the threshold to:
\begin{equation}\label{eq:A2006b}
\psi^*_{j,k} = \frac{n \cdot L_{k-j} \cdot \exp(K)}{d \epsilon}
\end{equation}
would guarantee meeting the error probability requirement. Note that tighter bounds can be obtained for specific structures of the metric, specifically when the metric is a product of single-letter metrics and $Q$ is i.i.d., by using the methods proposed by Feder \& Blits \todo{Missing ref to Navot}, and for these cases the factor $n$ in \eqref{eq:A2006b} could be avoided. Here we used the union bound on symbols, which is simpler and more general, but less tight.

\begin{figure}
\centering
\ifpdf
  \setlength{\unitlength}{1bp}%
  \begin{picture}(316.63, 200.98)(0,0)
  \put(0,0){\includegraphics{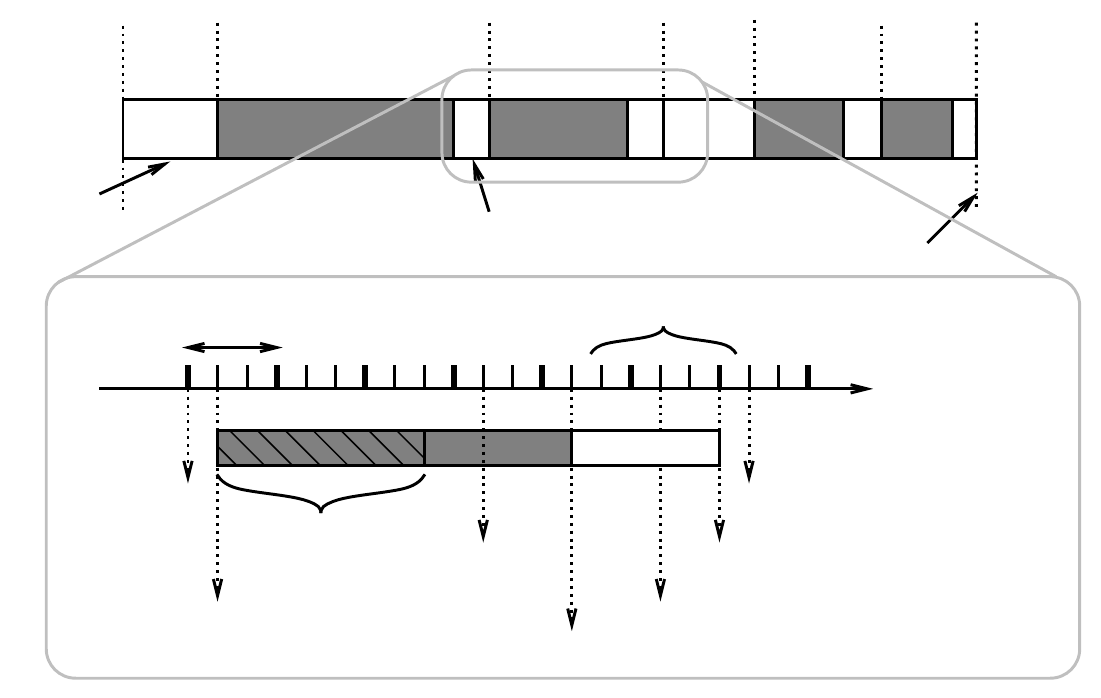}}
  \put(91.56,183.40){\fontsize{6.83}{8.19}\selectfont Block 2}
  \put(38.83,183.40){\fontsize{6.83}{8.19}\selectfont Block 1}
  \put(5.67,137.48){\fontsize{6.83}{8.19}\selectfont $< b_0 + 2d$}
  \put(137.48,133.23){\fontsize{6.83}{8.19}\selectfont $2 d - 1$}
  \put(153.64,183.40){\fontsize{6.83}{8.19}\selectfont Block 3}
  \put(193.61,183.40){\fontsize{6.83}{8.19}\selectfont Block 4}
  \put(228.47,183.40){\fontsize{6.83}{8.19}\selectfont Block 5}
  \put(256.54,183.40){\fontsize{6.83}{8.19}\selectfont Block 6}
  \put(256.54,123.87){\fontsize{6.83}{8.19}\selectfont Time $n$}
  \put(66.05,47.34){\fontsize{6.83}{8.19}\selectfont $b_0$ symbols}
  \put(59.24,104.31){\fontsize{6.83}{8.19}\selectfont $d$}
  \put(234.43,92.41){\fontsize{6.83}{8.19}\selectfont Time (symbols)}
  \put(59.24,19.28){\fontsize{6.83}{8.19}\selectfont Start of current block}
  \put(38.83,54.99){\fontsize{6.83}{8.19}\selectfont $j_b$}
  \put(103.46,36.28){\fontsize{6.83}{8.19}\selectfont First check point}
  \put(212.31,54.99){\fontsize{6.83}{8.19}\selectfont First symbol of next block}
  \put(200.41,36.28){\fontsize{6.83}{8.19}\selectfont Last symbol of current block}
  \put(185.10,20.98){\fontsize{6.83}{8.19}\selectfont Decoding decision made}
  \put(113.67,10.77){\fontsize{6.83}{8.19}\selectfont $k_b$ = last check point  in which decoding condition not met}
  \put(139.18,109.42){\fontsize{6.83}{8.19}\selectfont $2d-1$ unconstrained symbols}
  \put(21.83,112.82){\fontsize{8.54}{10.24}\selectfont Block 3 - a typical long block}
  \put(40.54,177.45){\fontsize{5.12}{6.15}\selectfont (short)}
  \put(94.96,177.45){\fontsize{5.12}{6.15}\selectfont (long)}
  \put(157.89,177.45){\fontsize{5.12}{6.15}\selectfont (long)}
  \put(229.32,177.45){\fontsize{5.12}{6.15}\selectfont (long)}
  \put(259.94,177.45){\fontsize{5.12}{6.15}\selectfont (long)}
  \put(202.11,177.45){\fontsize{5.12}{6.15}\selectfont (short)}
  \end{picture}%
\else
  \setlength{\unitlength}{1bp}%
  \begin{picture}(316.63, 200.98)(0,0)
  \put(0,0){\includegraphics{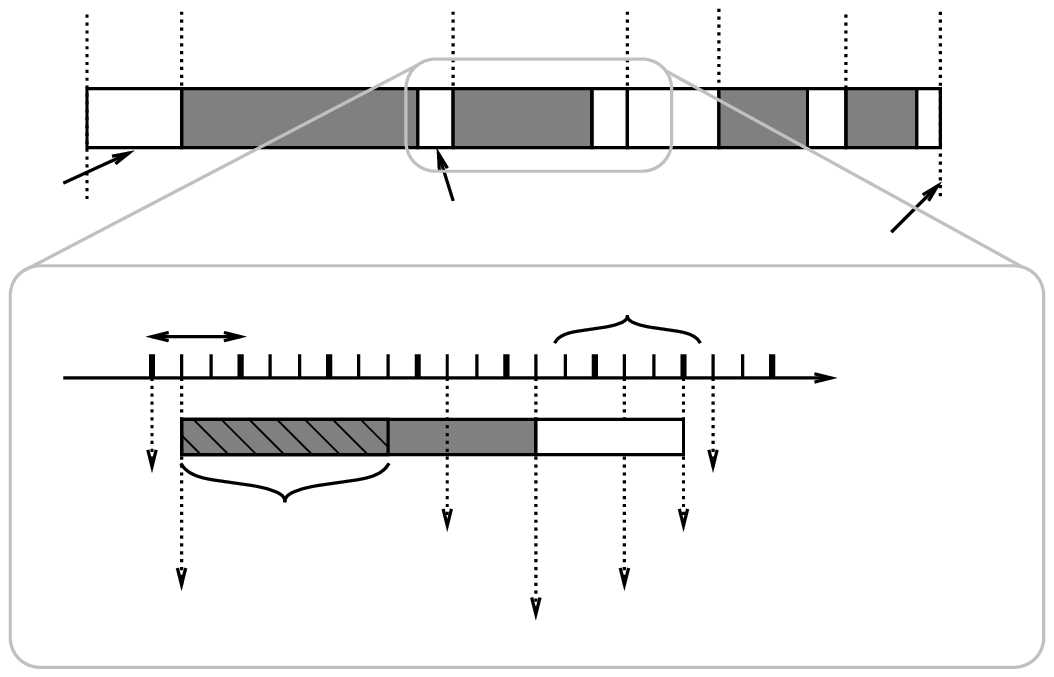}}
  \put(91.56,183.40){\fontsize{6.83}{8.19}\selectfont Block 2}
  \put(38.83,183.40){\fontsize{6.83}{8.19}\selectfont Block 1}
  \put(5.67,137.48){\fontsize{6.83}{8.19}\selectfont $< b_0 + 2d$}
  \put(137.48,133.23){\fontsize{6.83}{8.19}\selectfont $2 d - 1$}
  \put(153.64,183.40){\fontsize{6.83}{8.19}\selectfont Block 3}
  \put(193.61,183.40){\fontsize{6.83}{8.19}\selectfont Block 4}
  \put(228.47,183.40){\fontsize{6.83}{8.19}\selectfont Block 5}
  \put(256.54,183.40){\fontsize{6.83}{8.19}\selectfont Block 6}
  \put(256.54,123.87){\fontsize{6.83}{8.19}\selectfont Time $n$}
  \put(66.05,47.34){\fontsize{6.83}{8.19}\selectfont $b_0$ symbols}
  \put(59.24,104.31){\fontsize{6.83}{8.19}\selectfont $d$}
  \put(234.43,92.41){\fontsize{6.83}{8.19}\selectfont Time (symbols)}
  \put(59.24,19.28){\fontsize{6.83}{8.19}\selectfont Start of current block}
  \put(38.83,54.99){\fontsize{6.83}{8.19}\selectfont $j_b$}
  \put(103.46,36.28){\fontsize{6.83}{8.19}\selectfont First check point}
  \put(212.31,54.99){\fontsize{6.83}{8.19}\selectfont First symbol of next block}
  \put(200.41,36.28){\fontsize{6.83}{8.19}\selectfont Last symbol of current block}
  \put(185.10,20.98){\fontsize{6.83}{8.19}\selectfont Decoding decision made}
  \put(113.67,10.77){\fontsize{6.83}{8.19}\selectfont $k_b$ = last check point  in which decoding condition not met}
  \put(139.18,109.42){\fontsize{6.83}{8.19}\selectfont $2d-1$ unconstrained symbols}
  \put(21.83,112.82){\fontsize{8.54}{10.24}\selectfont Block 3 - a typical long block}
  \put(40.54,177.45){\fontsize{5.12}{6.15}\selectfont (short)}
  \put(94.96,177.45){\fontsize{5.12}{6.15}\selectfont (long)}
  \put(157.89,177.45){\fontsize{5.12}{6.15}\selectfont (long)}
  \put(229.32,177.45){\fontsize{5.12}{6.15}\selectfont (long)}
  \put(259.94,177.45){\fontsize{5.12}{6.15}\selectfont (long)}
  \put(202.11,177.45){\fontsize{5.12}{6.15}\selectfont (short)}
  \end{picture}%
\fi
\caption{\label{fig:Iterated_rateless_blocks_illustration_2}%
 An illustration of the constrained and unconstrained symbols}
\end{figure}

We now turn to analyze the rate. When $\vr X$ and $\vr y$ are given, and under the assumption that no decoding errors occurred, the decoding times are deterministic, and result in a deterministic rate. We denote by $B$ the number of blocks, including the last one which is potentially not decoded. The actual rate of the scheme satisfies
\begin{equation}\label{eq:A2028}
\Ract \geq \frac{(B-1) K}{n}
\end{equation}
We now use the summability condition to relate $\Ract$ and $\Remp$. Define by $j_b$ ($b=1,\ldots,B$) the end-time of the previous block for any of the blocks. A typical block, which is long enough, has the following time line: during the first $b_0$ symbols the decoding condition is not checked. The opportunities to send feedback are symbols $i \cdot d + 1$ in the block ($i=0,1,\ldots$). The decoding condition is checked for the first time in symbol $\lceil \frac{b_0}{d}  \rceil \cdot d + 1$ (the minimal $i \cdot d + 1$ satisfying $i \cdot d + 1 \geq b_0 + 1$). The condition may be met on this symbol, in which case the new block would begin $d$ symbols later. The block may be even terminated before this time, if time $n$ arrives. However in the typical case, as depicted in the bottom of Figure~\ref{fig:Iterated_rateless_blocks_illustration_2}, the condition is not met on this symbol, and then it is checked again each $d$ symbols, until it is finally met. For a block $b$, long enough, define $k_b$ as the last time, in which the decoding condition was checked (after location $b_0$) and did \emph{not} pass, i.e. the metric of none of the codewords, including the correct one, passed the threshold. Suppose that such a $k_b$ exists, then the decoding condition was met at time $k_b + d$, and a new block was started at time $k_b + 2d$. Therefore the length of the block is $l_b = k_b + 2 d - j_b - 1$. For a given block length $l_b$, the condition for the existence of $k_b$ is that the symbol number of this opportunity satisfies $k_b - j_b \geq b_0 + 1$, i.e. $l_b \geq b_0 + 2 d$. When this happens, the fact that the decoding condition failed at time $k_b$ yields an upper bound on the decoding metric, since we know that for the true codeword, we have:
\begin{equation}\label{eq:A2041}
\psi(\vr X^{k_b}, \vr y^{k_b}, j_b) < \psi^*_{k_b,j_b}
\end{equation}
Note that this yields a bound on the value of $\psi$ up to $2d-1$ symbols before the end of the block: after time $k_b$ there are $2d-1$ additional symbols which are not ``covered'' by this bound. For the shorter blocks, we do not have any bound on $\psi$.

We divide the $B$ blocks into a group $B_L$ of blocks whose length is at least $b_0 + 2d$ and a group $B_S$ of blocks whose length is smaller. The last block may be included in one group or the other. For the blocks in the first group, we define $j_b$ and $k_b$ as above, and we have the bound of \eqref{eq:A2041}. $(j_b,k_b)$ are interpreted as the ``segments'' referred to in the suitability condition. We effectively split the $n$ symbols into ``constrained'' symbols (contained in the segments $(j_b,k_b)$), for which we have a bound on $\psi$, and ``unconstrained'' symbols for which we do not have a bound. The summability condition allows us to relate the the overall metric $\psi_0^n$ to the values of the metric on the segments, and the number $m_0$ of ``unconstrained'' symbols. We now count the number of ``unconstrained'' symbols, i.e. those that are not covered by any segment. In each long block there are $2d-1$ unconstrained symbols, unless it is the last one, in which case there are at most $d$. And all the symbols of a short block, which are at most $b_0 +2d-1$, are unconstrained. Therefore the total number of unconstrained symbols is at most $m_0 = (2d-1) \cdot |B_L| + (b_0 + 2d - 1) \cdot |B_S|$. Substituting $|B_S| = B - |B_L|$ we may write $m_0$ as  $m_0 = (b_0 + 2d - 1) \cdot B - b_0 \cdot |B_L|$

Figure~\ref{fig:Iterated_rateless_blocks_illustration_2} illustrates the constrained and unconstrained symbols. The top of the figure shows the overall transmission time $1, \ldots, n$ divided into 6 blocks. Blocks 1,4 are short, and the rest are long. The dark parts denote the segments $(j_b,k_b)$ for which the constraint \eqref{eq:A2041} applies. The while parts denote unconstrained symbols, which occur on short blocks and at the last symbols of long blocks. The bottom of the figure illustrates the the time line of a long block, as was already discussed above.

Applying the summability condition \eqref{eq:A1945} we have:
\begin{equation}\label{eq:A1945b}
\log \psi_0^n - \sum_{b \in B_L} \log \psi(\vr X^{k_b}, \vr y^{k_b}, j_b) \leq f_0^{(n)}(\psi_0^n) \cdot m_0 = f_0^{(n)}(\psi_0^n) \cdot ((b_0 + 2d - 1) \cdot B - b_0 \cdot |B_L|)
\end{equation}

Substituting the threshold \eqref{eq:A1945b} we have:
\begin{equation}\begin{split}\label{eq:A2058}
\sum_{b \in B_L} \log \psi(\vr X^{k_b}, \vr y^{k_b}, j_b)
& \stackrel{\eqref{eq:A1945b}}{\leq}
\sum_{b \in B_L} \log \psi^*_{k_b,j_b}
\stackrel{\eqref{eq:A2006b}}{=}
\sum_{b \in B_L} \log \frac{n \cdot L_{k_b-j_b} \cdot \exp(K)}{d \epsilon}
\\& \leq
\sum_{b \in B_L} \log \frac{n \cdot L_n \cdot \exp(K)}{d \epsilon}
=
|B_L| \cdot \left( \log \frac{n \cdot L_n}{d \epsilon} + K \right)
\end{split}\end{equation}

Therefore
\begin{equation}\label{eq:A2072}
\log \psi_0^n
\stackrel{\eqref{eq:A1945b}}{\leq}
\underbrace{|B_L| \cdot \left( \log \frac{n \cdot L_n}{d \epsilon} + K \right)
+ f_0^{(n)}(\psi_0^n) \cdot ((b_0 + 2d - 1) \cdot B - b_0 \cdot |B_L|)}_{ \defeq \rho(|B_L|)}
\end{equation}
The above expression, denoted $\rho(|B_L|)$, is a linear function of $|B_L|$. Not knowing $|B_L|$, we may upper bound this expression by its maximum value $\max_{0 \leq |B_L| \leq B} \rho(|B_L|)$. Due to the linearity, the maximum is always obtained at the edges $|B_L| \in \{0,B\}$, therefore
\begin{equation}\label{eq:A2081}
\rho(|B_L|) \leq \max_{0 \leq |B_L| \leq B} \rho(|B_L|) =  \max_{|B_L| \in \{0,B\}} \rho(|B_L|) = \max(\rho(0), \rho(B)) =  \rho(0) + [\rho(B) - \rho(0)]^+
\end{equation}
where $[x]^+ \defeq \max(x,0)$. Substituting in \eqref{eq:A2072} we have:
\begin{equation}\begin{split}\label{eq:A2085}
\log \psi_0^n
&\leq
f_0^{(n)}(\psi_0^n) \cdot (b_0 + 2d - 1) \cdot B + B \cdot \left[ \log \frac{n \cdot L_n}{d \epsilon} + K - f_0^{(n)}(\psi_0^n) \cdot b_0 \right]^+
\\&\leq
f_0^{(n)}(\psi_0^n) \cdot (b_0 + 2d - 1) \cdot B + B \cdot \left[ \log \frac{n \cdot L_n}{d \epsilon} + K  \right]^+
\\& \stackrel{d \leq n, \epsilon \leq 1}{=}
\left( f_0^{(n)}(\psi_0^n) \cdot (b_0 + 2d - 1)  + \log \frac{n \cdot L_n}{d \epsilon} + K  \right) \cdot B
\end{split}\end{equation}
Extracting a lower bound on $B$ from \eqref{eq:A2085} we have:
\begin{equation}\begin{split}\label{eq:A2090}
\Ract
& \stackrel{\eqref{eq:A2028}}{\geq}
\frac{(B-1) K}{n}
\\& \geq
\left[ \frac{\log \psi_0^n}{f_0^{(n)}(\psi_0^n) \cdot (b_0 + 2d - 1) +  \log \frac{n \cdot L_n}{d \epsilon} + K} - 1 \right] \cdot \frac{K}{n}
\\& =
\frac{\frac{1}{n} \log \psi_0^n}{1 + \frac{1}{K} \left( \log \frac{n \cdot L_n}{d \epsilon}  + f_0^{(n)}(\psi_0^n) \cdot (b_0 + 2d - 1) \right)} - \frac{K}{n}
\\& =
\frac{\Remp}{1 + \frac{1}{K} \left( \underbrace{\log \frac{n \cdot L_n}{d \epsilon}}_{c_n}  + f_0^{(n)}(\exp(n \Remp)) \cdot \underbrace{(b_0 + 2d - 1)}_{b_1} \right)} - \frac{K}{n}
\end{split}\end{equation}
This proves the main claim of the theorem. Regarding the sufficient Markov-based CCDF condition \eqref{eq:A1933b}, it is easy to see that if \eqref{eq:A1933b} holds then the bound \eqref{eq:A1933} is obtained by applying Markov inequality \eqref{eq:Amarkov_bound}.
\endofproof

\textit{Proof of Corollary~\ref{corollary:framework_asymptote}: \\}
The proof is completely technical, by showing that under the conditions $F_n(t) \ntoinfty 0$. If $\frac{1}{n} \log L_n \ntoinfty 0$ then there exists a sequence $\Delta_n \in [0,1], \Delta_n \to 0$ such that  $\frac{1}{n \cdot \Delta_n} \log L_n \ntoinfty 0$. As an example we can choose $\Delta_n = \min \left( \sqrt{\frac{1}{n} \log L_n}, 1 \right)$. We choose $K = n \cdot \max \{\Delta_n, \delta_n, n^{-1/2} \}$. Then for all $t$, the term in the denominator of $F_n(t)$ \eqref{eq:A1962} satisfies:
\begin{equation}\begin{split}\label{eq:A1983}
\frac{c_n  + b_1 \cdot f_0^{(n)}(\exp(n t))}{K}
&=
\frac{\log \frac{n}{d \epsilon} + \log L_n  + b_1 \cdot f_0^{(n)}(\exp(n t))}{n \cdot \max \{\Delta_n, \delta_n, n^{-1/2} \}}
\\& \leq
\underbrace{    \frac{\log \frac{n}{d \epsilon}}{\sqrt{n}}  }_{\to 0}
+
\underbrace{    \frac{ \log L_n }{n \cdot \Delta_n} }_{\to 0}
+
b_1 \cdot \underbrace{    \frac{f_0^{(n)}(\exp(n t))}{n \cdot \delta_n} }_{\to 0}
\ntoinfty
0
\end{split}\end{equation}
and in addition
\begin{equation}\label{eq:A1998}
\frac{K}{n} = \max \{\Delta_n, \delta_n, n^{-1/2} \} \ntoinfty 0
\end{equation}
Therefore $F_n(t) \to t$, and by definition, $\Remp$ is asymptotically achievable. \endofproof

Note that the condition on $f_0$ is essentially that $\forall t: \frac{f_0^{(n)}(\exp(n t))}{n} \ntoinfty 0$, however it was defined by using a sequence $\delta_n$ since the convergence is not necessarily uniform in $t$, therefore it is not always possible to extract a sequence $\delta_n$ from $f_0^{(n)}$ itself (as we have done for the other overhead sequence $\frac{1}{n} \log L_n$).

\textit{Proof of Corollary~\ref{corollary:framework_redundancy}: \\}
Define $f_0^{(n)*} \defeq \max_{t \leq R_{\max}} f_0^{(n)}(\exp(n t))$, and bound $F_n(t)$ of \eqref{eq:A1962} for all $t \leq R_{\max}$ as
\begin{equation}\begin{split}\label{eq:A2013}
F_n(t)
& \geq
\frac{t}{1 + \frac{c_n  + b_1 \cdot f_0^{(n)*}}{K}} - \frac{K}{n}
\stackrel{k_n \defeq c_n  + b_1 \cdot f_0^{(n)*}}{=}
\frac{t}{1 + \frac{k_n}{K}} - \frac{K}{n}
\\& \stackrel{\frac{1}{1+x} \geq 1-x}{\geq}
t \cdot \left( 1 - \frac{k_n}{K} \right) - \frac{K}{n}
=
t - t \cdot \frac{k_n}{K} - \frac{K}{n}
\\& \geq
t - \underbrace{\left[ R_{\max} \cdot \frac{k_n}{K} + \frac{K}{n} \right]}_{\defeq \delta_n}
\end{split}\end{equation}

with $k_n = c_n  + b_1 \cdot f_0^{(n)*}$.

We choose the value of $K$ that minimizes the overhead term $\delta_n$ in the lower bound, using the following lemma:
\begin{lemma}\label{lemma:ab_bound}
For $a>0, b>0$ with $b \leq a$
\begin{equation}
r = \min_{k \in \mathbb{N}} \left[ \frac{a}{k} + b k \right] \leq 3 \sqrt{ab}
\end{equation}
\end{lemma}

\textit{Proof of the lemma:} It is easy to see by derivation that the minimizer over $x \in \mathbb{R}$ of $\frac{a}{x} + b x$ is $x^* = \sqrt{\frac{a}{b}}$. Choosing $k^* = \lceil x^* \rceil$ we have $k^* \in \mathbb{N}$ and since $\sqrt{\frac{a}{b}} \leq k^* \leq \sqrt{\frac{a}{b}} +1$:
\begin{equation}
\frac{a}{k^*} + b k^*
\leq
\frac{a}{\sqrt{\frac{a}{b}}} + b \left(\sqrt{\frac{a}{b}} +1 \right)
=
2\sqrt{ab} + b
=
2\sqrt{ab} + \sqrt{b \cdot b}
\stackrel{b \leq a} \leq
3\sqrt{ab}
\end{equation}
\endofproof

applying the lemma with $a = R_{\max} k_n, b = \frac{1}{n}$ we obtain
\begin{equation}\label{eq:A2052}
\delta_n \leq 3 \sqrt{\frac{R_{\max} k_n}{n}} = 3 \sqrt{\frac{R_{\max} \cdot (c_n  + b_1 \cdot f_0^{(n)*})}{n}}
\end{equation}

Since $k_n$ grows with $n$, asymptotically $a >> b$, and therefore the result in Lemma~\ref{lemma:ab_bound} is closer to $2 \sqrt{ab}$, and the factor in $\delta_n$ approaches $2$, however this coarse bound was chosen for its simplicity, as it doesn't change the order of magnitude. For large $n$ we can use the approximation $2 \sqrt{ab}$.
\endofproof

\textit{Proof of Corollary~\ref{corollary:framework_nonadaptive}:}
This stems directly from the CCDF condition, computed for $k=n, j=0$:
\begin{equation}\label{eq:A2417}
\placeunder{\Pr}{Q} \left\{ \psi(\vr X, \vr y, 0) \geq t \right\} \leq \frac{L_{n}}{t}
\end{equation}
Therefore the intrinsic redundancy \eqref{eq:Adef_intrinsic_redundancy} is
\begin{equation}\begin{split}\label{eq:A2421}
\mu_Q(\Remp)
&=
\sup_{\vr y,R \in \mathbb{R}} \left\{ \frac{1}{n} \log Q \{\Remp(\vr X, \vr y) \geq R\}  + R \right\}
\\&=
\sup_{\vr y,R \in \mathbb{R}} \left\{ \frac{1}{n} \log Q \{\psi(\vr X, \vr y, 0) \geq \exp(nR) \}  + R \right\}
\\&\leq
\sup_{\vr y,R \in \mathbb{R}} \left\{ \frac{1}{n} \log \frac{L_{n}}{\exp(nR)}  + R \right\}
\\&=
\frac{1}{n} \log L_{n}
\end{split}\end{equation}
\endofproof

\subsection{A conditional probability based empirical rate}\label{sec:Remp_adaptive_conditional}
In Section~\ref{sec:chernoff_cond_probability} we have shown that, asymptotically, all maximum attainable rate functions are of the form $\Remp(\vr x, \vr y) = \frac{1}{n} \log \frac{P(\vr x | \vr y)}{Q(\vr x)}$. We now present a specific ``causal'' structure for $P$ and show that with this structure $\Remp$ can also be \emph{adaptively} attained. The set of $P(\cdot|\cdot)$ we use is based on a ``causality'' condition.

\begin{definition}[causality]\label{def:D_causality}
A conditional probability distribution $P(\vr x | \vr y)$ defined over $\vr x \in \mathcal{X}^n, \vr y \in \mathcal{Y}^n$ is said to be $D$-causal (for some non-negative $D$), if for all $k \leq n$:
\begin{equation}\label{eq:A2292}
P(\vr x^k | \vr y) = P(\vr x^k | \vr y^{k+D})
\end{equation}
i.e. computing the conditional probability of a sub-vector only requires considering $D$ future symbols of $\vr y$.
\end{definition}

An equivalent condition is that $P(\vr x^n | \vr y)$ can be written as $P(\vr x^n | \vr y) = \prod_{i=1}^n P(x_i | \vr y^{i+D}, \vr x^{i-1})$. This is since we can always write $P(\vr x^n | \vr y) = \prod_{i=1}^n P(x_i | \vr y, \vr x^{i-1})$, and in this case $P(\vr x^k | \vr y) = \sum_{\vr x_{k+1}^n} P(\vr x^n | \vr y) = \prod_{i=1}^k P(x_i | \vr y, \vr x^{i-1})$, and the later should be a function of $\vr y^{k+D}$ for any $k$. Unfortunately, the causality we defined, and which is needed for the adaptive achievability, is the causality of the backward channel (from $\vr y$ to $\vr x$). Most channel models define a causal relation from $\vr x$ to $\vr y$, and if there is memory in the channel, the backward channel will not be causal, in general. To accommodate such cases we have allowed a dependence on $D$ future symbols of $\vr y$ (see example \unfinished{***}). Softer conditions can be defined instead of the strict equality in \eqref{eq:A2292} however this requirement is sufficient for our purposes.
\todo{mention that this is like "causal conditioning" defined by Kramer, after shifting $\vr y$ by $D+1$ samples (for $D=-1$ we obtain Kramer's definition). The references can be found in "to feed back or not to feed back" (from ISIT 2011) and \cite{FiniteStateChannelsPermuter09}}

\todo{maybe the following structure: put the definitions of causality (all of the above) at the beginning of the chapter, and here, only the theorem and the proof. Then, leading to the framework theorem, even before the theorem is presented, we can explain the system's operation for $P/Q$. This will also replace the paragraph after the theorem saying that $\psi=P/Q$ is a reasonable choice and will help rationalize it}

Given a $D$-causal distribution $P$, we define the following decoding metric:
\begin{equation}\label{eq:A2306}
\psi(\vr x^k, \vr y^k, j) = \frac{P(\vr x^{k-D} | \vr y^k)}{Q(\vr x^{k})} \cdot \left( \frac{P(\vr x^{j-D} | \vr y^j)}{Q(\vr x^{j})} \right)^{-1} = \frac{P \left(\vr x_{j+1-D}^{k-D} \big| \vr y^k, \vr x^{j-D} \right)}{Q \left(\vr x_{j+1}^{k} \big| \vr x^j \right)}
\end{equation}
Note that for $k \leq D$ $\vr x^{k-D}$ is simply the empty set and in this case we define $P(\vr x^{k-D} | \vr y^k) = 1$. The equality holds by using Bayes rule and since due to $D$-causality we can replace $P(\vr x^{j-D} | \vr y^j)$ by $P(\vr x^{j-D} | \vr y^k)$. Note that we can write (for $k \geq j$):
\begin{equation}\label{eq:A2312}
\psi(\vr x^k, \vr y^k, 0) =  \psi(\vr x^j, \vr y^j, 0) \psi(\vr x^k, \vr y^k, j)
\end{equation}
Note that the above is analogous to Bayes rule. We will assume that $\vr x$ is discrete and therefore $P(\vr x^k | \vr y) \leq 1$. Regarding the conditional distribution of the input we make the assumption that any symbol that has non-zero probability, has a probability of at least $q_{\min}$, i.e. for all $k$, $Q(x_k | \vr x^{k-1}) \in \{0\} \cup [q_{\min}, 1]$. Under these assumptions $\psi$ defined above satisfies the conditions of Theorem~\ref{theorem:framework} and its corollaries, and we have the following result:

\begin{theorem}\label{theorem:adaptive_causal_distribution}
Let $(\vr x, \vr y) \in \mathcal{X}^n \times \mathcal{Y}^n$ where $\vr x$ is discrete ($|\mathcal{X}| < \infty$), and let $Q(\vr x)$ be an input distribution that satisfies $\forall k: Q(x_k | \vr x^{k-1}) \in \{0\} \cup [q_{\min}, 1]$. Let $P(\vr x | \vr y)$ be a $D$-causal conditional distribution. Define the following rate function:
\begin{equation}\label{eq:A2323}
\Remp = \frac{1}{n} \log \frac{P(\vr x^n | \vr y^n)}{Q(\vr x^{n})}
\end{equation}
Then:
\begin{enumerate}
\item The scheme of Section~\ref{sec:rate_adaptive_scheme}, with $\psi$ defined in \eqref{eq:A2306} and $\psi^* = \frac{n \cdot |\mathcal{X}|^D \cdot \exp(K)}{d \epsilon}$, adaptively achieves $F_n(\Remp)$, where $F_n(t) = \left(1 + \frac{c_n  + (2\dfb - 1) \cdot \log q_{\min}^{-1}}{K} \right)^{-1} \cdot t - \frac{K}{n}$, with $c_n = \log \frac{n \cdot |\mathcal{X}|^D}{d \epsilon}$
\item $\Remp$ is adaptively achievable up to $\delta_n = 3 \sqrt{\frac{\log q_{\min}^{-1} \cdot (c_n  + (2\dfb - 1) \cdot \log q_{\min}^{-1})}{n}}$
\item $\Remp$ is asymptotically adaptively achievable
\end{enumerate}
\end{theorem}

Note: as in Corollary~\ref{corollary:framework_redundancy}, for small $\epsilon$ and large $n$, $\delta_n \approx 2 \sqrt{\frac{\log \frac{n}{\epsilon}}{n}}$.

\textit{Proof:}
What we will actually prove is the attainability of the rate function
\begin{equation}\label{eq:A2378}
\Remp' = \frac{1}{n} \log \psi_0^n = \frac{1}{n} \log \frac{P \left(\vr x^{n-D} \big| \vr y \right)}{Q \left(\vr x \right)}
\end{equation}
Since $P \left(\vr x^{n} \big| \vr y \right) = P \left(\vr x^{n-D} \big| \vr y \right) \cdot P \left(\vr x_{n-D+1}^n \big| \vr y \vr x^{n-D} \right) \leq  P \left(\vr x^{n-D} \big| \vr y \right)$, we have that $\Remp \leq \Remp'$ and therefore the achievability of $\Remp'$ shows the achievability of $\Remp$. The adaptive achievability of the rate function above is given by Theorem~\ref{theorem:framework}, when the conditions hold. Below we prove the conditions hold:

\textit{CCDF condition:} we use the Markov sufficient condition. By plugging the second form in \eqref{eq:A2306}:
\begin{equation}\begin{split}\label{eq:A2324}
\placeunder{\E}{Q} \left[ \psi(\vr X^k, \vr y^k, j)  | \vr x^j \right]
&=
\sum_{x_{j+1}^k \in \mathcal{X}^{k-j}} \psi(\vr x^k, \vr y^k, j) \cdot Q \left(\vr x_{j+1}^{k} \big| \vr x^j \right)
\\& \stackrel{\eqref{eq:A2306}}{=}
\sum_{x_{j+1}^k \in \mathcal{X}^{k-j}} P \left(\vr x_{j+1-D}^{k-D} \big| \vr y^k, \vr x^{j-D} \right)
\end{split}\end{equation}
If $k-j > D$, then we continue as follows:
\begin{equation}\begin{split}\label{eq:A2328}
\placeunder{\E}{Q} \left[ \psi(\vr X^k, \vr y^k, j)  | \vr x^j \right]
&=
\sum_{x_{k-D+1}^k \in \mathcal{X}^{D}} \sum_{x_{j+1}^{k-D}} P \left(\vr x_{j+1-D}^{k-D} \big| \vr y^k, \vr x^{j-D} \right)
=
\sum_{x_{k-D+1}^k \in \mathcal{X}^{D}} P \left(\vr x_{j+1-D}^{j} \big| \vr y^k, \vr x^{j-D} \right)
\\&\leq
\sum_{x_{k-D+1}^k \in \mathcal{X}^{D}} 1
=
|\mathcal{X}|^D
\end{split}\end{equation}
If $k-j \leq D$ then the same bound holds based on \eqref{eq:A2324} (the number of elements in the sum is at most $\mathcal{X}^{D}$). Therefore the condition is satisfied with $L_{k-j} = |\mathcal{X}|^D$ and $b_0 = 0$ (i.e. holds for any value of $k-j$).

\textit{Summability:}
Let $\{j_b, k_b\}_{b=1}^B$ be a set of segments as defined in the summability condition of Theorem~\ref{theorem:framework}, and let $\psi_b$ be as defined there. Let $A$ denote the set of indices not included in the segments, with $|A|=m_0$.
Using the condition on the input we have, for every sequence $\vr x$ with non-zero probability:
\begin{equation}\label{eq:A2319}
\psi(\vr x^k, \vr y^k, k-1) \stackrel{\eqref{eq:A2306}}{\leq} \frac{1}{Q \left(\vr x_k \big| \vr x^{k-1} \right)} \leq \frac{1}{q_{\min}}
\end{equation}
We recursively use \eqref{eq:A2312} to write $\psi_0^n$ as a product of $\psi$ over the segments and the $\psi(\vr x^i, \vr y^i, i-1)$ over the un-included symbols.
\begin{equation}\label{eq:A2325}
\psi_0^n = \psi(\vr x^n, \vr y^n, 0) = \prod_{b=1}^B \psi(\vr x^{k_b}, \vr y^{k_b}, j_b) \cdot \prod_{i \in A} \psi(\vr x^i, \vr y^i, i-1) \leq \prod_{b=1}^B \psi_b \cdot q_{\min}^{-|A|} = \prod_{b=1}^B \psi_b \cdot q_{\min}^{-m_0}
\end{equation}
Taking logarithm, we obtain the summability condition with $f_0^{(n)} = \log q_{\min}^{-1}$
\begin{equation}\label{eq:A2331}
\log \psi_0^n -  \sum_{b=1}^B \log  \psi_b  \leq  m_0 \cdot \underbrace{\log q_{\min}^{-1}}_{f_0^{(n)}}
\end{equation}

Property (1) in Theorem~\ref{theorem:adaptive_causal_distribution} is proven by directly plugging these values into Theorem~\ref{theorem:framework} and using $\Remp \leq \Remp'$. Further more, $\Remp'$ is upper bounded by $\Remp' \leq \log q_{\min}^{-1}$, due to the constraint on $Q$. This can be shown by definition but also derived from the summability condition with $B=0$ and hence $m_0=n$. Property (2) is shown by using Corollary~\ref{corollary:framework_redundancy} with $R_{\max} = \log q_{\min}^{-1}$. Property (3) can be shown by using Corollary~\ref{corollary:framework_asymptote}, or either of the previous properties.
\endofproof

Note that by \eqref{eq:A2324} for the case $D=0$, the CCDF condition holds also if the distribution is continuous (i.e. $P(\vr x | \vr y), Q(\vr x)$ are density functions and are not upper bounded by $1$), since the sum will be replaced by the integral of $P \left(\vr x_{j+1}^{k} \big| \vr y^k, \vr x^{j} \right)$ over $x_{j+1}^k \in \mathcal{X}^{k-j}$, which is one. The summability condition may hold with a different $f_0^{(n)}$, if $P(x_i | \vr x^{i-1}, \vr y)$ is bounded. Therefore in the general case if $P(\vr x | \vr y)$ is strictly causal (with $D=0$) $P(x_i | \vr x^{i-1}, \vr y)$ is upper bounded, and $Q(x_i | \vr x^{i-1})$ is lower bounded, the $\Remp$ of \eqref{eq:A2323} is asymptotically achievable.

It is worthwhile spending a few words on the limitation $Q(x_k | \vr x^{k-1}) \in \{0\} \cup [q_{\min}, 1]$. This limitation relates to the summability condition, where $f_0$ reflects the loss due to the fact we do not have a constraint on all the symbols as expressed in \eqref{eq:A1945b}. As an example, suppose that $\dfb=1$. We know that one symbol before the end of a block in the scheme, $\psi \leq \psi^*$. In the next symbol, the metric exceeds the threshold, but we do not have a bound by how much it exceeds it, and this gap is expressed by a loss with respect to the ideal rate function. If we let one of the values of $x_k$ have a very low a-priori probability, this symbol occurs, and has a high aposteriori probability $P(x_k | \vr y, \vr x_{k-1})$ after seeing $\vr y$, then the growth of the metric, $\frac{P(x_k | \vr y, \vr x_{k-1})}{Q(x_k |\vr x_{k-1})}$ may be unlimited. In \cite{YL_individual_full} we did not use this constraint but as a result (of this and other technical reasons), had to define a set of sequences $\vr x$ for which the assertions do not hold. This is further discussed in Section~\unfinished{***}. In the discrete case, this condition is plausible, and we can find a $q_{\min}$ for any $Q$ since there is always a minimum value to the non-zero probabilities. We will also use a similar constraint, from similar reasons, for the continuous case. In this case, the condition changes the distribution, but it can be considered a replacement to a ``failure set'' as was defined in \cite{YL_individual_full}, and its purpose is to prevent symbols which have unlimited contribution to the rate function.

\subsection{The ML rate function}\label{sec:rate_adaptivity_RempML}
We have presented the ML rate function in Section~\ref{sec:ML_rate_functions}:
\begin{equation}\label{eq:A2433}
\Rempname{\ML} = \max_{\theta \in \Theta} \frac{1}{n} \log \frac{P_{\theta}(\vr x | \vr y)}{Q(\vr x)}
=
\frac{1}{n} \log \frac{\hat p_{\ML}(\vr x | \vr y)}{Q(\vr x)}
\end{equation}
Where $P_{\theta}(\vr x | \vr y)$ be a family of conditional distributions, indexed by a parameter $\theta \in \Theta$, and $\hat p_{\ML}$ is the maximum likelihood conditional probability \eqref{eq:A1610c}. Our purpose is now to attain this rate function adaptively, up to overhead terms. We will present some general cases in which is it possible to do so. Note that achieving $\Rempname{\ML}$ adaptively also means achieving $\Rempname{\ML*}$ of \eqref{eq:A1696} adaptively.

\subsubsection{The discrete case based on a weight function}\label{sec:RempML_adaptive_redundancy_by_weighting}
The first case of interest is when there exists a weighting function over $\Theta$, denoted $w(\theta)$, with $\int_{\Theta} w(\theta) d \theta = 1$, and a constant $C_n$ such that
\begin{equation}\label{eq:A2444}
\hat p_{\ML}(\vr x | \vr y) = \max_{\theta \in \Theta} P_{\theta}(\vr x | \vr y) \leq C_n \cdot \int_{\theta \in \Theta} w(\theta) P_{\theta}(\vr x | \vr y) d \theta
\end{equation}
Where the constant $C_n$ grows sub-exponentially with $n$. The term $\int_{\theta \in \Theta} w(\theta) P_{\theta}(\vr x | \vr y) d \theta$ is sometimes termed the ``Bayesian mixture'' of the distributions $P_{\theta}$ with a prior $w(\theta)$. Mixtures of this type appear as solutions to the minimax redundancy problem \cite{FederMerhav98}\cite{Barron_MDL} (sometimes termed ``average regret''), seeking to minimize the maximum divergence $D(P^*||P_{\theta})$ between a universal distribution $P^*$ and the set of distributions $\{ P_{\theta} \}$, while we require the relation in \eqref{eq:A2444} to hold per point $(\vr x, \vr y)$. Our target in \eqref{eq:A2444} of upper bounding $\hat p_{\ML}(\vr x | \vr y)$ is related to the problem of minimax regret \cite{FederMerhav98} which was discussed in Section~\ref{sec:redundancy_of_Remp_ML}, i.e. the problem of finding a distribution $P^*(\vr x, \vr y)$ which is close to $\hat p_{\ML}(\vr x | \vr y)$ in the sense of minimizing the maximum regret $\max_{\vr x} \log \frac{\hat p_{\ML}(\vr x | \vr y)}{P^*(\vr x, \vr y)}$. For the class of conditionally memoryless distributions it was observed by Xie and Barron \cite{XieBarron00} that the the Dirichlet-$\half$ Bayesian mixture which is the asymptotically optimal solution to the minimax redundancy problem, also yields yields a nearly optimum maximum regret. See Section~\unfinished{**} for further details.

Suppose for example that the number of $\theta$-s achieving the maximum is sub-exponential. An example of such a case is when $P_{\theta}(\vr x | \vr y)$ is the memoryless distribution where $\theta$ is the single-letter conditional distribution. In this case the $\theta$ achieving the maximum is the empirical distribution (see Section~\ref{sec:ML_probability}), and the number of empirical distributions (conditional types) is bounded by $(n+1)^{|\mathcal{X}| \cdot |\mathcal{Y}|}$. Similarly, if $P_{\theta}(\vr x | \vr y)$ is defined by higher order conditional distributions, the number of maximizing $\theta$-s can be polynomially bounded. Denote by $\tilde{\Theta}$ the set of $\theta$-s that may achieve the maximum, and assume $|\tilde{\Theta}| \leq C_n$. Then \eqref{eq:A2444} holds in a straight-forward way by defining a uniform $w(\theta)$ over $\tilde{\Theta}$, i.e. use a discrete weighting that gives a weight $\frac{1}{|\tilde{\Theta}|}$ for every $\theta \in \tilde{\Theta}$ and zero otherwise. In this case:
\begin{equation}\label{eq:A2455}
\max_{\theta \in \Theta} P_{\theta}(\vr x | \vr y)
=
\max_{\theta \in \tilde{\Theta}} P_{\theta}(\vr x | \vr y)
\leq
\sum_{\theta \in \tilde{\Theta}} P_{\theta}(\vr x | \vr y)
\leq
C_n \cdot \sum_{\theta \in \tilde{\Theta}} \underbrace{\frac{1}{|\tilde{\Theta}|}}_{w(\theta)} P_{\theta}(\vr x | \vr y)
\end{equation}

However the number of maximizing $\theta$-s is a rather coarse bound, and a better bound may be obtained by assuming that $P_{\theta}(\vr x | \vr y)$ is smooth in $\theta$, and therefore if the $\theta$ achieving the maximum in the LHS of \eqref{eq:A2444} is $\theta^*$, the integral on the RHS includes a volume surrounding $\theta^*$ in which $P_{\theta}(\vr x | \vr y)$ is close to $P_{\theta^*}(\vr x | \vr y)$. Therefore, the integral in the RHS contains an integral over this volume of $w(\theta)$, rather than just the contribution of $w(\theta^*)$, and as a result the integral is larger than the simplistic bound of \eqref{eq:A2455}, and the coefficient $C_n$ can be reduced.

Supposing \eqref{eq:A2444} is satisfied, and assuming $P_{\theta}(\vr x | \vr y)$ is $D$-causal, then it is easy to see that the weighted distribution
\begin{equation}\label{eq:A2307}
P_w(\vr x | \vr y) = \int_{\Theta} w(\theta) P_{\theta}(\vr x | \vr y) d \theta
\end{equation}
is also $D$-causal, since $P_w(\vr x^k | \vr y)$ is only a function of $\vr y^k$:
\begin{equation}\label{eq:A2319b}
P_w(\vr x^k | \vr y) = \sum_{\vr x_{k+1}^n} P_w(\vr x | \vr y) = \int_{\Theta} w(\theta) \sum_{\vr x_{k+1}^n} P_{\theta}(\vr x | \vr y) d \theta =
\int_{\Theta} w(\theta) P_{\theta}(\vr x^k | \vr y) d \theta = \int_{\Theta} w(\theta) P_{\theta}(\vr x^k | \vr y^k) d \theta
\end{equation}
It is interesting to note that the fact $P_{\theta}(\vr x | \vr y)$ is $D$-causal, does not mean $\hat p_{\ML}(\vr x | \vr y)$ is $D$-causal (only that $P_w$ is, and $P_w$ can be used to upper bound $\hat p_{\ML}$). Therefore we have that
\begin{equation}\label{eq:A2433d}
\Rempname{\ML} = \frac{1}{n} \log \frac{\hat p_{\ML}(\vr x | \vr y)}{Q(\vr x)} \stackrel{\eqref{eq:A2444}}{\leq}  \frac{1}{n} \log \frac{C_n P_w(\vr x | \vr y)}{Q(\vr x)}
= \frac{1}{n} \log \frac{P_w(\vr x | \vr y)}{Q(\vr x)} + \frac{\log C_n}{n}
\end{equation}
if the conditions of Theorem~\ref{theorem:adaptive_causal_distribution} hold with respect to $P_{\theta}$ and $Q$, then $\frac{1}{n} \log \frac{P_w(\vr x | \vr y)}{Q(\vr x)}$ is adaptively achievable up to the factors given by Theorem~\ref{theorem:adaptive_causal_distribution}, and therefore $\Rempname{\ML}$ will be achievable up to these factors plus $\frac{\log C_n}{n}$ (which tends to zero with $n$ if $C_n$ is sub-exponential). The conclusions from this discussion are formalized in the following theorem:

\begin{theorem}\label{theorem:RempML_adaptive_redundancy_by_weighting}
Let $P_{\theta}(\vr x | \vr y)$ be a family of conditional distributions, indexed by a parameter $\theta \in \Theta$, and $\hat p_{\ML}$ be the maximum likelihood conditional probability \eqref{eq:A1610c}. If the conditions of Theorem~\ref{theorem:adaptive_causal_distribution} hold with respect to $P_{\theta}$ and $Q$, and \eqref{eq:A2444} holds, then $\Rempname{\ML}=\frac{1}{n} \log \frac{\hat p_{\ML}(\vr x | \vr y)}{Q(\vr x)}$ defined in \eqref{eq:A2433} is adaptively achievable up to $\delta_n' = \delta_n + \frac{\log C_n}{n}$, where $\delta_n$ is defined in Theorem~\ref{theorem:adaptive_causal_distribution}. If further $\frac{\log C_n}{n} \ntoinfty 0$, then $\Rempname{\ML}$ is asymptotically adaptively achievable.
\end{theorem}

Note: if \eqref{eq:A2444} is satisfied then the intrinsic redundancy of $\Rempname{\ML}$ satisfies
\begin{equation}\begin{split}\label{eq:A3076}
\mu_Q(\Rempname{\ML})
& \stackrel{\eqref{eq:Adef_L_Fn}, \eqref{eq:A756}}{\leq}
\frac{1}{n} \log \underset{Q}{\E} \left[ \exp(n \Rempname{\ML}(\vr X, \vr y)) \right]
=
\frac{1}{n} \log  \underset{Q}{\E} \left[ \frac{\hat p_{\ML}(\vr x | \vr y)}{Q(\vr x)} \right]
\\& \leq
\frac{1}{n} \log  \left\{ C_n \cdot \int_{\theta \in \Theta} w(\theta)  \underbrace{\underset{Q}{\E} \left[ \frac{P_{\theta}(\vr x | \vr y)}{Q(\vr x)} \right]}_{1} d \theta \right\}
=
\frac{1}{n} \log C_n
\end{split}\end{equation}
Therefore by Theorem~\ref{theorem:remp_achievability_upto} it is achievable (non adaptively) up to $\frac{\log \frac{1}{\epsilon}}{n}  + \frac{\log C_n}{n}$. The last term, which is related to the complexity of the parametric class is common to the adaptive and non-adaptive case. The first term increases from $\frac{\log \frac{1}{\epsilon}}{n}$ in the non-adaptive case to $\delta_n = \Theta \left(\sqrt{\frac{\log \frac{n}{\epsilon}}{n}} \right)$ of Theorem~\ref{theorem:adaptive_causal_distribution} in the adaptive case. I.e. the penalty payed for the error probability increases by a square root, and an additional redundancy of $\Theta \left(\sqrt{\frac{\log n}{n}} \right)$ is added. In many cases $\frac{\log C_n}{n}$ decays to $0$ like $\Theta \left(\frac{\log n}{n} \right)$, i.e. faster than $\delta_n$, and therefore the main overhead is due to the rate adaptivity scheme, and not for the complexity of the class.

\subsubsection{The conditionally memoryless discrete case}\label{sec:RempML_adaptive_redundancy_empirical}
In Theorem~\ref{theorem:RempML_adaptive_redundancy_by_weighting} we characterized the redundancy achieved by the adaptivity scheme using the factor $C_n$ from \eqref{eq:A2444}. The additional redundancy related to the parametric class according to Theorem~\ref{theorem:RempML_adaptive_redundancy_by_weighting} is $\frac{\log C_n}{n}$. We now give an expression for $C_n$ for the conditionally memoryless case, based on known results on minimax regret.

Let $\vr z \in \mathcal{Z}^n$ be a vector of states, which may have an arbitrary dependence on $\vr x$ and $\vr y$. In the simplest case $\vr z = \vr y$. Our parameter class $\Theta$ is the class of memoryless conditional distributions of $\vr x$ given $\vr z$, defined by the conditional probability function $\theta(x|z)$ with $x \in \mathcal{X}, z \in \mathcal{Z}$ and where $\sum_{x \in \mathcal{X}} \theta(x|z) = 1$. The probability of $\vr x$ is:
\begin{equation}\label{eq:A3439}
P_{\theta}(\vr x | \vr y) = \prod_{i=1}^n \theta(x_i | z_i)
\end{equation}
The functional dependence of $\vr z$ in $\vr x, \vr y$ is implicit in \eqref{eq:A3439}, i.e. for any value of $\vr x, \vr y$ we first calculate the vector $\vr z$ and apply it to \eqref{eq:A3439}. In order for $P_{\theta}(\vr x | \vr y)$ to be a probability (i.e. sum to unity over $\vr x \in \mathcal{X}^n$), we need to assume that $x_i$ does not affect $z_i$. Specifically, we restrict $z_i$ to depend only on the \emph{past} of $\vr x$, i.e. on $x_1^{i-1}$ and the entire $\vr y$. In this case it is easy to see that \eqref{eq:A3439} defines a legitimate probability (by summing first on $x_n$ and then on $x_{n-1}$, etc). This distribution was discussed in Section~\ref{sec:ML_probability} where it was shown that the maximum likelihood solution is the empirical conditional probability, and therefore the maximum likelihood probability $\hat p_{\ML}$ is the empirical conditional probability.

Since all $|\mathcal{X}| \cdot |\mathcal{Z}|$ elements of the conditional probability vectors are in $\left\{ \frac{i}{n} \right\}_{i=0}^n$, the maximum always occurs within a limited set $\tilde{\Theta}$ of at most $|\tilde{\Theta}| \leq (n+1)^{|\mathcal{X}| \cdot |\mathcal{Z}|}$ sequences, therefore as already mentioned in Section~\ref{sec:RempML_adaptive_redundancy_by_weighting}, a coarse bound on $C_n$ is $(n+1)^{|\mathcal{X}| \cdot |\mathcal{Z}|}$, which yields a redundancy of $\frac{\log C_n}{n} \approx |\mathcal{X}| \cdot |\mathcal{Z}| \frac{\log n}{n}$.

Xie and Barron \cite{XieBarron00} gave asymptotically tight expressions for the maximum regret associated with Bayesian mixtures in the memoryless case. We first state their results for the non-conditional case $\mathcal{Z} = \emptyset$. Define $\hat p_{\ML}(\vr x) = \max_{\theta(\cdot)} P_{\theta}(\vr x) = \max_{\theta(\cdot)}  \prod_{i=1}^n \theta(x_i)$, and $P_w(\vr x) = \int_{\Theta} P_{\theta}(\vr x) w(\theta) d \theta$. Their Lemma~1 states that when using the Diriclet-$\half$ prior for $w(\theta)$, i.e. $w(\theta) = \frac{c}{\sqrt{\prod_{x \in \mathcal{X}} \theta(x)}}$ (where $c$ is the normalizing factor), the regret satisfies (see Equation~(23) for an explicit bound):
\begin{equation}\label{eq:A3459}
\log \frac{\hat p_{\ML}(\vr x)}{P_w(\vr x)} \leq \frac{d}{2} \log \frac{n}{2 \pi} +  C_{\mathcal{X}} + \frac{|\mathcal{X}|}{2} \log e + o_n(1)
\end{equation}
where $d = |\mathcal{X}|-1$ is the number of free parameters, $o_n(1) = \frac{|\mathcal{X}|^2 \cdot \log e}{4 n} \ntoinfty 0$, and
\begin{equation}\label{eq:A3467}
C_{\mathcal{X}} = \log \frac{\Gamma \left( \half \right)^{|\mathcal{X}|}}{\Gamma \left( \frac{|\mathcal{X}|}{2} \right)}
\end{equation}
This observation is attributed to Shtarkov \cite{Shtarkov88} but was given a more explicit expression by Xie and Barron (see also in Cover and Thomas \cite[Section 13.2]{CoverThomas_InfoTheoryBook}, Cesa-Bianchy and Lugosi \cite[Remark 9.4]{Nicolo}).

Furthermore, they propose a slightly modified distribution $w(\theta)$ for which:
(Theorem~2):
\begin{equation}\label{eq:A3473}
\log \frac{\hat p_{\ML}(\vr x)}{P_w(\vr x)} \leq \frac{d}{2} \log \frac{n}{2 \pi} + C_{\mathcal{X}} + o_n(1)
\end{equation}
The term on the RHS of \eqref{eq:A3473} tends to the asymptotical minimax regret (i.e. the regret achieved by the NML), i.e. this weighting scheme asymptotically loses nothing with respect to the optimum regret. Note that the expressions in \eqref{eq:A3459} and \eqref{eq:A3467} both share the common factor $\frac{d}{2} \log n$, and the difference is an increase of the constant factor by $\frac{|\mathcal{X}|}{2} \log e$ in the Diriclet prior with respect to the minimax solution and the prior proposed by Xie and Barron. This weighting has the property that it depends on $n$, whereas the former Dirichlet mixture does not. They extend their results to the conditional case (see Section IX there), however the proofs are quite involved.

Below we show how the result regarding the Diriclet prior is extended to the conditional case. Although this extension is quite standard (and sub-optimal compared to Xie and Barron's extension), we present it explicitly here in order to show that the dependence between $\vr x, \vr y$ and $\vr z$ does not change the result. The parameters are now the set of $|\mathcal{X}| \cdot |\mathcal{Z}|$ values of the function $\theta(x|z)$ which have $(|\mathcal{X}| -1) \cdot |\mathcal{Z}|$ degrees of freedom (since $\forall z: \sum_x \theta(x|z) = 1$). The prior is simply the product of Diriclet priors assigned to each function $\theta(\cdot | z)$ for each value of $z$, i.e. for a probability vector $\vartheta = \theta(\cdot | z)$ let $w_0(\vartheta) = \frac{c}{\sqrt{\prod_{x \in \mathcal{X}} \vartheta(x)}}$ then the weight function is $w(\theta) = \prod_z w_0(\theta(\cdot | z)) = \frac{\tilde c}{\sqrt{\prod_{x \in \mathcal{X}, z \in \mathcal{Z}}\theta(x | z)}}$.

For each $z$, consider each sub-vector of $\vr x$ at the indices where $z_i = z$. The result is based on the fact that the parameters for each sub-vectors are separate, and therefor the problem can be reduced to the non-conditional case. In the maximum likelihood solution, each sub-vector has a set of variables independent of the other sub-vectors and therefore the maximum likelihood probability of the sub-vector depends only on the empirical distribution of $\vr x$ over the sub-vector. For the mixture distribution, the dependence on $\theta(\cdot|z)$ stems only from the elements of the sub-vector associated with $z$ and the integral can be separated into a set of weighted distributions on the sub-vectors, which are related to the maximum likelihood probabilities. The regret terms for each of the subvectors are accumulated, and bounded by a convexity argument. Rewrite the RHS of \eqref{eq:A3459} as $c_1 \log n + c_2$ to express explicitly the dependence on $n$, then:
\begin{equation}\begin{split}\label{eq:A3488}
\log P_w(\vr x | \vr y)
&=
\log \int w(\theta) \prod_{i=1}^n \theta(x_i | z_i) d \theta
\\&=
\log \int \prod_{z \in \mathcal{Z}} w_0(\theta(\cdot | z)) \cdot \prod_{z \in \mathcal{Z}} \prod_{i: z_i=z} \theta(x_i | z) d \theta
\\&=
\log \prod_{z \in \mathcal{Z}} \left[ \int w_0(\theta(\cdot | z)) \cdot \prod_{i: z_i=z} \theta(x_i | z) d \theta(\cdot|z) \right]
\\&=
\sum_{z \in \mathcal{Z}} \log \left[ \int w_0(\theta(\cdot | z)) \cdot \prod_{i: z_i=z} \theta(x_i | z) d \theta(\cdot|z) \right]
\\&\geq
\sum_{z \in \mathcal{Z}} \left[ \log \left( \max_{\theta(\cdot|z)} \prod_{i: z_i=z} \theta(x_i | z) \cdot \right) - c_1 \log \left( n \hat P_{\vr z}(z) \right) - c_2 \right]
\\&=
\log \hat p_{\ML}(\vr x | \vr y) - \sum_{z \in \mathcal{Z}} \left[ c_1 \log \left( n \hat P_{\vr z}(z) \right) + c_2 \right]
\\&=
\log \hat p_{\ML}(\vr x | \vr y) - |\mathcal{Z}| \cdot c_2 + |\mathcal{Z}| \cdot c_1 \cdot \sum_{z \in \mathcal{Z}} \frac{1}{|\mathcal{Z}|} \log \left( n \hat P_{\vr z}(z) \right)
\\&\stackrel{\text{Convexity}}{\geq}
\log \hat p_{\ML}(\vr x | \vr y) - |\mathcal{Z}| \cdot c_2 + |\mathcal{Z}| \cdot c_1 \cdot  \log \left( \sum_{z \in \mathcal{Z}} \frac{1}{|\mathcal{Z}|} n \hat P_{\vr z}(z) \right)
\\&=
\log \hat p_{\ML}(\vr x | \vr y) - |\mathcal{Z}| \cdot \left[ c_1 \cdot  \log \left( \frac{n}{|\mathcal{Z}|}  \right) + c_2 \right]
\end{split}\end{equation}
Therefore the regret becomes
\begin{equation}\label{eq:A3537}
r_n
=
|\mathcal{Z}| \cdot \left[ c_1 \cdot  \log \left( \frac{n}{|\mathcal{Z}|}  \right) + c_2 \right]
=
|\mathcal{Z}| \cdot \left[
\frac{|\mathcal{X}|-1}{2} \log \frac{n}{2 \pi |\mathcal{Z}|} +  C_{\mathcal{X}} + \frac{|\mathcal{X}|}{2} \log e + o_n(1) \right]
\end{equation}
It is important to note that $\vr x, \vr y$ and $\vr z$ are all constant throughout \eqref{eq:A3488}, and therefore the result is oblivious to any dependence between them. The modification of Xie and Barron's asymptotically optimal result to the conditional case results in a similar expression, i.e. $|\mathcal{Z}| \cdot \left[ c_1 \cdot  \log \left( \frac{n}{|\mathcal{Z}|}  \right) + c_2 \right]$, where $c_1, c_2$ are taken from \eqref{eq:A3473}. One way or the other, we have obtained a relation of the form:
\begin{equation}\label{eq:A3552}
\log \frac{\hat p_{\ML}(\vr x | \vr y)}{P_w(\vr x | \vr y)} \leq r_n
\end{equation}
I.e. \eqref{eq:A2444} holds with $C_n = \exp(r_n)$. Therefore the redundancy term $\frac{\log C_n}{n}$ of Theorem~\ref{theorem:RempML_adaptive_redundancy_by_weighting} is
\begin{equation}\label{eq:A3556}
\frac{\log C_n}{n} = \frac{r_n}{n}
\end{equation}
We summarize these results in the following theorem, which specializes Theorem~\ref{theorem:RempML_adaptive_redundancy_by_weighting} to the case of conditional memoryless distributions.

\begin{theorem}\label{theorem:RempML_adaptive_redundancy_empirical}
Let $\vr z \in \mathcal{Z}^n$ be a discrete vector of states, which is a function of $\vr x$ and $\vr y$, where $z_i$ may arbitrarily depend on $\vr x_1^{i-1}$ and $\vr y_1^{i+D}$ for some delay $D \geq 0$. Let $Q(\vr x)$ be an input distribution over a discrete set $\mathcal{X}$ that satisfies $\forall k: Q(x_k | \vr x^{k-1}) \in \{0\} \cup [q_{\min}, 1]$. Define the following rate function:
\begin{equation}\label{eq:A3565}
\Remp = \frac{1}{n} \log \frac{\hat p(\vr x | \vr z)}{Q(\vr x)}
\end{equation}
$\Remp$ is adaptively achievable up to $\delta_n' = \delta_n + \frac{1}{n} r_n$, where $\delta_n$ is defined in Theorem~\ref{theorem:adaptive_causal_distribution} and
\begin{equation}\label{eq:A3572}
r_n
=
|\mathcal{Z}| \cdot \left[
\frac{|\mathcal{X}|-1}{2} \log \frac{n}{2 \pi |\mathcal{Z}|} +  C_{\mathcal{X}} + \frac{|\mathcal{X}|}{2} \log e + o_n(1) \right]
\end{equation}
with $C_{\mathcal{X}}$ defined in \eqref{eq:A3467} and $o_n(1) = \frac{|\mathcal{X}|^2 \cdot \log e}{4 n} \ntoinfty 0$. Furthermore, this rate function has intrinsic redundancy $\mu_Q \leq \frac{1}{n} r_n$ and is achievable non adaptively, up to $\frac{1}{n} (r_n + \log \epsilon^{-1})$.
 \end{theorem}

\textit{Proof:} based on Theorem~\ref{theorem:RempML_adaptive_redundancy_by_weighting} and the discussion above. Note that for the conditionally memoryless class we defined, $\hat p_{\ML}(\vr x | \vr y) = \hat p(\vr x | \vr z)$. Theorem~\ref{theorem:RempML_adaptive_redundancy_by_weighting} requires that the conditions of Theorem~\ref{theorem:adaptive_causal_distribution} be satisfied with respect to $P_{\theta}(\vr x | \vr y)$ and $Q$. Specifically, $Q$ needs to be bounded from below, and $P_{\theta}(\vr x | \vr y)$ of \eqref{eq:A3439} is required to be $D$-causal which is obtained by allowing $z_i$ to depend only on the past of $\vr x$ and $D$ future samples of $\vr y$. In this case, in the conditionally memoryless model of \eqref{eq:A3439}, $P_{\theta}(x_i | \vr x^{i-1} \vr y) = P_{\theta}(x_i | \vr x^{i-1} \vr y^{i+D}) = \theta(x_i | z_i(\vr x^{i-1} \vr y^{i+D}))$, since $\vr x^{i-1}, \vr y^{i+D}$ completely define $z_i$ (see Definition~\ref{def:D_causality}).

The result in the non-adaptive case and the bound on the intrinsic redundancy follow from Lemma~\ref{lemma:remp_conditional_form} since we can write \eqref{eq:A3552}: $\Remp \leq \frac{1}{n} \log \frac{P_w(\vr x | \vr y)}{Q(\vr x)} + \frac{1}{n} r_n$, and by Lemma~\ref{lemma:remp_conditional_form} the first part has $\mu_Q \leq 0$ (see also the note following Theorem~\ref{theorem:RempML_adaptive_redundancy_by_weighting}).

Note that the additional redundancy $\frac{r_n}{n} \approx  \frac{|\mathcal{Z}| \cdot (|\mathcal{X}|-1)}{2} \cdot \frac{\log n}{n}$, is better than the redundancy $\approx |\mathcal{X}| \cdot |\mathcal{Z}| \frac{\log n}{n}$ which is obtained using the simple bound based on the number of types (Theorem~\ref{theorem:optimal_type_based}).

\subsubsection{The continuous and general case}\label{sec:ML_adaptivity_continuous}
\todo{This section is confusing. Consider removing it, passing the important things to the MIMO section, and merely saying that the example of the MIMO can be extended to handle other continuous cases. Perhaps move it to a the ``unfinished ideas'' document}

The discussion above was relevant for the discrete case only. For the continuous (or general) case, we need to consider additional constraints. We define the following decoding metric:
\begin{equation}\label{eq:A2497}
\psi(\vr x^k, \vr y^k, j) = \left( \frac{\max_{\theta} P_{\theta} \left(\vr x_{j+1}^{k} \big| \vr y^k, \vr x^{j} \right)}{Q \left(\vr x_{j+1}^{k} \big| \vr x^j \right)} \right)^{\gamma}
\end{equation}
where $\gamma \in (0,1)$ and we assume $P_{\theta}$ is strictly causal (with $D$=0). When this decoding metric meets the conditions of Theorem~\ref{theorem:framework}, the resulting rate function would be
\begin{equation}\label{eq:A2502}
\Remp = \frac{1}{n} \log \psi(\vr x, \vr y, 0) = \frac{1}{n} \cdot \gamma \cdot \log \left( \frac{\max_{\theta} P_{\theta} \left(\vr x \big| \vr y \right)}{Q \left(\vr x \right)} \right) = \gamma \Rempname{\ML}
\end{equation}
i.e. achieves $\Rempname{\ML}$ up to a multiplicative factor, which we would like to take to $1$ as $n \to \infty$.

We now analyze the conditions required for $\psi$. Unlike the discrete case in which we could easily characterize a set of rate functions which can be adaptively achieved by the scheme presented, in the general case we do not have such a simple characterization. Instead, we give below some analysis of the conditions.

The Markov sufficient condition for the CCDF requires bounding the following quantity:
\begin{equation}\label{eq:A2507}
\placeunder{\E}{Q} \left[ \psi(\vr X^k, \vr y^k, j)  | \vr x^j \right]
=
\int \left( \frac{\hat p_{\ML} \left(\vr x_{j+1}^{k} \big| \vr y^k, \vr x^{j} \right)}{Q \left(\vr x_{j+1}^{k} \big| \vr x^j \right)} \right)^{\gamma}  Q \left(\vr x_{j+1}^{k} \big| \vr x^j \right) d \vr x_{j+1}^{k}
=
\int \hat p_{\ML}^{\gamma} \left(\vr x_{j+1}^{k} \big| \vr y^k, \vr x^{j} \right) Q^{1-\gamma} \left(\vr x_{j+1}^{k} \big| \vr x^j \right) d \vr x_{j+1}^{k}
\end{equation}
Note that the same applies for discrete $\vr x$, replacing the integral with a sum. For $\gamma=0$ the value above is simply the integral of $Q$ and is therefore $1$ (and bounded), and therefore it is reasonable to assume that there exists a $0 < \gamma < 1$ for which the integral above is bounded. For $\gamma=1$ the above evaluates to the redundancy term in universal coding (see Section~\ref{sec:ML_rate_functions}), however this term may be infinite when the distribution is continuous.

The summability condition can be written as follows. Suppose that $\theta^* = \argmax{\theta} P_{\theta} \left(\vr x | \vr y \right)$, and as in the proof of Theorem~\ref{theorem:adaptive_causal_distribution}, let $\{j_b, k_b\}_{b=1}^B$ be a set of segments as defined in the summability condition of Theorem~\ref{theorem:framework}, and $A$ denote the set of indices not included in the segments (unconstrained symbols), with $|A|=m_0$.

We assume that $Q(x_i | \vr x^{i-1})$ is bounded from two sides, i.e.  $0 < q_{\min} \leq Q(x_i | \vr x^{i-1}) \leq q_{\max} < \infty$. For many distributions of interest (such as the Gaussian distribution), the lower bound $q_{\min}$ does not exist, and we need to ``enforce'' it by removing the tail of the distribution. In the current scheme it seems there is no way around this, since the scheme fails to attain $\Remp$ if an unconstrained symbol appears, which has a very small a-priori probability $Q$ and the posteriori probability (which is controlled by the channel), is not small, may increase $\Remp$ in an unbounded amount, which is not utilized by the scheme. \unfinished{See also comments in ****}.

We may expand the probability $P_{\theta^*} \left(\vr x \big| \vr y \right)$ by Bayes law:
\begin{equation}\label{eq:A2522}
P_{\theta^*} \left(\vr x \big| \vr y \right)
=
\prod_{b=1}^B P_{\theta^*} \left(\vr x_{j_b+1}^{k_b} \big| \vr y, \vr x^{j_b} \right) \cdot \prod_{i \in A}  P_{\theta^*} \left(\vr x_i \big| \vr y, \vr x^{i-1} \right)
\end{equation}
and similarly for $Q$:
\begin{equation}\label{eq:A2553}
Q \left(\vr x \right)
=
\prod_{b=1}^B Q \left(\vr x_{j_b+1}^{k_b} \big| \vr x^{j_b} \right) \cdot \prod_{i \in A}  Q \left(\vr x_i \big| \vr x^{i-1} \right)
\end{equation}

The terms in the first product in \eqref{eq:A2522} are bounded by the maximum likelihood value over the segment:
\begin{equation}\label{eq:A1895}
P_{\theta^*} \left(\vr x_{j_b+1}^{k_b} \big| \vr y, \vr x^{j_b} \right)
\stackrel{D=0\text{-Causality}}{=}
P_{\theta^*} \left(\vr x_{j_b+1}^{k_b} \big| \vr y^{k_b}, \vr x^{j_b} \right)
\leq
\max_{\theta} P_{\theta} \left(\vr x_{j_b+1}^{k_b} \big| \vr y^{k_b}, \vr x^{j_b} \right)
\end{equation}

The second product in \eqref{eq:A2522} relates to the ``unconstrained'' symbols (see the proof of Theorem~\ref{theorem:framework}). Regarding the terms in this product $P_{\theta^*} \left(\vr x_i \big| \vr y, \vr x^{i-1} \right)$ we do not have a general bound and they may be bounded in specific cases.

One simple case is when $P_{\theta}$ is globally upper bounded (i.e. $\forall \theta,\vr x, \vr y, i: P_{\theta} \left(\vr x_i \big| \vr y, \vr x^{i-1} \right) \leq c$), however this is a rare case, since if, for example, the parameter space enables scaling of $P_{\theta}$ and this scaling is not bounded, then it is possible to obtain unlimited values of $P_{\theta} \left(\vr x_i \big| \vr y, \vr x^{i-1} \right)$ by scaling. If $s$ denotes the shrinkage ratio between $\theta'$ and $\theta$ (applied, for example, for both $\vr x$ and $\vr y$), then $P_{\theta'} \left(\vr x_i \big| \vr y, \vr x^{i-1} \right) = s \cdot P_{\theta} \left(s \cdot \vr x_i \big| s \cdot \vr y, s \cdot \vr x^{i-1} \right)$, and we may obtain unbounded value by taking $s \to \infty$. As an example this occurs in the Gaussian case (see Section~\unfinished{***}) where the parameter $\theta$ is the covariance matrix.

A softer requirement is that the probability $P_{\theta}$ will be bounded per value of $\theta$: $\forall \theta,\vr x, \vr y, i: P_{\theta} \left(\vr x_i \big| \vr y, \vr x^{i-1} \right) \leq P_{\max}(\theta)$. In this case we may use the fact the gap in the summability condition depends on the value of $\psi_0^n$. In many cases we can draw a bound on $P_{\theta^*} \left(\vr x_i \big| \vr y, \vr x^{i-1} \right)$ from the knowledge of $\hat p_{\ML}(\vr x | \vr y)$. The reason is that $\theta = \hat \theta_{\ML}(\vr x | \vr y)$ maximizes the product of all $P_{\theta} \left(\vr x_i \big| \vr y, \vr x^{i-1} \right)$ (for $i=1,\ldots,n$), and therefore for many ``smooth'' distributions $\theta^*$ strikes a balance between the probabilities assigned to each symbol. In these cases the probability that any specific symbol may attain while the total probability is bounded, cannot grow indefinitely. Specifically in some cases of interest, including the Gaussian case, knowledge of $\hat p_{\ML}(\vr x | \vr y)$ yields an information on $\theta^*$, which can be used to upper-bound $P_{\theta} \left(\vr x_i \big| \vr y, \vr x^{i-1} \right)$, i.e. let
\begin{equation}\label{eq:A2587}
\Theta^{(ML)}(t) = \left\{ \hat \theta_{\ML}(\vr x | \vr y) :  \hat p_{\ML}(\vr x | \vr y) \leq t \right\}
\end{equation}
I.e. $\Theta^{(ML)}(t)$ is the range of possible values of the maximum likelihood estimator (over all $\vr x, \vr y$), for which the maximum likelihood probability is no more than $t$. For example in the Gaussian case \unfinished{(see ****) it is ***}. Now, since
\begin{equation}\label{eq:A2591}
\hat p_{\ML}(\vr x | \vr y) = Q(\vr x) \cdot (\psi_0^n)^{1/\gamma} \leq q_{\max}^n \cdot (\psi_0^n)^{1/\gamma}
\end{equation}
and recall that $\theta^* = \hat \theta_{\ML}(\vr x | \vr y)$, we may bound $P_{\theta^*}$ as:
\begin{equation}\label{eq:A2592}
P_{\theta^*} \left(\vr x_i \big| \vr y, \vr x^{i-1} \right) \leq \max_{\theta \in \Theta^{(ML)}(q_{\max}^n \cdot (\psi_0^n)^{1/\gamma})} P_{\max}(\theta) \defeq g_0(\psi_0^n)
\end{equation}
In other words, from $\psi_0^n$ we bound $\hat p_{\ML}(\vr x | \vr y)$, obtain a range of possible $\theta^*$-s and find the maximum single-symbol probability that may be assigned using these $\theta^*$-s. This bounding technique can be better understood by reviewing the example of the Gaussian case which is \unfinished{given in Section**}.

We summarize these conclusions in the following lemma:

\begin{lemma}\label{lemma:summability_continuous}
Let $\psi(\vr x^k, \vr y^k, j)$ be defined in \eqref{eq:A2497} where $\gamma \in (0,1)$ and we assume $P_{\theta}$ is strictly causal, and $Q(\vr x)$ is bounded by $Q(\vr x) \in \{0\} \cup [q_{\min}, q_{\max}]$ (where  $0 < q_{\min} < q_{\max} < \infty$). Let $\Theta^{(ML)}(t) = \left\{ \hat \theta_{\ML}(\vr x | \vr y) :  \hat p_{\ML}(\vr x | \vr y) \leq t \right\}$. If there exists $P_{\max}(\theta)$ such that $\forall \theta,\vr x, \vr y, i: P_{\theta} \left(\vr x_i \big| \vr y, \vr x^{i-1} \right) \leq P_{\max}(\theta)$, and $g_0(\psi_0^n) \defeq \max_{\theta \in \Theta^{(ML)}(q_{\max}^n \cdot (\psi_0^n)^{1/\gamma})} P_{\max}(\theta) < \infty$, then the summability condition in Theorem~\ref{theorem:framework} holds with $f_0(\psi_0^n) = \gamma \cdot \log (g_0(\psi_0^n) \cdot q_{\min}^{-1})$
\end{lemma}

\todo{consider move the discussion somewhere else}
Unfortunately, in the general case, the probability of a single symbol $P_{\theta^*} \left(\vr x_i \big| \vr y, \vr x^{i-1} \right)$ cannot be upper bounded even when $\hat \theta_{\ML}(\vr x | \vr y)$ is known (see Example \unfinished{***}). In this case the summability condition does not hold, and we cannot attain the rate function $\Rempname{\ML}$ using the scheme proposed here. The failure occurs with respect to the ``unconstrained'' symbols ($m_0$) in the summability condition. These symbols are related to the increase of the rate function at the symbol in which the decoding occurred. Therefore one might say that failure to obtain the rate function in these cases stems from the scheme and the fact that it does not ``use'' all the symbols. On the other hand, it is quite difficult to envision an adaptive scheme that does not have this limitation. If the rate is determined by negotiation between the encoder and the decoder, then an unlimited increase of the rate function $\Rempname{\ML}$ that occurs at the $n$-th symbol does not allow the system to adapt its rate (since the feedback for this symbol is not relevant). It's worth noting that in posterior matching scheme \cite{Ofer_Posterior_analysis} for the known memoryless channel (an extension of Horstein's scheme \cite{Horstein}), the rate for a given error probability $\epsilon$ can be determined by the decoder after reception (without coordination with the encoder, who always transmits the infinite sequence), however it is not trivial to extend this scheme to the individual case.

Assuming the above assumptions holds, we have:
\begin{equation}\begin{split}\label{eq:A2572}
\frac{1}{\gamma} \log \psi_0^n
&=
\log \frac{P_{\theta^*} \left(\vr x \big| \vr y \right)}{Q(\vr x)}
\\& \stackrel{\eqref{eq:A2522}}{=}
\sum_{b=1}^B \log \frac{P_{\theta^*} \left(\vr x_{j_b+1}^{k_b} \big| \vr y, \vr x^{j_b} \right)}{Q(\vr x_{j_b+1}^{k_b} \big| \vr y, \vr x^{j_b})} + \sum_{i \in A} \log \frac{P_{\theta^*} \left(\vr x_i \big| \vr y, \vr x^{i-1} \right)}{Q \left(\vr x_i \big| \vr x^{i-1} \right)}
\\& \stackrel{\eqref{eq:A1895}, \eqref{eq:A2592}}{\leq}
\sum_{b=1}^B \log \frac{\max_{\theta} P_{\theta} \left(\vr x_{j_b+1}^{k_b} \big| \vr y, \vr x^{j_b} \right)}{Q(\vr x_{j_b+1}^{k_b} \big| \vr y, \vr x^{j_b})} + \sum_{i \in A} \log \frac{g_0(\psi_0^n)}{q_{\min}}
\\& =
\frac{1}{\gamma} \sum_{b=1}^B \log \psi_b  + m_0 \cdot \log (g_0(\psi_0^n) \cdot q_{\min}^{-1})
\end{split}\end{equation}

Therefore the summability condition holds:
\begin{equation}\label{eq:A2631}
\log \psi_0^n - \sum_{b=1}^B \log \psi_b \leq m_0 \cdot \underbrace{\gamma \cdot \log (g_0(\psi_0^n) \cdot q_{\min}^{-1})}_{f_0(\psi_0^n)}
\end{equation}
with $f_0(\psi_0^n) = \gamma \cdot \log (g_0(\psi_0^n) \cdot q_{\min}^{-1})$.

To summarize, in the general case we do not have a general characterization of rate functions that are achieved by the scheme presented, and specifically there is no general claim that $\Rempname{\ML}$ can be adaptively achieved. In specific cases, we may use the techniques shown here: the CCDF condition requires bounding the value in \eqref{eq:A2507}. For $\Rempname{\ML}$, the summability condition holds in general with respect to the ``constrained'' segments, but particular treatment (per parametric family) is needed for the ``unconstrained'' symbols, possibly by Equations \eqref{eq:A2587}-\eqref{eq:A2592}. Furthermore, to obtain these bounds we need to constrain $Q$ by a minimum and a maximum value.

\subsubsection{Examples for the bound on the unconstrained symbols}
Below we give some examples to better illustrate the bounding technique presented above for the unconstrained symbols (Equations \eqref{eq:A2587}-\eqref{eq:A2592}), and its shortcomings.

\begin{example}[A gaussian model]\label{example:unconstrained_gausss}
Suppose that the model for $\vr x$ given $\vr y$ is an i.i.d. Gaussian model, where $X_i$ is Gaussian with mean $\alpha \cdot y_i$ and variance $\sigma^2_{x|y}$. There are two parameters $\theta = (\alpha,\sigma^2_{x|y})$, and the distribution is
\begin{equation}\label{eq:A2652}
P_{\theta}(\vr x | \vr y) = (2 \pi \sigma^2_{x|y})^{-n/2} e^{-\frac{1}{2 \sigma^2_{x|y}} \sum_{i=1}^n (x_i - \alpha \cdot y_i)^2}
\end{equation}
It is easy to check (e.g. by derivating $\log P_{\theta}(\vr x | \vr y)$, see also \unfinished{Section***}) that $\hat \alpha_{\ML} = \frac{\vr x^T \vr y}{\| \vr y \|^2}$, and $\hat \sigma^2_{x|y,ML} = \frac{1}{n} \| \vr x - \hat \alpha_{\ML} \vr y \|^2$, substituting we obtain
\begin{equation}\label{eq:A2657}
\hat p_{\ML}(\vr x | \vr y) = (2 \pi \sigma^2_{x|y})^{-n/2} e^{-\frac{1}{2 \sigma^2_{x|y}} \cdot n \sigma^2_{x|y}} =  (2 \pi \sigma^2_{x|y} e)^{-n/2}
\end{equation}
therefore \eqref{eq:A2587}:
\begin{equation}\label{eq:A2661}
\Theta^{(ML)}(t) = \left\{ \hat \theta_{\ML}(\vr x | \vr y) :  \hat p_{\ML}(\vr x | \vr y) \leq t \right\} = \left\{ (\alpha, \sigma^2_{x|y}) : (2 \pi \sigma^2_{x|y} e)^{-n/2} \leq t \right\}
\end{equation}
and the maximum of the single letter probability is
\begin{equation}\label{eq:A2666}
P_{\max}(\theta) = \max_{x_i,y_i} (2 \pi \sigma^2_{x|y})^{-1/2} e^{-\frac{1}{2 \sigma^2_{x|y}} (x_i - \alpha \cdot y_i)^2} = (2 \pi \sigma^2_{x|y})^{-1/2}
\end{equation}

by \eqref{eq:A2592}:
\begin{equation}\label{eq:A2667}
g_0(\psi_0^n)
=
\max_{\theta \in \Theta^{(ML)}(q_{\max}^n \cdot (\psi_0^n)^{1/\gamma})} P_{\max}(\theta)
=
\max_{\sigma^2_{x|y}: (2 \pi \sigma^2_{x|y} e)^{-n/2} \leq q_{\max}^n \cdot (\psi_0^n)^{1/\gamma}} (2 \pi \sigma^2_{x|y})^{-1/2}
=
q_{\max} e^{1/2} \cdot (\psi_0^n)^{\frac{1}{n \gamma}}
\end{equation}
and by \eqref{eq:A2631}:
\begin{equation}\label{eq:A2681}
f_0(\psi_0^n) = \gamma \cdot \log (g_0(\psi_0^n) \cdot q_{\min}^{-1}) = \gamma \cdot \log \left( \frac{q_{\max} e^{1/2}}{q_{\min}} \right) + \frac{1}{n} \cdot \log  (\psi_0^n)
\end{equation}
\end{example}

\begin{example}\label{example:unconstrained_exp}
As another example we consider the case is when $\vr X$ given $\vr y$ is modeled as i.i.d. where each symbol $X_i$ is conditionally distributed around $y_i$ with a scale factor proportional to $\theta$:
\begin{equation}\label{eq:A2689}
P_{\theta}(x_i | y_i) = \theta \cdot f \left( \theta \cdot (x_i - y_i) \right)
\end{equation}
where
\begin{equation}\label{eq:A3274}
f(t) = c \cdot e^{-|t|^p}
\end{equation}
$p \geq 1$ is a fixed parameter, and $c$ takes care of normalization so that $\int f(t) dt = 1$. This family includes as special cases the symmetric exponential distribution $p=1$ and the Gaussian distribution $p=2$. We have $P_{\theta}(\vr x | \vr y) = \theta^n c^n e^{-\theta^p \sum_i |x_i - y_i|^p}$. It is easy to check that
$\hat \theta_{\ML}  = \left( \frac{p}{n} \sum_i |x_i - y_i|^p \right)^{-1/p}$, and therefore $\hat p_{\ML} = \hat \theta_{\ML}^n c^n e^{-n/p}$. As before,  $\hat p_{\ML}$ and $\hat \theta_{\ML}$ are related, and we have: $\Theta^{(ML)}(t) = \left\{ \hat \theta_{\ML}(\vr x | \vr y) :  \hat p_{\ML}(\vr x | \vr y) \leq t \right\} = \left\{ \theta: \theta^n c^n e^{-n/p} \leq t \right\} = (-\infty, t^{1/n} e^{1/p} c^{-1}]$. In this case $P_{\max}(\theta) = \theta c$, and we have from \eqref{eq:A2592}:
\begin{equation}\label{eq:A2697}
g_0(\psi_0^n)
=
\max_{\theta \in \Theta^{(ML)}(q_{\max}^n \cdot (\psi_0^n)^{1/\gamma})} P_{\max}(\theta)
=
\max_{\theta \leq q_{\max} \cdot (\psi_0^n)^{\frac{1}{n \gamma}} e^{1/p} c^{-1}} P_{\max}(\theta)
=
q_{\max} e^{1/p} \cdot (\psi_0^n)^{\frac{1}{n \gamma}}
\end{equation}
\end{example}

\begin{example}[A general counter example]\label{example:unconstrained_counter1}
A rather general case where the summability condition does not hold is when on one hand the probability $P_{\theta} \left(\vr x_i \big| \vr y, \vr x^{i-1} \right)$ is not globally bounded, and on the other hand, the parameters $\theta$ contain a separate set of parameters for each value of $y_i$. In this case, if the value of $y_i$ on any symbol is unique and does not appear elsewhere, then the probability assigned to this symbol may grow indefinitely, while, with a suitable choice of the other symbols, the overall maximum-likelihood probability $\hat p_{\ML} (\vr x | \vr y)$ may remain bounded.
\end{example}

\begin{example}[The discrete case]\label{example:unconstrained_discrete}
Consider the discrete memoryless case where $\theta(x|y)$ is the conditional probability of symbol $x$ to appear when $y$ appears. In this case, the maximum likelihood estimator is the empirical distribution $\hat \theta_{\ML}(x|y) = \hat P_{\vr x | \vr y}(x | y)$, and the maximum likelihood probability is the empirical probability $\hat p_{\ML}(\vr x | \vr y) = \hat p(\vr x | \vr y) = \exp(-n \hat H(\vr x | \vr y))$ (see \eqref{eq:A1389}). $P_{\max}(\theta)$ in this case is simply $\max_{x,y} \theta(x|y)$. The empirical entropy related to the empirical probability, however there is an unknown factor which is the empirical distribution of $\vr y$. Since we are looking for a bound on $\theta(x|y)$ in terms of $\hat p_{\ML}(\vr x | \vr y)$, which holds for any $\vr x, \vr y$. Using the techniques of the previous section, we cannot do better than simply bound the probability by $1$, i.e. $g_0(\psi_0^n) = 1$ (see \eqref{eq:A2592}). This is because for a pair of random variables $X,Y$ it is possible to have a large conditional probability $\Pr(X|Y)$ with a small effect on the conditional entropy $H(X|Y)$ if $\Pr(Y)$ is small (tends to $0$). The actual implication is that if the value of $y_i$ on an``unconstrained'' symbol is unique (does not appear on the constrained symbols), the empirical probability of this symbol may be $1$, while the empirical probability of the rest of the sequence may vary arbitrarily.
\end{example}

\begin{example}[Another counter example]\label{example:unconstrained_counter2}
The counter example we gave above requires that $\theta$ contains a different set of parameters for each $\vr y$. However, we can show that much less is necessary in order to have an unlimited loss $g_n(\psi_0^n)$, and this may occur even for the simple case of a memoryless distribution with a single scale parameter. We argued in the previous section that the maximum likelihood solution tends to equalize the probabilities assigned to various symbols. The following example is based on creating a region in which the distribution decays rapidly to $0$. By letting one of the points reside in this region, the maximum likelihood solution gives a large part of the probability to this point.

We consider the same setting of Example~\ref{example:unconstrained_exp}, except the distribution $f$ is:
\begin{equation}\label{eq:A3336}
f(t) = \frac{c}{t^2} \cdot e^{-|t|^{-p}}
\end{equation}
Note that $f(t)$ is the probability density function of $1/Z$ where $Z$ is distributed according to the density $f(t)$ defined in Example~\ref{example:unconstrained_exp}, so we have just changed variables. Note also that $f(t)$ is upper bounded and therefore $P_{\theta}(x_i | y_i)$ is bounded for each value of $\theta$. $f(t)$ decays exponentially to $0$ for $t \to 0$ (due to the exponential term). We have
\begin{equation}\label{eq:A3342}
P_{\theta}(\vr x | \vr y)
=
\prod_{i=1}^n \left[ \theta \frac{c}{(\theta(x_i-y_i))^2} \cdot e^{-|(\theta(x_i-y_i))|^{-p}} \right]
=
\theta^{-n} c^n \frac{1}{\prod_{i=1}^n (x_i-y_i)^2} \cdot e^{-\theta^{-p} \sum_{i=1}^n |x_i-y_i|^{-p}}
\end{equation}
It is easy to check that
$\hat \theta_{\ML}  = \left( \frac{p}{n} \sum_{i=1}^n |x_i-y_i|^{-p} \right)^{1/p}$, however due to the term $\prod_{i=1}^n (x_i-y_i)^2$, $\hat p_{\ML}$ cannot be expressed via $\hat \theta_{\ML}$ alone, and $\hat \theta_{\ML}$ cannot be bounded given $\hat p_{\ML}$. We will now show a choice of $\vr x, \vr y$ for which the probability density of a single symbol $i=1$, $P_{\theta}(x_1 | y_1)$ tends to $\infty$ while the overall probability $\hat p_{\ML}$ tends to $0$. Let $x_1 - y_1 = \delta$, and $x_i - y_i \to \infty, i \geq 2$, then $\hat \theta_{\ML} \to \left( \frac{p}{n} \right)^{1/p} \delta^{-1}$, and
\begin{equation}\label{eq:A3355}
\hat p_{\ML}(\vr x | \vr y) = P_{\hat \theta_{\ML}}(\vr x | \vr y) \longrightarrow \const \cdot \delta^n \cdot 0 \cdot e^{-\frac{n}{p}} = 0
\end{equation}
while
\begin{equation}\label{eq:A3360}
P_{\hat \theta_{\ML}}(x_1 | y_1) = \hat \theta_{\ML}^{-1} c^n \frac{1}{(x_1-y_1)^2} \cdot e^{-\hat \theta_{\ML}^{-p} |x_1-y_1|^{-p}}
\longrightarrow
\const \cdot \delta \cdot \frac{1}{\delta^2} \cdot e^{-\frac{n}{p}} = \const \cdot \frac{1}{\delta}
\end{equation}
By taking $\delta \to 0$ we obtain $P_{\hat \theta_{\ML}}(x_1 | y_1) \to \infty$.
This demonstrates a distribution which is controlled by a simple scale parameter, where the summability condition does not hold.
\end{example}

\subsection{An infinite horizon adaptive scheme}\label{sec:rate_adaptive_inf_horizon}
The scheme of Section~\ref{sec:rate_adaptive_scheme} and Theorem~\ref{theorem:framework} is a finite-horizon scheme, i.e. the rate is measured at time $n$ and the scheme is aware of the value of $n$ and is designed to meet the promise of the theorem at this point. It is of interest to consider schemes that do not have this limitation, i.e. they are designed without knowing $n$ and still yield similar guarantees to the guarantees of Theorem~\ref{theorem:framework} for any $n$, and specifically, the convergence of the actual rate to the asymptotical rate function given by \eqref{eq:A1958}.

A straightforward modification of the scheme presented here to the infinite horizon case is difficult due to the inherent need to design the information contents of a single block, $K$, to keep the overheads small. As can be seen in Corollary~\ref{corollary:framework_redundancy} there is a balance between the overheads incurred at each block and the loss of the last block. One could change $K$ from block to block (e.g. according to the block index, the elapsed time $t$ or the value of the metric $\psi_0^{t}$), but an inherent difficulty occurs because the overhead term related to keeping the error probability small increases with time. If we have set a certain value of $K$ for the current block, and the block extends indefinitely (due to a very low value of $\psi$ or equivalently $\Remp$), then at some point the overhead for keeping the error probability low would become significant with respect to $K$. A possible solution is to stop the transmission at such a case, and re-start it with a larger value of $K$ but this complicates the scheme and its analysis.

We present here a simple, brute force, modification of the scheme to the indefinite horizon case by an extension termed ``the doubling trick'' and used in universal prediction as well \cite[Section 2.3]{Nicolo} to solve a similar problem of matching the scheme parameters to the block length. This scheme is certainly not the most efficient way to achieve the infinite horizon property, and is given here only in order to show that it is feasible to do so. To simplify, the result is particularized to the case where both $\Remp$  and $f_0^{(n)}$ are upper bounded by constants, and $L_n$ is subexponential in $n$ (these assumptions are correct for the cases \unfinished{XYZ ***}). The idea is to operate the scheme over epochs in time $n_i$ with increasing lengths. In each epoch, we design the scheme parameters to be optimal for the end of the epoch. If the observation time $n$ occurs before the end of the epoch, the parameters are slightly suboptimal but the loss is small. In the simplest form, each epoch is double the size of the previous one, hence the name ``doubling trick''.

The first step is to examine the loss incurred when the scheme's parameters are designed for time $h$ (where $h$ is the horizon for which the scheme is designed), while the actual performance is measured at time $n \leq h$. Considering again the proof of Theorem~\ref{theorem:framework}, we now make a distinction between the value of $n$ used for selecting the scheme's parameters (which is now termed $h$) and the value of $n$ which is the observation time, i.e. the time when the actual rate is measured and compared against the empirical rate function. It is easy to see by following the proof, that if the scheme is designed to yield an error of no more than $\epsilon$ up to any time $n \leq h$, then only the determination of the thresholds $\psi^*$ changes, and the rest of the analysis remains the same. The result is that if the scheme is not aware of $n$ and just given an horizon $h \geq n$, then the results of the theorem still hold with $c_n$ replaced by $c_h$ in \eqref{eq:A1962}. The next step is to choose $K$. Considering the proof of Corollary~\ref{corollary:framework_redundancy}, the value $k_n$ in \eqref{eq:A2013} is now replaced with $c_h + b_1 \cdot f_0^{(n)*}$, however because we assume that $f_0^{(n)}$ is upper bounded by a constant $f_0^{(n)} \leq f_0^{*}$, then this simply becomes a function of $h$, $k_h = c_h + b_1 \cdot f_0^{*}$, and by substituting in \eqref{eq:A2013}, we would have a redundancy of $\delta = \frac{R_{\max} k_h}{K} + \frac{K}{n}$. Note that the second factor is still a function of $n$ since the loss of $K$ bits of the last block is divided by the duration $n$ of the observation time. Choosing $K = \lceil \sqrt{h k_h R_{\max}} \rceil$ (optimized for $n=h$), we have
\begin{equation}\begin{split}\label{eq:A2218}
\delta
&\leq
\sqrt{\frac{R_{\max} k_h}{h}} + \frac{\sqrt{h k_h R_{\max}} + 1}{n}
\stackrel{\frac{1}{\sqrt{h}} \geq \frac{\sqrt{h}}{n}}{\leq}
\frac{2 \sqrt{h k_h R_{\max}} + 1}{n}
\\&=
\frac{1}{n} \left( 2 \sqrt{h \left(c_h + b_1 \cdot f_0^{*} \right) R_{\max}} + 1 \right)
=
\frac{1}{n} \left( 2 \sqrt{h \left(\log \frac{h \cdot L_h}{d \epsilon} + b_1 \cdot f_0^{*} \right) R_{\max}} + 1 \right)
\end{split}\end{equation}

We select the sequence of epoch lengths to be the power of $2$, $h_i = 2^i, i=1,2,\ldots$. Denote by $N_i$ the end time of the $i$-th epoch, i.e. $N_i = \sum_{j=1}^i h_i = 2^{i+1}-1$. We distinguish between the epochs themselves that do not depend on $n$, and the ``observed epoch'', which the part of the epoch which is included in the period of time $1,\ldots,n$ which we observe (and is an empty set of all epochs after time $n$). We denote by $j$ the index of the epoch that contains time $n$, i.e. $N_{j-1} < n \leq N_j$. We denote by $n_i$ the length of the observed epoch, i.e. $n_i = h_i$ for all epochs except the one containing symbol $n$, and is $n_j = n - N_{j-1}$ for this epoch. We denote by $\tilde{N}_i = \min(N_i, n)$ end of each observed epoch. In each epoch we design the scheme for a different error probability $\epsilon_i$ where the sequence of error probabilities satisfies $\sum_{i=1}^\infty \epsilon_i \leq \epsilon$. This guarantees an error probability at most $\epsilon$ no matter what the observation time is. Specifically we choose $\epsilon_i = \frac{\epsilon}{2 i^2}$ ($\sum_{n=1}^{\infty} \frac{1}{n^2} = 1 + \sum_{n=2}^{\infty} \frac{1}{n^2} \leq 1 + \sum_{n=2}^{\infty} \frac{1}{n(n-1)} =  1 + \sum_{n=2}^{\infty} \left[ \frac{1}{n-1} - \frac{1}{n} \right] =  1 + \left[ \frac{1}{2-1} - \frac{1}{\infty} \right] = 2$).

The scheme operated at each epoch uses the metric $\psi(\vr x^k, \vr y^k, j)$ to decode the blocks. This metric uses the entire history from time $1$, and therefore the scheme operation in each epoch is dependent of the value of $\vr x$ and $\vr y$ in previous epochs. We assume that the conditions of Theorem~\ref{theorem:framework} hold for any epoch with any length, and specifically the summability condition holds not only for periods of time starting at $1$ (in which case $\psi_{0}^n$ in the condition is replaced with $\psi(\vr x^{N_i}, \vr y^{N_i}, N_{i-1})$, for the observed epoch $[\tilde{N}_{i-1}+1, \tilde{N}_i]$). It is straightforward to modify the proof of Theorem~\ref{theorem:framework} to see that the rate function $\Remp_i = \frac{1}{n_i} \log \psi(\vr x^{\tilde{N}_i}, \vr y^{\tilde{N}_i}, \tilde{N}_{i-1})$ is obtained. From the derivation above \eqref{eq:A2218} we have that with our choice of $K$, it is obtained up to $\delta_i = \frac{1}{n_i} \left( 2 \sqrt{h_i \left(\log \frac{h_i \cdot L_{h_i}}{d \epsilon_i} + b_1 \cdot f_0^{*} \right) R_{\max}} + 1 \right)$, in other words the actual rate over the $i$-th observed epoch satisfies $\Ract \geq \Remp - \delta_i$. Since the number of bits transmitted in the $i$-th epoch satisfies $n_i \Ract$, we have that the total number of bits $k$ transmitted up to time $n$ satisfies:
\begin{equation}\begin{split}\label{eq:A2245}
k
&=
\sum_{i=1}^j n_i \Ract_i
\geq
\sum_{i=1}^j n_i \left(\Remp_i - \delta_i \right)
\\& \geq
\sum_{i=1}^j \log \psi(\vr x^{\tilde{N}_i}, \vr y^{\tilde{N}_i}, \tilde{N}_{i-1}) -  \underbrace{\sum_{i=1}^j  n_i \delta_i}_{\defeq n \delta(n)}
\geq
\log \log \psi_0^n - n \delta(n)
\end{split}\end{equation}
where the last inequality is due to the summability condition (note that here the segments cover the entire period $1,\ldots,n$ therefore $m_0=0$). Therefore with $\Remp = \frac{1}{n} \log \psi_0^n$ we have:
\begin{equation}\label{eq:A2260}
\Ract = \frac{k}{n} \geq \frac{1}{n} \log \log \psi_0^n - \delta(n) = \Remp - \delta(n)
\end{equation}

We now bound $\delta(n)$ to show $\delta(n) \ntoinfty 0$. By substituting $N_i = 2^{i+1}-1$ in $N_{j-1} < n$ we have that $n \geq 2^j$. Therefore none of the epochs $1,\ldots,j$ is larger than $n$: $h_i \geq h_j = 2^j \leq n$.

\begin{equation}\begin{split}\label{eq:A2261}
n \delta(n)
&=
\sum_{i=1}^j  n_i \delta_i
\\& \leq
\sum_{i=1}^j  \left( 2 \sqrt{h_i \left(\log \frac{h_i \cdot L_{h_i}}{d \epsilon_i} + b_1 \cdot f_0^{*} \right) R_{\max}} + 1 \right)
\\& \stackrel{h_i \leq n, \epsilon_i \geq \epsilon_j}{\leq}
\sum_{i=1}^j  \left( 2 \sqrt{h_i \left(\log \frac{n \cdot L_{n}}{d \epsilon_j} + b_1 \cdot f_0^{*} \right) R_{\max}} + 1 \right)
\\& =
j + 2 \sqrt{\left(\log \frac{n \cdot L_{n} \cdot j^2}{2 \epsilon d} + b_1 \cdot f_0^{*} \right) R_{\max}} \cdot \sum_{i=1}^j  \sqrt{h_i}
\\& =
j + 2 \sqrt{\left(\log \frac{n \cdot L_{n} \cdot j^2}{2 \epsilon d} + b_1 \cdot f_0^{*} \right) R_{\max}} \cdot  \frac{\sqrt{2}^j - 1}{\sqrt{2} - 1}
\\& \leq
\log_2(n) + 2 \sqrt{\left(\log \frac{n \cdot L_{n} \cdot (\log_2 (n))^2}{2 \epsilon d} + b_1 \cdot f_0^{*} \right) R_{\max}} \cdot   \frac{1}{\sqrt{2} - 1} \cdot \sqrt{n}
\end{split}\end{equation}
therefore $\delta(n) \ntoinfty 0$ under the assumption that $L_n$ is sub-exponential (i.e. $\log \frac{\log L_{n}}{n} \ntoinfty 0$).

\section{Examples}\label{chap:examples}
\subsection{Empirical mutual information}\label{sec:examples_eMI}
The empirical mutual information is probably the most intuitively appealing rate function. It was presented in \cite{YL_individual_full}, and revisited throughout the current \selector{paper}{work}. Below we review the main results regarding this rate function and discuss the overhead related to attaining it.

The alphabets $\mathcal{X},\mathcal{Y}$ are assumed to be discrete. We have:
\begin{equation}\label{eq:A3384}
\hat I(\vr x; \vr y) = \frac{1}{n} \log \frac{\hat p(\vr x | \vr y)}{\hat p(\vr x)} = \Rempname{\ML*} \leq \frac{1}{n} \log \frac{\hat p(\vr x | \vr y)}{Q(\vr x)} = \Rempname{\ML}
\end{equation}
In other words, $\hat I$ is of the $\Rempname{\ML*}$ form which is upper bounded by the $\Rempname{\ML}$ form (see \selector{Section}{Chapter}~\ref{chap:useful_constructions} and Section~\ref{sec:MLML_construction}). By definition, the respective $\Rempname{\ML}$ form guarantees this rate function asymptotically equals or exceeds the best reliably achievable rate (with the given prior) over any memoryless channel model (Section~\ref{sec:ML_rate_functions}), and since they are equivalent in high probability, $\Rempname{\ML*}=\hat I$ will asymptotically achieve this guarantee as well. In the case of the empirical mutual information it is easy to see this claim holds -- since for every memoryless model $\hat I(\vr x, \vr y)$ will tend to the statistical mutual information $I(X;Y)$,\footnote{by the law of large numbers the empirical probability tends to the letter probability, and the claim follows from the continuity of the mutual information} which upper bounds the attainable rate.

In Section~\ref{sec:eMI_optimality}, Lemma~\ref{lemma:maximum_iid_type_based_Remp} we saw that it is essentially, but not strictly speaking, the optimal rate function defined by zero-order statistics (asymptotically).

The redundancy of attaining $\hat I$ is upper bounded by Theorem~\ref{theorem:RempML_adaptive_redundancy_empirical} with $\vr z = \vr y$. In the non adaptive case, $\hat I$ is achievable up to $\delta \approx \frac{(|\mathcal{X}|-1) \cdot |\mathcal{Y}|}{2} \cdot \frac{\log n}{n}$ (this is the dominant term from Theorem~\ref{theorem:RempML_adaptive_redundancy_empirical}, assuming $\epsilon$ is constant). In the adaptive case, the dominant factor in the overhead becomes $\delta_n$ defined in Theorem~\ref{theorem:adaptive_causal_distribution}, which is $\delta_n \approx 2 \sqrt{\frac{\log \frac{n}{\epsilon}}{n}}$ (for large $n$).

A lower bound on the redundancy for the $\Rempname{\ML}$ form $\frac{1}{n} \log \frac{\hat p(\vr x | \vr y)}{Q(\vr x)}$ can be obtained via Lemma~\ref{lemma:remp_conditional_form_reverse} and the discussion in Section~\ref{sec:redundancy_of_Remp_ML}: writing
\begin{equation}\label{eq:A3418}
\Rempname{\ML}
=
\frac{1}{n} \log \frac{\hat p(\vr x | \vr y)}{Q(\vr x)}
\stackrel{\eqref{eq:A1717}}{=}
\frac{1}{n} \log \frac{c_{\NML} \cdot P_{\NML}(\vr x | \vr y)}{Q(\vr x)}
=
\frac{1}{n} \log \frac{P_{\NML}(\vr x | \vr y)}{Q(\vr x)} + \frac{1}{n} \log c_{\NML}
\end{equation}
The term $\log c_{\NML}$ is the minimax regret which in this case is known up to an additive factor to be $\log c_{\NML} \approx \frac{(|\mathcal{X}|-1) \cdot |\mathcal{Y}|}{2} \cdot \frac{\log n}{n}$ \cite[Section IX]{XieBarron00}. By Lemma~\ref{lemma:remp_conditional_form_reverse}, the first term in the RHS of \eqref{eq:A3418} requires redundancy of at least $\delta_0 \approx -\frac{\log n}{n}$. Therefore the redundancy in attaining $\Rempname{\ML}$ it at least $\delta_0 + \frac{1}{n} \log c_{\NML} \approx \frac{(|\mathcal{X}|-1) \cdot |\mathcal{Y}| - 2}{2} \cdot \frac{\log n}{n}$. The redundancy of the empirical mutual information itself can be bounded based on the method of types and Theorem~\ref{theorem:remp_achievability_upto}, but this bound is looser.

\subsection{Markov sources and stationary ergodic models}\label{sec:examples_markov}
The empirical mutual information is drawn from the $\Rempname{ML}$ construction with a memoryless model. Therefore it is not able to exploit memory in the channel. In a simple example where $y_i = x_{i-1}$ the empirical mutual information tends to $0$ while the capacity of the channel is $\log |\mathcal{X}|$.

An immediate extension is to replace the memoryless family of distributions with a Markov model. The simplest model could be one in which $X_i$ is a $k$-th order Markov process (the probability of $X_i$ is given as a function of $\vr X_{i-k}^{i-1}$), and the probability of $Y_i$ is given as a function of $X_i$ and the $k$-th order history $\vr X_{i-k}^{i-1}, \vr Y_{i-k}^{i-1}$. In this case, since the probability of $(X_i, Y_i)$ is given as a function of $\vr X_{i-k}^{i-1}, \vr Y_{i-k}^{i-1}$, the pair $(X_i, Y_i)$ is a $k$-th order Markov process. Unfortunately, $\vr Y_i$ alone is not a Markov process but a hidden Markov process (HMM) which has a more complex structure. As a result, the conditional distribution $P_{\theta}(\vr X^n | \vr Y^n)$ (where $\theta$ indexes a specific Markov model) does not have a simple closed form expression. Even values such as the the size of the conditional Markov type or the conditional entropy rate (which would be needed to characterize this rate function via Theorem~\ref{theorem:optimal_type_based}) are related to the entropy rate of HMM-s which does not have a closed form expression (see  for example \cite{SeroussiHMM04}).

To circumvent this problem using a more general characterization, suitable for stationary ergodic channels. Since $\Rempname{\ML}$ is based on modeling $P_{\theta}(\vr x^n | \vr y^n)$ we associate the parameters with the conditional distribution, by giving the probability of $X_i$ given the $D$ past letters of the input $\vr X_{i-D}^{i-1}$ and the past and future of the output $\vr y_{i-D}^{i+D}$. I.e.
\begin{equation}\label{eq:A3440}
P_{\theta}(\vr x^n | \vr y^n) = \prod_{i=1}^n \theta(x_i | \vr x_{i-D}^{i-1}, \vr y_{i-D}^{i+D})
\end{equation}
where $\theta(\cdot | \cdot): \mathcal{X}^{D+1} \times \mathcal{Y}^{2D+1} \to [0,1]$ is a set of conditional probability functions which is the parametric space. Regarding times $i \leq D$ in which the past $D$ samples are not defined, we may either define an arbitrary initial state, a special value (which effectively increases the $\mathcal{Y}$ alphabet size by one, and is equivalent to defining special probability functions for these times), or avoid communication during these times (treat them as a training sequence). To simplify the discussion below we adopt the first solution, although it is easy to modify it.

The probability $P_{\theta}(\vr x^n | \vr y^n)$ is $D$-causal (Definition~\ref{def:D_causality}). Defining the state variable $z_i = (\vr x_{i-D}^{i-1}, \vr y_{i-D}^{i+D})$, this distribution falls into the category of conditionally memoryless distributions. Hence, the maximum likelihood distribution equals the empirical distribution (and similarly for the entropies, see Section~\ref{sec:empirical_distributions_and_information_measures}). From the same model class we may extract a $D$-order Markov characterization of the probability of $\vr x$, therefore it makes sense to choose $Q$ as any $D$-order Markov distribution (note that this is only required for the inequality $\Rempname{\ML*} \leq \Rempname{\ML}$ which is needed for proving the achievability of $\Rempname{\ML*}$. Thus in this case we have the following information measures:
\begin{equation}\label{eq:A3456}
\Rempname{\ML} = \frac{1}{n} \log \frac{\hat p(\vr x | \vr z)}{Q(\vr x)} = \frac{1}{n} \log \frac{\hat p((x_i | \vr x_{i-D}^{i-1}, \vr y_{i-D}^{i+D})_{i=1}^n)}{Q(\vr x)} = \hat H_Q(\vr x) - \hat H(\vr x | \vr z)
\end{equation}
To write $\Rempname{\ML*}$ we split the state vector into $z_{x,i} = \vr x_{i-D}^{i-1}, z_{y,i} = \vr y_{i-D}^{i+D}$ and write:
\begin{equation}\label{eq:A3461}
\Rempname{\ML*} = \frac{1}{n} \log \frac{\hat p(\vr x | \vr z)}{\hat p(\vr x | \vr z_x)} = \hat H(\vr x | \vr z_x) - \hat H(\vr x | \vr z_x, \vr z_y) = \hat I(\vr x; \vr z_y | \vr z_x)
\end{equation}

These rate functions are adaptively achievable by Theorem~\ref{theorem:RempML_adaptive_redundancy_empirical}. The redundancy due to the complexity of the parametric family is $\frac{1}{n} r_n \approx  \frac{(|\mathcal{X}|-1)\cdot |\mathcal{Z}|}{2} \cdot \frac{\log n}{n} =  \frac{(|\mathcal{X}|-1)\cdot |\mathcal{X}|^{D} \cdot  |\mathcal{Y}|^{2D+1}}{2} \cdot \frac{\log n}{n}$ (this is the dominant term, the full expression appears in Theorem~\ref{theorem:RempML_adaptive_redundancy_empirical}), while the redundancy due to adaptation is $\delta_n = O \left( \sqrt{\frac{\log n}{n}} \right)$ (see Theorem~\ref{theorem:adaptive_causal_distribution}). Note that because of the delay $D$, the adaptive rate scheme is able to estimate the conditional probability of a symbol $x_i$ only after $y_{i+D}$ was received, and therefore the last $D$ input symbols of each block are ``wasted'' (at time $i$ the decoding metric considers only $\vr x_1^{i-D}$).

By definition, for any channel that satisfies the model $P_{\theta}(\vr x^n | \vr y^n)$, the maximum likelihood rate function yields an average rate which is at least as large as maximum attainable rate with the given input distribution. By taking $D \to \infty$, this model is able to account for all stationary ergodic channels, i.e. channels in which the joint distribution of the processes $\vr X, \vr Y$ is time invariant. Of course, in order that the redundancy still tends to $0$, $D$ can be taken to infinity only at a logarithmic rate, eg. $|\mathcal{X}|^{D} \cdot  |\mathcal{Y}|^{2D} \approx \sqrt{n} \Rightarrow D \approx \frac{\log n}{2 \log (|\mathcal{X}| \cdot  |\mathcal{Y}|^2)}$.

From another point of view, if the processes $\vr X, \vr Y$ are stationary ergodic, then
\begin{equation}\begin{split}\label{eq:A3479}
\Rempname{\ML*}(\vr X; \vr Y)
&=
\hat I(\vr X; \vr Z_y | \vr Z_x)
\arrowexpl{\mathrm{Prob.}}
I(X_i ; \vr Y_{i-D}^{i+D} | \vr X_{i-D}^{i-1})
\\& \arrowexpl{D \to \infty}
I(X_i ; \vr Y | \vr X_{1}^{i-1})
\arrowexpl{i \to \infty}
\overline I(\vr X; \vr Y)
\end{split}\end{equation}
where the convergence in probability is due to the law of large numbers (convergence of the empirical probability) and true for any $i \geq D$ (therefore we may take $i \to \infty$), and the last relation is due to $I(X_i ; \vr Y | \vr X_{1}^{i-1}) = H(X_i | \vr X_{1}^{i-1}) -  H(X_i | \vr X_{1}^{i-1},\vr Y) \arrowexpl{i \to \infty} \overline H(\vr X) -  H(\vr X_{1}^{i-1} | \vr Y)$ \cite[Section 4.2]{CoverThomas_InfoTheoryBook}. This shows that when the channel is indeed stationary ergodic, the rate function proposed tends to the mutual information rate of the channel, which upper bounds the achievable rate (with the given prior).

\subsection{Channel variation over time}\label{sec:examples_time_variation}
The stationary ergodic model does not cover all types of memory in the channel. Another type is a channel state that evolves irrespectively of the input (such as in fading channels). Note that in (static) Markov channels, i.e. when the state is just a function of the input, capacity does not improve with feedback. However if the state doesn't depend only on the input (but can also evolve randomly), then capacity improves with feedback  (since it improves the transmitter's guess as to the state) \cite{Viswanathan99}. While in channels of the first kind, we are able to reach the capacity, which is also the feedback capacity, with the ``individual channel'' model (and the above rate function, with the right prior), in channels of the second type, our model, in which the input distribution is determined a-priori will create an inherent limitation, since the best rate is achieved by modifying the input distribution.

However, if we target the mutual information (rather than the feedback capacity), a suitable rate function can be devised by modifying the model such that the conditional probabilities may slowly change with time. Naturally, the redundancy associated with such a model will not tend to $0$ with $n$, but behave like $\frac{d}{2} \frac{\log T}{T}$ where $d$ is the number of parameters and $T$ measures the coherence time (the typical referesh rate of the conditional distribution, e.g. the length of the fading block in a block fading model). This non-decreasing redundancy reflects the loss in rate from learning the channel in each coherence epoch (in a statistical setting this would be reflected by the difference between the known-channel mutual information $I(\vr X; \vr Y | \theta)$ and the unknown channel mutual information $I(\vr X; \vr Y)$).

It is easy to see the source of this factor, in a block fading model. The maximum likelihood probability of $\vr x$ given $\vr y$ is a product of maximum likelihood probabilities of each block. Each of these is related to a legitimate (NML) probability by the normalization factor $c_{\NML}(T)$ where for large $T$, $\log c_{\NML}(T) \approx \frac{d}{2} \log T$ (see \eqref{eq:A1759b} and Section \ref{sec:redundancy_of_Remp_ML}), therefore $\hat p_{\ML}(\vr x | \vr y)$ is related to a conditional probability function by $c_{\NML}(T)^{n/T}$, and this affects the overall redundancy in a factor of $\frac{1}{n} \log \left( c_{\NML}(T)^{n/T} \right) = \frac{1}{T} \log c_{\NML}(T) \approx \frac{d}{2 T} \log T$.

\subsection{The modulo additive channel}\label{sec:examples_modadditive}
Shayevitz and Feder's results \cite{Ofer_ModuloAdditive} for the modulo-additive channel (with $\mathcal{X}=\mathcal{Y}$) can be interpreted as asymptotic adaptive achievability of the rate function
\begin{equation}\label{eq:A3521}
\Remp = \log |\mathcal{X}| - \hat H(\vr y - \vr x)
\end{equation}
where $\vr y - \vr x$ refers to letter by letter modulo subtraction. This rate function is easily outperformed by the empirical mutual information when using a uniform i.i.d. input distribution \cite[Section TBD]{YL_individual_full}, since $\hat H(\vr x) \arrowexpl{\mathrm{Prob.}} \log |\mathcal{X}|$ while $\hat H(\vr x | \vr y) = \hat H(\vr y - \vr x | \vr y) \leq \hat H(\vr y - \vr x)$. On the other hand the redundancy of attaining this rate function (the part relating to the model complexity) is smaller due to the smaller number of parameters. This rate function can be identified with the maximum likelihood rate function, $\Rempname{\ML}$ with $Q(\vr x) = |\mathcal{X}|^{-n}$ (uniform) and where the noise sequence $\vr y - \vr x$ is modeled as an i.i.d. sequence, i.e. $P_{\theta}(\vr x | \vr y) = \prod_{i=1}^n \theta(x_i - y_i)$. The intrinsic redundancy will therefore be bounded by $\approx \frac{|\mathcal{X}|-1}{2} \cdot \frac{\log n}{n}$ (see Section~\ref{sec:redundancy_of_Remp_ML}). The actual redundancy of the adaptive scheme is again dominated by $\delta_n = O \left( \sqrt{\frac{\log n}{n}} \right)$ of Theorem~\ref{theorem:adaptive_causal_distribution}. However this convergence rate is significantly better than the attained by Shayevitz and Feder's scheme \cite[Section V.C, Table I]{Ofer_ModuloAdditive}, which is approximately $n^{-1/32}$.\footnote{This is the convergence rate of $\epsilon_2(n)$ according to the parameters chosen in Section V.C, with a target to only show convergence.}
\excluded{$\epsilon_1,\epsilon_3 = \const \Rightarrow \tau m = \const \sqrt{b} \log n$. $\epsilon_2 \geq O(\sqrt{\tau + m b^{-1} \log b}) = O(\sqrt{\const \cdot m^{-1} \sqrt{b} \log n + m b^{-1} \log b}) \stackrel{m^* = \sqrt{\sqrt{b} \log n / (b^{-1} \log b)}}{\geq} O((\sqrt{b} \log n \cdot b^{-1} \log b)^{1/4}) \geq O(( \log n / \sqrt{b})^{1/4}) \stackrel{b \leq \log n, \text{below (43)}}{\geq}  O(( \log n)^{1/8})$ ????
}
In a straightforward way, as done in Section~\ref{sec:examples_markov}), the rate function can be extended to $\Remp = \log |\mathcal{X}| - \hat \hat H(\vr y - \vr x | \vr z)$ where $\vr z$ denotes the past of the assumed noise sequence $z_i = (\vr x_{i-D}^{i-1} - \vr y_{i-D}^{i-1})$. In
Section~\ref{sec:examples_compression} below we extend this result further by replacing $\hat H(\vr y - \vr x | \vr z)$ by the normalized conditional compression length $\frac{1}{n} L(\vr x | \vr y)$ attached by any sequential compression scheme for the sequence $\vr x$ given $\vr y$ (and in particular the normalized compression length attained for the noise sequence $\vr y - \vr x$ by any compression scheme).
\todo{Can we show that this rate function is optimal in the sense of competition with any length decoder, in the same way that eMI is optimal in the sense of one symbol decoder?}

\subsection{Rate functions based on compression schemes}\label{sec:examples_compression}
A result generalizing the empirical mutual information and its stationary ergodic extensions (Section~\ref{sec:examples_markov}) for case of a uniform input distribution, as well as Shayevitz and Feder's result \cite{Ofer_ModuloAdditive} from Section~\ref{sec:examples_modadditive} is the asymptotic attainability of the following rate function:
\begin{equation}\label{eq:A3550}
\Remp = \log |\mathcal{X}| - \frac{1}{n} L(\vr x | \vr y)
\end{equation}
where $L(\vr x | \vr y)$ is the compression (output) length of the sequence $\vr x$ when the sequence $\vr y$ is given as side information. In the non adaptive case, this rate function is asymptotically attainable for every uniquely decodable code, while for the adaptive case we need to assume the compressor is ``sequential'' (which will be formalized below).

\subsubsection{Attainability}\label{sec:examples_compression_attanability}
In the non adaptive case this directly stems from Kraft's inequality $\sum_{\vr x} \exp(-L(\vr x | \vr y)) \leq 1$ -- we can write $\Remp = \frac{1}{n} \log \frac{f(\vr x | \vr y)}{Q(\vr x)}$ where $f(\vr x | \vr y) = c(\vr y) \exp(-L(\vr x | \vr y))$ is a legitimate conditional probability with $c(\vr y) \leq 1$. Formally, using the Markov/Chernoff bound (Section~\ref{sec:chernoff_application})
\begin{equation}\begin{split}\label{eq:A3562}
\mu_Q(\Remp)
&\stackrel{\eqref{eq:A756}}{\leq}
\frac{1}{n} \log L_{F=t,n}
=
\frac{1}{n} \log \underset{Q}{\E} \left[ \exp(n \Remp(\vr X, \vr y)) \right]
\\&=
\frac{1}{n} \log \sum_{\vr x} \underbrace{Q(\vr x) \cdot |\mathcal{X}|^n}_{=1} \exp(-L(\vr x | \vr y))
\leq
0
\end{split}\end{equation}
Another way to prove the same result is by using the fact there are at most $\exp(T)$ sequences with $L(\vr x | \vr y) \leq T$, and that the total probability of these sequences is therefore at most $\frac{\exp(T)}{|\mathcal{X}|^n}$, and therefore $Q(\Remp \geq R) = Q(L(\vr x | \vr y) \leq n (\log |\mathcal{X}| - R)) \leq \frac{\exp[n (\log |\mathcal{X}| - R)]}{|\mathcal{X}|^n} = \exp(-nR)$, therefore by definition \eqref{eq:Adef_intrinsic_redundancy} $\mu_Q(\Remp) \leq 0$. The fact that we obtained a lower intrinsic redundancy than the one of  Section~\ref{sec:examples_modadditive} is not surprising, since some of the redundancy is hidden in the compression length itself.

For the rate adaptive case additional assumptions are needed. We assume the sequential compression scheme receives $x_i$ and $y_i$ sequentially (for $i=1,2,\ldots$, and occasionally outputs encoded bits representing $\vr x$. There is an additional input causing the machine to terminate (i.e. declaring the input pair as the end of the block), in which case it may emit additional bits that terminate the encoded block. The decoder is required to be able to reconstruct $\vr x$ (not necessarily sequentially) when $\vr y$ and the encoded bits are given.

Define $L_S(\vr x | \vr y)$ as the unterminated coding length, i.e. the length of the output of the encoder after the input $\vr x, \vr y$ has been fed, but the sequence has not been terminated (i.e. the encoder is expecting additional input), and $L_T(\vr x  | \vr y) = L(\vr x  | \vr y)$ as the terminated coding length, i.e. the length of encoding the complete sequence. The sequence $\vr x$ is uniquely decodable from the $L_T(\vr x  | \vr y)$ bits of the terminated code, but not necessarily from the $L_S(\vr x  | \vr y)$ bits of the unterminated one. The difference $L_T(\vr x  | \vr y) - L_S(\vr x  | \vr y) \geq 0$ is the information stored in the encoder which has not been output yet. We require that:
\begin{enumerate}
\item The difference between the terminated and unterminated lengths is bounded by an asymptotically negligible value: $\frac{1}{n} (L_T(\vr x  | \vr y) - L_S(\vr x  | \vr y)) \leq \ \frac{1}{n} \Delta_L(n) \arrowexpl{n \to \infty} 0$ \\
This can be considered an embodiment of the limitation to ``sequential'' encoders and precludes encoders that need to process the entire sequence in order to produce outputs.
\item The encoding length does not decrease when the sequence is extended: $L_T(\vr x_1^i  | \vr y_1^i) \geq L_T(\vr x_1^{i-1}  | \vr y_1^{i-1})$
\end{enumerate}

Consider the system of Section~\ref{sec:rate_adaptive_scheme} with the decoding metric $\psi(\vr x^k, \vr y^k, j)$ defined by:
\begin{equation}\label{eq:A3598}
\log \psi(\vr x^k, \vr y^k, j) = (k-j) \cdot \log |\mathcal{X}| - (L_T(\vr x^k | \vr y^k) - L_T(\vr x^j | \vr y^j))
\end{equation}
I.e. the metric compares the encoding length accumulated from $j$ to $k$ with the encoding length of a random sequence. If this difference is large, then $\vr x_{j+1}^k$ is assumed to be related to $\vr y$. We denote $\Delta_L^{*}(n) = \max \{\Delta_L(m) \}_{m=1}^n$.

We begin by evaluating the CCDF condition of Theorem~\ref{theorem:framework}. In order to bound $\placeunder{\Pr}{Q} \left\{ \psi(\vr X^k, \vr y^k, j) \geq t | \vr x^j\right\}$ we need to bound the number of sequences $x_{j+1}^k$ that satisfy this condition for given $\vr y^k$ and $\vr x^j$. Suppose that we insert $\vr x^j, \vr y^j$ and then further append them by  $\vr x_{j+1}^k, \vr y_{j+1}^k$ and terminate the encoding. Consider the length $L_T(\vr x^k | \vr y^k) - L_S(\vr x^j | \vr y^j)$. This is the number of bits emitted by the machine between times $j$ and $k$, and these bits uniquely encode the sequence $\vr x_{j+1}^k$ (i.e. it is possible to reconstruct $\vr x_{j+1}^k$ from $\vr x^k, \vr y$ and this bit sequence). Therefore the number of sequences that are encoded by less than $T$ bits is at most $\exp(T)$, and therefore their probability (over $Q(\vr x_{j+1}^k | \vr x^j)$) is at most $\frac{\exp(T)}{|\mathcal{X}|^{k-j}}$. I.e.
\begin{equation}\label{eq:A3629}
\placeunder{\Pr}{Q} \left\{ L_T(\vr x^k | \vr y^k) - L_S(\vr x^j | \vr y^j) \leq T | \vr x^j\right\} \leq \frac{\exp(T)}{|\mathcal{X}|^{k-j}}
\end{equation}
Therefore
\begin{equation}\begin{split}\label{eq:A3608}
\placeunder{\Pr}{Q} \left\{ \psi(\vr X^k, \vr y^k, j) \geq t \big| \vr x^j\right\}
& =
\placeunder{\Pr}{Q} \left\{  L_T(\vr x^k | \vr y^k) - L_T(\vr x^j | \vr y^j) \leq (k-j) \cdot \log |\mathcal{X}|  - \log t \big| \vr x^j\right\}
\\ & \stackrel{\text{Assumption (1)}: L_T \leq L_S + \Delta}{\leq}
\placeunder{\Pr}{Q} \left\{  L_T(\vr x^k | \vr y^k) - L_S(\vr x^j | \vr y^j) \leq (k-j) \cdot \log |\mathcal{X}|  - \log t + \Delta_L^{*}(n) \big| \vr x^j\right\}
\\& \stackrel{\eqref{eq:A3629}}{\leq}
\frac{\exp((k-j) \cdot \log |\mathcal{X}| - \log t + \Delta_L^{*}(n))}{|\mathcal{X}|^{k-j}}
\\&=
\frac{\exp(\Delta_L^{*}(n))}{t}
\end{split}
\end{equation}
which satisfies the CCDF condition of Theorem~\ref{theorem:framework} with $L_{m} = \exp(\Delta_L^{*}(n))$ (this holds for all $m$ therefore $b_0=0$).

The summability condition is satisfied using the assumptions above: given a set of segments $\{j_b, k_b\}_{b=1}^B$ as defined in Theorem~\ref{theorem:framework} with $\sum_{b=1}^B (k_b-j_b) = n-m_0$, we extend the sequence by defining $j_{B+1} = n$, and write:
\begin{equation}\begin{split}\label{eq:A3652}
\sum_{b=1}^B \log \psi_b
&=
\sum_{b=1}^B \left[ (k_b-j_b) \cdot \log |\mathcal{X}| - (L_T(\vr x^{k_b} | \vr y^{k_b}) - L_T(\vr x^{j_b} | \vr y^{j_b})) \right]
\\& \stackrel{\text{Assumption (2)}, j_{b+1} \geq k_b}{\geq}
(n-m_0) \cdot \log |\mathcal{X}| - \sum_{b=1}^B \left[ L_T(\vr x^{j_{b+1}} | \vr y^{j_{b+1}}) - L_T(\vr x^{j_b} | \vr y^{j_b}) \right]
\\&=
(n-m_0) \cdot \log |\mathcal{X}| - \left[ L_T(\vr x^{n} | \vr y^{n}) - L_T(\vr x^{j_1} | \vr y^{j_1}) \right]
\\& \geq
\left[ n \cdot \log |\mathcal{X}| - L_T(\vr x^{n} | \vr y^{n}) \right] - m_0 \cdot \log |\mathcal{X}|
\\&=
\log \psi_0^n - m_0 \cdot \log |\mathcal{X}|
\end{split}\end{equation}
Therefore the summability condition of Theorem~\ref{theorem:framework} is met with $f_0(\psi_0^n) = \log |\mathcal{X}|$. The values $c_n,b_1$ of Theorem~\ref{theorem:framework} evaluate to $c_n = \log \frac{n \cdot L_n}{\dfb \epsilon} = \log \frac{n}{\dfb \epsilon} + \Delta^*_L(n)$ and $b_1 = b_0 + 2\dfb - 1 = 2\dfb - 1$. Since our rate function is upper bounded by $R_{\max} = \log |\mathcal{X}|$, and $f_0$ is constant, we obtain the following result by substitution in Corollary~\ref{corollary:framework_redundancy}:

\begin{theorem}\label{theorem:adaptive_L}
Given a sequential source coding scheme with input symbols from alphabet $\mathcal{X}$ that satisfies assumptions (1,2), and assigns a codeword length of $L(\vr x | \vr y)$ to the sequence $\vr x \in \mathcal{X}^n$ given $\vr y \in \mathcal{Y}^n$, then the following rate function
is adaptively achievable
\begin{equation}\label{eq:A3551}
\Remp = \log |\mathcal{X}| - \frac{1}{n} L(\vr x | \vr y)
\end{equation}
up to $\delta_n$, where
\begin{equation}\label{eq:A3663}
\delta_n = 3 \sqrt{\frac{\log |\mathcal{X}|}{n} \cdot \left(\log \frac{n}{\dfb \epsilon} + \Delta^*_L(n)  + (2\dfb - 1) \cdot \log |\mathcal{X}| \right)} \ntoinfty 0
\end{equation}
and $\Delta_L^{*}(n) = \max \{\Delta_L(m) \}_{m=1}^n$.
\end{theorem}

\begin{figure}
\centering
\ifpdf
  \setlength{\unitlength}{1bp}%
  \begin{picture}(249.87, 231.41)(0,0)
  \put(0,0){\includegraphics{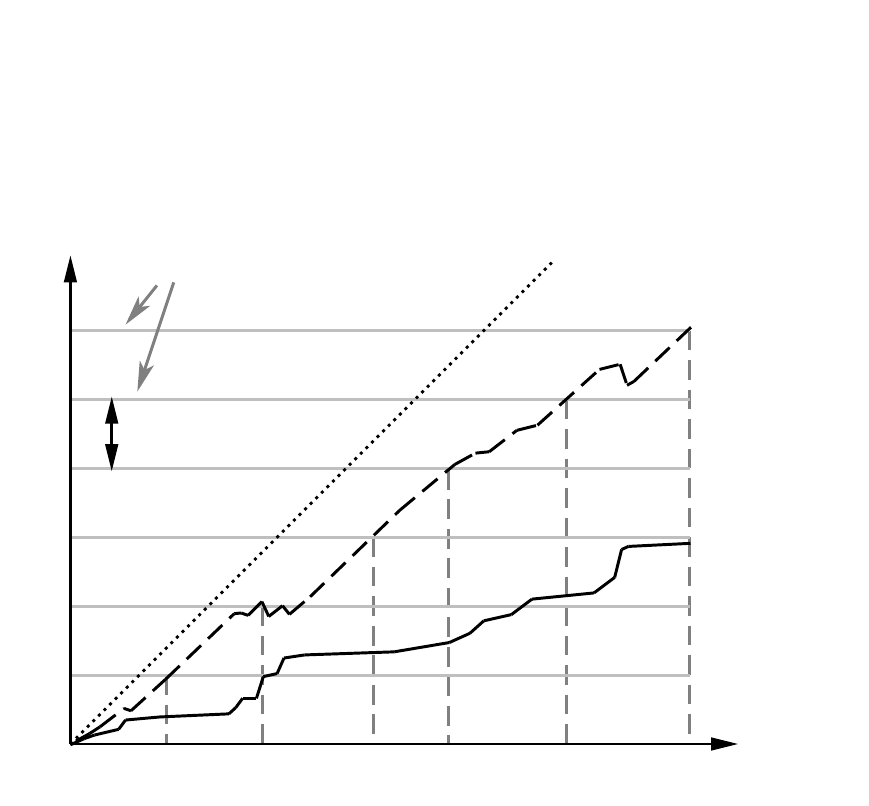}}
  \put(79.67,86.32){\rotatebox{45.00}{\fontsize{9.96}{11.95}\selectfont \smash{\makebox[0pt][l]{$k \cdot \log |\mathcal{X}|$}}}}
  \put(84.22,61.78){\rotatebox{45.00}{\fontsize{9.96}{11.95}\selectfont \smash{\makebox[0pt][l]{$N_k = k \cdot \log |\mathcal{X}| - L(\vr x^k | \vr y^k)$}}}}
  \put(141.32,65.36){\fontsize{9.96}{11.95}\selectfont $L(\vr x^k | \vr y^k)$}
  \put(143.30,7.81){\fontsize{9.96}{11.95}\selectfont  $k$ - time index}
  \put(13.45,107.59){\rotatebox{90.00}{\fontsize{9.96}{11.95}\selectfont \smash{\makebox[0pt][l]{Number of bits}}}}
  \put(40.12,154.65){\fontsize{9.96}{11.95}\selectfont \textcolor[rgb]{0.50196, 0.50196, 0.50196}{Decoding thresholds}}
  \put(36.15,105.04){\fontsize{9.96}{11.95}\selectfont $K$}
  \end{picture}%
\else
  \setlength{\unitlength}{1bp}%
  \begin{picture}(249.87, 231.41)(0,0)
  \put(0,0){\includegraphics{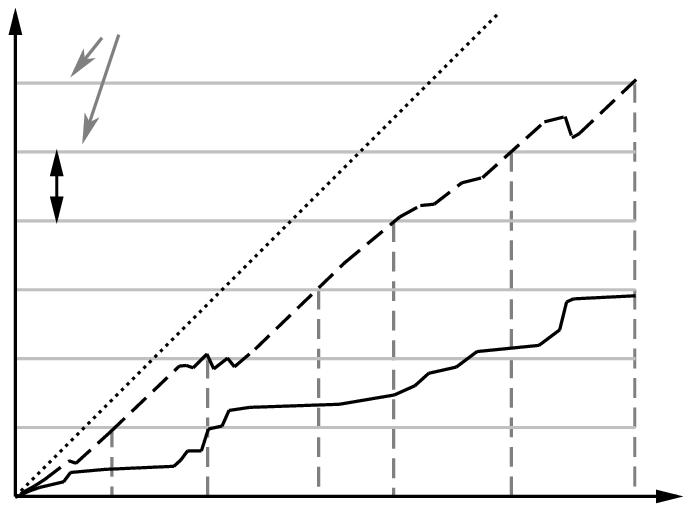}}
  \put(79.67,86.32){\rotatebox{45.00}{\fontsize{9.96}{11.95}\selectfont \smash{\makebox[0pt][l]{$k \cdot \log |\mathcal{X}|$}}}}
  \put(84.22,61.78){\rotatebox{45.00}{\fontsize{9.96}{11.95}\selectfont \smash{\makebox[0pt][l]{$N_k = k \cdot \log |\mathcal{X}| - L(\vr x^k | \vr y^k)$}}}}
  \put(141.32,65.36){\fontsize{9.96}{11.95}\selectfont $L(\vr x^k | \vr y^k)$}
  \put(143.30,7.81){\fontsize{9.96}{11.95}\selectfont  $k$ - time index}
  \put(13.45,107.59){\rotatebox{90.00}{\fontsize{9.96}{11.95}\selectfont \smash{\makebox[0pt][l]{Number of bits}}}}
  \put(40.12,154.65){\fontsize{9.96}{11.95}\selectfont \textcolor[rgb]{0.50196, 0.50196, 0.50196}{Decoding thresholds}}
  \put(36.15,105.04){\fontsize{9.96}{11.95}\selectfont $K$}
  \end{picture}%
\fi
\caption{\label{fig:rate_adaptive_nRemp_over_time}%
 Illustration of the decoding rule of the rate adaptive system. $L(\vr x^k | \vr y^k)$ is the compression length. Decoding thresholds with respect to $N_K=k \cdot \log |\mathcal{X}| - L(\vr x^k | \vr y^k)$ are depicted by horizontal lines.}
\end{figure}

Note that the decoding metric \eqref{eq:A3598} in this case is a difference of two values of the form $N_k = k \cdot \log|\mathcal{X}| - L(\vr x^k | \vr y_k)$ that can be interpreted as the ``incompressibility'' of the sequence up to time $k$ (the gap between the compressibility of the hypothetical noise sequence, and the compressibility of a random sequence). It is interesting to give an interpretation of the rate adaptive scheme of Section~\ref{sec:rate_adaptive_scheme} using $N_k$. Recall that to terminate a block, the decoder compares the decoding metric against a threshold. Ignoring the overhead terms this threshold is approximately $\exp(K)$ (see $\psi^*$ in Theorem~\ref{theorem:framework}), therefore the termination condition may be interpreted as decoding when the value of $N_k$ increases by $K$ from the start of the current block. For random sequences (the codewords that were not transmitted), $N_k$ is not expected to increase (the compression length is approximately $\log |\mathcal{X}|$ per symbol), and $K$ reflects the value of the threshold needed to make sure the probability of a random sequence appearing to be ``compressible'' is small. When $N_k$ increased by $K$, the termination condition is satisfied, and we begin a new block, therefore is a correspondence between the increase in $N_k$ and the number of blocks and bits that are transmitted, i.e. the termination condition can be approximately interpreted as $N_k \geq K (b+1)$ where $b$ is the number of blocks so far. Therefore assuming by time $n$, $B$ blocks were transmitted, the number of transmitted bits is $K \cdot B \approx N_n = n \cdot \log|\mathcal{X}| - L(\vr x^n | \vr y^n)$. This is depicted in Figure~\ref{fig:rate_adaptive_nRemp_over_time}, where the horizontal axis is the time $k$. The solid line presents $L(\vr x^k | \vr y^k)$, and the dashed line $N_k$. The decoding thresholds $K b$ ($b=1,2,\ldots$) are depicted as horizontal lines, while the vertical lines depict the decoding times. We can see that a decoding occurs whenever $N_k$ crossed a threshold.

\subsubsection{The modulo additive case}\label{sec:examples_compression_modadditive}
A specific case of the rate function proposed here is obtained for the modulo-additive channel when using a non-conditional source encoder operating over the (hypothesized) noise sequence $\vr z = \vr y - \vr x$, i.e.
\begin{equation}\label{eq:A3675}
\Remp = \log |\mathcal{X}| - \frac{1}{n} L(\vr y - \vr x)
\end{equation}
In this case $\frac{1}{n} L(\vr y - \vr x)$ can be considered a generalization of the notion of empirical entropy, and therefore generalizes the rate function \eqref{eq:A3521} presented previously for this channel.

It is specifically interesting to consider an application of the Lempel-Ziv algorithms (LZ77 \cite{LZ77} or LZ78 \cite{LZ78}), since their compression rate asymptotically reaches the finite state compressibility of the noise sequence $\rho(\vr z)$, which surpasses empirical entropies of any order. This substitution can be used to prove the universality of the system of Section~\ref{sec:rate_adaptive_scheme} attaining \eqref{eq:A3675}, over any finite-block length system operating over the modulo additive \onlypaper{channel \cite{YL_UnivModuloAdditive}.}\onlyphd{channel. See Section~***.}

We need to show that LZ77 \cite{LZ77} and LZ78 \cite{LZ78} fulfil the assumptions of Theorem~\ref{theorem:adaptive_L}. Both algorithms operate by creating a dictionary from previous symbols in the string, compressing a new substring to a tuple containing its location in the dictionary, plus, possibly one additional symbol. In LZ77 the dictionary consists of all substrings that begin in a window of specified length before the first symbol that was not encoded yet. LZ78 parses the string $\vr z$ into phrases. Each phrase is a substring which is not a prefix of any previous phrase, but can be generated from concatenating a previous phrase with one additional symbol. The dictionary contains all phrases.

It is easy to make sure that $L_T$ is monotonous (Assumption~(2) of Theorem~\ref{theorem:adaptive_L}). This depends on the way the last phrase in the string is treated (and does not affect the asymptotical performance), since this phrase may be an incomplete substring of a string in the dictionary, and therefore does not naturally terminate and produce a tuple. If, for example, the last phrase is sent without coding, then $L_T$ will not be monotonous (since adding more symbols to $\vr z$ that will terminate the phrase will result in a shorter compression). A simple treatment is to encode the last phrase similarly to other phrases - refer to one of the phrases in the dictionary which is a prefix of the remaining substring, and always give the length of the last substring (or the length of the block) at the end. This way the compression length associated with the last substring does not decrease when the substring is extended.

In order to bound $L_T(\vr z) - L_S(\vr z)$ (Assumption~(1)), we need to bound the tuple which encodes the last  phrase. In LZ78 this tuple carries an index to a previous phrase, plus a new symbol. The number of previous phrases is bounded by $n$ (a coarse bound, but sufficient for our purpose), and therefore \cite[Lemma 13.5.1]{CoverThomas_InfoTheoryBook} its encoding will be of length $\log n + \log \log n + 1$, and the length of the tuple will be $\log n + \log \log n + c$ (where $c$ is a constant accounting also for rounding, encoding of the additional symbol, etc). Therefore, if we end the block with an indication of its length we have total $\Delta_{LZ78}(n) \leq 2 \log n + 2 \log \log n + c$. In LZ77 this tuple carries a pointer to the window and a length (i.e. two numbers bounded to $\{1,\ldots,n\}$). Therefore after adding an indication of the length at the termination we would have $\Delta_{LZ77}(n) \leq 3 \log n + 3 \log \log n + c$. In both cases $\Delta_{LZ}(n) = O(\log n)$ and the requirement is satisfied.

\subsubsection{A converse for the modulo additive case}
An interesting thing to note is that all rate functions that depend only on the noise sequence $\Remp(\vr x, \vr y) = R(\vr z)$ ($\vr z = \vr y - \vr x$), can be written in the form $R(\vr z) = \log |\mathcal{X}| - \tfrac{1}{n} L(\vr z)$, where $L$ a compression length.

Two way to see this is by using the achievability of $R(\vr z)$ to bound the maximum number of sequences with $R(\vr z) > R$, which then bounds the number of sequences with $L(\vr z) < n \log |\mathcal{X}| - n R(\vr z)$, and we can show that Kraft inequality is met. Since $R(\vr z)$ can always be written as
\begin{equation}\label{eq:3392}
R(\vr z) = \log |\mathcal{X}| - \tfrac{1}{n} L(\vr z)
,
\end{equation}
the purpose is now to prove that for any achievable $R(\vr z)$, $L(\vr z)$ satisfies Kraft's inequality. Rounding issues are ignored as their effect is at most 1 bit, so $L(\vr z)$ is allowed to be non-integer. The input distribution $Q(\vr x)$ is not limited to be the uniform distribution. Choose a fixed $\vr y$ and define the random variable $\vr Z = \vr X - \vr y$. Then, taking any $\gamma < 1$, the necessary condition of Lemma~\ref{lemma:chernoff_tightness_lemma} yields:
\begin{equation}\label{eq:3374}
\E \left[ \exp(n \gamma R(\vr Z)) \right] \leq \frac{1}{(1-\epsilon)(1-\gamma)}
,
\end{equation}
Because the above holds for any $\vr y$, the same inequality holds for $\vr Y$ generated randomly and uniformly over $\mathcal{X}^n$. In this case, irrespective of the distribution of $\vr x$, $\vr Z$ becomes uniformly distributed as well. Therefore:
\begin{equation}\label{eq:3374c}
\underset{\vr Z \sim \unif(\mathcal{X}^n)}{\E} \left[ \exp(n \gamma R(\vr Z)) \right]
=
\frac{1}{|\mathcal{X}|^n} \sum_{\vr z} \exp(n \gamma R(\vr z))
\leq
\frac{1}{(1-\epsilon)(1-\gamma)}
.
\end{equation}
This can be written as:
\begin{equation}\label{eq:3410}
\sum_{\vr z} \exp(n [\gamma R(\vr z) - \log |\mathcal{X}|] + \log (1-\epsilon) + \log(1-\gamma))
\leq
1
\end{equation}
I.e. the following encoding lengths
\begin{equation}\begin{split}\label{eq:3416}
L'(\vr z)
&=
n \log |\mathcal{X}| - n \gamma R(\vr z)  - \log (1-\epsilon) - \log(1-\gamma)
\\& \stackrel{\eqref{eq:3392}}{=}
\gamma L(\vr z) + n (1-\gamma) \log |\mathcal{X}| - \log (1-\epsilon) - \log(1-\gamma)
\end{split}\end{equation}
satisfy Kraft's inequality $\sum_{\vr z} \exp(-L'(\vr z)) \leq 1$. Since $\gamma L(\vr z)$ is shorter (better) than $L(\vr z)$, $\gamma$ is chosen to minimize the overhead terms the second and fourth terms of \eqref{eq:3416}). The optimal $\gamma$ is $\gamma = 1 - \frac{1}{n \ln |\mathcal{X}|}$, which when substituted above yields:
\begin{equation}\label{eq:3429}
L'(\vr z) = \gamma L(\vr z) + \underbrace{\log \left(\frac{n e \ln |\mathcal{X}|}{1-\epsilon} \right)}_{\defeq \delta_L}
\leq
L(\vr z) + \delta_L
.
\end{equation}
To make $L'(\vr z)$ feasible encoding lengths one may have to add an overhead of $1$ bit. This is summarized in the following theorem:
\begin{theorem}\label{theorem:Rz_converse}
If $\Remp(\vr x, \vr y) = \log |\mathcal{X}| - \tfrac{1}{n} L(\vr x - \vr y)$ is an achievable rate function (with $\epsilon, Q(\vr x)$), then $\lceil L(\vr z) + \delta_L \rceil$ are feasible compression lengths (i.e. satisfy Kraft's inequality) where $\delta_L = \log \left(\frac{n e \ln |\mathcal{X}|}{1-\epsilon} \right)$.
\end{theorem}
Note that the overhead $\delta_L$ satisfies $\frac{1}{n} \delta_L \ntoinfty 0$ and is therefore asymptotically negligible. Combining this with the positive result of Section~\ref{sec:examples_compression_modadditive}, implies that every rate function which is a function of only the noise sequence $\vr z = \vr y - \vr x$, is asymptotically bounded by the form $\log |\mathcal{X}| - \tfrac{1}{n} L(\vr z)$ (for some compression lengths $L(\vr z)$).

Another interesting way of proof is to generate a compression scheme from the encoder and decoder: suppose we use the decoder to decode the message from $\vr y$, re-encode it to obtain $\hat {\vr x}$, and calculate an estimate of the noise $\hat {\vr z} = \vr y - \hat {\vr x}$. Suppose we run all combinations of $n R(\vr z)$ bits as inputs to the encoder, then take the output and pass it through the channel with a specific noise sequence $\vr z$. Then we obtained $2^{n r(\vr z)}$ different sequences $\vr y$, $1-\epsilon_s$ of which will be mapped by the previous machine to $\vr z$ ($s$ denotes the common randomness, and we know that on average $E_s \epsilon_s \leq \epsilon$). If we generate $\vr y$ at random (uniformly), the probability of the machine to output $\vr z$ is at least $\frac{2^{n r(\vr z)}}{|\mathcal{X}|^{n}} = 2^{-n (\log_2 |\mathcal{X}| - R(\vr z))}$. Now to encode, we generate for each coded sequence, in each length (i.e. $'0', '1', '00', '01', ...$, which to be a prefix code needs to be added a length indication) a random choice of a $\vr y$ sequence, and pass it to the previous machine to generate a $\vr z$ sequence.  The encoding of a sequence $\vr z$ is done by taking the first coded sequence which generates $\vr z$ in the generated codebook. Since we have at least $2^m$ sequence until exhausting all combinations up to length $m$, and the probability of each one to produce $\vr z$ is at least $2^{-n (\log_2 |\mathcal{X}| - R(\vr z))}$, we can see that this probability will be high if $L(\vr z) = m$ is slightly larger than $n (\log_2 |\mathcal{X}| - R(\vr z))$. More accurately, the probability that the length will be higher than $m$, i.e. that all words up to length $m$ will not produce $\vr z$ is $\left(1 - 2^{-n (\log_2 |\mathcal{X}| - R(\vr z))} \right)^{2^m} \approx e^{-2^{m - n (\log_2 |\mathcal{X}| - R(\vr z))}}$, so it decays very quickly after this point.

\todo{
Just for proving no need to create such a complex codebook: we may say that to encode with $L_0$ bits, we produce a codebook of $2^{nL}$ choices of $\vr y$. Then, if $L(\vr z) < L_0$ we are likely to find a match in our codebook. From this fixed-length scheme it is possible to move to a variable length scheme. However note that because of the codebook randomization and because of the error probability $\epsilon$, we can only create random source coding schemes. These schemes cannot be converted to non-random coding schemes (since the errors that occur with probability $\epsilon$ for each sequence, for any fixed encoder, either happen or not happen for each sequence). So the proof using the converse (we should actually use the Chernoff bound formulation) is preferred.
}

\subsubsection{The conditional Lempel-Ziv}\label{sec:examples_compression_condLZ}
We now consider another interesting substitution in $L(\vr x | \vr y)$ for the general (non modulo additive) case, which is the conditional Lempel-Ziv algorithm, described e.g. by Ooi \cite[Section 4.3.1]{Ooi}. This algorithm based on LZ78 \cite{LZ78} performs Lempel-Ziv incremental parsing of the combined sequence $(x_i, y_i)$. With this parsing each $\vr x$ phrase is associated with a $\vr y$ phrase. Then for each phase the algorithm sends the last letter of the phrase, plus the index of the phrase obtained by removing the last letter, out of all phrases with the same value of $\vr y$. The assumptions of Theorem~\ref{theorem:adaptive_L} are met in the same way as they are for the non-conditional case (the output phrases are of same or smaller length).

Note that the metric that results from using the conditional LZ we $L(\vr x | \vr y)$ is similar to the metric used by Ziv \cite{ZivUniversal} in order to construct a universal decoder that attains the maximum likelihood error exponent for all finite state channels. Ziv's metric which was later termed the conditional LZ complexity \cite{Uyematsu03} (see \eqref{eq:A3796}) refers directly to the number of phrases generated for each $\vr y$-phrase, and can be shown to be asymptotically close to the $L(\vr x | \vr y)$. Furthermore the conditional LZ algorithm was used by Ooi \cite{OOi} for constructing a universal communication scheme for finite state channels based on iterative compression.

The results known for the non-conditional LZ such as Ziv's lemma \cite{CoverThomas_InfoTheoryBook} can be extended to the conditional case \cite{Uyematsu03}, and therefore for every stationary ergodic channel with a stationary ergodic input, the compression rate tends asymptotically (for $n \to \infty$ almost surely) to the conditional entropy rate $\frac{1}{n} L(\vr X | \vr Y) \to \overline H(\vr X|\vr Y)$ \cite[Theorem 2]{Uyematsu03}, and hence our rate function tends to the mutual information.

The probability $\hat P_{LZ}(\vr x | \vr y) = \exp(-L(\vr x | \vr y))$ assigned by the conditional LZ to an input sequence, asymptotically surpasses (up to vanishing factors) the probability that can be assigned to the sequence by any finite state machine operating on the sequences $\vr x, \vr y$. Since we have not found an explicit derivation of this result we show this explicitly in \onlyphd{Appendix~\ref{sec:cond_LZ_performance}}\onlypaper{\cite{YL_PhdThesis}}. Therefore considering the setting of Section~\ref{sec:rate_of_given_metric}, using this rate function we can compete with the performance of every maximum likelihood decoder using a finite state characterization of the channel (this is not surprising given Ziv's results \cite{ZivUniversal}, and especially related to his Lemma~1). Therefore the current result gives us another angle on Ziv's result regarding the finite state channel: while Ziv considered competing systems operating at the same rate, and showed that the system using the conditional LZ complexity as a decoding metric achieves the same error exponent universally, here we may compare against systems operating at different rates (tuned to specific FS channels), and show that the rate adaptive system attains at least the rate obtained by any of these systems (however we have a suboptimal error exponent).

Another possible candidate for $L(\vr x | \vr y)$ with similar properties (but possibly better convergence rate) is the conditional version of the context tree weighting algorithm \cite{HaixiaoCai06}.

\subsubsection{Kolmogorov complexity?!}

\subsection{Second order rate function for the MIMO channel}\label{sec:GaussianMIMO}
In the \selector{previous paper}{initial work} \cite{YL_individual_full} we presented the rate function $\half \log \frac{1}{1-\hat \rho^2}$ where $\rho$ is the empirical correlation factor for the real valued channel $\mathbb{R} \to \mathbb{R}$ and showed it is asymptotically adaptively achievable. In this section we extend this result in several directions: we consider a MIMO channel with $t$ transmit and $r$ receive antennas, where the components may be real or complex numbers (i.e. $\mathbb{R}^t \to \mathbb{R}^r$ or $\mathbb{C}^t \to \mathbb{C}^r$), and where the correlation matrix or alternatively the covariance matrix may be used to define the rate (the difference being in subtracting the mean before taking second moments). The non-adaptive attainability of the rate function for the real-valued MIMO channel was shown in a conference paper \cite{YL_MIMO_ITW2010} on the subject.

We have altogether four cases (complex/real, covariance/correlation), for which the results and the techniques are very similar. In order to avoid duplication, we will prove them together (and apologize for the additional complication caused). For that purpose, we define $d$ as the dimensionality of the input, i.e. $1$ for real valued and $2$ for complex input, and $u$ as an indicator whether the mean is subtracted, i.e. $u=0$ for correlation matrices, and $u=1$ for covariance matrices. The input and output alphabets are denoted $\mathcal{X} = \mathbb{B}^t, \mathcal{Y}=\mathbb{B}^r$, where $\mathbb{B} \defeq \begin{cases} \mathbb{R} & d=1 \\ \mathbb{C} & d=2 \end{cases}$.
For a matrix $A$, $A^*$ denotes the conjugate-transpose of $A$. We use $\vr 1$ to denote a column vector of $1$-s, whose dimension is implicit.

We collect the input vectors over $n$ symbols into the $n \times t$ matrix $\mt X$ and similarly the $n \times r$ matrix $\mt Y$ denotes the output. The rate function is given as a function of $\mt X, \mt Y$. We denote sub-matrices similarly to sub-vectors, i.e. $\mt X_j^k$ denotes the matrix composed of rows $j$ to $k$ of $\mt X$.

Although the result here is stronger, the proof in the conference paper \cite{YL_MIMO_ITW2010} is more intuitive than here. Here we use similar techniques but the proof is more complex due to the need to show adaptive achievability and the other generalizations mentioned, and some of the intuition may be lost.

\begin{table}
  \centering
  \begin{tabular}{|c||c|c||c|c|}
    \hline
    Item & $d=0$ & $d=1$ & $u=0$ & $u=1$ \\ \hline
    Input alphabet & $\mathbb{R}^t$ & $\mathbb{C}^t$ & - & - \\
    Output alphabet & $\mathbb{R}^r$ & $\mathbb{C}^r$ & - & - \\
    $\hat C_X$ & - & - & $=\frac{1}{n} \mt X^* \mt X$ & $=\frac{1}{n} (\mt X - \vr 1 \cdot \mu)^* \mt (\mt X - \vr 1 \cdot \mu), \mu = \frac{1}{n} \cdot 1^T \cdot \mt X$ \\
    Gaussian Family & Real valued & Complex & Zero mean & Non zero mean \\
    \hline
  \end{tabular}
  \caption{Main differences between the 4 cases defined for the MIMO channel}\label{tbl:mimo_4_cases}
\end{table}

\subsubsection{The Gaussian parametric family and the maximum likelihood distribution}\label{sec:ML_gaussian_family}
The rate function we present is based on the maximum likelihood construction \eqref{eq:A1674} relating to the Gaussian i.i.d. family of distributions. In this section we present the distribution and its associated maximum likelihood probability. The parametric family defining the joint distribution of $\vr x$ and $\vr y$ is the family of Gaussian or complex Gaussian i.i.d. distributions:
\begin{equation}\label{eq:A3089}
\Theta = \begin{cases}
    \Normal(\mu_{XY}, \Lambda_{XY})^n, \qquad \mu_{XY} \in \mathbb{R}^{t+r}, \Lambda_{XY} \in \mathbb{R}^{(t+r) \times (t+r)}      &   u=1, d=1 \\
    \Normal(0, \Lambda_{XY})^n, \qquad \Lambda_{XY} \in \mathbb{R}^{(t+r) \times (t+r)}      &   u=0, d=1 \\
    \mathcal{CN}(\mu_{XY}, \Lambda_{XY})^n, \qquad \mu_{XY} \in \mathbb{C}^{t+r}, \Lambda_{XY} \in \mathbb{C}^{(t+r) \times (t+r)}      &   u=1, d=2 \\
    \mathcal{CN}(0, \Lambda_{XY})^n, \qquad \Lambda_{XY} \in \mathbb{C}^{(t+r) \times (t+r)}      &   u=0, d=2 \\
\end{cases}
\end{equation}
Using the maximum likelihood rate function \eqref{eq:A1674} over this family, guarantees attaining the mutual information for every Gaussian memoryless MIMO channel (where the input and output are jointly Gaussian).

We would like to find the maximum likelihood probabilities for the families above. We start with the non-conditional case, i.e. the maximum likelihood probability of a vector (which we denote by $\vr x$, but it may be a concatenation of $\vr x, \vr y$).
In the non-conditional form, each of the $n$ rows of $\mt X$ is modeled as a Gaussian random vector $\Normal(\mu_{1 \times t}, \Lambda_{t \times t})$, independent of the others. The probability density of a single row $\vr x$ (a row vector) in the real valued case is:
\begin{equation}\label{eq:A3708}
P_{\vr \mu, \Lambda}(\vr x) = \left| 2 \pi \Lambda \right|^{-\half} e^{-\half (\vr x - \vr \mu) \Lambda^{-1} (\vr x - \vr \mu)^T} \qquad \vr x \in \mathbb{R}^t
\end{equation}
In the complex-valued case, we have instead:\footnote{It is easy to produce this distribution by taking a complex Gaussian vector who's real and imaginary parts are i.i.d. distributed $\Normal(0,\half)$ and multiply it by $\Lambda^{\half}$}
\begin{equation}\label{eq:A3713}
P_{\vr \mu, \Lambda}(\vr x) = \left| \pi \Lambda \right|^{-1} e^{- (\vr x - \vr \mu) \Lambda^{-1} (\vr x - \vr \mu)^*} \qquad \vr x \in \mathbb{C}^t
\end{equation}
Where in both cases $\mu = \E \vr x$ and $\Lambda = \E (\vr x - \vr u)^* (\vr x - \vr u)$. $\Lambda$ is non-negative definite. Note that in the complex case, the power of each component of $\vr x$ is split between the real and imaginary components). In general we can write:
\begin{equation}\label{eq:A3720}
P_{\vr \mu, \Lambda}(\vr x) = \left| d \pi \Lambda \right|^{-d/2} e^{-\frac{d}{2} (\vr x - \vr \mu) \Lambda^{-1} (\vr x - \vr \mu)^*} \qquad \vr x \in \mathbb{B}^t
\end{equation}
To obtain the rate function based on correlation matrices ($u=0$) we will degenerate this family by fixing $\mu=0$.
For brevity, in the rest of the section, we will use the word ``Gaussian'' to refer to both Gaussian and complex Gaussian vectors.

Considering the $n \times t$ matrix $\mt X = \left( \vr x_1^T, \ldots, \vr x_t^T \right)^T$ where the rows are i.i.d. and distributed according to \eqref{eq:A3720}, we have the following distribution for the matrix:
\begin{equation}\begin{split}\label{eq:A3731}
P_{\vr \mu, \Lambda}(\mt X)
&=
\prod_{i=1}^{n} P_{\vr \mu, \Lambda}(\vr x_i)
=
\left| d \pi \Lambda \right|^{-\dalf n} e^{-\dalf \sum_{i=1}^{n} \vr (\vr x_i - \vr \mu) \Lambda^{-1} (\vr x_i - \vr \mu)^*}
\\&=
\left| d \pi \Lambda \right|^{-\dalf n} e^{-\dalf \tr \left( (\mt X - \vr 1 \cdot \vr \mu) \Lambda^{-1} (\mt X - \vr 1 \cdot \vr \mu)^* \right) }
\stackrel{\tr AB = \tr BA}{=}
\left| d \pi \Lambda \right|^{-\dalf n} e^{-\dalf \tr \left(  (\mt X - \vr 1 \cdot \vr \mu)^* (\mt X - \vr 1 \cdot \vr \mu) \Lambda^{-1} \right) }
\end{split}\end{equation}

We would now like to find the find the ML estimate of $\vr \mu$ and $\Lambda$ given $\mt X$. For $u=0$ we fix $\vr \mu=0$ and optimize \eqref{eq:A3731} with respect to $\Lambda$. It is intuitively clear that for $u=1$, $\hat \mu_{\ML}$ is just the empirical mean $\hat \mu_{\ML} = \frac{1}{n} \vr 1^T \cdot \mt X$, and that $\hat \Lambda_{\ML}$ is the empirical covariance ($u=1$) or correlation matrix ($u=0$) $\hat \Lambda_{\ML} = \frac{1}{n} (\mt X - \vr 1 \cdot \hat \mu_{\ML})^* (\mt X - \vr 1 \cdot \hat \mu_{\ML})$ (where for $u=0$ we just take $\hat \mu_{\ML}=0$).

To prove this, we first maximize \eqref{eq:A3731} with respect to $\mu$, which implies minimizing $\tr \left(  (\mt X - \vr 1 \cdot \vr \mu)^* (\mt X - \vr 1 \cdot \vr \mu) \Lambda^{-1} \right)$. Defining
\begin{equation}\label{eq:A3753}
\mt X_c = \mt X - \vr 1 \cdot \vr {\hat \mu_{\ML}}
\end{equation}
we have that $\vr 1^T \cdot \mt X_c = 0$ and therefore:
\begin{equation}\begin{split}\label{eq:A3754}
\tr \left(  (\mt X - \vr 1 \cdot \vr \mu)^* (\mt X - \vr 1 \cdot \vr \mu) \Lambda^{-1} \right)
&=
\tr \left(  (\mt X_c + \vr 1 \cdot (\vr {\hat \mu_{\ML}} - \vr \mu))^* (\mt X_c + \vr 1 \cdot (\vr {\hat \mu_{\ML}} - \vr \mu)) \Lambda^{-1} \right)
\\&=
\tr \left( \mt X_c^* \mt X_c \Lambda^{-1} \right) + \tr \left(  \vr 1 \cdot (\vr {\hat \mu_{\ML}} - \vr \mu)^* (\vr {\hat \mu_{\ML}} - \vr \mu) \vr 1^T \Lambda^{-1} \right)
\end{split}\end{equation}
The second term is non-negative and is minimized for $\vr \mu = \vr {\hat \mu_{\ML}}$.

Substituting $\vr \mu = \vr {\hat \mu_{\ML}}$ in \eqref{eq:A3731} we obtain
\begin{equation}\label{eq:A3766}
\max_{\vr \mu} P_{\vr \mu, \Lambda}(\mt X)
=
\left| d \pi \Lambda \right|^{-\dalf n} e^{-\dalf \tr \left(  \mt X_c^* \mt X_c \Lambda^{-1} \right) }
\end{equation}
Where $\mt X_c$ is defined by \eqref{eq:A3753} (where for $u=0$ we fix $\hat \mu_{\ML}=0$). It remains to maximize the above with respect to $\Lambda$. We change optimization variable by defining $\mt A = \mt X_c^T \mt X_c \Lambda^{-1}$; The determinants of the two matrices are related by $\ln | \mt A | = \ln \left| \mt X_c^T \mt X_c \right| - \ln \left| \Lambda \right| = \const - \ln \left| \Lambda \right|$ so taking the logarithm of \eqref{eq:A3766} and removing constants, it remains to maximize:
\begin{equation}\label{eq:A3789}
n \ln | \mt A | - \tr \mt A
\end{equation}
with respect to $\mt A$. By Hadamard inequality since $\mt A$ is non-negative definite, $ | \mt A | \leq \prod_{i=1}^t \mt A_{ii} $ (with equality iff $\mt A$ is diagonal), therefore \eqref{eq:A3789} is upper bounded by $\sum_{i=1}^t \left( n \ln \mt A_{ii} - \mt A_{ii} \right)$, which is maximized for $\mt A_{ii} = n$. The upper bound can be met by choosing a diagonal $\mt A$, and therefore we have $\mt A = n \cdot I_{t \times t}$. Changing variables we obtain the ML estimate of $\Lambda$ is the empirical covariance/correlation:
\begin{equation}
\hat \Lambda_{\ML} (\mt X) = \mt X_c^T \mt X_c \cdot \mt A^{-1} = \frac{1}{n} \mt X_c^T \mt X_c
\end{equation}

Substituting the result into the probability density we obtain:
\begin{equation}\begin{split}\label{eq:A3794}
\hat p_{\ML} (\mt X)
&=
P_{\hat \mu(\mt X), \hat \Lambda (\mt X)}(\mt X)
=
\left| d \pi \frac{1}{n} \mt X_c^T \mt X_c \right|^{-\dalf n} e^{-\dalf \tr \left( \mt X_c^T \mt X_c \left(\frac{1}{n} \mt X_c^T \mt X_c \right)^{-1} \right)}
\\&=
\left| d \pi \frac{1}{n} \mt X^T \mt X \right|^{-\dalf n} e^{-\dalf n \cdot t}
\\&=
\left| \frac{d \pi e}{n} \mt X_c^T \mt X_c \right|^{-\dalf n}
\end{split}\end{equation}

Note that $\hat p_{\ML} (\mt X)$ diverges when the columns of $\mt X_c$ are linearly dependent.

We now discuss the conditional case. Assume $[\vr x, \vr y]$ are jointly Gaussian row vectors of sizes $t,r$ respectively, with means $[\vr \mu_x, \vr \mu_y]$ and covariances $\Lambda_{xx},\Lambda_{yy},\Lambda_{xy}$. Then the conditional distribution is known to be Gaussian as well with:
\begin{equation}\label{eq:A3822}
P_{\vr \mu_x, \vr \mu_y, \Lambda_{xx},\Lambda_{yy},\Lambda_{xy}}(\vr x | \vr y) = \left| d \pi \Lambda_{x|y} \right|^{-\dalf} e^{-\dalf (\vr x - \mu_{x|y}(\vr y)) \Lambda_{x|y}^{-1} (\vr x - \mu_{x|y}(\vr y))^*}
\end{equation}
where
\begin{equation}\label{eq:A3825cond_mu_lambda}
\mu_{x|y}(\vr y) = \vr \mu_x + (\vr y - \vr \mu_y)\Lambda_{yy}^{-1}\Lambda_{yx}  \hspace{5ex}
\Lambda_{x|y} = \Lambda_{xy}\Lambda_{yy}^{-1} \Lambda_{xy}^*
\end{equation}
For our purposes, it will be convenient to define the conditional distribution by a different set of parameters. We write:
\begin{equation}\label{eq:A3832}
P_{\theta}(\vr x | \vr y) = \left| d \pi \Lambda_{x|y} \right|^{-\dalf} e^{-\dalf (\vr x -  \vr y \mt A - \vr b) \Lambda_{x|y}^{-1} (\vr x - \vr y \mt A - \vr b)^*}
\end{equation}
Where $\theta = [\mt A_{[r \times t]}, \vr b_{[1 \times t]}, {\Lambda_{x|y}}_{[t \times t]}]$ is the vector of new parameters. $\vr y \mt A + \vr b$ is the MMSE estimator $\E \left[ \vr x | \vr y \right]$. For the case $u=0$ we fix $\vr b =0$.

For matrices $\mt X, \mt Y$ whose rows are distributed i.i.d. based on the distribution above, we have:
\begin{equation}\begin{split}\label{eq:A3839}
P_{\theta}(\mt X | \mt Y)
&=
\prod_{i=1}^n P_{\theta}(\vr x_i | \vr y_i)
=
\left| d \pi \Lambda_{x|y} \right|^{-\dalf n} e^{-\dalf \sum_{i=1}^n (\vr x_i - \vr y_i \mt A - \vr b) \Lambda_{x|y}^{-1} (\vr x_i - \vr y_i \mt A - \vr b)^*}
\\&=
\left| d \pi \Lambda_{x|y} \right|^{-\dalf n} e^{-\dalf \tr \left[ (\mt X - \mt Y \mt A - \vr 1 \cdot \vr b) \Lambda_{x|y}^{-1} (\mt X - \mt Y \mt A - \vr 1 \cdot \vr b)^* \right]}
=
\left| d \pi \Lambda_{x|y} \right|^{-\dalf n} e^{-\dalf \tr \left[ (\mt X - \mt Y \mt A - \vr 1 \cdot \vr b)^* (\mt X - \mt Y \mt A - \vr 1 \cdot \vr b) \Lambda_{x|y}^{-1} \right]}
\end{split}\end{equation}

To find the ML estimator, we begin by maximizing with respect to $\mt A, \vr b$. This is a simple quadratic problem, but the algebra can be avoided, by considering it as an estimation problem. Consider the matrix $\Lambda_{\epsilon} = \frac{1}{n} (\mt X - \mt Y \mt A - \vr 1 \cdot \vr b)^* (\mt X - \mt Y \mt A - \vr 1 \cdot \vr b)$. This matrix can be considered as the mean estimation error covariance matrix in the following scenario: there is a linear estimator $\hat {\vr x} = \vr y \mt A + \vr b$ is sought, and the matrix above is the estimation error covariance matrix, when $(\vr x, \vr y)$ are selected from the $i$-th row of $[\mt X, \mt Y]$ and $i \sim \unif\{1,\ldots,n\}$. In other words, when one seeks a linear estimator, which given a randomly selected row in $\mt Y$ will produce an estimate of the respective row in $\mt X$. The LMMSE estimator brings the matrix $\Lambda_{\epsilon}$ to minimum (in the matrix sense) and therefore would bring $P_{\theta}$ to maximum. In this scenario, the covariances and means of $(\vr x, \vr y)$ are the empirical covariances and means (since the rows are selected uniformly). Therefore the optimal linear estimator is
\begin{equation}\label{eq:A3866}
\vr y \mt A + \vr b = \vr {\hat \mu}_{\mt X} +  (\vr y - \vr {\hat \mu}_{\mt Y}) \hat {\mt C}_{\mt Y \mt Y}^{-1} \hat {\mt C}_{\mt Y \mt X}
\end{equation}
where
\begin{eqnarray*}
  \vr {\hat \mu}_{\mt X} &=& \hat E_i \vr x_i = \frac{1}{n} \vr 1^T \mt X \\
  \vr {\hat \mu}_{\mt Y} &=& \hat E_i \vr y_i = \frac{1}{n} \vr 1^T \mt Y \\
  \hat {\mt C}_{\mt Y \mt X} &=& \hat E_i (\vr y_i - \vr {\hat \mu}_{\mt Y})^T (\vr x_i - \vr {\hat \mu}_{\mt X}) = \frac{1}{n} (\mt Y - \vr 1 \cdot \vr {\hat \mu}_{\mt Y})^* (\mt X - \vr 1 \cdot \vr {\hat \mu}_{\mt X}) \\
  \hat {\mt C}_{\mt Y \mt Y} &=& \hat E_i (\vr y_i - \vr {\hat \mu}_{\mt Y})^T (\vr y_i - \vr {\hat \mu}_{\mt Y}) = \frac{1}{n} (\mt Y - \vr 1 \cdot \vr {\hat \mu}_{\mt Y})^* (\mt Y - \vr 1 \cdot \vr {\hat \mu}_{\mt Y}) \\
\end{eqnarray*}
Furthermore, after substituting $\mt A, \vr b$ from \eqref{eq:A3866} we will obtain in the exponent of \eqref{eq:A3839} the LMMSE error matrix (of the aforementioned scenario) which is:
\begin{equation}\label{eq:A3879}
\Lambda_{\epsilon}^{LMMSE} = \hat {\mt C}_{\mt X \mt X} - \hat {\mt C}_{\mt Y \mt X}^* \hat {\mt C}_{\mt Y \mt Y}^{-1} \hat {\mt C}_{\mt Y \mt X} \defeq \hat {\mt C}_{\mt X|\mt Y}
\end{equation}
where $\hat {\mt C}_{\mt X \mt X}$ is defined similarly $\hat {\mt C}_{\mt Y \mt Y}$. Substituting $\Lambda_{\epsilon}$ in \eqref{eq:A3839} we have:
\begin{equation}\label{eq:A3885}
\max_{\mt A, \vr b} P_{\theta}(\mt X | \mt Y)
=
\left| d \pi \Lambda_{x|y} \right|^{-\dalf n} e^{-\dalf \tr \left[ n \cdot \hat {\mt C}_{\mt X|\mt Y} \Lambda_{x|y}^{-1} \right]}
\end{equation}
This can be also verified by direct substitution of \eqref{eq:A3866} in \eqref{eq:A3839}. In the case of $u=0$, where we have $\vr b=0$, we are limited to linear estimators of the form $\hat {\vr x} = \vr y \mt A$. The solution in this case is to replace $\vr {\hat \mu}_{\mt X}, \vr {\hat \mu}_{\mt Y}$ with zeros, and $\hat {\mt C}_{\cdot,\cdot}$ with the respective correlation matrices (i.e. obtained without removing the mean). The proof is technical and appears in Appendix~\ref{sec:MMSE_no_mean_appendix}.

We remain with the problem of maximizing with respect to $\Lambda_{x|y}$, which is identical to the non-conditional case \eqref{eq:A3766}, where $\frac{1}{n} \mt X_c^* \mt X_c$ is replaced with $\hat {\mt C}_{\mt X|\mt Y}$. Therefore the maximum in \eqref{eq:A3885} will be attained for $\Lambda_{x|y} = \hat {\mt C}_{\mt X|\mt Y}$, and the maximum likelihood distribution is:
\begin{equation}\begin{split}\label{eq:A3900}
\hat p_{\ML} (\mt X | \mt Y)
&=
\max_{\theta} P_{\theta}(\mt X | \mt Y)
=
\left| d \pi \hat {\mt C}_{\mt X|\mt Y} \right|^{-\dalf n} e^{-\dalf \tr \left[ n \cdot \hat {\mt C}_{\mt X|\mt Y} \hat {\mt C}_{\mt X|\mt Y}^{-1} \right]}
\\&=
\left| d \pi \hat {\mt C}_{\mt X|\mt Y} \right|^{-\dalf n} e^{-\dalf n t}
=
\left| d \pi e \hat {\mt C}_{\mt X|\mt Y} \right|^{-\dalf n}
\end{split}\end{equation}
where $\hat {\mt C}_{\mt X|\mt Y}$ is a function of $\mt X, \mt Y$ defined by \eqref{eq:A3879}.

Note that if the columns of $\mt Y$ are linearly dependent, or are linearly dependent on the $\vr 1$ vector (in the case $u=1$), the value of \eqref{eq:A3866} is not defined. In this case, return to \eqref{eq:A3839} and observe that the result of $\hat p_{\ML}$ only depends on the subspace spanned by the columns of $\mt Y$ (plus the vector $\vr 1$) since this determines the values that $\mt Y \mt A - \vr 1 \cdot \vr b$ can attain. Therefore, removing linearly dependent columns from $\mt Y$ does not change the result (and it does not matter which columns are removed).

We summarize the results of this sub-section in the following Lemma:
\begin{lemma}\label{lemma:Gaussian_pML}
Let the matrix $\mt X$ be defined by an i.i.d. Gaussian $\Normal(\mu,\Lambda)$ distribution ($d=1$) or a complex Gaussian $\mathcal{CN}(\mu,\Lambda)$ distribution ($d=2$) on its rows, as defined in \eqref{eq:A3720}. Then the maximum likelihood probability, which is obtained by maximizing \eqref{eq:A3720} with respect to $\mu,\Lambda$ (in the case $u=1$) or with respect to $\Lambda$ for $\mu=0$ (in the case $u=0$) is:
\begin{equation}\label{eq:A3922}
\hat p_{\ML} (\mt X) = \left| d \pi e \hat {\mt C}_{\mt X \mt X} \right|^{-\dalf n}
\end{equation}
where $\hat {\mt C}_{\mt X \mt X}$ is defined below. When $\mt X$ is defined by a conditional i.i.d. distribution on its rows, conditioned on the respective rows of $\mt Y$, as defined in \eqref{eq:A3822} or \eqref{eq:A3839}, then the maximum likelihood probability, obtained by maximizing with respect to \eqref{eq:A3839} to $\theta = [\mt A_{[r \times t]}, \vr b_{[1 \times t]}, {\Lambda_{x|y}}_{[t \times t]}]$ (where for $u=0$, $\vr b=0$ and is excluded from $\theta$), is:
\begin{equation}\label{eq:A3931}
\hat p_{\ML} (\mt X | \mt Y) = \left| d \pi e \hat {\mt C}_{\mt X|\mt Y} \right|^{-\dalf n}
\end{equation}
where the covariance matrices are defined as follows:
\begin{eqnarray}
\vr {\hat \mu}(\mt Z) &=& \begin{cases} \vr 0 & u = 0 \\ \frac{1}{n} \vr 1^T \cdot \mt Z & u = 1 \end{cases} \\
\hat {\mt C}_{\mt Z \mt W} &=& \frac{1}{n} (\mt Z - \hat \mu(\mt Z))^* (\mt W - \hat \mu(\mt W)) \\
\hat {\mt C}_{\mt X|\mt Y} &=& \hat {\mt C}_{\mt X \mt X} - \hat {\mt C}_{\mt Y \mt X}^* \hat {\mt C}_{\mt Y \mt Y}^{-1} \hat {\mt C}_{\mt Y \mt X} \\
\end{eqnarray}
where $\mt Z, \mt W$ are generic matrices which are replaced with $\mt X$ or $\mt Y$ as appropriate. If $\hat {\mt C}_{\mt Y \mt Y}$ is singular, the result is obtained by removing columns of $\mt Y$ until the columns are linearly in-dependent of each other (and the $\vr 1$ vector, in case of $u=1$).
\end{lemma}

\subsubsection{The maximum likelihood rate function}
The input distribution is based on the the i.i.d. Gaussian distribution $\Normal(0, \Lambda_X)^n$ or $\mathcal{CN}(0, \Lambda_X)^n$ (we always use mean zero even if $u=1$). We define $\tilde Q$ as the ideal distribution $\Normal(0, \Lambda_X)^n$:
\begin{equation}\label{eq:A3299}
\tilde Q(\mt X) \stackrel{\eqref{eq:A3731}}{=} \left| d \pi \Lambda_X \right|^{-\dalf n} e^{-\dalf \tr \left(  \mt X^* \mt X \Lambda_X^{-1} \right) }
\end{equation}
Since $\tilde Q(\mt X)$ is unbounded from below (for non-degenerate $\mt X$, taking $\alpha \to \infty$ yields $\tilde Q(\alpha \mt X) \to 0$), the actual input distribution will be a trimmed Gaussian $\mt Q(\mt X)$ which will be defined in the sequel. However the rate function will be defined with respect to the ideal $\tilde Q$.

As in Section~\ref{sec:Remp_ML_as_entropy_diff} we can define the rate function by the empirical and quazi-empirical entropies:
\begin{equation}\label{eq:A3304}
\hat H_{\tilde Q}(\mt X) = -\frac{1}{n} \log {\tilde Q}(\mt X)
=
\dalf \log \left| d \pi \Lambda_X \right| + \dalf \cdot \log e \cdot \tr \left(  \frac{1}{n} \mt X^* \mt X \cdot \Lambda_X^{-1} \right)
=
\dalf \log \left| d \pi e \Lambda_X \right| + \dalf \cdot \log e \cdot \tr \left(  \frac{1}{n} \mt X^* \mt X \cdot \Lambda_X^{-1} - \mt I \right)
\end{equation}
\begin{equation}\label{eq:A3307}
\hat H_{\ML}(\mt X) = -\frac{1}{n} \log \hat p_{\ML}(\mt X)
\stackrel{\eqref{eq:A3922}}{=}
\dalf \cdot \log \left| d \pi e \hat {\mt C}_{\mt X \mt X} \right|
\end{equation}
Note the similarity to the expression for the entropy of a Gaussian random vector.

\begin{equation}\label{eq:A3315}
\hat H_{\ML}(\mt X | \mt Y) = -\frac{1}{n} \log \hat p_{\ML}(\mt X | \mt Y)
\stackrel{\eqref{eq:A3931}}{=}
\dalf \cdot \log \left| d \pi e \hat {\mt C}_{\mt X | \mt Y} \right|
\end{equation}
and the rate functions:
\begin{equation}\label{eq:A3321}
\Rempname{\ML} \stackrel{\eqref{eq:A1674}}{=} \frac{1}{n} \log \frac{\hat p_{\ML}(\mt X | \mt Y)}{{\tilde Q}(\mt X)} \stackrel{\eqref{eq:A1666}}{=} \hat H_{{\tilde Q}}(\mt X) - \hat H_{\ML}(\mt X | \mt Y) = \dalf \log \frac{\left| \Lambda_X \right|}{\left| \hat {\mt C}_{\mt X | \mt Y} \right|} +  \dalf \cdot \log e \cdot \tr \left(  \frac{1}{n} \mt X^* \mt X \cdot \Lambda_X^{-1} - \mt I \right)
\end{equation}
\begin{equation}\label{eq:A3324}
\Rempname{\ML*} \stackrel{\eqref{eq:A1696}}{=}  \frac{1}{n} \log \frac{\hat p_{\ML}(\mt X | \mt Y)}{\hat p_{\ML}(\mt X)} \stackrel{\eqref{eq:A1670}}{=}  \hat H_{\ML}(\mt X) - \hat H_{\ML}(\mt X | \mt Y) = \dalf \cdot \log \frac{\left| \hat {\mt C}_{\mt X \mt X} \right|}{\left| \hat {\mt C}_{\mt X | \mt Y} \right|}
\end{equation}

where $\hat {\mt C}_{\mt X \mt X}, \hat {\mt C}_{\mt X | \mt Y}$ are as defined in Lemma~\ref{lemma:Gaussian_pML}. Note the similarity of the maximum likelihood empirical entropies to the entropies of gaussian random vectors where the true covariance is replaced with the empirical covariance (or correlation) matrices (the entropy of $\vr Z \sim \Normal(0,\Lambda_Z)$ is $\half \log | 2 \pi e \Lambda_Z |$). Regarding the quazi-empirical entropy $\hat H_{\tilde Q}(\mt X)$, it is composed of two parts: the first is the true (statistical) entropy of the channel input $\vr x$, and the second part is a measure for the similarity between the empirical correlation matrix of the input and the average one. For typical $\mt X$, $\frac{1}{n} \mt X^* \mt X \approx \Lambda_X$ and the second part tends to $0$. By definition (since ${\tilde Q}$ belongs to the parametric family $\Theta$), we have $\hat p_{\ML}(\mt X) \geq {\tilde Q}(\mt X)$ and $\hat H_{\tilde Q}(\mt X) \geq \hat H_{\ML}(\mt X)$.

The parametric class we defined is separable in the sense discussed in Section~\ref{sec:def_emp_entropy} (Equations \eqref{eq:A1430}, \eqref{eq:A1442}), i.e. the joint Gaussian distribution of the vectors $\vr x, \vr y$ (defined by the joint mean and covariance) can be equivalently defined by the mean and covariance of $\vr y$, and parameters defining the conditional mean and covariance of $\vr x$ given $\vr y$ (or equivalently, the matrices $\Lambda_{x|y}$, $\mt A$ and the vector $\vr b$ as in \eqref{eq:A3832}). Therefore \eqref{eq:A1430}, \eqref{eq:A1442} hold with equality, i.e. we can write:
\begin{equation}\label{eq:A3330}
\hat H_{\ML}(\mt X | \mt Y) = \hat H_{\ML}(\mt X, \mt Y) -  \hat H_{\ML}(\mt Y)
=
\dalf \cdot \log \left| d \pi e \hat {\mt C}_{(\mt X \mt Y)(\mt X \mt Y)} \right| - \dalf \cdot \log \left| d \pi e \hat {\mt C}_{\mt Y \mt Y} \right|
\end{equation}
Where $\hat {\mt C}_{(\mt X \mt Y)(\mt X \mt Y)}$ is the empirical covariance/correlation matrix of the matrix $[\mt X, \mt Y]$. Alternatively, this relation can be obtained by using Leibnitz formula
\begin{equation}\label{eq:A3337}
\left[ \begin{array}{cc}  A & B \\  C & D  \end{array} \right] =
   \left[ \begin{array}{cc}  A & 0 \\  C & I  \end{array} \right] \cdot
   \left[ \begin{array}{cc}  I & A^{-1}B \\  0 & D - C A^{-1} B  \end{array} \right]
\end{equation}
To obtain the relation:
\begin{equation}\begin{split}\label{eq:A3343}
\left| \hat {\mt C}_{(\mt X \mt Y)(\mt X \mt Y)} \right|
&=
\left| \begin{array}{cc}  \hat {\mt C}_{\mt X \mt X} & \hat {\mt C}_{\mt X \mt Y} \\  \hat {\mt C}_{\mt Y \mt X} & \hat {\mt C}_{\mt Y \mt Y}  \end{array} \right|
=
\left| \begin{array}{cc}  \hat {\mt C}_{\mt Y \mt Y} & \hat {\mt C}_{\mt Y \mt X} \\  \hat {\mt C}_{\mt X \mt Y} & \hat {\mt C}_{\mt X \mt X}  \end{array} \right|
\\&=
   \left| \begin{array}{cc}  \hat {\mt C}_{\mt Y \mt Y} & 0 \\  \hat {\mt C}_{\mt X \mt Y} & I  \end{array} \right| \cdot
   \left| \begin{array}{cc}  I & \hat {\mt C}_{\mt Y \mt Y}^{-1} \hat {\mt C}_{\mt Y \mt X} \\  0 & \hat {\mt C}_{\mt X \mt X} - \hat {\mt C}_{\mt X \mt Y} \hat {\mt C}_{\mt Y \mt Y}^{-1} \hat {\mt C}_{\mt Y \mt X}  \end{array} \right|
\\&=
\left|  \hat {\mt C}_{\mt Y \mt Y}  \right| \cdot
   \left|  \hat {\mt C}_{\mt X \mt X} - \hat {\mt C}_{\mt X \mt Y} \hat {\mt C}_{\mt Y \mt Y}^{-1} \hat {\mt C}_{\mt Y \mt X} \right|
=
\left|  \hat {\mt C}_{\mt Y \mt Y}  \right| \cdot
   \left|  \hat {\mt C}_{\mt X | \mt Y} \right|
\end{split}\end{equation}
Plugging into \eqref{eq:A3330} and noting that the factors $d \pi e$ are canceled out due to the matching sizes of the matrices, proves the relation.

Using this equality we can alternatively write $\Rempname{\ML*}$ in a symmetrical form \eqref{eq:A1675}:
\begin{equation}\label{eq:A3371}
\Rempname{\ML*} = \hat H_{\ML}(\mt X) + \hat H_{\ML}(\mt Y)  - \hat H_{\ML}(\mt X, \mt Y) = \dalf \cdot \log \frac{\left| \hat {\mt C}_{\mt X \mt X} \right| \cdot \left| \hat {\mt C}_{\mt Y \mt Y} \right|}{\left| \hat {\mt C}_{(\mt X \mt Y)(\mt X \mt Y)} \right|}
\end{equation}
This form was presented in a previous paper \cite{YL_MIMO_ITW2010} for the case $d=1,u=0$ and was proven to be asymptotically attainable (non adaptively). \onlypaper{In that paper, the rate function was justified based on different considerations, of convergence to the mutual information for Gaussian channels}

% phdonly

\subsubsection{Achievability of the rate function}
In the Gaussian case, the parametric class is continuous, and $\hat p_{\ML}(\mt X | \mt Y)$ may take unbounded values (when the matrices are highly correlated). Therefore the achievability proof is quite involved and uses the tools developed in Section~\ref{sec:ML_adaptivity_continuous}. We will use the metric defined in \eqref{eq:A2497} with a parameter $\gamma \in (0,1)$, which, using Theorem~\ref{theorem:framework} and Lemma~\ref{lemma:summability_continuous}, can achieve adaptively the rate function $\gamma \Rempname{\ML}$, and then take $\gamma \to 1$.

The main parts which are specific to the Gaussian case and need to be proven are:
\begin{enumerate}
\item We need to bound $Q$: $0 < q_{\min} \leq Q(x_i | \vr x^{i-1}) \leq q_{\max} < \infty$. This is done by trimming the input probability.
\item For the CCDF condition, we need to bound the quantity appearing in \eqref{eq:A2507}
\item For the summability condition, calculate $g_0(\psi_0^n)$ from \eqref{eq:A2592} related to the unconstrained symbols.
\end{enumerate}

We first state the result. The proof is partially followed in the next sub-sections, while the more tedious parts are in the appendix.

\begin{theorem}\label{theorem:GaussianMIMO}
Consider the channel $\mathcal{X} \to \mathcal{Y}$, where the input and output are vectors of size $t,r$ respectively $\mathcal{X} = \mathbb{B}^t, \mathcal{Y}=\mathbb{B}^r$, where each element is either real or complex valued $\mathbb{B} \defeq \begin{cases} \mathbb{R} & d=1 \\ \mathbb{C} & d=2 \end{cases}$. Let the $n \times t$ matrix $\mt X$ and the $n \times r$ matrix $\mt Y$ denote the channel input and output respectively.

Let the input distribution $Q$ be defined by an i.i.d. generation of each symbol $\vr x_i$ (row of $\mt X$) according to the following distribution:
\begin{equation}\label{eq:A4035}
Q(\vr x_i) =  c \cdot \Ind(\vr x_i^*  \Lambda_X^{-1} \vr x_i \leq \Omega^2) \cdot e^{-\frac{d}{2} \vr x_i \Lambda_X^{-1} \vr x_i^*}
\end{equation}
Where $\Lambda_X$ is a chosen positive semidefinite matrix, $\Omega$ is a chosen radius, and $c$ is a normalization factor chosen such that $\int_{\mathbb{R}^t} Q(\vr x) d\vr x = 1$. When $\Omega \to \infty$, $Q(\vr x)$ tends to the Gaussian or complex Gaussian distribution with zero mean and covariance matrix $\Lambda_X$. Consider the following rate functions:
\begin{equation}\label{eq:A4043}
\Rempname{\ML} = \dalf \log \frac{\left| \Lambda_X \right|}{\left| \hat {\mt C}_{\mt X | \mt Y} \right|} +  \dalf \cdot \log e \cdot \tr \left(  \frac{1}{n} \mt X^* \mt X \cdot \Lambda_X^{-1} - \mt I \right)
\end{equation}
\begin{equation}\label{eq:A4048}
\Rempname{\ML*} = \dalf \cdot \log \frac{\left| \hat {\mt C}_{\mt X \mt X} \right|}{\left| \hat {\mt C}_{\mt X | \mt Y} \right|}
= \dalf \cdot \log \frac{\left| \hat {\mt C}_{\mt X \mt X} \right| \cdot \left| \hat {\mt C}_{\mt Y \mt Y} \right|}{\left| \hat {\mt C}_{(\mt X \mt Y)(\mt X \mt Y)} \right|} \leq \Rempname{\ML}
\end{equation}
where $\hat {\mt C}_{\mt X \mt X}, \hat {\mt C}_{\mt X | \mt Y}$ are either empirical correlation matrices (for $u=0$) or covariance matrices (for $u=1$), defined in Lemma~\ref{lemma:Gaussian_pML}.
Then:
\begin{enumerate}
\item $F(\Rempname{\ML})$ and $F(\Rempname{\ML*})$ are adaptively achievable, where:
\begin{equation}\label{eq:A4054}
F(t) = \frac{\eta \cdot t}{1 + \alpha t} - \delta
\end{equation}
where $\eta, \alpha, \delta$ are defined as a function of the transmission length $n$, $\Omega$, the feedback delay $\dfb$, the number of bits per block $K$ (a chosen parameter), and $\gamma \in (0,1)$ (a chosen parameter) as follows:
\begin{eqnarray*}
\eta &=& \gamma \left(1 + \frac{B_{n,\gamma}}{K} \right)^{-1} \\
\alpha &=& \frac{A_{n,\gamma}}{K + B_{n,\gamma}} \\
\delta &=& a_0 + \frac{K}{n} \\
A_{n,\gamma} &=& \gamma \left( \frac{a_3}{1-\gamma} + a_4 \right) \\
B_{n,\gamma} &=& \log n + a_1 + a_2 \log \frac{1}{1-\gamma} + \left( \frac{a_3}{1-\gamma} + a_4 \right) \cdot \gamma \cdot a_5 \\
a_0 &=& \log \frac{1}{1 - \delta_{\Omega}} \\
a_1 &=& a_0 + \log \frac{1}{\dfb \epsilon} + a_2 \log(e) \\
a_2 &=& \frac{d}{4} \left(t + 1 + 2r + 2u \right) \cdot t \\
a_3 &=& t+1+r+u \\
a_4 &=& 2 \dfb - 1 \\
a_5 &=& \dalf (t + \Omega^2) \cdot \log (e) \\
\delta_{\Omega} &=& \frac{\Gamma \left( \frac{dt}{2}, \frac{d \Omega^2}{2} \right)}{\Gamma \left( \frac{dt}{2} \right)} \\
\end{eqnarray*}

\item $\Rempname{\ML}$ and $\Rempname{\ML*}$ are asymptotically adaptively achievable with a sequence of priors defined by $Q$ above \eqref{eq:A4035} with $\Omega \ntoinfty \infty$ (i.e. with the input distribution tending to Gaussian)
\end{enumerate}
\end{theorem}

The proof is organized as follows: in the subsections below we discuss the modified input probability and the summability condition. The computation of the CCDF condition which is rather involved appears in the appendix (Section~\ref{sec:GaussianMIMO_CCDF}). The final calculations that combine these results together also appear in the appendix (Section~\ref{sec:GaussianMIMO_Wrapup}). Finally, we show in Section~\ref{sec:GaussianMIMO_opt} a Lemma (which can be considered a corollary to Theorem~\ref{theorem:GaussianMIMO}), which gives a way to choose the parameters $\gamma,K$ that guarantees a bounded loss within a specified region.

\begin{figure}
\center
  \includegraphics[width=8cm]{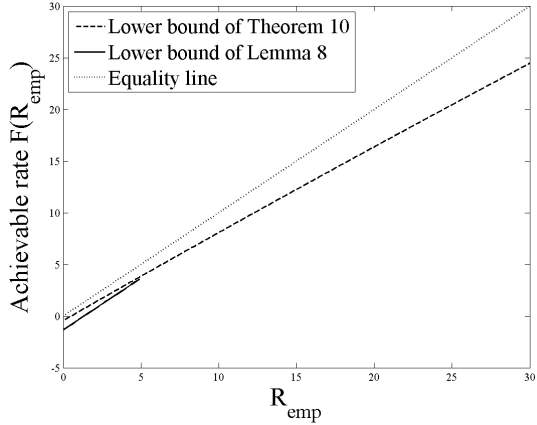}
  \caption{Illustration of $\Remp$ lower bound of Theorem~\ref{theorem:GaussianMIMO} and of Lemma~\ref{lemma:GaussianMIMO_opt}. The achieved rate is plotted against $\Rempname{\ML}$ for $n=100,000, r=t=2$. The full list of parameters appears in table \ref{table:GaussianMIMO_params} in the appendix.}\label{fig:GaussianMIMO}
\end{figure}

Figure~\ref{fig:GaussianMIMO} illustrates the lower bounds of Theorem~\ref{theorem:GaussianMIMO} and of Lemma~\ref{lemma:GaussianMIMO_opt}. The achieved rate is plotted against $\Rempname{\ML}$ for $n=100,000, r=t=2$. The full list of parameters appears in table \ref{table:GaussianMIMO_params} in the appendix. Due to the choice $R_0=5$ The bound of the lemma applies only for $\Remp \leq 5$. A comparison between Theorem~\ref{theorem:GaussianMIMO} when specialized to the SISO real valued case $t=1,r=1,u=0,d=1$ and the looser results obtained for the same setting in our \selector{previous paper}{initial work} \cite{YL_individual_full} appears in \selector{\cite{YL_PhdThesis}}{Section~\ref{sec:GaussianMIMO_comparison}}. Note that with mild values of $\Omega$, very small values of $\delta_{\Omega}$ are obtained, and thus the resulting input distribution is very close to the desired Gaussian distribution.

\todo{note that this rate function has infinite $\mu_Q$ and therefore cannot be achieved up to a factor, but only with a multiplicative loss. Prove this in appendix based on the proof in the previous paper with other side of equality}

A result on non-adaptive achievability stems as a byproduct of CCDF condition required for the proof of Theorem~\ref{theorem:GaussianMIMO}:
\begin{lemma}\label{lemma:GaussianMIMO_nonadaptive}
Under the definitions of Theorem~\ref{theorem:GaussianMIMO}, for any $\gamma \leq 1 - \frac{t+1+r+u}{n}$, the rate function $\gamma \Rempname{\ML}$ has an intrinsic redundancy:
\begin{equation}\label{eq:A5425}
\mu_Q(\gamma \Rempname{\ML})
\leq
\frac{1}{n} \log \left( \frac{1}{ 1 - \delta_{\Omega}} \right) + \frac{1}{n} \cdot \frac{d}{4} \left(t + 1 + 2r + 2u \right) \cdot t \cdot \log  \left( \frac{e}{1-\gamma} \right)
\end{equation}
and therefore by Theorem~\ref{theorem:remp_achievability_upto}, $\Remp = \gamma \Rempname{\ML} - \left(\mu_Q + \frac{\log \epsilon^{-1}}{n} \right)$ is achievable.
\end{lemma}
The proof of the lemma appears at the end of Section~\ref{sec:GaussianMIMO_CCDF}.

\subsubsection{The trimmed input probability}
As noted, the distribution $\tilde Q$ is unbounded:
\begin{equation}\label{eq:A3447}
\tilde Q(\vr x_i | \vr x^{i-1}) = \tilde Q(\vr x_i)
\stackrel{\eqref{eq:A3720}}{=}
\left| d \pi \Lambda_X \right|^{-d/2} e^{-\frac{d}{2} \vr x \Lambda_X^{-1} \vr x^*}
\end{equation}
When $\vr x \to \infty$ (in almost every direction), $\tilde Q(\vr x) \to 0$, therefore it is not bounded from below as required by the conditions of Section~\ref{sec:ML_adaptivity_continuous}. To meet the condition we define the trimmed distribution which limits $\vr x$ into a an ellipse define by a radius $\Omega$:
\begin{equation}\label{eq:A3455}
B_{\Omega} \defeq \left\{ \vr x : \vr x^*  \Lambda_X^{-1} \vr x \leq \Omega^2 \right\}
\end{equation}
$Q$ is the conditional density of $\vr x$ given that it belongs to $B_{\Omega}$:
\begin{equation}\label{eq:A3455b}
Q(\vr x) =  \frac{\Ind(\vr x \in B_{\Omega})}{ \tilde Q \left\{ B_{\Omega} \right\}} \cdot {\tilde Q}(\vr x)
\end{equation}
In the case of a white input $\Lambda_X = \mt I_{t \times t}$, this bounds the peak power of each input vector (which makes sense from a practical point of view). $\tilde Q \left\{ B_{\Omega} \right\}$ can be easily evaluated. Since according to $\tilde Q$, $d \cdot \vr x^*  \Lambda_X^{-1} \vr x$ is distributed $\chi^2$ with $d \cdot t$ degrees of freedom (it is the power of the white vector $\sqrt{d} \cdot \Lambda_X^{-1/2} \vr x$, which has Gaussian i.i.d. entries, where the factor $\sqrt{d}$ for the complex case normalizes the variance of the real and imaginary parts to $1$ rather than $\half$)
\begin{equation}\label{eq:A3460}
\tilde Q \left\{ B_{\Omega} \right\}
=
1 - \underset{\tilde Q}{\Pr} \left\{ d \vr x^*  \Lambda_X^{-1} \vr x \geq d \Omega^2 \right\}
=
1 - \underbrace{\frac{\Gamma \left( \frac{dt}{2}, \frac{d \Omega^2}{2} \right)}{\Gamma \left( \frac{dt}{2} \right)}}_{\defeq \delta_{\Omega}}
=
1 - \delta_{\Omega}
\end{equation}
where $\Gamma(t)$ is the gamma function, and $\Gamma(t,s)$ is the upper incomplete gamma function. $\delta_{\Omega}$ decays exponentially to $0$ when $\Omega \to \infty$. Therefore we have:
\begin{equation}\label{eq:A3477}
Q(\vr x) = \Ind(\vr x \in B_{\Omega}) \cdot \frac{1}{ 1 - \delta_{\Omega}} \cdot {\tilde Q}(\vr x)
\end{equation}

Below we address some properties of $Q$ and differences that arise from substituting $Q$ instead of $\tilde Q$. For the trimmed distribution $Q$ we have that $Q(\vr x) \in \{0 \} \cup [q_{\min}, q_{\max}]$ where:
\begin{equation}\label{eq:A3482}
q_{\min} = \min_{\vr x \in B_{\Omega}} Q(\vr x) = \frac{1}{1 - \delta_{\Omega}} \left| d \pi \Lambda_X \right|^{-d/2} \min_{\vr x \in B_{\Omega}} e^{-\frac{d}{2} \vr x \Lambda_X^{-1} \vr x^*} = \frac{1}{1 - \delta_{\Omega}} \left| d \pi \Lambda_X \right|^{-d/2} e^{-\frac{d}{2} \Omega^2}
\end{equation}
\begin{equation}\label{eq:A3486}
q_{\max} = \max_{\vr x \in B_{\Omega}} Q(\vr x) = \frac{1}{1 - \delta_{\Omega}} \left| d \pi \Lambda_X \right|^{-d/2} \max_{\vr x \in B_{\Omega}} e^{-\frac{d}{2} \vr x \Lambda_X^{-1} \vr x^*} = \frac{1}{1 - \delta_{\Omega}} \left| d \pi \Lambda_X \right|^{-d/2}
\end{equation}

We defined the quazi empirical entropy $\hat H_{\tilde Q}$ \eqref{eq:A3304} and the rate function in \eqref{eq:A3321} using $\tilde Q$, but the results of Lemma~\ref{lemma:summability_continuous} and Theorem~\ref{theorem:framework} apply to rate functions defined using the true input distribution $Q$. However since for $\vr x \in B_{\Omega}$ we have $Q(\vr x) \geq \tilde Q(\vr x)$, we have:
\begin{equation}\label{eq:A3506}
\hat H_{Q}(\mt X)
=
-\frac{1}{n} \log {Q}(\mt X)
=
-\frac{1}{n} \log \left[ \frac{1}{(1 - \delta_{\Omega})^n} {\tilde Q}(\mt X) \right]
=
\log (1 - \delta_{\Omega}) + \hat H_{\tilde Q}(\mt X)
\end{equation}
And therefore there is a loss of $\log (1 - \delta_{\Omega})$ in the rate.

In the sequel, we compute the expected value in the Markov CCDF condition of Theorem~\ref{theorem:framework}. It is convenient for the sake of this calculation to assume $\mt X \sim \tilde Q$ (i.e. is Gaussian) rather than $\mt X \sim Q$. There is a simple relation between the expected values in this case. For every non-negative function $g(\vr x)$:
\begin{equation}\label{eq:A3478}
\underset{Q}{\E} g(\vr x)
=
\int_{\vr x \in B_{\Omega}} Q(\vr x) g(\vr x) d \vr x
=
\frac{1}{ 1 - \delta_{\Omega}} \int_{\vr x \in B_{\Omega}} \tilde Q(\vr x) g(\vr x) d \vr x
\leq
\frac{1}{ 1 - \delta_{\Omega}} \int_{\vr x \in \mathbb{B}^t} \tilde Q(\vr x) g(\vr x) d \vr x
=
\frac{1}{ 1 - \delta_{\Omega}} \underset{\tilde Q}{\E} g(\vr x)
\end{equation}

\subsubsection{The summability condition}
We use Lemma~\ref{lemma:summability_continuous} in order to prove the summability condition.
In our case $\theta = [\mt A, \vr b, \Lambda_{x|y}]$ (see Section~\ref{sec:ML_gaussian_family}). As we saw, the ML estimate of $\Lambda_{x|y}$ is $\hat {\mt C}_{\mt X|\mt Y}$ and $\hat p_{\ML}(\mt X | \mt Y) \stackrel{\eqref{eq:A3931}}{=} \left| d \pi e \hat {\mt C}_{\mt X|\mt Y} \right|^{-\dalf n}$. On the other hand, the per-letter probability satisfies \eqref{eq:A3832}:
\begin{equation}\begin{split}\label{eq:A3816}
P_{\max}(\theta)
&=
\max_{\vr x, \vr y} P_{\theta}(\vr x | \vr y)
=
\max_{\vr x, \vr y}  \left| d \pi \Lambda_{x|y} \right|^{-\dalf} e^{-\dalf (\vr x -  \vr y \mt A - \vr b) \Lambda_{x|y}^{-1} (\vr x - \vr y \mt A - \vr b)^*}
\\&=
\left| d \pi \Lambda_{x|y} \right|^{-\dalf}
\end{split}\end{equation}
where $\vr x, \vr y$ are single rows of $\mt X, \mt Y$ (single symbols). We can observe that knowing $\hat p_{\ML}(\mt X | \mt Y)$ determines $\left| \hat \Lambda_{x|y} \right|$ and this relates to $P_{\max}(\theta)$.

Referring to Lemma~\ref{lemma:summability_continuous} we have:
\begin{equation}\label{eq:A3829}
\Theta^{(ML)}(t) = \left\{ \hat \theta_{\ML}(\vr x | \vr y) :  \hat p_{\ML}(\vr x | \vr y) \leq t \right\} =
\left\{ \hat \theta_{\ML}(\vr x | \vr y) :  \left| d \pi e \hat \Lambda_{x|y} \right|^{-\dalf n} \leq t \right\}
=
\left\{ \theta: \left| d \pi e \Lambda_{x|y} \right|^{-\dalf n} \leq t \right\}
\end{equation}

\begin{equation}\begin{split}\label{eq:A3826}
g_0(\psi_0^n)
&=
\max_{\theta \in \Theta^{(ML)}(q_{\max}^n \cdot (\psi_0^n)^{1/\gamma})} P_{\max}(\theta)
=
\max_{\left| d \pi e \Lambda_{x|y} \right|^{-\dalf n} \leq q_{\max}^n \cdot (\psi_0^n)^{1/\gamma}} \left| d \pi \Lambda_{x|y} \right|^{-\dalf}
\\&=
\max_{\left| d \pi \Lambda_{x|y} \right|^{-\dalf}  \leq e^{\dalf t} q_{\max} \cdot (\psi_0^n)^{\frac{1}{n \gamma}}} \left| d \pi \Lambda_{x|y} \right|^{-\dalf}
=
e^{\dalf t} q_{\max} \cdot (\psi_0^n)^{\frac{1}{n \gamma}}
\end{split}\end{equation}

Therefore by the lemma, the summability condition in Theorem~\ref{theorem:framework} holds with
\begin{equation}\begin{split}\label{eq:A3849}
f_0(\psi_0^n)
&=
\gamma \cdot \log \left(g_0(\psi_0^n) \cdot q_{\min}^{-1} \right)
=
\dalf t \gamma \cdot \log (e)
+
\gamma \cdot \log \frac{q_{\max}}{q_{\min}}
+
\gamma \cdot \log \left( (\psi_0^n)^{\frac{1}{n \gamma}} \right)
\\& \stackrel{\eqref{eq:A3482},\eqref{eq:A3486}}{=}
\dalf t \gamma \cdot \log (e)
+
\dalf \Omega^2 \gamma \cdot \log (e)
+
\frac{1}{n} \cdot \log (\psi_0^n)
=
\dalf (t + \Omega^2) \gamma \cdot \log (e)
+
\frac{1}{n} \cdot \log (\psi_0^n)
\end{split}\end{equation}

The proof of Theorem~\ref{theorem:GaussianMIMO} is finalized in the appendix (Section~\ref{sec:GaussianMIMO_Wrapup}).

\subsubsection{Selection of parameters for finite $n$ by approximate optimization}\label{sec:GaussianMIMO_opt}
The rate $\Remp$ defined in Theorem~\ref{theorem:GaussianMIMO} has a rather complex expression and it is not clear how to select the parameters. Below, we present a coarse way to choose these parameters by trying to minimize the main loss factors. We assume $\Omega$ is fixed, and so are the overheads related to it, and focus on $K,\gamma$. For various values of $K,\gamma$ we obtain different curves, none of which is uniformly better than others. The loss with respect to $\Rempname{\ML}$ determined by \eqref{eq:A3965} increases with $\Rempname{\ML}$, therefore it makes sense to optimize for all rates up to a certain value $\Rempname{\ML} = R_0$. In the appendix (Section~\ref{sec:GaussianMIMO_optproof}), we develop a coarse bound for the rate loss in the region $0 \leq \Rempname{\ML} \leq R_0$, and minimize the bound. This results in the following Lemma:
\begin{lemma}\label{lemma:GaussianMIMO_opt}
Under the definitions of Theorem~\ref{theorem:GaussianMIMO}, let $R_0 \geq 0$, and select $\gamma = 1 - \sqrt{\frac{a_6}{K}}$, $K = \lceil \left( n \cdot \sqrt{a_6} \cdot R_0 \right)^{\frac23} \rceil$, where $a_6 = \log n + a_1 + a_2 + (a_3 + a_4)\left( R_0 + a_5 \right)$, then
\begin{equation}\label{eq:A4085}
\forall t \in [0,R_0]: F(t) \geq t - \delta_0 - a_0
\end{equation}
where $\delta_0 = 3 n^{-\frac13} a_6^{\frac13} R_0^{\frac23} + \frac{1}{n}$
\end{lemma}

\onlyphd{
\subsection{Achieving capacity for probabilistic and semi-probabilistic channels}\label{sec:examples_probabilistic}
\todo{Give the examples from the paper for the state averaged model, etc.
Also, try to see if we can use our results to compute an error exponent (in the sense of computing the probability of $\Remp$ to fall below a threshold) (yes - Shannon did for $\Remp=i$ and is it tight near capacity), and the error exponent for random transmission time (in the sense of dependence of error probability in average transmission time).}
}

\section{Comments \& further research}\label{chap:comments_research}
\onlypaper{
\subsection{Comparison with previous results and techniques}\label{sec:comparison_with_prev_paper}
The asymptotic adaptive and non-adaptive achievability of the empirical mutual information and the second order rate function of Theorem~\ref{theorem:GaussianMIMO} (when particularized to the real valued SISO case $t=r=1,d=1,u=0$) was shown in the \selector{previous paper}{initial work} \cite{YL_individual_full}. The current results are improved in many senses compared to the previous results (although are also inferior in other aspects). Due to space limits, the reader is referred to \cite{YL_PhdThesis} for a detailed comparison.
} % only paper

\onlypaper{
\section*{Acknowledgment}
This work was partially supported by Weinstein institution. \& Feder prise. \todo{copy from papers}
}

\onlypaper{
\appendix{}
}

\subsection{Proof of the properties of intrinsic redundancy}\label{sec:int_redundancy_properties}
In this section we prove the two properties of intrinsic redundancy presented in Section~\ref{sec:irp524}.

\textit{Proof of property 1:} The intrinsic redundancy increases linearly when an offset $\delta \in \mathbb{R}$ is added to (or subtracted from) the rate function:
\begin{equation}\begin{split}\label{eq:A490}
\mu_Q(\Remp + \delta)
&=
\sup_{\vr y,R} \left\{ \frac{1}{n} \log Q \{\Remp(\vr X, \vr y) + \delta \geq R\}  + R \right\}
\\&\stackrel{R'=R - \delta}{=}
\sup_{\vr y,R'} \left\{ \frac{1}{n} \log Q \{\Remp(\vr X, \vr y) \geq R' \}  + R' + \delta \right\}
\\&=
\mu_Q(\Remp) + \delta
\end{split}\end{equation}

\textit{Proof of property 2:}  by the union bound:
\begin{equation}\label{eq:A545}
Q \{\max_{k \in \{1,\ldots,K\}} \Remp_k  > R\}
=
Q \left\{\bigcup_{k \in \{1,\ldots,K\}} \left( \Remp_k  > R\ \right) \right\}
\leq
\sum_{k=1}^K Q \{\Remp_k  > R\}
\leq
K \max_{k \in \{1,\ldots,K\}} Q \{\Remp_k  > R\}
\end{equation}

\begin{equation}\begin{split}\label{eq:A544}
\mu_Q \left(\max_{k \in \{1,\ldots,K\}} \Remp_k \right)
&=
\sup_{\vr y,R} \left\{ \frac{1}{n} \log Q \{\max_{k \in \{1,\ldots,K\}} \Remp_k  > R\}  + R \right\}
\\&\leq
\sup_{\vr y,R} \left\{ \frac{1}{n} \log \left[ K \max_{k \in \{1,\ldots,K\}} Q \{\Remp_k  > R\} \right]  + R \right\}
\\&=
\sup_{\vr y,R,k} \left\{ \frac{1}{n} \log \left[ K  Q \{\Remp_k  > R\}  \right]  + R \right\}
\\&=
\sup_{\vr y,R,k \in \{1,\ldots,K\}} \left\{ \frac{1}{n} \log \left[ Q \{\Remp_k  > R\} \right]  + R  \right\} + \frac{\log(K)}{n}
\\&=
\max_{k \in \{1,\ldots,K\}} \mu_Q(\Remp_k)  + \frac{\log(K)}{n}
\end{split}\end{equation}
\endofproof

% PhdOnly

\subsection{Achievability of good-put functions for rate adaptive systems [UNSFINISHED]}\label{sec:good_put_bound_rateadaptive}
In Section~\ref{sec:good_put_bound} it was shown that good-put functions (defined therein) for fixed-rate systems, are asymptotically achievable rate functions. Here, the result is extended to good-put functions of rate adaptive systems. Notice that it is not shown that these functions are \emph{adaptively} achievable.

The same derivation of Section~\ref{sec:good_put_bound} is followed, while conditioning on $R_\tsubs{sys}$. Consider the conditional form:
\begin{equation}\label{eq:A1040d}
R_\tsubs{good}(\vr x, \vr y | R_\tsubs{sys}) \defeq \E \left[ (1 - \epsilon_\tsubs{sys}) R_\tsubs{sys} \Big| \vr x, \vr y, R_\tsubs{sys} \right]
.
\end{equation}
Since $R_\tsubs{sys} = R_\tsubs{sys}(S, \vr y)$, and $\vr y$ is considered constant, this conditioning only affects the distribution of $S$ (and not of $\vr X$ and $\msg$). Thus, it still holds that \eqref{eq:A743}:
\begin{equation}\label{eq:A743g}
\exp(-n R_\tsubs{sys}) = \sum_{\vr x} \frac{R_\tsubs{good}(\vr x, \vr y | R_\tsubs{sys})}{R_\tsubs{sys}} \Pr(\vr X = \vr x | R_\tsubs{sys})
.
\end{equation}
Or, in other words:
\begin{equation}\label{eq:A743h}
\E \left[ R_\tsubs{good}(\vr x, \vr y | R_\tsubs{sys}) \Big| R_\tsubs{sys} \right] = R_\tsubs{sys} \exp(-n R_\tsubs{sys})
\end{equation}

\todo{Need to complete the proof for the rate adaptive case ?? should work through chernoff. in the above I tried to use the concavity of $x e^{-x}$ but this is true only for $x < 2$}

\subsection{A binary on/off channel}\label{sec:binary_onoff_example}

In Section~\ref{sec:ML_rate_functions} we mentioned the binary on/off channel as an example for a non-ergodic channel, where the rate that can be achieved on average (adaptively) is larger than the rate that is achieved in worst case (the Han-Verd\'u capacity). Here we complete the example by analyzing the information density of this channel.

The channel may be in one of two states, which are determined by a single random drawing with equal probabilities -- either the output equals the input for $j=1,\ldots,n$, or it is independent of the input. The information density of this channel, for uniform i.i.d. input, is a random variable taking values close to $0,1$ [bits] with equal probabilities, as shown below.

\begin{eqnarray}\label{eq:A1641}
\Pr(\vr X) &=& \frac{1}{2^n} \\
\Pr(\vr Y | \vr X) &=& \half \delta_{\vr X, \vr Y} + \half \cdot \frac{1}{2^n} \\
\Pr(\vr Y) &=& \frac{1}{2^n} \\
\end{eqnarray}

\begin{equation}\begin{split}\label{eq:A1460}
i
&=
\frac{1}{n} \log \frac{\Pr(\vr Y | \vr X)}{\Pr(\vr Y)}
=
\frac{1}{n} \log \left( \half 2^n \delta_{\vr X, \vr Y} + \half \right) \delta_{\vr X, \vr Y}
\\&=
\begin{cases} 1 & \Pr=\half \cdot 1 + \half \cdot \frac{1}{2^n} \\ 0 & \text{o.w.} \end{cases}
=
\begin{cases} \frac{1}{n} \log \left( \half 2^n \cdot 1 + \half \right) & \Pr=\half \left( 1 + \frac{1}{2^n} \right) \\ \frac{1}{n} \log \left( \half \right) & \text{o.w.} \end{cases}
\\&=
-\frac{1}{n} + \begin{cases} \frac{1}{n} \log \left( 2^n + 1 \right) & \Pr=\half \left( 1 + \frac{1}{2^n} \right) \\ 0 & \text{o.w.} \end{cases}
\\&\approx
\begin{cases} 1 & \Pr=\half  \\ 0 & \Pr=\half \end{cases}
\end{split}\end{equation}
Therefore the liminf in probability of $i$ is $0$, and therefore we see also by Han-Verd\'u formula that the Shannon capacity of this channel is $0$ (which is clear from operational perspective).

Note: the reason that $E(i) \leq \half$ is that some information is lost due to not knowing the channel state $I(\vr X; \vr Y) < I(\vr X; \vr Y | \text{State}) = \half$.

\subsection{Proof of Lemma~\ref{lemma:remp_conditional_form_reverse}}\label{sec:proof_remp_conditional_form_reverse}
Assume $\Rempname{*}(\vr x, \vr y) = \frac{1}{n} \log \frac{f(\vr x | \vr y)}{Q(\vr x)} - \delta$ is achievable. For every $\gamma \in (0,1)$, by  Lemma~\ref{lemma:chernoff_tightness_lemma}, one has
\begin{equation}\label{eq:A4290}
\underset{Q}{\E} \left[ \exp(n \gamma \Rempname{*}(\vr X, \vr y)) \right] \leq \frac{1}{(1-\epsilon)(1-\gamma)}
\end{equation}
On the other hand
\begin{equation}\begin{split}\label{eq:A1264}
\underset{Q}{\E} \left[ \exp(n \gamma \Rempname{*}(\vr X, \vr y)) \right]
&=
\exp(-n \gamma \delta) \underset{Q}{\E} \left[ \exp(n \Rempname{*}(\vr X, \vr y)) \cdot \exp(- n (1-\gamma) \Rempname{*}(\vr X, \vr y))  \right]
\\& \geq
\exp(-n \gamma \delta) \underbrace{\underset{Q}{\E} \left[ \frac{f(\vr x | \vr y)}{Q(\vr x)} \right]}_{=1} \cdot \exp(- n (1-\gamma) R_{\max})
\\& =
\exp(-n \gamma \delta - n (1-\gamma) R_{\max})
\end{split}\end{equation}
Combining with \eqref{eq:A4290} we have:
\begin{equation}\label{eq:A1273}
\exp(-n \gamma \delta - n (1-\gamma) R_{\max}) \leq \frac{1}{(1-\epsilon)(1-\gamma)}
\end{equation}
Which yields after rearrangement:
\begin{equation}\label{eq:A1276}
\delta \geq \frac{\log (1-\epsilon) + \log (1-\gamma) -  n (1-\gamma) R_{\max}}{n \gamma}
\end{equation}
To approximately maximize the RHS with respect to $\gamma$ (in fact, to maximize $\log (1-\gamma) -  n (1-\gamma) R_{\max}$) we set $\gamma  = 1 - \frac{1}{n R_{\max}}$ and obtain:
\begin{equation}\label{eq:A1282}
\delta \geq \frac{\log (1-\epsilon) - \log (n R_{\max}) -  1}{n - R_{\max}^{-1}} = - \frac{\log (n) + \log \frac{e \cdot R_{\max}}{1-\epsilon}}{n - R_{\max}^{-1}}
\end{equation}
which proves the Lemma.
\endofproof

\subsection{Completion of the proofs for the Gaussian MIMO case}
In the below we give the detailed derivations to complete the proofs of Theorem~\ref{theorem:GaussianMIMO}, and some related results that appear in Section~\ref{sec:GaussianMIMO}.

\subsubsection{Optimal linear estimator without an additive factor}\label{sec:MMSE_no_mean_appendix}
In Section~\ref{sec:ML_gaussian_family} we presented a conditional probability density for the Gaussian family \eqref{eq:A3839}, which includes a linear estimator of the form $\mt A \vr y + \vr b$. The maximization of \eqref{eq:A3839} over $\mt A, \vr b$ was solved using an LMMSE estimator \eqref{eq:A3866}. For the case where $\vr b=0$ ($u=0$), i.e. the estimator is required to be of the form $\mt A \vr y$, we claimed the same solution holds, where $\vr {\hat \mu}_{\mt X}, \vr {\hat \mu}_{\mt Y}$ are replaced with zeros. Here we provide a proof of this claim (which follows the same proof as the optimality of MMSE estimator).
\begin{lemma}\label{lemma:quadratic_minimizer}
The matrix $\mt A$ minimizing $(\mt X - \mt Y \mt A)^* (\mt X - \mt Y \mt A)$ (in matrix sense) is
\begin{equation}\label{eq:A3209}
\mt A = (\mt Y^* \mt Y)^{-1} \mt Y^* \mt X
\end{equation}
\end{lemma}
\textit{proof:}
The matrix $\mt A$ defined above satisfies the orthogonality criterion:
\begin{equation}\label{eq:A3214}
\mt Y^* (\mt X - \mt Y \mt A) = 0
\end{equation}
Consider a different matrix $\mt {\tilde A}$ and write:
\begin{equation}\begin{split}\label{eq:A3214b}
(\mt X - \mt Y \mt {\tilde A})^* (\mt X - \mt Y \mt {\tilde A})
&=
\left[ (\mt X - \mt Y \mt A )+ \mt Y (\mt A - \mt {\tilde A}) \right]^* \left[ (\mt X - \mt Y \mt A )+ \mt Y (\mt A - \mt {\tilde A}) \right]
\\&\stackrel{\eqref{eq:A3214}}{=}
(\mt X - \mt Y \mt A)^* (\mt X - \mt Y \mt A)
+
(\mt A - \mt {\tilde A})^* \mt Y^* \mt Y (\mt A - \mt {\tilde A})
\\&\geq
(\mt X - \mt Y \mt A)^* (\mt X - \mt Y \mt A)
\end{split}\end{equation}
\endofproof

\subsubsection{The CCDF condition}\label{sec:GaussianMIMO_CCDF}
Based on Section~\ref{sec:ML_adaptivity_continuous} let:
\begin{equation}\label{eq:A3536}
\psi(\mt X^k, \mt Y^k, j)  =
\left( \frac{\hat p_{\ML}(\mt X_{j+1}^k | \mt Y_{j+1}^k)}{Q \left(\mt X_{j+1}^{k} \right)} \right)^{\gamma}
=
\psi(\mt X_{j+1}^k, \mt Y_{j+1}^k, 0)
\end{equation}
Note that $\psi$ is of the form \eqref{eq:A2497}, where some dependencies were removed due to the i.i.d. nature of the distribution $P_{\theta}$. Note that $\psi(\mt X^k, \mt Y^k, j)$ (recall: the metric at time $k$ for the block which started at time $j+1$) is dependent only on $\mt X_{j+1}^k, \mt Y_{j+1}^k$, i.e. the values of the channel input and output inside the block. The Markov sufficient condition of Theorem~\ref{theorem:framework} is:
\begin{equation}\label{eq:A3545}
\placeunder{\E}{Q} \left[ \psi(\mt X^k, \mt Y^k, j)  | \mt X^j \right]
=
\placeunder{\E}{Q} \left[ \psi(\mt X_{j+1}^k, \mt Y_{j+1}^k, 0) \right] \leq L_{k-j}
\end{equation}
For brevity we define $m=k-j$, and the  matrices $\mt {\tilde X} = \mt X_{j+1}^k, \mt {\tilde Y} = \mt Y_{j+1}^k$ of sizes $m \times t, m \times r$ respectively. We have $\placeunder{\E}{Q} \left[ \psi(\mt X_{j+1}^k, \mt Y_{j+1}^k, 0) \right] = \placeunder{\E}{Q} \left[ \psi(\mt {\tilde X}, \mt {\tilde Y}, 0) \right]$. Using \eqref{eq:A3478}, we bound, instead, the following value:
\begin{equation}\label{eq:A3553}
\tilde L_m
=
\placeunder{\E}{\tilde Q} \left[ \psi(\mt {\tilde X}, \mt {\tilde Y}, 0) \right]
=
\placeunder{\E}{\tilde Q} \left[ \left( \frac{\hat p_{\ML}\left(\mt {\tilde X} | \mt {\tilde Y} \right)}{Q \left(\mt {\tilde X} \right)} \right)^{\gamma} \right]
\end{equation}
and therefore for the rest of this section we assume $\mt {\tilde X}$ has a Gaussian distribution.

\todo{The rest should be moved to appendix}

We define $\mt V = \mt {\tilde X} \Lambda_X^{-1/2}$ as the whitened version of $\mt {\tilde X}$: the elements of $\mt V_{m \times t}$ are independent unit variance Gaussian (/complex Gaussian) random variables. To calculate $L_m$ it is convenient to present $\hat p_{\ML}\left(\mt {\tilde X} | \mt {\tilde Y} \right)$ by a way of sequential projection of the columns of $\mt V$ on the subspaces created by $\mt {\tilde Y}$ and the previous columns. The concept is the same as was used in the conference paper \cite{YL_MIMO_ITW2010}, but the details slightly differ mainly due to the different rate function ($\Rempname{\ML}$ rather than $\Rempname{\ML*}$).

We define the combined matrix $\mt Z_{m \times (u+t+r)} \defeq [\vr 1_{u}, \mt {\tilde Y}, \mt V]$, where $\vr 1_{u} \defeq \begin{cases} \vr 1_{m \times 1} & u=1 \\ [ \emptyset ] & u=0 \end{cases}$, i.e. for $u=0$ the vector $\vr 1_{u}$ is an empty vector and is excluded from $\mt Z$. By QR decomposition we can write $\mt Z = \mt Q_z \cdot \mt R_z$ with $\mt Q_z^* \mt Q_z = \mt I$ and $\mt R_z$ upper triangular. As a reminder, QR decomposition is performed by Gram-Schmidt process. We start from the left column of $\mt Z$ and work our way to the last one. At each time we take a column of $\mt Z$ and split it to the part which can be represented by a linear combination of the columns to the left of it (equivalently, to the columns of $\mt Q_z$ that were already generated), and the "innovation", i.e. the part which is orthogonal to the subspace generated by the previous columns. The vector representing the innovation is normalized, and becomes the respective column of $\mt Q_z$, and its power becomes the diagonal element in $\mt R_z$. The coefficients representing the part of the vector which is in the subspace of previous columns become the elements of $\mt R_z$ above the diagonal. Another important property of QR decomposition is that the determinant of $ \mt Z^* \mt Z$ can be written in terms of the diagonal elements in $\mt R_z$:  $\left| \mt Z^* \mt Z \right| = \left| \mt R_z^* \mt Q_z^* \mt Q_z \mt R_z \right| = \left| \mt R_z^* \mt R_z \right| = \left| \mt R_z \right|^2 = \prod_{i=1}^k \left| {R_{Z}}_{ii} \right|^2$. For this equality to be correct in the complex case we define the operation $\left| \cdot \right|$ to imply absolute-determinant.

We may split the matrices $\mt Q_z, \mt R_z$ into several parts, matching the separate matrices $\vr 1_u, \mt {\tilde Y}, \mt V$ as follows:
\begin{equation}\label{eq:A3577}
\mt Z = [\vr 1_{u}, \mt {\tilde Y}, \mt V] =
\left[ \begin{array}{c|c|c} \mt Q_1 & \mt Q_{y|1} & \mt Q_{v|y1} \end{array} \right] \cdot
    \left[ \begin{array}{c|c|c} \sqrt{m} & \vr r_{y|1} & \vr r_{v|1} \\ \hline 0 & \mt R_y & \mt R_{v|y} \\ \hline 0 & 0 & \mt R_v \end{array} \right]
\end{equation}
Where the blocks dividing the matrices $\mt Q_z, \mt R_z$ have sizes $u, r, t$ (respectively), $\mt R_v$ and $\mt R_y$ are upper triangular and $\mt Q_1 = \begin{cases} \frac{1}{\sqrt{m}} \cdot \mt 1 & u=1 \\ [ \emptyset ] & u=0 \end{cases}$ is just the normalization of the vector $1_{u}$ (when $u=0$ the first row and column of the RHS of \eqref{eq:A3577} are absent). The matrices $\mt Q_1,\mt Q_{y|1},\mt Q_{v|y1}$ contain orthogonal columns. The meaning of \eqref{eq:A3577} is that each column of $\mt V$ is represented by its projection on $\vr 1_u$ (which is the mean of the rows, up to a constant), it's projection on the subspace defined by the rows of $\mt {\tilde Y}$ and on the previous columns of $\mt V$, and finally by a new element which is orthogonal to the previous subspaces. We can write:
\begin{equation}\label{eq:A3586}
\mt {\tilde Y} = \vr 1_{u} \frac{1}{\sqrt{m}} \vr r_{y|1} + \mt Q_{y|1} \mt R_y
\end{equation}
\begin{equation}\label{eq:A3587}
\mt V = \vr 1_{u} \frac{1}{\sqrt{m}} \vr r_{v|1} + \mt Q_{y|1} \mt R_{v|y1} + \mt Q_{v|y1} \mt R_{v}
\end{equation}
\begin{equation}\label{eq:A3588}
\mt {\tilde X} = \mt V  \Lambda_X^{\half} = \vr 1_{u} \frac{1}{\sqrt{m}} \vr r_{v|1} \Lambda_X^{\half} + \mt Q_{y|1} \mt R_{v|y1} \Lambda_X^{\half} + \mt Q_{v|y1} \mt R_{v} \Lambda_X^{\half}
\end{equation}

We would like to show that $\hat p_{\ML}(\mt {\tilde X} | \mt {\tilde Y})$ can be written as a function of the diagonal elements in $\mt R_{v}$ alone. This can be proven in a technical form simply by substitution of \eqref{eq:A3586},\eqref{eq:A3588} into the expressions in Lemma~\ref{lemma:Gaussian_pML}, but an alternative proof that shows the fundamental reason for that is by recalling that $\hat p_{\ML}(\mt {\tilde X} | \mt {\tilde Y})$ maximizes $P_{\theta}$ given by \eqref{eq:A3839}. In maximizing $P_{\theta}$ we first find the best linear approximation of $\mt {\tilde X}$ by $\mt {\tilde Y}$ and $\vr 1_{u}$, and then the covariance matrix of the remainder (error). Clearly the best approximation of $\mt {\tilde X}$ by $\mt {\tilde Y}$ and $\vr 1_{u}$ is in the subspace spanned by $\vr 1_{u}, \mt Q_{y|1}$, which is described by the first two elements in \eqref{eq:A3588} and therefore the error is the remainder $\mt Q_{v|y1} \mt R_{v} \Lambda_X^{\half}$. \onlypaper{we obtain
\begin{equation}\label{eq:A3618}
\hat p_{\ML}(\mt {\tilde X} | \mt {\tilde Y}) = \left| \frac{d \pi e}{m} \Lambda_X \right|^{-\dalf m} \cdot \left| \mt R_{v} \right|^{-d m}
\end{equation}
} % only

% only

Substituting into \eqref{eq:A3553} we have:
\begin{equation}\begin{split}\label{eq:A3623}
\tilde L_m
& \stackrel{\eqref{eq:A3299},\eqref{eq:A3618}}{=}
\placeunder{\E}{\tilde Q} \left[ \left( \frac{\left| \frac{d \pi e}{m} \Lambda_X \right|^{-\dalf m} \cdot \left| \mt R_{v} \right|^{-d m}}{\left| d \pi \Lambda_X \right|^{-\dalf m} e^{-\dalf \tr \left(  \mt {\tilde X}^* \mt {\tilde X} \Lambda_X^{-1} \right) }} \right)^{\gamma} \right]
=
\left( \frac{e}{m} \right)^{-\dalf \gamma t m} \cdot \placeunder{\E}{\tilde Q} \left[  \left| \mt R_{v} \right|^{-\gamma d m} e^{\dalf \gamma \tr \left(  \mt V^* \mt V \right)}  \right]
\\&=
\left( \frac{e}{m} \right)^{-\dalf \gamma t m} \cdot \placeunder{\E}{\tilde Q} \left[  \prod_{i=1}^t {\mt R_v}_{ii}^{-\gamma d m} e^{\dalf \gamma \| \vr v_i \|^2} \right]
=
\placeunder{\E}{\tilde Q} \left[  \prod_{i=1}^t \underbrace{\left( \frac{e}{m} \right)^{-\dalf \gamma m} {\mt R_v}_{ii}^{-\gamma d m} e^{\dalf \gamma \| \vr v_i \|^2}}_{\defeq D_i} \right]
\end{split}\end{equation}
where $\vr v_i$ is the $i$-th column of $\mt V$. Since $\vr v_i$ are independent is are isotropically distributed (since their elements are Gaussian i.i.d.), the innovation norms ${\mt R_v}_{ii}$ are independent. Recall that ${\mt R_v}_{ii}$ is the norm of the innovation of $\vr v_i$ with respect to the subspace spanned by $\vr 1_u, \mt {\tilde Y}$ and $\vr v_1, \ldots, \vr v_{i-1}$, however because $\vr v_i$ is isotropically distributed, this power is independent of the specific subspace in question, and only depends on the dimensions of the subspace. Formally, consider the squared norm of the innovation of a $m \times 1$ vector of Gaussian (/complex Gaussian) i.i.d. random variables $\vr v$ with respect to a $k$ dimensional subspace spanned by the unitary matrix $\mt U_{m \times k}$, i.e. $p = \| \vr v - \mt U \mt U^* \vr v \|^2$. Completing $\mt U$ to an orthonormal basis $\mt {\tilde U}_{m \times m}$, and defining $\vr w = \mt {\tilde U}^* \cdot \vr v$, we have that $\mt U^* \vr v = \vr w_1^k$, and $\mt U \mt U^* \vr v = \mt U \vr w_1^k = \mt {\tilde U} \left[ \substack{ \vr w_1^k \\ \vr 0_{m-k \times 1}} \right]$. Therefore the innovation norm can be written as
\begin{equation}\label{eq:A3662}
p = \left\| \mt {\tilde U} \left(\vr w - \left[ \substack{ \vr w_1^k \\ \vr 0_{m-k \times 1}} \right] \right) \right\|^2 = \left\| \left[ \substack{ \vr 0_{k \times 1} \\ \vr w_{k+1}^m} \right] \right\|^2 = \left\| \vr w_{k+1}^m \right\|^2
\end{equation}
Since $\vr w$ has the same distribution of $\vr v$, the distribution of $p$ does not depend on $\mt U$. Furthermore $p \cdot d$ is distributed $\chi^2$ with $d \cdot (m-k)$ degrees of freedom (the multiplication with $d$ is needed in order to normalize the real and imaginary to unit power). Therefore ${\mt R_v}_{ii}^2$ are independent and are distributed $\chi^2_{d \cdot (m-i)}$. Furthermore, $\| \vr v_i \|^2$ in \eqref{eq:A3623} can be replaced by $\| \vr w_i \|^2$ (where $\vr w_i$ is the vector $\vr v_i$ rotated according to the same subspace), which are also independent. Therefore the expected value in \eqref{eq:A3623} can be written as the product of expected values
\begin{equation}\label{eq:A3682}
\tilde L_m = \E \left[ \prod_{i=1}^t D_i \right] = \prod_{i=1}^t \E \left[ D_i \right]
\end{equation}

It remains to bound this expected value. The $i$-th column of $\mt V$ that generates ${\mt R_v}_{ii}$ is projected into a $k = (i-1) + r + u$ dimensional subspace ($i-1$ previous columns of $\mt V$, $r$ columns of $\mt {\tilde Y}$ and an all-ones vector if $u=1$). In the below we take $\vr w$ to be the rotated version of $\vr v_i$:
\begin{equation}\begin{split}\label{eq:A3668}
\left( \frac{e}{m} \right)^{\dalf \gamma m}  \E \left[ D_i \right]
&=
\E \left[ {\mt R_v}_{ii}^{-\gamma d m} e^{\dalf \gamma \| \vr v_i \|^2} \right]
=
\E \left[ \left\| \vr w_{i+r+u}^m \right\|^{-\gamma d m} e^{\dalf \gamma \| \vr w \|^2} \right]
\\&=
\E \left[ \left\| \vr w_{i+r+u}^m \right\|^{-\gamma d m} e^{\dalf \gamma \left\| \vr w_{i+r+u}^m \right\|^2} \cdot  e^{\dalf \gamma \left\| \vr w_{1}^{i+r+u-1} \right\|^2} \right]
\\&=
\underset{\substack{S=d \left\| \vr w_{i+r+u}^m \right\|^2 \\ \sim \chi^2_{d(m-i-r-u+1)}}}{\E} \left[ \left( \frac{S}{d} \right)^{-\half \gamma d m} e^{\half \gamma S} \right] \cdot
\underset{\substack{S=d \left\| \vr w_{1}^{i-1+r+u} \right\|^2 \\ \sim \chi^2_{d \cdot (i+r+u-1)}}}{\E} \left[ e^{\half \gamma S} \right]
\end{split}\end{equation}

for general $k, \alpha$:
\begin{equation}\begin{split}\label{eq:A3686}
\underset{S \sim \chi^2_{k}}{\E} \left[ S^{-\alpha} e^{\half \gamma S} \right]
&=
\int_{s=0}^\infty s^{-\alpha} e^{\half \gamma s} \cdot \frac{s^{\frac{k}{2} - 1} e^{-\frac{s}{2}}}{2^{k/2} \Gamma \left(\frac{k}{2} \right)} ds
\\&=
\frac{1}{2^{k/2} \Gamma \left(\frac{k}{2} \right)} \int_{s=0}^\infty s^{\frac{k}{2} - 1 - \alpha} \cdot e^{-\half (1-\gamma) s} ds
\\& \stackrel{h=\half (1-\gamma) s}{=}
\frac{\left( \half (1-\gamma) \right)^{\alpha - \frac{k}{2}}}{2^{k/2} \Gamma \left(\frac{k}{2} \right)} \int_{h=0}^\infty h^{\frac{k}{2} - 1 - \alpha} \cdot e^{-h} dh
\\& \stackrel{(*)}{=}
\frac{\Gamma \left(\frac{k}{2} - \alpha \right)}{2^{\alpha} \cdot (1-\gamma)^{\frac{k}{2} - \alpha} \cdot \Gamma \left(\frac{k}{2} \right)}
\end{split}\end{equation}
where (*) is by definition $\Gamma(z) \defeq \int_{h=0}^\infty h^{z-1} \cdot e^{-h} dh$, and in order for the integral to exist (near $h = 0$) we need to assume $\frac{k}{2} - 1 - \alpha > -1$, i.e. $\alpha < \frac{k}{2}$.

Substituting this in \eqref{eq:A3668} ($\alpha=\half \gamma d m, k=d(m-i+1-r-u)$ for the first expression and $\alpha=0, k=d \cdot (i-1+r+u)$ for the other) we have:
\begin{equation}\begin{split}\label{eq:A3700}
\E \left[ D_i \right]
&=
\left( \frac{e}{m} \right)^{-\dalf \gamma m}  \left( \frac{1}{d} \right)^{-\half \gamma d m} \frac{\Gamma \left(\frac{d(m-i+1-r-u)}{2} - \half \gamma d m \right)}{2^{\half \gamma d m} \cdot (1-\gamma)^{\frac{d(m-i+1-r-u)}{2} - \half \gamma d m} \cdot \Gamma \left(\frac{d(m-i+1-r-u)}{2} \right)} \cdot
\frac{1}{(1-\gamma)^{\frac{d \cdot (i-1+r+u)}{2}}}
\\&=
\left( \frac{d m}{2 e} \right)^{\dalf \gamma m}  \frac{\Gamma \left(\frac{d((1-\gamma) m-(i-1+r+u))}{2} \right)}{(1-\gamma)^{\frac{dm(1-\gamma)}{2}} \cdot \Gamma \left(\frac{d(m-i+1-r-u)}{2} \right)}
\end{split}\end{equation}
Where to meet the condition $\alpha < \frac{k}{2}$ we need to require $\half \gamma d m < \half d(m-i+1-r-u) \Rightarrow \gamma < 1 - \frac{i-1+r+u}{m}$. Since this must hold for any $i=1,\ldots,t$, this implies $\gamma < 1 - \frac{t+r+u-1}{m}$. Recall that in order to have a decreasing redundancy in Theorem~\ref{theorem:framework} (see for example Corollary~\ref{corollary:framework_asymptote}) we need to have $\frac{1}{n} \log L_n \to 0$, which implies in our case $\frac{1}{m} \log \E \left[ D_i \right] \to 0$. This is not immediately clear from \eqref{eq:A3700}. We use Stirling's approximation for Gamma function:
\begin{equation}\label{eq:A3739}
\Gamma(z) = \sqrt{2 \pi} z^{z-\half} e^{-z + \frac{\eta}{12 z}}, \qquad 0 < \eta < 1
\end{equation}

For brevity we define $z_1 = \frac{d m}{2}$, $z_2 = (1-\gamma) z_1$, $z_3 = \frac{d(i-1+r+u)}{2} = z_1 \cdot \frac{i-1+r+u}{m}$. Under our assumptions, $z_1 > z_2 > z_3$. We further assume $z_2 - z_3 \geq 1$.
\begin{equation}\begin{split}\label{eq:A3741}
\frac{\Gamma \left(\frac{d((1-\gamma) m-(i-1+r+u))}{2} \right)}{\Gamma \left(\frac{d(m-i+1-r-u)}{2} \right)}
& =
\frac{\Gamma \left(z_2 - z_3 \right)}{\Gamma \left( z_1 - z_3 \right)}
\\& \leq
\frac{\sqrt{2 \pi} (z_2 - z_3)^{(z_2 - z_3)-\half} e^{-(z_2 - z_3) + \frac{1}{12} (z_2 - z_3)}}{\sqrt{2 \pi} (z_1 - z_3)^{(z_1 - z_3)-\half} e^{-(z_1 - z_3) }}
\leq
\frac{(z_2 - z_3)^{(z_2 - z_3)-\half} e^{z_1-z_2} e^{\frac{1}{12}}}{(z_1 - z_3)^{(z_1 - z_3)-\half}}
\\&=
\frac{(z_1 - z_3)^{(z_2 - z_3)-\half} \cdot \left( \frac{(z_2 - z_3)}{(z_1 - z_3)} \right)^{(z_2 - z_3)-\half} e^{z_1-z_2} e^{\frac{1}{12}}}{(z_1 - z_3)^{(z_1 - z_3)-\half}}
\\& \stackrel{(a)}{\leq}
(z_1 - z_3)^{z_2 -z_1} \cdot \left( \frac{z_2}{z_1} \right)^{(z_2 - z_3)-\half} e^{z_1-z_2} e^{\frac{1}{12}}
\\&=
\left(\frac{d m}{2} \left( 1 - \frac{i-1+r+u}{m} \right) \right)^{-\gamma \frac{d m}{2}} \cdot (1-\gamma)^{(1-\gamma)\frac{d m}{2} -  \frac{d(i-1+r+u+1)}{2}} e^{\gamma \frac{d m}{2}} e^{\frac{1}{12}}
\\&=
\left(\frac{d m}{2 e} \right)^{- \frac{d}{2} \gamma  m}  \cdot (1-\gamma)^{\frac{d m(1-\gamma)}{2}} \cdot \left( 1 - \frac{i-1+r+u}{m} \right)^{- \frac{d}{2} \gamma  m}
 \cdot
(1-\gamma)^{- \frac{d(i+r+u)}{2}}
e^{\frac{1}{12}}
\end{split}\end{equation}
Where in the last inequality (a) we used $\frac{z_2 - z_3}{z_1 - z_3} \leq \frac{z_2}{z_1}$ (which stems from $z_1 > z_2$), and under the assumption $z_2-z_3 \geq 1$ the exponent $(z_2 - z_3)-\half$ is positive. This condition implies $(1-\gamma) m  \geq \frac{2}{d} + i-1+r+u$, so it is sufficient that $(1-\gamma) m  \geq i+1+r+u$. Note that the two first terms cancel out respective terms in \eqref{eq:A3700} and the last two terms are independent of $m$. The term $\left( 1 - \frac{i-1+r+u}{m} \right)^{- \frac{d}{2} \gamma  m}$ tends to $e^{(i-1+r+u) \frac{d}{2} \gamma}$ as $m \to \infty$. For finite $m$, using $\ln(1+x) \geq \frac{x}{1+x}$ we have:
\begin{equation}\begin{split}\label{eq:A3795}
\ln \left[ \left( 1 - \frac{i-1+r+u}{m} \right)^{- \frac{d}{2} \gamma  m} \right]
& \leq
- \frac{d}{2} \gamma  m \frac{- \frac{i-1+r+u}{m} }{1 - \frac{i-1+r+u}{m}}
=
\frac{d}{2} \gamma (i-1+r+u) \frac{1}{1 - \frac{i-1+r+u}{m}}
\\& \stackrel{\eqref{eq:A3700}: \gamma < 1 - \frac{i-1+r+u}{m}}{<}
\frac{d}{2} (i-1+r+u)
\end{split}\end{equation}

substituting \eqref{eq:A3741} and \eqref{eq:A3795} in \eqref{eq:A3700},
\begin{equation}\label{eq:A3804}
\E \left[ D_i \right]
<
e^{\frac{d}{2} (i-1+r+u)}  \cdot (1-\gamma)^{- \frac{d(i+r+u)}{2}} e^{\frac{1}{12}}
\end{equation}
Where we have assumed $(1-\gamma) m  \geq i+1+r+u$. Substituting into \eqref{eq:A3682} we obtain:
\begin{equation}\begin{split}\label{eq:A3771}
\tilde L_m
&=
\prod_{i=1}^t \E \left[ D_i \right]
\leq
e^{\frac{d}{2} \sum_{i=1}^t (i-1+r+u)}  \cdot (1-\gamma)^{- \frac{d \sum_{i=1}^t (i+r+u)}{2}} e^{\frac{t}{12}}
\\&=
e^{\frac{d}{2} \left(\half (t-1) +r+u \right) t}  \cdot (1-\gamma)^{- \frac{d \left(\half(t+1)+r+u \right)t}{2}} e^{\frac{t}{12}}
\\& \leq
e^{\frac{d}{4} \left(t + 1 + 2r + 2u \right) t}  \cdot (1-\gamma)^{- \frac{d \left(t+1+2r+2u \right)t}{4}}
=
\left( \frac{e}{1-\gamma} \right)^{ \frac{d}{4} \left(t + 1 + 2r + 2u \right) \cdot t}
\end{split}\end{equation}
Note that we obtained a constant bound on $L_m$ that does not grow with $m$.

Returning to \eqref{eq:A3545} (recall that $m=k-j$, $\mt {\tilde X} = \mt X_{j+1}^k, \mt {\tilde Y} = \mt Y_{j+1}^k$):
\begin{equation}\begin{split}\label{eq:A3778}
\placeunder{\E}{Q} \left[ \psi(\mt X^k, \mt Y^k, j)  | \mt X^j \right]
&=
\placeunder{\E}{Q} \left[ \psi(\mt {\tilde X}, \mt {\tilde Y}, 0) \right]
\stackrel{\eqref{eq:A3478}}{\leq}
\frac{1}{ 1 - \delta_{\Omega}} \underset{\tilde Q}{\E} \left[ \psi(\mt {\tilde X}, \mt {\tilde Y}, 0) \right]
\\&=
\frac{1}{ 1 - \delta_{\Omega}} \tilde L_m
\leq
\frac{1}{ 1 - \delta_{\Omega}} \left( \frac{e}{1-\gamma} \right)^{ \frac{d}{4} \left(t + 1 + 2r + 2u \right) \cdot t}
\defeq L_m
\end{split}\end{equation}
\eqref{eq:A3778} defines $L_m$ under which the CCDF condition of Theorem~\ref{theorem:framework} holds, and $L_m$ is non-decreasing as required. To satisfy the assumption $(1-\gamma) m  \geq i+1+r+u$ for all $i \leq t$, we define $b_0 = \frac{t+1+r+u}{1-\gamma}$ as the minimal symbol for which the bound holds (see the definitions of Theorem~\ref{theorem:framework}).

The CCDF condition directly yields the result of Lemma~\ref{lemma:GaussianMIMO_nonadaptive}: from the CCDF condition we have that the intrinsic redundancy of $\gamma \Rempname{\ML}$ satisfies:
\begin{equation}\begin{split}\label{eq:A5425b}
\mu_Q(\gamma \Rempname{\ML})
& \stackrel{\eqref{eq:A756}}{\leq}
\frac{1}{n} \log L_{\gamma t,n}
=
\frac{1}{n} \log \underset{Q}{\E} \left[ \exp(n \gamma \Remp(\vr X, \vr Y)) \right]
\\& =
\frac{1}{n} \log \underset{Q}{\E} \left[ \psi(\mt X^n, \mt Y^n, 0) \right]
\leq
\frac{1}{n} \log L_n
\\& =
\frac{1}{n} \log \left[ \frac{1}{ 1 - \delta_{\Omega}} \left( \frac{e}{1-\gamma} \right)^{ \frac{d}{4} \left(t + 1 + 2r + 2u \right) \cdot t} \right]
\\& =
\frac{1}{n} \log \left( \frac{1}{ 1 - \delta_{\Omega}} \right) + \frac{1}{n} \cdot \frac{d}{4} \left(t + 1 + 2r + 2u \right) \cdot t \cdot \log  \left( \frac{e}{1-\gamma} \right)
\end{split}\end{equation}
The condition on $\gamma$ is obtained by the requirement to satisfy the conditions of \eqref{eq:A3778} for $m=n$.

\subsubsection{Proof of Theorem~\ref{theorem:GaussianMIMO}}\label{sec:GaussianMIMO_Wrapup}
In this section we wrap up the proof of Theorem~\ref{theorem:GaussianMIMO} by combining the results together. From \eqref{eq:A3778} we have that the CCDF condition holds with $L_m = \frac{1}{ 1 - \delta_{\Omega}} \left( \frac{e}{1-\gamma} \right)^{ \frac{d}{4} \left(t + 1 + 2r + 2u \right) \cdot t}$ and $b_0 = \frac{t+1+r+u}{1-\gamma}$. Substituting this and the summability condition with $f_0$ defined in \eqref{eq:A3849} in Theorem~\ref{theorem:framework}, we have that the following rate function is adaptively achievable:
\begin{equation}\label{eq:A3888}
\Remp =  \left(1 + \frac{c_n  + b_1 \cdot f_0^{(n)}(\psi_0^n)}{K} \right)^{-1} \cdot \frac{1}{n} \log (\psi_0^n) - \frac{K}{n}
\end{equation}
with $c_n = \log \frac{n \cdot L_n}{\dfb \epsilon}$ and $b_1 = b_0 + 2\dfb - 1$. We have
\begin{equation}\begin{split}\label{eq:A3872}
\frac{1}{n} \cdot \log (\psi_0^n)
& \stackrel{\eqref{eq:A3536}}{=}
\gamma \left[ \hat H_{Q} (\mt X) - \hat H_{\ML} (\mt X | \mt Y) \right]
\stackrel{\eqref{eq:A3506}}{=}
\gamma \left[ \hat H_{\tilde Q} (\mt X) - \hat H_{\ML} (\mt X | \mt Y) \right] + \gamma \log (1 - \delta_{\Omega})
\\&=
\gamma \Rempname{\ML} + \gamma \log (1 - \delta_{\Omega})
\leq
\gamma \Rempname{\ML}
\end{split}\end{equation}
Where $\Rempname{\ML}$ is defined in \eqref{eq:A3321}.

Substituting we obtain:
\begin{equation}\label{eq:A3892}
f_0^{(n)}(\psi_0^n)
\stackrel{\eqref{eq:A3849}, \eqref{eq:A3872}}{\leq}
\dalf (t + \Omega^2) \gamma \cdot \log (e) + \gamma \Rempname{\ML}
\end{equation}
\begin{equation}\label{eq:A3889}
c_n = \log \frac{n \cdot L_n}{\dfb \epsilon} = \log \frac{n}{\dfb \epsilon (1 - \delta_{\Omega})}
+
\frac{d}{4} \left(t + 1 + 2r + 2u \right) \cdot t \cdot \log \left( \frac{e}{1-\gamma} \right)
\end{equation}
\begin{equation}\begin{split}\label{eq:A3888b}
c_n  + b_1 \cdot f_0^{(n)}(\psi_0^n)
& \leq
\underbrace{\log \frac{n}{\dfb \epsilon (1 - \delta_{\Omega})}
+
\frac{d}{4} \left(t + 1 + 2r + 2u \right) \cdot t \cdot \log \left( \frac{e}{1-\gamma} \right)}_{c_n}  +
\underbrace{\left( \frac{t+1+r+u}{1-\gamma} + 2 \dfb - 1 \right)}_{b_1}
\\ & \qquad \cdot \underbrace{\left(\dalf (t + \Omega^2) \gamma \cdot \log (e) + \gamma \Rempname{\ML} \right)}_{\geq f_0^{(n)}(\psi_0^n)}
\\& \stackrel{(*)}{=}
\log n + a_1 + a_2 \log \frac{1}{1-\gamma} +
\left( \frac{a_3}{1-\gamma} + a_4 \right)
\cdot \gamma \left( \Rempname{\ML} + a_5 \right) = A_{n,\gamma} \cdot \Rempname{\ML} + B_{n,\gamma}
\end{split}\end{equation}
where
\begin{eqnarray}
a_0 &=& \log \frac{1}{1 - \delta_{\Omega}} \\
a_1 &=& a_0 + \log \frac{1}{\dfb \epsilon} + a_2 \log(e) \\
a_2 &=& \frac{d}{4} \left(t + 1 + 2r + 2u \right) \cdot t \\
a_3 &=& t+1+r+u \\
a_4 &=& 2 \dfb - 1 \\
a_5 &=& \dalf (t + \Omega^2) \cdot \log (e) \\
A_{n,\gamma} &=& \gamma \left( \frac{a_3}{1-\gamma} + a_4 \right) \\
B_{n,\gamma} &=& \log n + a_1 + a_2 \log \frac{1}{1-\gamma} + \left( \frac{a_3}{1-\gamma} + a_4 \right) \cdot \gamma \cdot a_5 \\
\end{eqnarray}

We may lower bound the achievable rate $\Remp$ \eqref{eq:A3888} by:
\begin{equation}\begin{split}\label{eq:A3965}
\Remp
& \stackrel{\eqref{eq:A3872}, \eqref{eq:A3888b}}{\geq}
\left(1 + \frac{A_{n,\gamma} \cdot \Rempname{\ML} + B_{n,\gamma}}{K} \right)^{-1} \cdot \left[ \gamma \Rempname{\ML} + \gamma \log (1 - \delta_{\Omega}) \right] - \frac{K}{n}
\\ & \geq
\left[ \left(1 + \frac{B_{n,\gamma}}{K} \right) \left(1 + \frac{A_{n,\gamma} \cdot \Rempname{\ML}}{K + B_{n,\gamma}} \right) \right]^{-1} \cdot \gamma \cdot \Rempname{\ML} - a_0 - \frac{K}{n}
=
\frac{\eta \cdot \Rempname{\ML}}{1 + \alpha \cdot \Rempname{\ML}} - \delta
\end{split}\end{equation}
where
\begin{eqnarray}
\eta &=& \gamma \left(1 + \frac{B_{n,\gamma}}{K} \right)^{-1} \\
\alpha &=& \frac{A_{n,\gamma}}{K + B_{n,\gamma}} \\
\delta &=& a_0 + \frac{K}{n}
\end{eqnarray}
This shows the main results of the theorem.

In order to show asymptotic achievability we need to show there exists a choice of $\gamma, \Omega$ and $K$ as functions of $n$ such that $\eta \ntoinfty 1, \alpha,\delta \ntoinfty 0$. This requires that $\frac{K}{n} \ntoinfty 0$, $\gamma \ntoinfty 1$ and $a_0 \ntoinfty 0$ while $\frac{A_{n,\gamma}}{K},\frac{B_{n,\gamma}}{K} \ntoinfty 0$. Examining these expression we observe it is sufficient that $\frac{\Omega^2}{(1-\gamma) K} \ntoinfty 0$. A possible choice is $K=\lceil n^{1/4} \rceil, \gamma = 1 - n^{-1/4}, \Omega^2 = n^{1/4}$.
\endofproof

\subsubsection{Proof of Lemma~\ref{lemma:GaussianMIMO_opt}}\label{sec:GaussianMIMO_optproof}
Using $\log(x) < x$ (this is true for $\log$ of base larger than $e^{1/e}=1.44$ and results from $\ln(x)/x \leq e^{-1}$ which can be proven by derivation) and assuming $\Rempname{\ML} \leq R_0$ we may coarsely bound $A_{n,\gamma} \cdot \Rempname{\ML} + B_{n,\gamma}$ in \eqref{eq:A3965} by:
\begin{equation}\begin{split}\label{eq:A3961}
A_{n,\gamma} \cdot \Rempname{\ML} + B_{n,\gamma}
&\leq
\log n + a_1 + a_2 \frac{1}{1-\gamma} +
\left( \frac{a_3 + a_4}{1-\gamma}  \right)
\cdot \left( \Rempname{\ML} + a_5 \right)
\\& \leq
\frac{\log n + a_1 + a_2 + (a_3 + a_4)\left( R_0 + a_5 \right) }{1-\gamma}
\\&\defeq
\frac{a_6}{1-\gamma}
\end{split}\end{equation}

Using $\frac{1}{1+x} \geq 1-x$ and \eqref{eq:A3965} we write (for $\Rempname{\ML} \leq R_0$):
\begin{equation}\label{eq:A3993}
\Remp
\geq
\left(1 - \frac{a_6}{(1-\gamma) K} \right) \cdot \gamma \cdot \Rempname{\ML} - a_0 - \frac{K}{n}
\geq
\Rempname{\ML} - \underbrace{\left[ (1-\gamma) \cdot R_0 + \frac{a_6}{(1-\gamma) K} \cdot R_0  + \frac{K}{n}\right]}_{\delta_0} - a_0
\end{equation}
We now choose $\gamma,K$ that minimize $\delta_0$. To minimize $(1-\gamma) \cdot R_0 + \frac{a_6}{(1-\gamma) K} \cdot R_0$ we choose $(1-\gamma) = \sqrt{\frac{a_6}{K}}$ (see Lemma~\ref{lemma:ab_bound}), and obtain $(1-\gamma) \cdot R_0 - \frac{a_6}{(1-\gamma) K} \cdot R_0 = 2 \sqrt{\frac{a_6}{K}} \cdot R_0$. Following, $K$ is chosen to minimize $2 \sqrt{\frac{a_6}{K}} \cdot R_0 + \frac{K}{n}$ which yields $K = \left( n \cdot \sqrt{a_6} \cdot R_0 \right)^{\frac23}$. This value is rounded up to an integer value, incurring an additional loss of at most $\frac{1}{n}$. Substituting we have $2 \sqrt{\frac{a_6}{K}} \cdot R_0 + \frac{K}{n} = 3 n^{-\frac13} a_6^{\frac13} R_0^{\frac23}$. Accounting for the additional loss of $\frac{1}{n}$ due to rounding $K$, we have $\delta_0 \leq 3 n^{-\frac13} a_6^{\frac13} R_0^{\frac23} + \frac{1}{n}$
\endofproof

\begin{table}[h]
\centering
\caption{Parameters of the rate adaptive scheme for MIMO (Section~\ref{sec:GaussianMIMO}), for Figure~\ref{fig:GaussianMIMO}}\label{table:GaussianMIMO_params}
\begin{tabular}{p{12cm}}
\textbf{Parameters of the scheme used for Figure~\ref{fig:GaussianMIMO}} \\ \hline
Basic parameters: $n=1e+005, t=2, r=2, d=2, u=1, \epsilon=0.001, \Omega=5, \dfb=1$ \\
Parameters of Lemma~\ref{lemma:GaussianMIMO_opt}: $R_0=5, a_6=356 \Rightarrow K=4.5e+004, \gamma=0.911$ \\
Intermediate parameters of Theorem~\ref{theorem:GaussianMIMO}: $a_0=0, a_1=23, a_2=9, a_3=6, a_4=1, a_5=39, A_{n,\gamma}=62, B_{n,\gamma}=2.5e+003$ \\
Final parameters of Theorem~\ref{theorem:GaussianMIMO}: $\delta=0.45, \alpha=0.0013, \eta=0.863, \delta_{\Omega}=3.17e-019$ \\
Final parameters of Lemma~\ref{lemma:GaussianMIMO_opt}: $\delta_0=1.3$ \\
Saturation (limit) of lower bound for $\Rempname{\ML} \to \infty$: $\frac{\eta}{\alpha}-\delta = 654.56$\\
\end{tabular}
\end{table}

\subsubsection{The intrinsic redundancy}\label{sec:ir_of_siso_proof}
In Example~\ref{example:ir_of_siso} we claimed that the SISO version of the rate function $\Remp = \half \log \frac{1}{1 - \hat\rho^2}$ has an intrinsic redundancy $\mu_Q(\Remp) = \infty$. This implies of course that also the MIMO rate function has an infinite intrinsic redundancy (since the SISO rate function can be attained as a particular case by zeroing some of the inputs and outputs). This results from the fact that $\Pr(\Remp \geq R) \approx \exp(-(n-1)R)$ (instead of $\exp(-nR)$ as required to satisfy the necessary or sufficient condition of Theorem~\ref{theorem:remp_ublb}). This exponent is already implied by Lemma~4 in the \selector{previous paper}{initial work} \cite{YL_individual_full}, but Lemma~4 is an upper bound and to prove that $\mu_Q(\Remp) = \infty$ a lower bound on the probability $\Pr(\Remp \geq R)$ is required. Below we prove the claim of Example~\ref{example:ir_of_siso} using such a lower bound.

We use the same technique and notation of the proof of Lemma~4 the \selector{previous paper}{initial work} \cite{YL_individual_full}. There we showed that
\begin{equation}\label{eq:A712}
\Pr(|\hat \rho| \geq t) = \Pr \left( X_1^2 \geq \frac{t^2}{1-t^2} \lVert \vr X_2^n \rVert^2  \right)
\end{equation}
where $\vr x$ is a Gaussian normal vector of length $n$, $\vr X \sim \Normal^n(0,1)$. $\lVert \vr X_2^n \rVert^2$ is distributed Chi-square with $k=n-1$ degrees of freedom. For a random variable $V \sim \chi^2_k$ (Chi square with $k$ degrees of freedom), one has:
\begin{equation}\label{eq:A784t}
\Pr(V \leq v) = \int_{t=0}^v \underbrace{\frac{1}{2^{k/2} \Gamma(k/2)}}_{c_1(k)} t^{k/2-1} e^{-t/2} dt \geq c_1(k) \int_{t=0}^v t^{k/2-1} e^{-v/2} dt = \underbrace{\frac{c_1(k)}{k/2}}_{c_2(k)} v^{k/2} e^{-v/2}
\end{equation}
In our case:
\begin{equation}\begin{split}\label{eq:A789}
\Pr(|\hat \rho| \geq t)
&=
\E \left[ \Pr \left( \lVert \vr X_2^n \rVert^2  \leq \frac{1-t^2}{t^2} X_1^2  \bigg| X_1 \right)\right]
\geq
\E \left[ c_2(n-1) \left( \frac{1-t^2}{t^2} X_1^2 \right)^{\frac{n-1}{2}} e^{-half \left( \frac{1-t^2}{t^2} X_1^2 \right)} \right]
\\&=
\int_{-\infty}^{\infty} c_2(n-1) \left( \frac{1-t^2}{t^2} x_1^2 \right)^{\frac{n-1}{2}} e^{-half \left( \frac{1-t^2}{t^2} x_1^2 \right)} (2 \pi)^{-\frac{n-1}{2}} e^{-\half x_1^2} d x_1
=
\underbrace{c_3(n)}_{=c_2(n-1)(2 \pi)^{-\frac{n-1}{2}}} \int_{-\infty}^{\infty} \left( \frac{1-t^2}{t^2} x^2 \right)^{\frac{n-1}{2}} e^{-half \left( \frac{1}{t^2} x^2 \right)} d x
\\&\stackrel{z=x/t}{=}
t (1-t^2)^{\frac{n-1}{2}} \underbrace{c_3(n)  \int_{-\infty}^{\infty} z^{n-1} e^{-half z^2}  \cdot d z}_{c_4(n)}
=
c_4(n) t (1-t^2)^{\frac{n-1}{2}}
\end{split}\end{equation}
Therefore
\begin{equation}\label{eq:A806}
\Pr(\Remp \geq R)
=
\Pr \left\{ |\hat \rho| \geq \sqrt{1-\exp(-2R)} \right\}
\geq
c_4(n) \sqrt{1-\exp(-2R)} \exp(-(n-1)R)
\end{equation}
and
\begin{equation}\label{eq:A817}
\mu_Q(\Remp) \defeq
\sup_{\vr y,R \in \mathbb{R}} \left\{ \frac{1}{n} \log \Pr \{\Remp \geq R\}  + R \right\}
\geq
\sup_{R \in \mathbb{R}} \left\{ \frac{1}{n} \log c_4(n) + \frac{1}{n} \log \sqrt{1-\exp(-2R)} - \frac{n-1}{n} R  + R \right\}
\geq
\frac{1}{n} \log c_4(n) + \lim_{R \to \infty} \left\{ \frac{1}{n} \log \sqrt{1-\exp(-2R)} + \frac{1}{n} R \right\}
=
\infty
\end{equation}
The limit diverges because $\lim_{R \to \infty} \log \sqrt{1-\exp(-2R)} = \log 1 = 0$.
\endofproof
The geometric interpretation of Lemma~4 given in the appendix of the paper \cite{YL_individual_full} may also be used to prove the same claim.

\subsection{The conditional Lempel-Ziv and probability assignments implemented by FSM-s}\label{sec:cond_LZ_performance}
Below we prove the claim from Section~\ref{sec:examples_compression_condLZ} that the probability $\hat P_{LZ}(\vr x | \vr y) = \exp(-L(\vr x | \vr y))$ assigned by the conditional LZ to an input sequence, asymptotically surpasses (up to vanishing factors) the probability that can be assigned to the sequence by any finite state machine operating on the sequences $\vr x, \vr y$. For simplicity of notation we will use $\vr x, \vr y$ to denote phrases, and the full sequences will be denoted $\vr x^n, \vr y^n$. Although this claim is straightforward and similar claims appear in \cite{ZivUniversal}\cite{Uyematsu03}, we did not find the exact claim, and therefore we prove it below.

The state machine with $S$ states. At each symbol it receives $y_i, x_i$, assigns a probability for $x_i$ and moves to a next state based on $y_i, x_i$. The total probability is the product of (conditional) probabilities assigned to the letters. It is required of course that the sum of the probabilities assigned to different $x_i$-s (and as a consequence different sequences $\vr x$) will be $1$.

Let $(\vr x_l, \vr y_l)$ denote the $l$-th phrase out of $c$ phrases in the joint parsing of $\vr x, \vr y$. Suppose the state of the state machine at the beginning of this phrase is $s_l$. The cumulative probability assigned by the state machine to the phrase can be written as function of $\vr x_l, \vr y_l, s_l$. Denote the probability assigned to a phrase $\vr x$ given the phrase $\vr y$ with the initial state $s$ as $P(\vr x | \vr y, s)$ (this function characterizes the state machine, and must satisfy $\sum_{\vr x} P(\vr x | \vr y, s)$), then the overall probability assigned by the state machine is:
\begin{equation}\label{eq:A3744}
P(\vr x^n | \vr y^n) = \prod_{l=1}^c P(\vr x_l | \vr y_l, s_l)
\end{equation}
let $c_l(\vr x | \vr y)$ count the number of different $\vr x_l$ that appear jointly with $\vr y_l$, and $c_l(\vr x | \vr y, s)$  the number of different $\vr x_l$ that appear jointly with $\vr y_l$ with $s_l = s$ (i.e. $c_l(\vr x | \vr y, s) = \sum_{l: \vr y_l = \vr y, s_l = s} 1$), then looking at the part of the product above associated with specific $\vr y_l$ and $s_l$ we have:
\begin{equation}\begin{split}\label{eq:A3750}
\log \prod_{l: \vr y_l = \vr y, s_l = s} P(\vr x_l | \vr y_l, s_l)
&=
c_l(\vr x | \vr y, s) \cdot \frac{1}{c_l(\vr x | \vr y, s)} \sum_{l: \vr y_l = \vr y, s_l = s} \log P(\vr x_l | \vr y_l, s_l)
\\& \leq
c_l(\vr x | \vr y, s) \cdot \log \left( \frac{1}{c_l(\vr x | \vr y, s)} \sum_{l: \vr y_l = \vr y, s_l = s} P(\vr x_l | \vr y_l, s_l) \right)
\\& \leq
c_l(\vr x | \vr y, s) \cdot \log \left( \frac{1}{c_l(\vr x | \vr y, s)}  \right)
\end{split}\end{equation}
where $\sum_{l: \vr y_l = \vr y, s_l = s} P(\vr x_l | \vr y_l, s_l) \leq 1$ since no phrase $x_l$ can appear twice. Hence
\begin{equation}\begin{split}\label{eq:A3762}
\log P(\vr x^n | \vr y^n)
&=
\log \prod_{\vr y, s} \prod_{l: \vr y_l = \vr y, s_l = s} P(\vr x_l | \vr y_l, s_l)
\leq
\sum_{\vr y, s} c_l(\vr x | \vr y, s) \cdot \log \left( \frac{1}{c_l(\vr x | \vr y, s)}  \right)
\\&=
\sum_{\vr y} c_l(\vr x | \vr y) \underbrace{\sum_{s} \frac{c_l(\vr x | \vr y, s)}{c_l(\vr x | \vr y)} \cdot \log \left( \frac{c_l(\vr x | \vr y)}{c_l(\vr x | \vr y, s)} \right)}_{\leq \log S}
-
\sum_{\vr y} c_l(\vr x | \vr y) \cdot \log c_l(\vr x | \vr y)
\\& \stackrel{(a)}\leq
\sum_{\vr y} c_l(\vr x | \vr y) \cdot \log S
-
\sum_{\vr y} c_l(\vr x | \vr y) \cdot \log c_l(\vr x | \vr y)
=
c \cdot \log S - \sum_{\vr y} c_l(\vr x | \vr y) \cdot \log c_l(\vr x | \vr y)
\end{split}\end{equation}
where (a) is because the braced expression can be interpreted as the entropy of the probability over $s$ $p(s) = \frac{c_l(\vr x | \vr y, s)}{c_l(\vr x | \vr y)}$ and is therefore bounded by the entropy of a uniform distribution over $s=1,\ldots,S$. The value $\sum_{\vr y} c_l(\vr x | \vr y) \cdot \log c_l(\vr x | \vr y)$ is the conditional LZ complexity. Therefore we have that for any conditional probability $P(\vr x^n | \vr y^n)$ implemented by a finite state machine with no more than $S$ states, one has:
\begin{equation}\label{eq:A3793}
\log P(\vr x^n | \vr y^n) \leq c \cdot \log S - C_{LZ}(\vr x | \vr y)
\end{equation}
where
\begin{equation}\label{eq:A3796}
C_{LZ}(\vr x | \vr y) = \sum_{\vr y} c_l(\vr x | \vr y) \cdot \log c_l(\vr x | \vr y) = \sum_{l=1}^c \log c_l(\vr x | \vr y)
\end{equation}
is the conditional LZ complexity and $c_l(\vr x | \vr y)$ is defined above, and $c$ is the number of phrases in joint parsing of $\vr x, \vr y$. The number of phrases $c$ is bounded by $\approx \frac{n \log (|\mathcal{X}| \cdot |\mathcal{Y}|)}{\log n}$ \cite[Eq.(6)]{LZ78}. Therefore when considering $\frac{1}{n} \log P(\vr x^n | \vr y^n)$ the first term in the RHS of \eqref{eq:A3793} yields an asymptotically vanishing factor $\frac{c \cdot \log S}{n} \ntoinfty 0$.

Next we connect $C_{LZ}(\vr x | \vr y)$ with $L(\vr x | \vr y)$ obtained by the conditional LZ algorithm. Since the index this algorithm sends for each phrase $l$ encodes $\vr x_l$ by sending the last letter plus the index of the phrase composed of the other letters out of the $c_l(\vr x|\vr y)$ phrases with the same $\vr y$, this requires at most $\log |\mathcal{X}| + \log c_l(\vr x|\vr y) + r_n$ where $r_n$ accounts for the additional overhead due to rounding, and the need to encode the length of $c_l(\vr x|\vr y)$ (since $c_l(\vr x|\vr y) \leq n$ the length of its encoding, i.e. the number of bits $\log c_l(\vr x|\vr y)$ is at most $\log \log n$). Therefore
\begin{equation}\label{eq:A3805}
L(\vr x^n | \vr y^n)
\leq
\sum_{l} \left[ \log |\mathcal{X}| + \log c_l(\vr x|\vr y) + r_n \right]
=
C_{LZ}(\vr x | \vr y) + c \cdot (\log |\mathcal{X}| + r_n)
\end{equation}

Therefore
\begin{equation}\label{eq:A3818}
\frac{1}{n} L(\vr x^n | \vr y^n) \leq \frac{1}{n} C_{LZ}(\vr x | \vr y) + \frac{c}{n} \cdot (\log |\mathcal{X}| + r_n)
\leq
- \log P(\vr x^n | \vr y^n) + \underbrace{\frac{c}{n} \cdot (\log |\mathcal{X}| + r_n + \log S)}_{\delta_n}
\end{equation}
where the factor $\delta_n$ in the RHS vanishes with $n$. Plugging this into the rate function \eqref{eq:A3550} we obtain
\begin{equation}\label{eq:A3824}
\Remp =  \log |\mathcal{X}| - \frac{1}{n} L(\vr x^n | \vr y^n) \geq \frac{1}{n} \log \frac{P(\vr x^n | \vr y^n)}{Q(\vr x^n)} - \delta_n
\end{equation}
I.e. this rate function surpasses up to $\delta_n$ all rate functions defined by any $P(\vr x^n | \vr y^n)$ that can be implemented by a finite state machine.

% \bibliographystyle{IEEEtran} %plain
% \bibliography{../Globals/Master}

\end{document}